\documentclass[aip,reprint,floatfix]{revtex4-2}

\usepackage[utf8]{inputenc}
\usepackage[T1]{fontenc}
\usepackage{amsmath,amssymb}
\usepackage{graphicx}
\usepackage{booktabs}
\usepackage{xcolor}
\usepackage{tikz}
\usetikzlibrary{positioning,calc,arrows.meta,fit,decorations.pathreplacing,backgrounds}
\usepackage[hidelinks]{hyperref}

\usepackage{listings}
\usepackage[caption=false]{subfig}
\lstdefinestyle{prompt}{basicstyle=\ttfamily\scriptsize,breaklines=true,breakatwhitespace=false,breakindent=0pt,columns=fullflexible,keepspaces=true,frame=none,escapeinside={(*@}{@*)}}

\definecolor{storagecol}{HTML}{FFF2CC}
\definecolor{gatecol}{HTML}{F8CECC}
\definecolor{junccol}{HTML}{D5E8D4}
\definecolor{storageline}{HTML}{D6B656}
\definecolor{gateline}{HTML}{B85450}
\definecolor{juncline}{HTML}{82B366}
\definecolor{expfill}{HTML}{DAE8FC}
\definecolor{expline}{HTML}{6C8EBF}
\definecolor{qubitcol}{HTML}{0050EF}   %
\definecolor{qubitline}{HTML}{001DBC}
\definecolor{qgatecol}{HTML}{E51400}   %
\definecolor{qgateline}{HTML}{B20000}

\tikzset{
  storage/.style={circle,draw=storageline,fill=storagecol,inner sep=0pt,minimum size=4.6mm,font=\scriptsize},
  gatev/.style  ={circle,draw=gateline,fill=gatecol,inner sep=0pt,minimum size=4.6mm,font=\scriptsize},
  junc/.style   ={circle,draw=juncline,fill=junccol,inner sep=0pt,minimum size=4.6mm,font=\scriptsize},
  edge/.style   ={draw=black,line width=0.5pt},
  smallv/.style ={circle,draw=black,inner sep=0pt,minimum size=4.0mm,font=\tiny},
  sstorage/.style={smallv,fill=storagecol,draw=storageline},
  sgate/.style   ={smallv,fill=gatecol,draw=gateline},
  sjunc/.style   ={smallv,fill=junccol,draw=juncline},
  bsmallv/.style ={circle,draw=black,inner sep=0pt,minimum size=2.5mm,font={\fontsize{3.6}{4}\selectfont}},
  bstorage/.style={bsmallv,fill=storagecol,draw=storageline},
  bgate/.style   ={bsmallv,fill=gatecol,draw=gateline},
  bjunc/.style   ={bsmallv,fill=junccol,draw=juncline},
  brep/.style    ={bsmallv,fill=junccol,draw=juncline,densely dashed},
}

\newcommand{\appref}[1]{\hyperref[#1]{Appendix~\ref*{#1}}}

\makeatletter\def\p@subfigure#1{\thefigure(#1)}\makeatother

\newtheorem{example}{Example}
\makeatletter
\def\@begintheorem#1#2{\trivlist\item[\hskip\labelsep{\bfseries #1\ #2.}]\normalfont\hyper@makecurrent{example}\Hy@raisedlink{\hyper@anchorstart{\@currentHref}\hyper@anchorend}}
\def\@opargbegintheorem#1#2#3{\trivlist\item[\hskip\labelsep{\bfseries #1\ #2\ (#3).}]\normalfont\hyper@makecurrent{example}\Hy@raisedlink{\hyper@anchorstart{\@currentHref}\hyper@anchorend}}
\makeatother

\makeatletter
\renewcommand{\fnum@figure}{Fig.\ \thefigure}
\renewcommand{\fnum@table}{Tab.\ \thetable}
\makeatother

\makeatletter
\def\jk@tabletype{table}
\AtBeginDocument{%
  \long\def\@makecaption#1#2{%
    \par
    \ifx\@captype\jk@tabletype \vskip 2\p@ \else \vskip\abovecaptionskip \fi
    \begingroup\small\rmfamily
      \parindent\z@ \flushing
      \@make@capt@title{#1}{#2}\par
    \endgroup
    \vskip\belowcaptionskip}%
}
\makeatother

\begin{document}

\title{Efficient LLM-Generated Shuttling Compilers for Complex Trapped-Ion Architectures}

\author{Fabian Kreppel}
\email{f.kreppel@uni-mainz.de}
\author{Reza Salkhordeh}
\email{rsalkhor@uni-mainz.de}
\affiliation{Institute of Computer Science, Johannes Gutenberg University, Mainz, Germany}

\author{Ferdinand Schmidt-Kaler}
\email{fsk@uni-mainz.de}
\affiliation{Institute of Physics, Johannes Gutenberg University, Mainz, Germany}

\author{André Brinkmann}
\email{andre.brinkmann@uni-saarland.de}
\affiliation{Department of Computer Science, Saarland University, Saarbrücken, Germany}

\begin{abstract}
Trapped-ion quantum computers rely on shuttling compilers, which cast an input algorithm into a sequence of ion-qubit movements within a given architecture. We present the first study in which a single frontier large language model (LLM), Claude Opus 4.7, generates and iteratively refines the Python code of shuttling compilers from written specifications. Starting with a compiler for (i) a linear segmented trap, we extend it to (ii) a trap with junctions, and finally achieve efficient compilation for (iii) a broad class of connected trap graphs. The compilers for the more general cases are seeded with code from the previous ones. We benchmark the LLM-generated compilers against state-of-the-art hand-crafted ones using a common suite of quantum circuits. The number of shuttling timesteps is reduced by up to 72\,\% for (i) and up to 43\,\% for (ii). For the broad case (iii) of freely connected architectures, we find large variations in the required number of shuttling timesteps, depending on the connectivity. A densely connected, junction-rich architecture yields up to an order-of-magnitude reduction in shuttling timesteps compared to a corridor-like one. Repeating the complete generation and evaluation with a second frontier LLM, Claude Fable 5, reproduces these findings, with the Fable 5 compilers surpassing the hand-crafted ones more often on the largest circuits. Our results show that an unmodified frontier LLM can produce working, correct, and competitive shuttling compilers without additional manual algorithmic engineering. Generating compilers this way reduces the development time for new architectures from several months to a few days.
\end{abstract}


\maketitle %

\section{Introduction}
\label{sec:introduction}
Trapped-ion quantum computing (QC) comes with the advantage that ion qubits can be moved between different sections of a scalable architecture \cite{KielpinskiMW02, Seidelin+06, Leibrandt+09, BowlerGLTHJHLW12, MuraliDBM20, KaushalLSHPSBMSP20, PinoDFGMABFHMRN21, LeeJPJKC21, MosesB+23, Ransford+26} to execute an algorithm efficiently. Such QC architectures are based on linear segmented traps with qubit storage and operation zones \cite{KaushalLSHPSBMSP20, PinoDFGMABFHMRN21, BowlerGLTHJHLW12, LeeJPJKC21}, racetrack loops \cite{MosesB+23}, or junctions \cite{Ransford+26}. Shuttling compilers \cite{Wagner22, DurandauWMBSPB23, DurandauBSPMB26, AshSakiTG22, WuW26, KreppelMWHPSB24, WebberHWH20, DaiBR24, ChangJCHL25, SchoenbergerHSW25, SchoenbergerHBW25, SchoenbergerW25, SchmaleTBPKDOWB22} require substantial engineering effort by domain experts for each specific architecture as the QC hardware grows in complexity and connectivity.

We investigate whether this per-layout effort can be replaced by code generation. Rather than having domain experts hand-craft each compiler, a state-of-the-art general-purpose large language model (LLM), Claude Opus 4.7 \cite{Anthropic26_2}, writes the compiler's Python code from a specification given in a single prompt. We demonstrate our novel approach on multiple architectures of strictly increasing generality, shown in \autoref{fig:pipeline}: (i) a linear segmented shuttling architecture with a central qubit operation zone, (ii) an architecture with multiple junctions, and (iii) a broad class of connected trap graphs, including dense architectures with multiple ion transport options. Follow-up prompts then refine each emitted compiler toward fewer shuttling timesteps for a given algorithm, which shortens the algorithm's run time on the QC, and toward shorter compile time and lower memory use. Generating a compiler this way is far more time-efficient than writing one by hand. A hand-crafted shuttling compiler takes several months to develop, whereas writing the specification and generating the compiler with LLM support cuts this effort to a few days per architecture.

\begin{figure*}[t]
\centering
\resizebox{\textwidth}{!}{%
\begin{tikzpicture}[
  font=\small,>=Latex,
  gen/.style={draw=expline,rounded corners,minimum height=14mm,text width=30mm,align=center,fill=expfill,thick},
  io/.style={align=center},
  lab/.style={align=center},
  optlab/.style={align=center,text width=21mm},
  ar/.style={-{Latex[length=2.4mm]},thick},
  loop/.style={-{Latex[length=1.8mm]},gray!70},
]
\node[io]  (seed) at (0,0.4)      {empty\\seed};
\node[gen] (e1)   at (0,-1.7)   {\textbf{Linear architecture}\\[1pt]{\footnotesize\autoref{sec:compiler:linear}}};
\node[gen] (e2)   at (0,-4.5)   {\textbf{Branched architecture}\\[1pt]with junctions and stacks\\[1pt]{\footnotesize\autoref{sec:compiler:branched}}};
\node[gen] (e3)   at (0,-9.5)  {\textbf{General architecture graph}\\[1pt]{\footnotesize\autoref{sec:compiler:general}}};
\node[io]  (out)  at (0,-12.8)  {general\\compiler};
\draw[ar] (seed) -- node[lab,right=1pt]{prompt} (e1);
\draw[ar] (e1) -- node[lab,right=1pt]{final code\\+\,new prompt} (e2);
\draw[ar] (e2) -- node[lab,right=1pt]{final code\\+\,new prompt} (e3);
\draw[ar] (e3) -- (out);
\foreach \n in {e1,e2,e3}{
  \draw[loop] ([yshift=3mm]\n.west) to[out=160,in=200,looseness=5] ([yshift=-3mm]\n.west);
}
\node[optlab] at ([xshift=-19mm]e1.west) {follow-up\\optimizations};
\node[optlab] at ([xshift=-19mm]e2.west) {follow-up\\optimizations};
\node[optlab] at ([xshift=-19mm]e3.west) {follow-up\\optimizations};

\node (d1) at (8.4,-1.7) {%
\begin{tikzpicture}[x=7.9mm,y=7.9mm,>=Latex,edge/.style={draw=black,line width=1pt},storage/.append style={minimum size=5.5mm},gatev/.append style={minimum size=5.5mm},junc/.append style={minimum size=5.5mm}]
  \foreach \i in {0,1,2,3,4,5,6,8,9,10,11,12,13,14} {\node[storage] (n\i) at (\i,0) {\i};}
  \node[gatev] (n7) at (7,0) {7};
  \foreach \a/\b in {0/1,1/2,2/3,3/4,4/5,5/6,6/7,7/8,8/9,9/10,10/11,11/12,12/13,13/14} {\draw[edge] (n\a) -- (n\b);}
\end{tikzpicture}};

\node (d2) at (8.4,-4.5) {%
\begin{tikzpicture}[x=7.9mm,y=7.9mm,>=Latex,edge/.style={draw=black,line width=1pt},storage/.append style={minimum size=5.5mm},gatev/.append style={minimum size=5.5mm},junc/.append style={minimum size=5.5mm}]
  \foreach \i in {0,1,3,4,6,8,10,11,13,14} {\node[storage] (m\i) at (\i,0) {\i};}
  \node[junc] (m2) at (2,0) {2}; \node[junc] (m5) at (5,0) {5};
  \node[junc] (m9) at (9,0) {9}; \node[junc] (m12) at (12,0) {12};
  \node[gatev] (m7) at (7,0) {7};
  \foreach \a/\b in {0/1,1/2,2/3,3/4,4/5,5/6,6/7,7/8,8/9,9/10,10/11,11/12,12/13,13/14} {\draw[edge] (m\a) -- (m\b);}
  \node[storage] (m15) at (2,1) {15};  \node[storage] (m16) at (2,2)  {16};
  \node[storage] (m17) at (5,1) {17};  \node[storage] (m18) at (5,2)  {18};
  \node[storage] (m19) at (9,1) {19};  \node[storage] (m20) at (9,2)  {20};
  \node[storage] (m21) at (12,1){21};  \node[storage] (m22) at (12,2) {22};
  \draw[edge] (m2)--(m15); \draw[edge] (m15)--(m16);
  \draw[edge] (m5)--(m17); \draw[edge] (m17)--(m18);
  \draw[edge] (m9)--(m19); \draw[edge] (m19)--(m20);
  \draw[edge] (m12)--(m21); \draw[edge] (m21)--(m22);
  \draw[decorate,decoration={brace,amplitude=3pt}] (12.6,2.4) -- (12.6,0.6)
      node[midway,right=2pt,align=center]{stack\\height};
  \draw[decorate,decoration={brace,amplitude=3pt,mirror}] (2.5,-0.5) -- (5.5,-0.5)
      node[midway,below=1pt]{junction distance};
  \draw[decorate,decoration={brace,amplitude=3pt,mirror}] (8.5,-0.5) -- (11.5,-0.5)
      node[midway,below=1pt]{junction distance};
\end{tikzpicture}};

\node (d3) at (8.4,-9.5) {\resizebox{11.5cm}{!}{%
\begin{tikzpicture}
\node at (0,0) {\begin{tikzpicture}[edge/.style={draw=black,line width=.4pt}]
  \def\R{12mm}
  \foreach \i in {1,2,3,4,5,6,7,8,9,10} {\node[sstorage] (b\i) at ({90-360*\i/11}:\R) {\i};}
  \node[sgate] (b0) at ({90}:\R) {0};
  \foreach \a/\b in {0/1,1/2,2/3,3/4,4/5,5/6,6/7,7/8,8/9,9/10,10/0} {\draw[edge] (b\a) -- (b\b);}
\end{tikzpicture}};
\node at (4.2,0) {\begin{tikzpicture}[edge/.style={draw=black,line width=.4pt}]
  \def\R{4.5mm}
  \node[sjunc]    (t0) at ($(-13.5mm,0)+({  0}:\R)$) {0};
  \node[sstorage] (t1) at ($(-13.5mm,0)+({ 90}:\R)$) {1};
  \node[sstorage] (t2) at ($(-13.5mm,0)+({180}:\R)$) {2};
  \node[sstorage] (t3) at ($(-13.5mm,0)+({-90}:\R)$) {3};
  \foreach \a/\b in {0/1,1/2,2/3,3/0} {\draw[edge] (t\a) -- (t\b);}
  \node[sjunc]    (t4) at ($(13.5mm,0)+({180}:\R)$) {4};
  \node[sstorage] (t5) at ($(13.5mm,0)+({ 90}:\R)$) {5};
  \node[sstorage] (t6) at ($(13.5mm,0)+({  0}:\R)$) {6};
  \node[sstorage] (t7) at ($(13.5mm,0)+({-90}:\R)$) {7};
  \foreach \a/\b in {4/5,5/6,6/7,7/4} {\draw[edge] (t\a) -- (t\b);}
  \node[sstorage] (t8)  at (-4.5mm,0) {8}; \node[sgate] (t9) at (0,0) {9}; \node[sstorage] (t10) at (4.5mm,0) {10};
  \draw[edge] (t0)--(t8); \draw[edge] (t8)--(t9); \draw[edge] (t9)--(t10); \draw[edge] (t10)--(t4);
\end{tikzpicture}};
\node at (8.4,0) {\begin{tikzpicture}[edge/.style={draw=black,line width=.4pt}]
  \node[sjunc] (h0) at (-13mm,0) {0}; \node[sjunc] (h1) at (13mm,0) {1};
  \node[sstorage] (h2) at (-7mm,8mm) {2}; \node[sgate] (h3) at (0,8mm) {3}; \node[sstorage] (h4) at (7mm,8mm) {4};
  \draw[edge] (h0)--(h2); \draw[edge] (h2)--(h3); \draw[edge] (h3)--(h4); \draw[edge] (h4)--(h1);
  \node[sstorage] (h5) at (-7mm,0) {5}; \node[sstorage] (h6) at (0,0) {6}; \node[sstorage] (h7) at (7mm,0) {7};
  \draw[edge] (h0)--(h5); \draw[edge] (h5)--(h6); \draw[edge] (h6)--(h7); \draw[edge] (h7)--(h1);
  \node[sstorage] (h8) at (-7mm,-8mm) {8}; \node[sstorage] (h9) at (0,-8mm) {9}; \node[sstorage] (h10) at (7mm,-8mm) {10};
  \draw[edge] (h0)--(h8); \draw[edge] (h8)--(h9); \draw[edge] (h9)--(h10); \draw[edge] (h10)--(h1);
\end{tikzpicture}};
\node at (12.6,0) {\begin{tikzpicture}[edge/.style={draw=black,line width=.4pt}]
  \def\R{6mm}
  \node[sjunc] (f0) at (0,0) {0};
  \node[sstorage] (f1) at ($(-6mm,0)+({60}:\R)$) {1};
  \node[sstorage] (f2) at ($(-6mm,0)+({120}:\R)$) {2};
  \node[sgate]    (f3) at ($(-6mm,0)+({180}:\R)$) {3};
  \node[sstorage] (f4) at ($(-6mm,0)+({240}:\R)$) {4};
  \node[sstorage] (f5) at ($(-6mm,0)+({300}:\R)$) {5};
  \foreach \a/\b in {0/1,1/2,2/3,3/4,4/5,5/0} {\draw[edge] (f\a) -- (f\b);}
  \foreach \i/\k in {6/1,7/2,8/3,9/4,10/5,11/6} {\node[sstorage] (f\i) at ($(6mm,0)+({180-360*\k/7}:\R)$) {\i};}
  \foreach \a/\b in {0/6,6/7,7/8,8/9,9/10,10/11,11/0} {\draw[edge] (f\a) -- (f\b);}
\end{tikzpicture}};
\node at (0,-2.1)    {\mbox{\texttt{cycle}}};
\node at (4.2,-2.1)  {\mbox{\texttt{two\_cycles\_bridged}}};
\node at (8.4,-2.1)  {\mbox{\texttt{theta}}};
\node at (12.6,-2.1) {\mbox{\texttt{figure8}}};
\node at (0,-4.5) {\begin{tikzpicture}[edge/.style={draw=black,line width=.4pt}]
  \def\R{12mm}
  \foreach \i in {1,3,4,5,8,9,10,12} {\node[sstorage] (q\i) at ({90-360*\i/13}:\R) {\i};}
  \node[sgate] (q0) at ({90}:\R) {0};
  \node[sjunc] (q2) at ({90-360*2/13}:\R) {2}; \node[sjunc] (q6) at ({90-360*6/13}:\R) {6};
  \node[sjunc] (q7) at ({90-360*7/13}:\R) {7}; \node[sjunc] (q11) at ({90-360*11/13}:\R) {11};
  \foreach \a/\b in {0/1,1/2,2/3,3/4,4/5,5/6,6/7,7/8,8/9,9/10,10/11,11/12,12/0} {\draw[edge] (q\a) -- (q\b);}
  \draw[edge] (q2) -- (q6); \draw[edge] (q7) -- (q11);
\end{tikzpicture}};
\node at (4.2,-4.5) {\begin{tikzpicture}[edge/.style={draw=black,line width=.4pt}]
  \node[sjunc] (k0) at (-12mm,12mm) {0}; \node[sjunc] (k1) at (12mm,12mm) {1};
  \node[sjunc] (k2) at (12mm,-12mm) {2}; \node[sjunc] (k3) at (-12mm,-12mm) {3};
  \node[sstorage] (n4) at (-6mm,12mm) {4}; \node[sgate] (n5) at (0,12mm) {5}; \node[sstorage] (n6) at (6mm,12mm) {6};
  \draw[edge] (k0)--(n4); \draw[edge] (n4)--(n5); \draw[edge] (n5)--(n6); \draw[edge] (n6)--(k1);
  \node[sstorage] (n9) at (-12mm,4mm) {9}; \node[sstorage] (n10) at (-12mm,-4mm) {10};
  \draw[edge] (k0)--(n9); \draw[edge] (n9)--(n10); \draw[edge] (n10)--(k3);
  \node[sstorage] (n11) at (12mm,4mm) {11}; \node[sstorage] (n12) at (12mm,-4mm) {12};
  \draw[edge] (k1)--(n11); \draw[edge] (n11)--(n12); \draw[edge] (n12)--(k2);
  \node[sstorage] (n15) at (4mm,-12mm) {15}; \node[sstorage] (n16) at (-4mm,-12mm) {16};
  \draw[edge] (k2)--(n15); \draw[edge] (n15)--(n16); \draw[edge] (n16)--(k3);
  \node[sstorage] (n7) at (-4mm,4mm) {7}; \node[sstorage] (n8) at (4mm,-4mm) {8};
  \draw[edge] (k0)--(n7); \draw[edge] (n7)--(n8); \draw[edge] (n8)--(k2);
  \node[sstorage] (n13) at (4mm,4mm) {13}; \node[sstorage] (n14) at (-4mm,-4mm) {14};
  \draw[edge] (k1)--(n13); \draw[edge] (n13)--(n14); \draw[edge] (n14)--(k3);
\end{tikzpicture}};
\node at (8.4,-4.5) {\begin{tikzpicture}[edge/.style={draw=black,line width=.4pt}]
  \node[sjunc] (qq0) at (-13mm,-13mm) {0}; \node[sjunc] (qq1) at (13mm,-13mm) {1};
  \node[sjunc] (qq2) at (-13mm,13mm) {2}; \node[sjunc] (qq3) at (13mm,13mm) {3};
  \node[sjunc] (qq4) at (-6mm,-6mm) {4}; \node[sjunc] (qq5) at (6mm,-6mm) {5};
  \node[sjunc] (qq6) at (-6mm,6mm) {6}; \node[sjunc] (qq7) at (6mm,6mm) {7};
  \node[sstorage] (n8) at (-6mm,-13mm) {8}; \node[sgate] (n9) at (0,-13mm) {9}; \node[sstorage] (n10) at (6mm,-13mm) {10};
  \draw[edge] (qq0)--(n8); \draw[edge] (n8)--(n9); \draw[edge] (n9)--(n10); \draw[edge] (n10)--(qq1);
  \node[sstorage] (n11) at (-13mm,-4mm) {11}; \node[sstorage] (n12) at (-13mm,4mm) {12};
  \draw[edge] (qq0)--(n11); \draw[edge] (n11)--(n12); \draw[edge] (n12)--(qq2);
  \node[sstorage] (n13) at (13mm,0) {13}; \draw[edge] (qq1)--(n13); \draw[edge] (n13)--(qq3);
  \node[sstorage] (n14) at (0,13mm) {14}; \draw[edge] (qq2)--(n14); \draw[edge] (n14)--(qq3);
  \node[sstorage] (n15) at (0,-6mm) {15}; \draw[edge] (qq4)--(n15); \draw[edge] (n15)--(qq5);
  \node[sstorage] (n16) at (-6mm,0) {16}; \draw[edge] (qq4)--(n16); \draw[edge] (n16)--(qq6);
  \node[sstorage] (n17) at (6mm,0) {17}; \draw[edge] (qq5)--(n17); \draw[edge] (n17)--(qq7);
  \node[sstorage] (n18) at (0,6mm) {18}; \draw[edge] (qq6)--(n18); \draw[edge] (n18)--(qq7);
  \node[sstorage] (n19) at (-9.5mm,-9.5mm) {19}; \draw[edge] (qq0)--(n19); \draw[edge] (n19)--(qq4);
  \node[sstorage] (n20) at (9.5mm,-9.5mm) {20}; \draw[edge] (qq1)--(n20); \draw[edge] (n20)--(qq5);
  \node[sstorage] (n21) at (-9.5mm,9.5mm) {21}; \draw[edge] (qq2)--(n21); \draw[edge] (n21)--(qq6);
  \node[sstorage] (n22) at (9.5mm,9.5mm) {22}; \draw[edge] (qq3)--(n22); \draw[edge] (n22)--(qq7);
\end{tikzpicture}};
\node at (12.6,-4.5) {\begin{tikzpicture}[edge/.style={draw=black,line width=.4pt}]
  \node[bjunc] (b0)  at (-18mm, 12.5mm) {0};  \node[bjunc] (b1) at (-18mm, 4.2mm) {1};
  \node[bjunc] (b2)  at (-18mm, -4.2mm) {2};  \node[bjunc] (b3) at (-18mm,-12.5mm) {3};
  \node[bjunc] (b4)  at ( -6mm, 12.5mm) {4};  \node[bjunc] (b5) at ( -6mm, 4.2mm) {5};
  \node[bjunc] (b6)  at ( -6mm, -4.2mm) {6};  \node[bjunc] (b7) at ( -6mm,-12.5mm) {7};
  \node[bjunc] (b8)  at (  6mm, 12.5mm) {8};  \node[bjunc] (b9) at (  6mm, 4.2mm) {9};
  \node[bjunc] (b10) at (  6mm, -4.2mm) {10}; \node[bjunc] (b11) at ( 6mm,-12.5mm) {11};
  \node[brep] (r0) at (18mm, 12.5mm) {0};  \node[brep] (r1) at (18mm, 4.2mm) {1};
  \node[brep] (r2) at (18mm, -4.2mm) {2};  \node[brep] (r3) at (18mm,-12.5mm) {3};
  \node[bstorage] (s12) at (-15mm,12.5mm) {12}; \node[bgate] (s13) at (-12mm,12.5mm) {13}; \node[bstorage] (s14) at (-9mm,12.5mm) {14};
  \draw[edge] (b0)--(s12); \draw[edge] (s12)--(s13); \draw[edge] (s13)--(s14); \draw[edge] (s14)--(b4);
  \node[bstorage] (s15) at (-9.6mm,6.69mm) {15}; \draw[edge] (b0)--(s15); \draw[edge] (s15)--(b5);
  \node[bstorage] (s18) at (-14.4mm,6.7mm) {18}; \draw[edge] (b1)--(s18); \draw[edge] (s18)--(b4);
  \node[bstorage] (s19) at (-12mm,4.2mm) {19}; \draw[edge] (b1)--(s19); \draw[edge] (s19)--(b5);
  \node[bstorage] (s22) at (-12mm,-4.2mm) {22}; \draw[edge] (b2)--(s22); \draw[edge] (s22)--(b6);
  \node[bstorage] (s23) at (-14.4mm,-6.69mm) {23}; \draw[edge] (b2)--(s23); \draw[edge] (s23)--(b7);
  \node[bstorage] (s26) at (-14.4mm,-10.0mm) {26}; \draw[edge] (b3)--(s26); \draw[edge] (s26)--(b6);
  \draw[edge] (b3)--(b7);
  \foreach \a/\b in {4/8,4/10,5/9,5/11,6/8,6/10,7/9,7/11} {\draw[edge] (b\a)--(b\b);}
  \node[bstorage] (s16) at (12mm,12.5mm) {16}; \draw[edge] (b8)--(s16); \draw[edge] (s16)--(r0);
  \node[bstorage] (s17) at (10.2mm,7.1mm) {17}; \draw[edge] (b9)--(s17); \draw[edge] (s17)--(r0);
  \node[bstorage] (s20) at (13.8mm,7.1mm) {20}; \draw[edge] (b8)--(s20); \draw[edge] (s20)--(r1);
  \node[bstorage] (s21) at (12mm,4.2mm) {21}; \draw[edge] (b9)--(s21); \draw[edge] (s21)--(r1);
  \node[bstorage] (s24) at (12mm,-4.2mm) {24}; \draw[edge] (b10)--(s24); \draw[edge] (s24)--(r2);
  \node[bstorage] (s25) at (10.2mm,-9.6mm) {25}; \draw[edge] (b11)--(s25); \draw[edge] (s25)--(r2);
  \draw[edge] (b10)--(r3); \draw[edge] (b11)--(r3);
\end{tikzpicture}};
\node at (0,-6.5)    {\mbox{\texttt{cycle\_with\_two\_chords}}};
\node at (4.2,-6.5)  {\mbox{\texttt{complete\_graph}}};
\node at (8.4,-6.5)  {\mbox{\texttt{hypercube}}};
\node at (12.6,-6.5) {\mbox{\texttt{butterfly}}};
\end{tikzpicture}}};
\end{tikzpicture}}
\caption{Overview of the approach. The three shuttling compilers are generated in a chained sequence of increasing architectural generality: from top to bottom, the linear architecture of \autoref{sec:compiler:linear}, the branched architecture with junctions and stacks of \autoref{sec:compiler:branched}, and the general architecture of \autoref{sec:compiler:general}. Each subsequent compiler is seeded with the previous one's optimized code together with a written specification of what must change, while the linear-architecture compiler starts from scratch. Follow-up optimization prompts then refine each emitted compiler, targeting timesteps, compile time, and memory use. Beside each stage, we draw the corresponding architecture, in which qubit segments (yellow) connect to gate segments (red) and junction segments (green). In the \texttt{butterfly} architecture, vertices 0 to 3 appear on both sides and denote the same four vertices.}
\label{fig:pipeline}
\end{figure*}

A QC encodes each qubit in a single ion, so we will use ``qubit'' and ``ion'' interchangeably. The ions are confined in a segmented microchip trap \cite{SchulzPSS06}, where control voltages applied to its electrodes define localized potential wells. Changing these control voltages reshapes the wells and \emph{shuttles} ions from one segment to the next. In a similar way, ions can be passed through junctions. We represent the trap as a graph whose vertices are its segments and whose edges connect segments between which an ion can shuttle. Apart from junctions, a segment is either a \emph{qubit segment} or a \emph{gate segment}, at which single- and two-qubit operations are executed. Each segment holds at most two qubits in an ordered chain, and a gate is applied by laser or microwave pulses once a gate segment holds exactly the qubits the gate acts on.

Commercial quantum charge-coupled device (QCCD) processors have executed gates simultaneously in several gate zones for several years \cite{PinoDFGMABFHMRN21, MosesB+23, Ransford+26}. In this work, we nevertheless choose a single gate segment for architectures (i) to (iii). This choice isolates the shuttling problem from the scheduling of simultaneous gates, which we leave to future work. A single gate segment is also the hardest shuttling regime, since the qubits of every gate must be routed to it, with no second gate segment to share the load, and this choice allows a direct benchmark against the hand-crafted reference compilers \cite{Wagner22, KreppelMWHPSB24}.

The different layouts are distinguished by their connectivity. In a linear array \cite{KaushalLSHPSBMSP20, PinoDFGMABFHMRN21, BowlerGLTHJHLW12, LeeJPJKC21} or a racetrack loop \cite{MosesB+23}, every segment has at most two neighbors, so the qubit count rises only with the physical length of the chip. Junctions \cite{HensingerOSHYADMR06, BlakestadOVABLW09, BlakestadODWBLW11, BurtonEHPVP23} raise this limit by allowing a segment to connect to several others \cite{AminiUWSBBLODW10, MoehringHSFHTB11, WrightAFVDHPLDKSH13, ShuVBNVSB14}. For example, the Quantinuum Helios processor realizes a single $\mathsf{X}$ junction connecting a ring-shaped storage region to two linear segmented sections \cite{Ransford+26}. Adding multiple junctions drives the architecture toward two-dimensional QCCD designs \cite{LekitschWFMDWH17}. Our choice of architectures follows this increase in connectivity.

The compilation can be divided into two steps: First, a circuit compiler \cite{KreppelMOWHPSB23} expresses the quantum circuit in the QC's native gate set, fixing a gate order (the program order) and applying other transpilation optimizations. In the following step, a shuttling compiler identifies which ions should encode which qubits \cite{UpadhyayATG22, OvideCA24, Wagner22, DurandauWMBSPB23, DurandauBSPMB26, ChangJCHL25} and how to move and rearrange the ions in the trap architecture to execute the gate sequence as efficiently as possible. Thus, this compiler derives a \emph{shuttling schedule}, an ordered sequence of ion movements consistent with the trap's connectivity. The schedule is assembled from four reconfiguration primitives:
\begin{itemize}
	\item \textsc{Translate} moves the qubits in a segment to an adjacent segment.
	\item \textsc{Separate} splits the two qubits in a segment, sending one to each of its neighbors.
	\item \textsc{Merge} brings one qubit into a segment from each of its neighbors.
	\item \textsc{Swap} reverses the order of the two qubits in a segment.
\end{itemize}
Separation, merge, and swap can be performed only at the gate segment, as they require precise calibration of the trap potentials. Gate and shuttling operations are never executed in parallel. At every step of compilation, our compilers choose the next gate from the set of \emph{ready gates}: gates for which every predecessor on each of their qubits, if any, has already been executed. Choosing the gate order at this stage, rather than adopting the program order determined upstream by the circuit compiler \cite{KreppelMOWHPSB23}, lets the shuttling compiler exploit the current ion placement, which is unavailable earlier. It thereby relaxes the otherwise strict compilation boundary.

Shuttling matters physically because its primitives are slow and are themselves a source of errors. In segmented microchip traps, a single shuttling operation takes on the order of tens of microseconds, comparable to or longer than a gate operation \cite{KaushalLSHPSBMSP20}. Since a circuit needs several shuttling timesteps per gate, shuttling dominates the execution time of a circuit, during which the qubits dephase. In addition, shuttling excites the ions' motion, which degrades subsequent gate fidelities. Thus, every timestep a compiler saves shortens the run time and reduces these error sources.

Shuttling compilers are also a tool for designing trap architectures before they are built. Numerical simulation is already an established step in trap design. Segmented traps are laid out with electrostatic field simulations, which experiments confirm at the percent level, for example for the secular frequencies of a monolithic segmented three-dimensional trap \cite{BruckeWFLMNSTVWHKH26}. Shuttling operations, including transport through junctions, are designed with numerical toolchains that combine field solvers, waveform optimization, and ion-motion simulations \cite{ContaBKSP26}. A compiler adds the next level to this toolset: it determines the shuttling schedule of a complete algorithm on a whole trap. To evaluate a candidate layout, it suffices to describe it as a graph of segments. The numbers of timesteps and operations in the resulting schedule then quantify the run time and the motional excitation that the layout imposes, without fabricating the trap or testing it with ions. Since such a compiler is generated within days, many candidate layouts can be compared this way, as we do for ten architectures in \autoref{sec:compiler:general}.

Shuttling compilers have so far been written by hand for linear segmented traps \cite{Wagner22, DurandauWMBSPB23, AshSakiTG22, WuW26}, for architectures with junctions \cite{KreppelMWHPSB24}, and for other layouts \cite{WebberHWH20, DaiBR24, ChangJCHL25, SchoenbergerHBW25, SchoenbergerHSW25, SchoenbergerW25}. Recent approaches pursue architecture independence either through a hand-designed graph abstraction paired with a generic routing heuristic \cite{BachSY26, RussonBYS26, ZhuWWW25} or through an attempt to teach an LLM to output a shuttling schedule directly \cite{KreppelSSB25}, though in the latter case complete schedules were obtained only for circuits with few qubits.

Our approach employs an unmodified frontier LLM and lets it write Python code from prompts that define the specifications. The LLM is consulted only to build the Python shuttling compiler, which then compiles any quantum circuit without further calls to the LLM. Generating code rather than schedules also lets us inspect and test the compiler like any other source code. The approach needs no fine-tuning data, compiles every benchmark circuit in full, with the largest reaching 50 qubits, has no inherent circuit-size limit, and extends beyond the linear and branched architectures to a general class of connected trap graphs. These properties make the proposed approach well suited for shuttling compilation. With a second frontier model, Claude Fable 5 \cite{Anthropic26}, we then check whether these properties depend on the particular LLM. Finally, we separate the effects of the initial mapping and the routing by letting each compiler route from the initial mapping of every other compiler and from random mappings.

The paper is organized as follows. Prior work on shuttling compilation and learning-based routing is reviewed in \autoref{sec:related}, and the prompting protocol and the benchmark setup are described in \autoref{sec:setup}. Next, \autoref{sec:compiler:linear}, \autoref{sec:compiler:branched}, and \autoref{sec:compiler:general} present the linear-, branched-, and general-architecture compilers, each from its prompt and emitted compiler through optimization to its benchmark results, with the last also comparing the trap architectures. The Opus 4.7 generation run is then compared with its Fable 5 replication in \autoref{sec:comparison}, and \autoref{sec:mapping} separates the effect of each compiler's initial mapping from that of its routing. The paper concludes in \autoref{sec:conclusion}.

\section{Related work}
\label{sec:related}

In a quantum circuit, any qubit may eventually interact with any other. No platform achieves this directly at scale, and a general all-to-all coupling would in any case lead to increased gate and memory errors. Shuttling-based trapped-ion and optical-tweezer architectures can adapt the connectivity between qubits as the circuit runs, by physically moving the qubits between operation sites. For such architectures, the circuit is first expressed and possibly optimized in software such as Qiskit \cite{JavadiAbhariTKWLGMNBCJG24}, t$\vert$ket$\rangle$ \cite{SivarajahDCSED20}, Cirq \cite{CirqDevelopers25}, or PennyLane \cite{BergholmISG+22}, and this output is then passed to the shuttling compiler, which assigns qubits to ions and produces the schedule of ion movements. A single gate then generally requires an entire sequence of operations drawn from the four distinct shuttling primitives, and these operations are further constrained by the limited capacity of each segment. Due to the large search space, heuristic approaches have been developed to minimize the number of shuttling operations.

Typically, such compilers work with one specific architecture \cite{PinoDFGMABFHMRN21, MosesB+23, Ransford+26, Wagner22, DurandauWMBSPB23, DurandauBSPMB26, AshSakiTG22, WuW26, KreppelMWHPSB24, WebberHWH20, DaiBR24, ChangJCHL25, SchoenbergerHSW25, SchoenbergerHBW25, SchoenbergerW25, SchmaleTBPKDOWB22, OvideAC26}. One of our compilers instead targets general architectures, a goal shared by several recent approaches. SHAW and SHAPER \cite{BachSY26} aim for a single compiler that handles many architectures. They abstract the operation sites and movement paths of a trap as a \emph{position graph} and transfer the SWAP-based SABRE heuristic \cite{LiDX19} from superconducting-qubit routing to shuttled ions. A later variant scales the search of SHAW and SHAPER to larger circuits by caching repeated computations \cite{RussonBYS26}. Similarly, S-SYNC \cite{ZhuWWW25} treats the whole QCCD device as a static connectivity graph and turns shuttling into SWAP-gate insertion. Compilers can also be generated automatically, from a description of the device as a state machine in a domain-specific language, by instantiating a parametric mapping and routing algorithm \cite{MolaviXCTA26}. In all of these approaches, the routing is implemented by hand, whereas the code of our compilers is written by an LLM from a written specification.

A different hardware model is addressed by the MQT IonShuttler, which schedules shuttling on a two-dimensional QCCD grid with dedicated processing zones, exactly via Boolean satisfiability for small instances \cite{SchoenbergerHBW24} and heuristically for larger ones \cite{SchoenbergerHBW25, SchoenbergerW25}. Each memory site holds a single ion, and the ions are reordered by rotating them around the loops of the grid, so separation, merge, and swap operations are not required.

Instead of being designed by hand, the routing of shuttled ions can also be learned. An agent trained by reinforcement learning in a simulation of the chip moves ions between segments and through junctions \cite{SchierRSSBHR26}, also without separation, merge, or swap operations.

Another direction targets the parallelism of grouped ion transport, with the FluxTrap framework \cite{RuanZFLCHHHHPD25} aggregating SIMD-style intra-trap shifts and junction transfers to parallelize ion movement on two-dimensional trapped-ion machines. For QCCD devices in which every segment holds several ions and can execute gates, one compiler reduces the number of transports between the segments \cite{AshSakiTG22}, and another runs operations in parallel across the segments \cite{OvideAC26}. Compilation has also been extended to multi-SIMD platforms, where regions communicate by teleportation \cite{HeckeyPJHKBFCM15}, and to distributed architectures, whose traps exchange qubits over photonic links \cite{WuZWW25}.

Routing on fixed-connectivity platforms such as superconducting processors takes a different form, in which the qubits remain stationary on a coupling graph and routing reduces to inserting SWAP gates. Learning-based methods have been explored extensively for this setting \cite{PalerSFA23, PozziHSM22, FanGL22, PascoalFA24, TangDKFKS24, SinhaAS22, RussoPPAC25, SangHH26, ZeynaliB25}. For distributed and multi-core architectures \cite{EscofetOBPVFAAA25}, the qubit assignment has also been formulated as a network-flow problem \cite{CuomoCKTAPC23} or solved as a quadratic assignment problem \cite{SundaramGR25}. Unlike these platforms, shuttling architectures change their connectivity as the ions move.

\begin{figure*}[t]
\centering
\resizebox{\textwidth}{!}{%
\begin{tikzpicture}[
  fbox/.style={draw, rounded corners, text width=84mm, align=center, font=\small, inner sep=4pt},
  cbox/.style={draw, rounded corners, text width=52mm, align=center, font=\small, inner sep=4pt},
  prov/.style={fill=junccol, draw=juncline},
  mod/.style={fill=expfill, draw=expline, very thick},
  arr/.style={-Latex, thick},
  darr/.style={-Latex, semithick, densely dashed, draw=black!60},
  dline/.style={semithick, densely dashed, draw=black!60},
  ttl/.style={font=\small\bfseries},
  lbl/.style={font=\small}
]
\node[fbox,prov] (parse) {\texttt{parse\_input()}\\reads and parses the input\\\textbf{Input:} circuit file, architecture, gate vertex\\\textbf{Output:} \texttt{TrapGraph}, list of \texttt{QuantumGate}};
\node[fbox,mod] (place) [below=6mm of parse] {\texttt{initial\_mapping()}\\assigns qubits to vertices\\\textbf{Input:} \texttt{TrapGraph}, qubits, list of \texttt{QuantumGate}\\\textbf{Output:} initial mapping};
\node[fbox,mod] (route) [below=6mm of place] {\texttt{compile()}\\routes gates, groups moves into timesteps\\\textbf{Input:} initial mapping, \texttt{TrapGraph}, qubits, list of \texttt{QuantumGate}\\\textbf{Output:} shuttling schedule};
\node[fbox,prov] (val) [below=6mm of route] {\texttt{validate\_and\_output()}\\checks the shuttling schedule and writes it out\\\textbf{Input:} shuttling schedule, initial mapping, \texttt{TrapGraph}, qubits, list of \texttt{QuantumGate}\\\textbf{Output:} output file};
\node[ttl, anchor=south] at (parse.north) {compilation pipeline};
\draw[arr] (parse) -- (place);
\draw[arr] (place) -- (route);
\draw[arr] (route) -- (val);
\node[cbox,prov,anchor=north west] (tg) at ([xshift=35.25mm]parse.north east) {\texttt{TrapGraph}\\vertices \\adjacency of vertices\\distances between vertices\\shortest paths between vertices\\qubit vertices\\gate vertices\\junction vertices\\\mbox{}\hfil$\vdots$\hfil\mbox{}};
\node[cbox,prov,anchor=north west] (qg) at ([xshift=-0.75mm,yshift=-22mm]tg.south west) {\texttt{QuantumGate}\\operation\\qubits the gate acts on\\parameters\\gate id};
\node[align=center, lbl, anchor=south, inner sep=0pt] (qghead) at ([xshift=0.75mm,yshift=1mm]qg.north) {list of \texttt{QuantumGate}\\in program order};
\coordinate (shc) at ([xshift=1.5mm,yshift=-1.5mm]qg.south east);
\begin{scope}[on background layer]
  \node[draw, rounded corners, inner sep=3mm, fill=storagecol, draw=storageline, fit=(qghead)(qg)(shc)] (qgbox) {};
  \draw[rounded corners, fill=junccol, draw=juncline] ([xshift=1.5mm,yshift=-1.5mm]qg.north west) rectangle ([xshift=1.5mm,yshift=-1.5mm]qg.south east);
  \draw[rounded corners, fill=junccol, draw=juncline] ([xshift=0.75mm,yshift=-0.75mm]qg.north west) rectangle ([xshift=0.75mm,yshift=-0.75mm]qg.south east);
\end{scope}
\node[ttl, anchor=south] at (tg.north) {data classes};
\coordinate (bf) at ([xshift=11mm]parse.east);
\draw[dline] (parse.east) -- node[above,lbl]{builds} (bf);
\draw[darr] (bf) to[out=0,in=180] ([yshift=-5mm]tg.north west);
\draw[darr] (bf) to[out=0,in=180] ([yshift=-4mm]qgbox.north west);
\draw[decorate, decoration={brace,amplitude=6pt}, draw=black!60] ([xshift=2mm]place.north east) -- ([xshift=2mm]val.south east) coordinate[midway](brc);
\coordinate (mf) at ([xshift=15mm]brc);
\draw[dline] ([yshift=6mm]tg.south west) to[out=180,in=40] (mf);
\draw[dline] ([yshift=4mm]qgbox.south west) to[out=180,in=-40] (mf);
\draw[darr] (mf) -- node[above,lbl]{used by} ([xshift=3mm]brc);
\end{tikzpicture}%
}
\caption{Structure of the single-file compiler \texttt{ion\_trap\_compiler.py}. Boxes in green are the provided components that the LLM may not change: the outer functions \texttt{parse\_input} and \texttt{validate\_and\_output} and the data classes \texttt{TrapGraph} and \texttt{QuantumGate}. Boxes in blue are the two functions written by the LLM: \texttt{initial\_mapping} and \texttt{compile}. The shaded frame groups the \texttt{QuantumGate} objects into the list the pipeline passes along. Solid arrows follow the compilation pipeline, and dashed lines show which data-class objects each function builds or uses. Of \texttt{TrapGraph}'s query methods, the figure shows only those shared by all three compilers.}
\label{fig:compiler-structure}
\end{figure*}

\section{Setup}
\label{sec:setup}

This section sets out the prompting protocol and benchmark setup shared by the three compilers. The compilers themselves are then presented in turn in \autoref{sec:compiler:linear}, \autoref{sec:compiler:branched}, and \autoref{sec:compiler:general}.

\subsection{Prompting protocol}
\label{sec:compiler:setup}

The compilers in this paper are generated using Claude Code, Anthropic's command-line agent, running Claude Opus 4.7 \cite{Anthropic26_2} as the underlying LLM. The LLM is used unmodified, without fine-tuning or retrieval augmentation, and its only tools are those built into Claude Code for reading and editing files and for running shell commands, with which it tests its code. Each compiler is generated through a single conversation. The first prompt provides the LLM with a written specification together with the circuit and architecture files for testing and asks for a complete Python implementation in a single file. None of the prompts leaves the algorithmic design to the LLM. Each sets out a complete algorithm, based on the hand-crafted compilers \cite{Wagner22, KreppelMWHPSB24}, as the required default that the LLM implements and later refines. The linear- and branched-architecture prompts additionally invite further strategies for the initial mapping, provided they perform no worse than the default, and for the gate scheduling. Fixing the algorithm this way is necessary, since Opus 4.7 did not produce a working compiler in preliminary tests without one. For an unbiased comparison, the Fable 5 replication run of \autoref{sec:comparison} keeps the same prompts, although Fable 5 might not share this limitation, and prompts leaving more of the design open to it could yield better results. The full prompt for each compiler appears in Appendices \ref{app:linear-prompt}, \ref{app:branched-prompt}, and \ref{app:general-prompt}. Each compiler's first prompt is always issued at medium reasoning effort. In every follow-up prompt we ask the LLM to identify code-level optimizations that reduce shuttling timesteps, compile time, and memory use, and then to apply them in place. These follow-up prompts use medium or extra-high reasoning effort to expose the trade-off between optimization depth and the LLM's own run time.

Every prompt builds on a Python file named \texttt{ion\_trap\_compiler.py}, which the LLM adapts and extends. Its structure is illustrated in \autoref{fig:compiler-structure}. The compilation pipeline runs through four functions, from the input files to the shuttling schedule. \texttt{parse\_input} reads and parses the input, constructing the objects that every later step uses: a single \texttt{TrapGraph} for the whole architecture and a list holding one \texttt{QuantumGate} per circuit gate in program order. \texttt{initial\_mapping} assigns the qubits to vertices, and \texttt{compile} then routes the qubits of each gate to the gate segment and groups the resulting operations into timesteps (the steps a circuit takes to run, each holding either one gate or shuttling operations executed in parallel) to produce the shuttling schedule. Finally, \texttt{validate\_and\_output} checks this schedule and writes it out. The two outer functions \texttt{parse\_input} and \texttt{validate\_and\_output}, together with the two data classes \texttt{TrapGraph} and \texttt{QuantumGate}, are provided and may not be changed by the LLM. \texttt{TrapGraph} also exposes many query methods that the compilers use. The LLM writes only the two functions in between, \texttt{initial\_mapping} and \texttt{compile}, whose interfaces and full specification are given in every prompt. When one compiler is seeded from the previous one, these two functions are the only parts the LLM adapts and extends.

The three compilers are generated in a chain, each building on its predecessor. Starting from scratch, the LLM produces the linear-architecture compiler from the first prompt and the provided code alone. Each later compiler is then seeded with the previous one's optimized Python file, in which the provided functions and data classes are swapped for their new-architecture versions while the LLM's \texttt{initial\_mapping} and \texttt{compile} are carried over unchanged. Because this file sits in the working directory, Claude Code reads and edits it directly, and the prompt asks the LLM to adapt and extend only those two functions in place. Each adaptation prompt treats the seed as the source of truth for everything that stays unchanged and describes only what must change. Because it names some unchanged features from the earlier specification rather than from the seed's code, it can list a feature the seed lacks as one to keep, and the LLM then carries that gap forward rather than filling it.

Every first prompt also contains a working example and an acceptance-test matrix, both built from given input files in the working directory. The working example specifies one circuit, one architecture, and its gate vertex, together with the exact bound the schedule must meet. The matrix itself is a list of circuit-and-graph pairs that the compiler must run successfully, drawn from the benchmark circuits and a few small hand-written checks. Some carry a tight bound, such as zero swaps for a chain-like circuit. The full matrices for the linear- and branched-architecture compilers are listed in \autoref{sec:linear:spec} and \autoref{sec:branched:spec}. For the general-architecture compiler, whose evaluation spans many layouts, the matrix graphs are not the evaluation instances themselves but are built from the same construction rules, as described in \autoref{sec:general:spec}. Each schedule is checked by the provided validator functions, which the LLM is not allowed to modify. The LLM runs these tests itself and debugs its code against them during generation. Every schedule of the generated compilers reported in this paper passes the same validation.

The three compilers are presented in a common structure in \autoref{sec:compiler:linear}, \autoref{sec:compiler:branched}, and \autoref{sec:compiler:general}. Each section opens with a specification that summarizes the corresponding prompt and states the task we set for the LLM, with the full prompt reproduced in the appendix. The section then reports the emitted compiler, noting where it follows or departs from that specification, and presents its optimizations. These come from the follow-up prompts we issue, which ask only for code-level improvements. The LLM itself decides which changes to make, and we then describe them. Each section closes with the compiler's evaluation.

\subsection{Evaluation setup}
\label{sec:eval:setup}

\begin{figure}[t]
	\centering
	\includegraphics{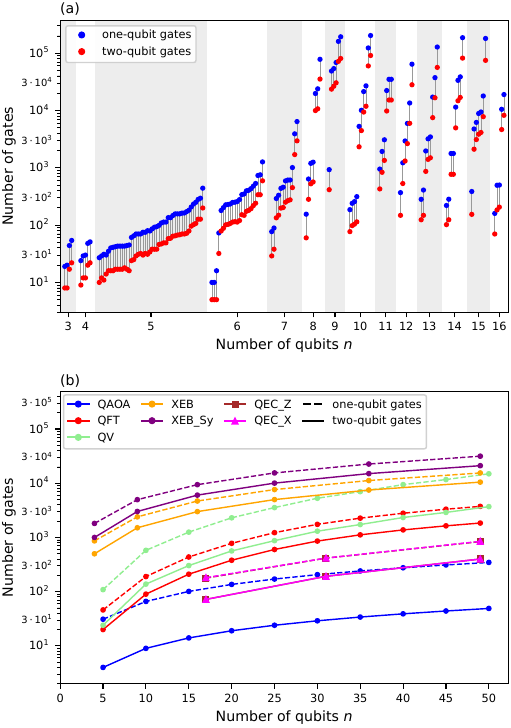}
	\caption{Numbers of one-qubit and two-qubit gates in the benchmark circuits after transpilation into the native gate set \cite{KreppelMOWHPSB23}. (a) The 153 library circuits, grouped by the qubit count $n$ and sorted by size within each group, with the two gate counts of each circuit joined by a vertical line. (b) The scalable families against the qubit count $n$.}
	\label{fig:circuit-stats}
\end{figure}

We benchmark every generated compiler on two circuit sets. The first is a library of 153 quantum circuits \cite{Zhou19} with 3 to 16 qubits. The second comprises seven scalable families with up to 50 qubits: Quantum Approximate Optimization Algorithm (QAOA) \cite{Moll+18}, Quantum Fourier Transform (QFT) \cite{NielsenC10}, Quantum Volume (QV) \cite{CrossBSNG19}, Cross-Entropy Benchmarking (XEB) \cite{BoixoISBDJBMN18}, a Sycamore-specialized XEB variant (XEB\_Sy) \cite{AruteAB+19}, and two families of quantum error correction (QEC) circuits. The latter are memory experiments of the rotated surface code \cite{FowlerMMC12, TomitaS14}, in which the logical qubit is prepared and measured in the Z basis (QEC\_Z) or in the X basis (QEC\_X). A circuit generator \cite{TomeshC21} produces the QAOA, QFT, XEB, and XEB\_Sy circuits, and Qiskit \cite{JavadiAbhariTKWLGMNBCJG24} the QV and QEC circuits. In the XEB and XEB\_Sy circuits, the qubits form a square lattice, and every circuit runs 1,000 cycles of random single-qubit gates and two-qubit gates between neighboring qubits. Each QEC circuit, of code distance $c \in \{3, 4, 5\}$, acts on $2c^2-1$ qubits and runs $c$ rounds of syndrome extraction. Every circuit of both sets is then transpiled into a trapped-ion native gate set by a circuit compiler \cite{KreppelMOWHPSB23}, and \autoref{fig:circuit-stats} shows the resulting gate counts. The library circuits of \autoref{fig:circuit-stats}(a) range from 15 to about 294,000 gates, largely independently of the qubit count, with up to 203,000 one-qubit and 91,000 two-qubit gates. In the scalable families of \autoref{fig:circuit-stats}(b), by contrast, the gate count grows steadily with the qubit count. At the largest qubit counts, the families range from about 400 gates for QAOA, of which 49 are two-qubit gates, to about 53,000 for XEB\_Sy, of which about 21,000 are two-qubit gates.

Each circuit of $n$ qubits is compiled on an architecture with $2n+1$ non-junction segments. The linear and branched architectures follow a previously published construction \cite{KreppelMWHPSB24}. In the linear architecture, the gate segment sits at the center. The branched architectures are mirror-symmetric about that single central gate segment and are parameterized by a stack height $h$, the number of segments in each side-stack, and a junction distance $d$, the spacing between consecutive junctions along the main axis. Both parameters are shown in \autoref{fig:pipeline}. Each branched architecture is built outward from the gate segment on both sides: first a qubit segment for separation and merge, then a stack, then a further stack after every $d$ segments of the main axis, with the qubit segments split equally between the two sides. Once a side has $h$ or fewer qubit segments left, no further stack is added, and the remaining qubit segments extend the main axis until none is left. We vary $h$ from 1 to 10 and $d$ from 1 to 4 on the circuit library. On the scalable families, both run from 1 to 10 and then in steps of 5, $h$ up to 45 and $d$ up to 50. Once $d > n - 1 - 2h$, only one stack fits per side, so all larger distances yield the same architecture and are evaluated once. A stack of more than $n-2$ segments does not fit, so such heights are skipped. We call each remaining pair of $h$ and $d$ a \emph{tested configuration}.

The general-architecture compiler of \autoref{sec:compiler:general} is evaluated on ten layouts: the eight cyclic families of \autoref{fig:pipeline} together with the linear and branched architectures. Three of the cyclic families are built from classical interconnection networks, though not in their original form: a complete graph on four vertices, the three-dimensional hypercube \cite{Leighton91}, and the two-dimensional butterfly network \cite{Leighton91}. The vertices of these networks are predominantly junctions, in which qubits cannot rest, since the electric fields there are designed for ions to pass through \cite{BlakestadODWBLW11, WrightAFVDHPLDKSH13, MokhberiSW17}. Therefore, the edges are subdivided by paths of non-junction segments, as shown in \autoref{fig:pipeline}. This subdivision also lets the networks grow with the circuit, since segments are inserted until each instance has the $2n+1$ non-junction segments required for an $n$-qubit circuit.

Two state-of-the-art hand-crafted compilers serve as references, one for the linear architecture \cite{Wagner22} and one for the branched one \cite{KreppelMWHPSB24}. Each is a heavily optimized heuristic for exactly the layout its LLM-generated counterpart targets, and operates under the same hardware limits as our compilers, namely a single gate segment, two qubits per segment, and the confinement of separation, merge, and swap to the gate segment. Both were evaluated on the same benchmark circuits, so the comparison is direct. In contrast to the generated compilers, which select a gate from the ready set at each step, the baselines execute a circuit's gates in the input order. The branched-architecture baseline further uses its junctions to push translation, separation, and merge counts below the linear figures on most configurations and to eliminate physical ion swaps outright on its targets, which makes it an especially stringent reference. It is also limited to a single stack per junction, whereas the generated compiler handles junctions that carry several stacks.

For every schedule we record the number of timesteps, our primary objective, together with the translation, separation, merge, and swap counts. All schedules are computed with one constant Python hash seed. The \emph{timestep reduction factor} is the timestep count of the reference divided by that of the optimized compiler, where the reference is either a hand-crafted baseline or the emitted compiler. Therefore, a value above 1 means the optimized LLM-based compiler outputs fewer timesteps. In the text, we summarize a circuit set by the median of its per-circuit reduction factors. The ratio of the summed timesteps would be dominated by the few largest circuits, and the mean of the factors would be sensitive to single outlying circuits. Each compiler section also shows the absolute counts behind these factors, as totals over the circuit library, whose circuits differ too much in size at equal qubit count to be grouped by qubit count, and per circuit for the scalable families. Since every gate occupies a timestep of its own at the single gate segment, the number of gates is a hard lower bound on the number of timesteps, and the timesteps a compiler needs beyond this bound are its shuttling overhead on that architecture.

Besides the timesteps, we record the compile time and the peak memory use of every compilation. For all compilers, the compile time covers only the initial mapping, the routing, and, for the generated compilers, the post-processing. The baselines need no separate post-processing, since each operation they emit already forms a timestep of its own. As peak memory use, we take the largest amount of main memory the process occupies, which includes the Python interpreter. Both are measured over ten runs of each compilation, of which we report the mean. All runs take place on the MOGON NHR cluster at Johannes Gutenberg University Mainz, each as a separate single-threaded Python 3.12 process on one core of an AMD EPYC 7713 processor with up to 16\,GB of reserved memory.

\section{Linear shuttling architecture}
\label{sec:compiler:linear}

\subsection{Specification}
\label{sec:linear:spec}

\begin{figure*}[t]
\centering
\resizebox{\textwidth}{!}{%
\begin{tikzpicture}[
  every node/.style={font=\small},
  gdot/.style={circle, fill=black, minimum size=4pt, inner sep=0pt},
  dagnode/.style={circle, draw, minimum size=16pt, inner sep=0pt, font=\scriptsize},
  qnode/.style={circle, draw, minimum size=16pt, inner sep=0pt},
]
\begin{scope}[xshift=0.72cm]
\foreach \q [count=\i from 0] in {0,1,2,3} {
\pgfmathsetmacro{\y}{1.2-0.8*\i}
\draw (-0.6,\y) -- (3.4,\y);
\node[anchor=east] at (-0.72,\y) {$q_\q$};
}
\node[gdot] (g1t) at (0.0,0.4) {};
\node[gdot] (g1b) at (0.0,-1.2) {};
\draw (g1t) -- (g1b);
\node[above] at (0.0,1.45) {$g_1$};
\node[gdot] (g2t) at (0.6,1.2) {};
\node[gdot] (g2b) at (0.6,-0.4) {};
\draw (g2t) -- (g2b);
\node[above] at (0.6,1.45) {$g_2$};
\node[gdot] (g3t) at (1.4,1.2) {};
\node[gdot] (g3b) at (1.4,-1.2) {};
\draw (g3t) -- (g3b);
\node[above] at (1.4,1.45) {$g_3$};
\node[gdot] (g4t) at (2.2,0.4) {};
\node[gdot] (g4b) at (2.2,-1.2) {};
\draw (g4t) -- (g4b);
\node[above] at (2.2,1.45) {$g_4$};
\node[gdot] (g5t) at (3.0,1.2) {};
\node[gdot] (g5b) at (3.0,0.4) {};
\draw (g5t) -- (g5b);
\node[above] at (3.0,1.45) {$g_5$};
\node[anchor=base west] at (-1.25,2.25) {(a)};
\node[anchor=base] at (1.13,2.25) {example circuit};
\end{scope}
\begin{scope}[xshift=6.93cm]
\node[dagnode] (G1) at (0,0.9) {$g_1$};
\node[dagnode] (G2) at (0,-0.9) {$g_2$};
\node[dagnode] (G3) at (1.3,-0.2) {$g_3$};
\node[dagnode] (G4) at (2.6,0.9) {$g_4$};
\node[dagnode] (G5) at (3.9,-0.2) {$g_5$};
\draw[->] (G1) -- (G3);
\draw[->] (G2) -- (G3);
\draw[->] (G1) -- (G4);
\draw[->] (G3) -- (G4);
\draw[->] (G4) -- (G5);
\draw[->] (G3) -- (G5);
\node[anchor=base west] at (-0.45,2.25) {(b)};
\node[anchor=base] at (1.95,2.25) {circuit-DAG};
\end{scope}
\begin{scope}[xshift=14.7cm]
\node[qnode] (Q1) at (0,1.2) {$q_1$};
\node[qnode] (Q3) at (2,1.2) {$q_3$};
\node[qnode] (Q0) at (2,-1.2) {$q_0$};
\node[qnode] (Q2) at (0,-1.2) {$q_2$};
\draw (Q1) -- (Q3) node[midway, above] {$\frac{4}{3}$};
\draw (Q3) -- (Q0) node[midway, right] {$\frac{1}{2}$};
\draw (Q0) -- (Q2) node[midway, below] {$1$};
\draw (Q1) -- (Q0) node[pos=0.62, above right=-2pt] {$\frac{1}{4}$};
\node[anchor=base west] at (-1.4,2.25) {(c)};
\node[anchor=base] at (1.2,2.25) {qubit-interaction graph};
\end{scope}
\draw[-Latex, thick] (4.5,0) -- (6.0,0);
\draw[-Latex, thick] (11.88,0) -- (13.38,0);
\end{tikzpicture}%
}\\[4mm]
\resizebox{\textwidth}{!}{%
\begin{tabular}{@{}c@{\hspace{14mm}}c@{}}
\sbox0{\begin{tikzpicture}[x=11.5mm,y=11.5mm,
  vtx/.style={minimum size=8.0mm,semithick},
  qb/.style={circle,draw=qubitline,fill=qubitcol,text=white,inner sep=0pt,minimum size=3.7mm,font={\fontsize{6.5}{6.5}\selectfont}},
  qg/.style={qb,draw=qgateline,fill=qgatecol},
  ed/.style={draw=black,thick},
  idx/.style={font=\small},
  op/.style={font=\small,anchor=east,align=right}]
\draw[ed] (0.000,0.000) -- (1.000,0.000);
\draw[ed] (1.000,0.000) -- (2.000,0.000);
\draw[ed] (2.000,0.000) -- (3.000,0.000);
\draw[ed] (3.000,0.000) -- (4.000,0.000);
\draw[ed] (4.000,0.000) -- (5.000,0.000);
\draw[ed] (5.000,0.000) -- (6.000,0.000);
\draw[ed] (6.000,0.000) -- (7.000,0.000);
\draw[ed] (7.000,0.000) -- (8.000,0.000);
\node[storage,vtx] at (0.000,0.000) {};
\node[idx] at (0.000,-0.591) {0};
\node[storage,vtx] at (1.000,0.000) {};
\node[idx] at (1.000,-0.591) {1};
\node[storage,vtx] at (2.000,0.000) {};
\node[idx] at (2.000,-0.591) {2};
\node[storage,vtx] at (3.000,0.000) {};
\node[idx] at (3.000,-0.591) {3};
\node[gatev,vtx] at (4.000,0.000) {};
\node[idx] at (4.000,-0.591) {4};
\node[idx] at (4.000,0.591) {GS};
\node[storage,vtx] at (5.000,0.000) {};
\node[idx] at (5.000,-0.591) {5};
\node[storage,vtx] at (6.000,0.000) {};
\node[idx] at (6.000,-0.591) {6};
\node[storage,vtx] at (7.000,0.000) {};
\node[idx] at (7.000,-0.591) {7};
\node[storage,vtx] at (8.000,0.000) {};
\node[idx] at (8.000,-0.591) {8};
\node[qg] at ($(4.000,0.000)+(-1.95mm,0)$) {$q_{1}$};
\node[qg] at ($(4.000,0.000)+(1.95mm,0)$) {$q_{3}$};
\node[qb] at ($(6.000,0.000)+(-1.95mm,0)$) {$q_{0}$};
\node[qb] at ($(6.000,0.000)+(1.95mm,0)$) {$q_{2}$};
\end{tikzpicture}}%
\makebox[\wd0]{\fontsize{10.69}{12.8}\selectfont (d)\hfill linear architecture\hfill\phantom{(d)}}
& \sbox0{\begin{tikzpicture}[x=11.5mm,y=11.5mm,
  vtx/.style={minimum size=8.0mm,semithick},
  qb/.style={circle,draw=qubitline,fill=qubitcol,text=white,inner sep=0pt,minimum size=3.7mm,font={\fontsize{6.5}{6.5}\selectfont}},
  qg/.style={qb,draw=qgateline,fill=qgatecol},
  ed/.style={draw=black,thick},
  idx/.style={font=\small},
  op/.style={font=\small,anchor=east,align=right}]
\draw[ed] (0.000,0.000) -- (1.000,0.000);
\draw[ed] (1.000,0.000) -- (2.000,0.000);
\draw[ed] (2.000,0.000) -- (3.000,0.000);
\draw[ed] (3.000,0.000) -- (4.000,0.000);
\draw[ed] (4.000,0.000) -- (5.000,0.000);
\draw[ed] (5.000,0.000) -- (6.000,0.000);
\draw[ed] (6.000,0.000) -- (7.000,0.000);
\draw[ed] (7.000,0.000) -- (8.000,0.000);
\draw[ed] (1.000,0.000) -- (1.000,1.000);
\draw[ed] (2.000,0.000) -- (2.000,1.000);
\draw[ed] (6.000,0.000) -- (6.000,1.000);
\draw[ed] (7.000,0.000) -- (7.000,1.000);
\node[storage,vtx] at (0.000,0.000) {};
\node[idx] at (0.000,-0.591) {0};
\node[junc,vtx] at (1.000,0.000) {};
\node[idx] at (1.000,-0.591) {1};
\node[junc,vtx] at (2.000,0.000) {};
\node[idx] at (2.000,-0.591) {2};
\node[storage,vtx] at (3.000,0.000) {};
\node[idx] at (3.000,-0.591) {3};
\node[gatev,vtx] at (4.000,0.000) {};
\node[idx] at (4.000,-0.591) {4};
\node[idx] at (4.000,0.591) {GS};
\node[storage,vtx] at (5.000,0.000) {};
\node[idx] at (5.000,-0.591) {5};
\node[junc,vtx] at (6.000,0.000) {};
\node[idx] at (6.000,-0.591) {6};
\node[junc,vtx] at (7.000,0.000) {};
\node[idx] at (7.000,-0.591) {7};
\node[storage,vtx] at (8.000,0.000) {};
\node[idx] at (8.000,-0.591) {8};
\node[storage,vtx] at (1.000,1.000) {};
\node[idx] at (1.000,1.591) {9};
\node[storage,vtx] at (2.000,1.000) {};
\node[idx] at (2.000,1.591) {10};
\node[storage,vtx] at (6.000,1.000) {};
\node[idx] at (6.000,1.591) {11};
\node[storage,vtx] at (7.000,1.000) {};
\node[idx] at (7.000,1.591) {12};
\node[qg] at ($(4.000,0.000)+(-1.95mm,0)$) {$q_{3}$};
\node[qg] at ($(4.000,0.000)+(1.95mm,0)$) {$q_{1}$};
\node[qb] at ($(2.000,1.000)+(-1.95mm,0)$) {$q_{0}$};
\node[qb] at ($(2.000,1.000)+(1.95mm,0)$) {$q_{2}$};
\end{tikzpicture}}%
\makebox[\wd0]{\fontsize{10.69}{12.8}\selectfont (e)\hfill branched architecture, $h=d=1$\hfill\phantom{(e)}}
\\[2mm]
\begin{tikzpicture}[x=11.5mm,y=11.5mm,
  vtx/.style={minimum size=8.0mm,semithick},
  qb/.style={circle,draw=qubitline,fill=qubitcol,text=white,inner sep=0pt,minimum size=3.7mm,font={\fontsize{6.5}{6.5}\selectfont}},
  qg/.style={qb,draw=qgateline,fill=qgatecol},
  ed/.style={draw=black,thick},
  idx/.style={font=\small},
  op/.style={font=\small,anchor=east,align=right}]
\draw[ed] (0.000,0.000) -- (1.000,0.000);
\draw[ed] (1.000,0.000) -- (2.000,0.000);
\draw[ed] (2.000,0.000) -- (3.000,0.000);
\draw[ed] (3.000,0.000) -- (4.000,0.000);
\draw[ed] (4.000,0.000) -- (5.000,0.000);
\draw[ed] (5.000,0.000) -- (6.000,0.000);
\draw[ed] (6.000,0.000) -- (7.000,0.000);
\draw[ed] (7.000,0.000) -- (8.000,0.000);
\node[storage,vtx] at (0.000,0.000) {};
\node[idx] at (0.000,-0.591) {0};
\node[storage,vtx] at (1.000,0.000) {};
\node[idx] at (1.000,-0.591) {1};
\node[storage,vtx] at (2.000,0.000) {};
\node[idx] at (2.000,-0.591) {2};
\node[storage,vtx] at (3.000,0.000) {};
\node[idx] at (3.000,-0.591) {3};
\node[gatev,vtx] at (4.000,0.000) {};
\node[idx] at (4.000,-0.591) {4};
\node[idx] at (4.000,0.591) {GS};
\node[storage,vtx] at (5.000,0.000) {};
\node[idx] at (5.000,-0.591) {5};
\node[storage,vtx] at (6.000,0.000) {};
\node[idx] at (6.000,-0.591) {6};
\node[storage,vtx] at (7.000,0.000) {};
\node[idx] at (7.000,-0.591) {7};
\node[storage,vtx] at (8.000,0.000) {};
\node[idx] at (8.000,-0.591) {8};
\node[qg] at ($(4.000,0.000)+(-1.95mm,0)$) {$q_{1}$};
\node[qg] at ($(4.000,0.000)+(1.95mm,0)$) {$q_{3}$};
\node[qb] at ($(6.000,0.000)+(-1.95mm,0)$) {$q_{0}$};
\node[qb] at ($(6.000,0.000)+(1.95mm,0)$) {$q_{2}$};
\end{tikzpicture}
& \begin{tikzpicture}[x=11.5mm,y=11.5mm,
  vtx/.style={minimum size=8.0mm,semithick},
  qb/.style={circle,draw=qubitline,fill=qubitcol,text=white,inner sep=0pt,minimum size=3.7mm,font={\fontsize{6.5}{6.5}\selectfont}},
  qg/.style={qb,draw=qgateline,fill=qgatecol},
  ed/.style={draw=black,thick},
  idx/.style={font=\small},
  op/.style={font=\small,anchor=east,align=right}]
\draw[ed] (0.000,0.000) -- (1.000,0.000);
\draw[ed] (1.000,0.000) -- (2.000,0.000);
\draw[ed] (2.000,0.000) -- (3.000,0.000);
\draw[ed] (3.000,0.000) -- (4.000,0.000);
\draw[ed] (4.000,0.000) -- (5.000,0.000);
\draw[ed] (5.000,0.000) -- (6.000,0.000);
\draw[ed] (6.000,0.000) -- (7.000,0.000);
\draw[ed] (7.000,0.000) -- (8.000,0.000);
\draw[ed] (1.000,0.000) -- (1.000,1.000);
\draw[ed] (2.000,0.000) -- (2.000,1.000);
\draw[ed] (6.000,0.000) -- (6.000,1.000);
\draw[ed] (7.000,0.000) -- (7.000,1.000);
\node[storage,vtx] at (0.000,0.000) {};
\node[idx] at (0.000,-0.591) {0};
\node[junc,vtx] at (1.000,0.000) {};
\node[idx] at (1.000,-0.591) {1};
\node[junc,vtx] at (2.000,0.000) {};
\node[idx] at (2.000,-0.591) {2};
\node[storage,vtx] at (3.000,0.000) {};
\node[idx] at (3.000,-0.591) {3};
\node[gatev,vtx] at (4.000,0.000) {};
\node[idx] at (4.000,-0.591) {4};
\node[idx] at (4.000,0.591) {GS};
\node[storage,vtx] at (5.000,0.000) {};
\node[idx] at (5.000,-0.591) {5};
\node[junc,vtx] at (6.000,0.000) {};
\node[idx] at (6.000,-0.591) {6};
\node[junc,vtx] at (7.000,0.000) {};
\node[idx] at (7.000,-0.591) {7};
\node[storage,vtx] at (8.000,0.000) {};
\node[idx] at (8.000,-0.591) {8};
\node[storage,vtx] at (1.000,1.000) {};
\node[idx] at (1.000,1.591) {9};
\node[storage,vtx] at (2.000,1.000) {};
\node[idx] at (2.000,1.591) {10};
\node[storage,vtx] at (6.000,1.000) {};
\node[idx] at (6.000,1.591) {11};
\node[storage,vtx] at (7.000,1.000) {};
\node[idx] at (7.000,1.591) {12};
\node[qg] at ($(4.000,0.000)+(-1.95mm,0)$) {$q_{3}$};
\node[qg] at ($(4.000,0.000)+(1.95mm,0)$) {$q_{1}$};
\node[qb] at ($(2.000,1.000)+(-1.95mm,0)$) {$q_{0}$};
\node[qb] at ($(2.000,1.000)+(1.95mm,0)$) {$q_{2}$};
\end{tikzpicture}
\end{tabular}%
}
\caption{Initial mapping of an example circuit. (a) The circuit, (b) its circuit-DAG, and (c) its qubit-interaction graph, from which the emitted compilers compute the initial placements on (d) the linear and (e) the branched architecture with $h=d=1$. The gate vertex (GS) is red and the junctions are green. The qubits of the first gate $g_1$ are drawn in red, the others in blue.}
\label{fig:graph-construction}
\end{figure*}

This specification summarizes the prompt, which is reproduced in full in \appref{app:linear-prompt}. The prompt defines the architecture as a linear chain of vertices, each either a qubit vertex or a gate vertex, and provides the four shuttling operations of \autoref{sec:introduction}, together with a gate-execution operation that runs the corresponding one- or two-qubit gate when a gate vertex holds exactly the required qubits. The architecture files place a single gate vertex at the center of the chain, as drawn in \autoref{fig:pipeline}. Building on these definitions, the prompt describes the compiler in three parts.

The \emph{initial mapping} places each of the $n$ qubits on a vertex and follows three rules inherited from the linear baseline \cite{Wagner22}. At most two qubits sit on a vertex, exactly $\lceil \frac{n}{2} \rceil$ vertices hold qubits, and all remaining vertices stay empty. To compute the mapping, two graphs are derived from the circuit, which \autoref{fig:graph-construction}(b) and \autoref{fig:graph-construction}(c) show for the example circuit of \autoref{fig:graph-construction}(a). The first is a directed acyclic graph (DAG), the \emph{circuit-DAG}, whose nodes are the gates of the circuit and whose edges encode the program-order dependency between two consecutive gates that share a qubit. A gate's depth in the circuit-DAG is the length of the longest directed path from any source node to it. The second is the \emph{qubit-interaction graph}, an undirected graph whose nodes are the qubits. Its edge weight between two qubits $q$ and $q'$ is the sum, over every two-qubit gate that acts on $q$ and $q'$, of a depth-based weight $\frac{1}{1 + \delta/\tau}$, where $\delta$ is the gate's circuit-DAG depth and $\tau \approx \frac{\mathrm{total\_dag\_depth}}{3}$. The prompt describes $\mathrm{total\_dag\_depth}$ both as the largest gate depth in the circuit-DAG and as the number of layers of the circuit, which is the largest gate depth plus one. This depth-weighting biases the mapping toward interactions that occur early in the circuit, where the initial mapping has the largest effect on schedule length. The qubits are then paired by a greedy matching on the qubit-interaction graph. The heaviest edge whose two qubits are both still unmatched is picked, those two qubits are matched into a pair, and the procedure repeats until no such edge remains. 

When the number of qubits is odd, one qubit cannot be matched and is left as a \emph{singleton}. The prompt specifies that the singleton is the qubit with the lowest total interaction weight across all its edges. A \emph{chain extraction} then orders these pairs and the singleton by greedily appending the matched pair with the heaviest connecting weight to the running chain, and the chain is placed on consecutive vertices in that order. The one exception is the first gate to execute. If it is a two-qubit gate, its two qubits are placed directly on the gate vertex, so that this gate runs with no preceding shuttling, and the gate vertex's two neighbors stay empty, so that a subsequent \textsc{Separate} has room to split the pair.

\begin{example}[Initial mapping]
In the circuit of \autoref{fig:graph-construction}(a), gates $g_1$ and $g_2$ have depth 0, $g_3$ has depth 1, $g_4$ has depth 2, and $g_5$ has depth 3, as the circuit-DAG of \autoref{fig:graph-construction}(b) shows. With $\tau = 1$, taken from the largest depth, their depth-based weights are 1, 1, $\frac{1}{2}$, $\frac{1}{3}$, and $\frac{1}{4}$. Summed per pair, they yield the edge weights of the qubit-interaction graph in \autoref{fig:graph-construction}(c), $1 + \frac{1}{3} = \frac{4}{3}$ for $q_1\text{-}q_3$ from $g_1$ and $g_4$, $1$ for $q_0\text{-}q_2$ from $g_2$, $\frac{1}{2}$ for $q_0\text{-}q_3$ from $g_3$, and $\frac{1}{4}$ for $q_0\text{-}q_1$ from $g_5$. Therefore, the greedy matching pairs $(q_1,q_3)$ and $(q_0,q_2)$, leaving the lighter $q_0\text{-}q_3$ and $q_0\text{-}q_1$ edges unused. Chain extraction then starts with the heavier pair $(q_1,q_3)$ and appends $(q_0,q_2)$. Since the first gate $g_1$ acts on $q_1$ and $q_3$, this pair occupies the gate vertex 4, and both of its neighbors stay empty. The emitted compiler places $(q_0,q_2)$ on the next vertex beyond, vertex 6, as \autoref{fig:graph-construction}(d) shows. With $\tau = \frac{4}{3}$, taken from the four layers, the weights change, but the matching and the placement stay the same.
\end{example}

Once the initial mapping has placed every qubit on its starting vertex, the \emph{routing} begins, which iteratively brings each gate's qubits together at the gate vertex. At every scheduling step the next gate is chosen from the set of ready gates. Among them, the compiler prefers a gate that acts on the same qubit set as the one just executed, so a run of gates on one pair completes without re-shuttling. Otherwise it takes the gate whose qubits lie closest to a gate vertex, with ties broken in favor of the gate that appears earliest in the input circuit.

For a single-qubit gate, the routine is simple. If the target qubit shares its vertex with another qubit, the pair is first moved to the gate vertex and split with a \textsc{Separate}, which sends each qubit to a neighboring vertex. The target qubit then moves back onto the gate vertex, where the gate executes.

For a two-qubit gate, the qubits sitting strictly between its two \emph{endpoints} on the current layout form an ordered set of \emph{intermediates} that must be cleared before the gate can execute. The intermediates are removed one at a time by a \textsc{Merge}-\textsc{Swap}-\textsc{Separate} cycle on the gate vertex. Each cycle advances the endpoint that the compiler estimates to be cheaper to move, and this choice may be re-evaluated every round. Before each cycle, that endpoint and its adjacent intermediate must each sit alone on a vertex. If either of them shares its vertex with another qubit, the pair is first moved to the gate vertex and split with a \textsc{Separate}. The cycle then proceeds in three steps. A \textsc{Merge} combines the endpoint and the intermediate at the gate vertex, a \textsc{Swap} reverses that chain, and a \textsc{Separate} sends them back out with their positions exchanged. Each cycle moves the endpoint one position closer to its partner and the intermediate one position farther away, so the intermediate set shrinks by one. In the special case where the endpoint already shares its vertex with its adjacent intermediate, the \textsc{Merge} is skipped, and moving the pair to the gate vertex followed by a \textsc{Swap} and a \textsc{Separate} completes the cycle. After $k$ cycles, one for each of the $k$ initial intermediates, a final \textsc{Merge} brings the two endpoints together at the gate vertex, and the gate executes.

\begin{figure}[t]
\centering
\resizebox{\columnwidth}{!}{\begin{tikzpicture}[x=11.5mm,y=11.5mm,
  vtx/.style={minimum size=8.0mm,semithick},
  qb/.style={circle,draw=qubitline,fill=qubitcol,text=white,inner sep=0pt,minimum size=3.7mm,font={\fontsize{6.5}{6.5}\selectfont}},
  qg/.style={qb,draw=qgateline,fill=qgatecol},
  ed/.style={draw=black,thick},
  idx/.style={font=\small},
  op/.style={font=\small,anchor=east,align=right}]
\node[op] at (-0.550,0.000) {(1)~initial state};
\draw[ed] (0.000,0.000) -- (1.000,0.000);
\draw[ed] (1.000,0.000) -- (2.000,0.000);
\draw[ed] (2.000,0.000) -- (3.000,0.000);
\draw[ed] (3.000,0.000) -- (4.000,0.000);
\draw[ed] (4.000,0.000) -- (5.000,0.000);
\draw[ed] (5.000,0.000) -- (6.000,0.000);
\node[storage,vtx] at (0.000,0.000) {};
\node[idx] at (0.000,-0.591) {0};
\node[storage,vtx] at (1.000,0.000) {};
\node[idx] at (1.000,-0.591) {1};
\node[storage,vtx] at (2.000,0.000) {};
\node[idx] at (2.000,-0.591) {2};
\node[gatev,vtx] at (3.000,0.000) {};
\node[idx] at (3.000,-0.591) {3};
\node[idx] at (3.000,0.591) {GS};
\node[storage,vtx] at (4.000,0.000) {};
\node[idx] at (4.000,-0.591) {4};
\node[storage,vtx] at (5.000,0.000) {};
\node[idx] at (5.000,-0.591) {5};
\node[storage,vtx] at (6.000,0.000) {};
\node[idx] at (6.000,-0.591) {6};
\node[qg] at ($(2.000,0.000)+(0mm,0)$) {$q_{0}$};
\node[qb] at ($(4.000,0.000)+(0mm,0)$) {$q_{2}$};
\node[qg] at ($(5.000,0.000)+(0mm,0)$) {$q_{1}$};
\node[op] at (-0.550,-1.650) {(2)~\textsc{Merge}};
\draw[ed] (0.000,-1.650) -- (1.000,-1.650);
\draw[ed] (1.000,-1.650) -- (2.000,-1.650);
\draw[ed] (2.000,-1.650) -- (3.000,-1.650);
\draw[ed] (3.000,-1.650) -- (4.000,-1.650);
\draw[ed] (4.000,-1.650) -- (5.000,-1.650);
\draw[ed] (5.000,-1.650) -- (6.000,-1.650);
\node[storage,vtx] at (0.000,-1.650) {};
\node[idx] at (0.000,-2.241) {0};
\node[storage,vtx] at (1.000,-1.650) {};
\node[idx] at (1.000,-2.241) {1};
\node[storage,vtx] at (2.000,-1.650) {};
\node[idx] at (2.000,-2.241) {2};
\node[gatev,vtx] at (3.000,-1.650) {};
\node[idx] at (3.000,-2.241) {3};
\node[idx] at (3.000,-1.059) {GS};
\node[storage,vtx] at (4.000,-1.650) {};
\node[idx] at (4.000,-2.241) {4};
\node[storage,vtx] at (5.000,-1.650) {};
\node[idx] at (5.000,-2.241) {5};
\node[storage,vtx] at (6.000,-1.650) {};
\node[idx] at (6.000,-2.241) {6};
\node[qg] at ($(3.000,-1.650)+(-1.95mm,0)$) {$q_{0}$};
\node[qb] at ($(3.000,-1.650)+(1.95mm,0)$) {$q_{2}$};
\node[qg] at ($(5.000,-1.650)+(0mm,0)$) {$q_{1}$};
\node[op] at (-0.550,-3.300) {(3)~\textsc{Swap}};
\draw[ed] (0.000,-3.300) -- (1.000,-3.300);
\draw[ed] (1.000,-3.300) -- (2.000,-3.300);
\draw[ed] (2.000,-3.300) -- (3.000,-3.300);
\draw[ed] (3.000,-3.300) -- (4.000,-3.300);
\draw[ed] (4.000,-3.300) -- (5.000,-3.300);
\draw[ed] (5.000,-3.300) -- (6.000,-3.300);
\node[storage,vtx] at (0.000,-3.300) {};
\node[idx] at (0.000,-3.891) {0};
\node[storage,vtx] at (1.000,-3.300) {};
\node[idx] at (1.000,-3.891) {1};
\node[storage,vtx] at (2.000,-3.300) {};
\node[idx] at (2.000,-3.891) {2};
\node[gatev,vtx] at (3.000,-3.300) {};
\node[idx] at (3.000,-3.891) {3};
\node[idx] at (3.000,-2.709) {GS};
\node[storage,vtx] at (4.000,-3.300) {};
\node[idx] at (4.000,-3.891) {4};
\node[storage,vtx] at (5.000,-3.300) {};
\node[idx] at (5.000,-3.891) {5};
\node[storage,vtx] at (6.000,-3.300) {};
\node[idx] at (6.000,-3.891) {6};
\node[qb] at ($(3.000,-3.300)+(-1.95mm,0)$) {$q_{2}$};
\node[qg] at ($(3.000,-3.300)+(1.95mm,0)$) {$q_{0}$};
\node[qg] at ($(5.000,-3.300)+(0mm,0)$) {$q_{1}$};
\node[op] at (-0.550,-4.950) {(4)~\textsc{Separate}};
\draw[ed] (0.000,-4.950) -- (1.000,-4.950);
\draw[ed] (1.000,-4.950) -- (2.000,-4.950);
\draw[ed] (2.000,-4.950) -- (3.000,-4.950);
\draw[ed] (3.000,-4.950) -- (4.000,-4.950);
\draw[ed] (4.000,-4.950) -- (5.000,-4.950);
\draw[ed] (5.000,-4.950) -- (6.000,-4.950);
\node[storage,vtx] at (0.000,-4.950) {};
\node[idx] at (0.000,-5.541) {0};
\node[storage,vtx] at (1.000,-4.950) {};
\node[idx] at (1.000,-5.541) {1};
\node[storage,vtx] at (2.000,-4.950) {};
\node[idx] at (2.000,-5.541) {2};
\node[gatev,vtx] at (3.000,-4.950) {};
\node[idx] at (3.000,-5.541) {3};
\node[idx] at (3.000,-4.359) {GS};
\node[storage,vtx] at (4.000,-4.950) {};
\node[idx] at (4.000,-5.541) {4};
\node[storage,vtx] at (5.000,-4.950) {};
\node[idx] at (5.000,-5.541) {5};
\node[storage,vtx] at (6.000,-4.950) {};
\node[idx] at (6.000,-5.541) {6};
\node[qb] at ($(2.000,-4.950)+(0mm,0)$) {$q_{2}$};
\node[qg] at ($(4.000,-4.950)+(0mm,0)$) {$q_{0}$};
\node[qg] at ($(5.000,-4.950)+(0mm,0)$) {$q_{1}$};
\node[op] at (-0.550,-6.600) {(5)~two translation\\timesteps};
\draw[ed] (0.000,-6.600) -- (1.000,-6.600);
\draw[ed] (1.000,-6.600) -- (2.000,-6.600);
\draw[ed] (2.000,-6.600) -- (3.000,-6.600);
\draw[ed] (3.000,-6.600) -- (4.000,-6.600);
\draw[ed] (4.000,-6.600) -- (5.000,-6.600);
\draw[ed] (5.000,-6.600) -- (6.000,-6.600);
\node[storage,vtx] at (0.000,-6.600) {};
\node[idx] at (0.000,-7.191) {0};
\node[storage,vtx] at (1.000,-6.600) {};
\node[idx] at (1.000,-7.191) {1};
\node[storage,vtx] at (2.000,-6.600) {};
\node[idx] at (2.000,-7.191) {2};
\node[gatev,vtx] at (3.000,-6.600) {};
\node[idx] at (3.000,-7.191) {3};
\node[idx] at (3.000,-6.009) {GS};
\node[storage,vtx] at (4.000,-6.600) {};
\node[idx] at (4.000,-7.191) {4};
\node[storage,vtx] at (5.000,-6.600) {};
\node[idx] at (5.000,-7.191) {5};
\node[storage,vtx] at (6.000,-6.600) {};
\node[idx] at (6.000,-7.191) {6};
\node[qb] at ($(1.000,-6.600)+(0mm,0)$) {$q_{2}$};
\node[qg] at ($(2.000,-6.600)+(0mm,0)$) {$q_{0}$};
\node[qg] at ($(4.000,-6.600)+(0mm,0)$) {$q_{1}$};
\node[op] at (-0.550,-8.250) {(6)~\textsc{Merge},\\gate $g(q_0,q_1)$};
\draw[ed] (0.000,-8.250) -- (1.000,-8.250);
\draw[ed] (1.000,-8.250) -- (2.000,-8.250);
\draw[ed] (2.000,-8.250) -- (3.000,-8.250);
\draw[ed] (3.000,-8.250) -- (4.000,-8.250);
\draw[ed] (4.000,-8.250) -- (5.000,-8.250);
\draw[ed] (5.000,-8.250) -- (6.000,-8.250);
\node[storage,vtx] at (0.000,-8.250) {};
\node[idx] at (0.000,-8.841) {0};
\node[storage,vtx] at (1.000,-8.250) {};
\node[idx] at (1.000,-8.841) {1};
\node[storage,vtx] at (2.000,-8.250) {};
\node[idx] at (2.000,-8.841) {2};
\node[gatev,vtx] at (3.000,-8.250) {};
\node[idx] at (3.000,-8.841) {3};
\node[idx] at (3.000,-7.659) {GS};
\node[storage,vtx] at (4.000,-8.250) {};
\node[idx] at (4.000,-8.841) {4};
\node[storage,vtx] at (5.000,-8.250) {};
\node[idx] at (5.000,-8.841) {5};
\node[storage,vtx] at (6.000,-8.250) {};
\node[idx] at (6.000,-8.841) {6};
\node[qb] at ($(1.000,-8.250)+(0mm,0)$) {$q_{2}$};
\node[qg] at ($(3.000,-8.250)+(-1.95mm,0)$) {$q_{0}$};
\node[qg] at ($(3.000,-8.250)+(1.95mm,0)$) {$q_{1}$};
\end{tikzpicture}}
\caption{Routing of the qubits of a two-qubit gate $g(q_0,q_1)$ on the linear architecture, with the gate vertex 3 (GS, red) at the center. Both gate qubits are drawn in red, all others in blue. Each of the states (1) to (6) shows the architecture after the operation named on its left.}
\label{fig:routing-linear}
\end{figure}

\begin{example}[Routing]
In \autoref{fig:routing-linear}, the intermediate $q_2$ lies between the endpoints $q_0$ and $q_1$. One \textsc{Merge}-\textsc{Swap}-\textsc{Separate} cycle at the gate vertex exchanges the positions of $q_0$ and $q_2$. Translations then bring $q_0$ and $q_1$ onto the two neighbors of the gate vertex. A final \textsc{Merge} combines them on the gate vertex, where the gate executes.
\end{example}

Once the gate has run, the prompt asks the compiler to consider a one-step \emph{post-execution lookahead} \textsc{Swap} at the gate vertex. While the chain order is irrelevant for gate execution, it influences the routing cost of the next gate touching either qubit. For each of the two qubits, the prompt defines its \emph{preferred side} of the gate vertex as the side that currently holds its next two-qubit-gate partner, with no preferred side when there is no such partner. The compiler inserts a \textsc{Swap} when the resulting flip places more qubits on their preferred side than the no-\textsc{Swap} arrangement does.

The \emph{post-processing} exists because the routing emits a plain sequential stream of operations. The hand-crafted compilers \cite{Wagner22, KreppelMWHPSB24} determine directly during routing which operations run in parallel within each translation phase between two non-translation operations. Splitting routing and grouping into two steps keeps the prompt simpler for the LLM, and once the full stream is known, unnecessary operations can be eliminated as well. Therefore, the post-processing applies two reduction passes followed by a grouping pass. Roundtrip elimination removes a \textsc{Translate} $A\!\to\!B$ followed by $B\!\to\!A$ when no operation between them touches $A$ or $B$. Consecutive cancellation likewise removes a \textsc{Separate} and a \textsc{Merge} in either order and a double \textsc{Swap} at the gate vertex. These two passes are alternated until neither removes any further operation, at which point a final grouping pass assembles the operations into timesteps. The prompt fixes only which operations may share a timestep, not how the grouper fills them. Independent operations may share one, and translations may also chain: a \textsc{Translate} $A\!\to\!B$ and a \textsc{Translate} $B\!\to\!C$ fit in one timestep, since $B$ is vacated in the same step it is filled. In effect, this combines independent and chained \textsc{Translate} moves into one translation timestep, while non-translation operations each occupy their own timestep, since any two of them share the single gate vertex.

\begin{figure}[t]
\centering
\resizebox{\columnwidth}{!}{%
\begin{tikzpicture}[
  op/.style={font=\small,anchor=west,inner sep=1.5pt},
  gone/.style={op,text=black!45},
  ts/.style={draw=expline,fill=expfill,rounded corners=2pt,font=\small,align=left,anchor=west,
             inner sep=3pt,text width=29mm,minimum height=5.5mm},
  link/.style={-{Stealth[length=1.8mm]},draw=expline,semithick},
  head/.style={font=\small\bfseries,anchor=west}]
\node[head] at (0,0.75) {emitted operations};
\foreach \i/\s/\t in {
    1/op/{\textsc{Merge} at 3: $q_0$, $q_2$},
    2/op/{\textsc{Swap} at 3: $q_0$, $q_2$},
    3/gone/{\textsc{Translate} $q_1$: $5\!\to\!6$},
    4/gone/{\textsc{Translate} $q_1$: $6\!\to\!5$},
    5/op/{\textsc{Separate} at 3: $q_2\!\to\!2$, $q_0\!\to\!4$},
    6/op/{\textsc{Translate} $q_2$: $2\!\to\!1$},
    7/op/{\textsc{Translate} $q_0$: $4\!\to\!3$},
    8/op/{\textsc{Translate} $q_0$: $3\!\to\!2$},
    9/op/{\textsc{Translate} $q_1$: $5\!\to\!4$},
    10/op/{\textsc{Merge} at 3: $q_0$, $q_1$}}
  \node[\s] (o\i) at (0,-0.62*\i+0.62) {(\i)~\t};
\draw[black!45,thick] (o3.west) -- (o3.east);
\draw[black!45,thick] (o4.west) -- (o4.east);
\draw[decorate,decoration={brace,amplitude=3pt,mirror},black!60] ([xshift=-1mm]o3.north west) -- ([xshift=-1mm]o4.south west)
  node[midway,left=2pt,font=\small,text=black!60,align=right] {roundtrip\\removed};
\node[head] at (6.6,0.75) {timesteps};
\node[ts] (t1) at (6.6,0) {1: \textsc{Merge}};
\node[ts] (t2) at (6.6,-0.62) {2: \textsc{Swap}};
\node[ts] (t3) at (6.6,-2.48) {3: \textsc{Separate}};
\node[ts,minimum height=10mm] (t4) at (6.6,-3.41) {4: translations\\\phantom{4: }(6), (7)};
\node[ts,minimum height=10mm] (t5) at (6.6,-4.65) {5: translations\\\phantom{5: }(8), (9)};
\node[ts] (t6) at (6.6,-5.58) {6: \textsc{Merge}};
\foreach \o/\t in {1/1, 2/2, 5/3, 6/4, 7/4, 8/5, 9/5, 10/6}
  \draw[link] (o\o.east) to[out=0,in=180] (t\t.west);
\end{tikzpicture}}
\caption{Post-processing of the operations of \autoref{fig:routing-linear}. The left column lists the operations in the order the routing emits them, the right column the timesteps into which the grouping assembles them. The struck-through roundtrip is removed before the grouping.}
\label{fig:postprocessing}
\end{figure}

\begin{example}[Post-processing]
The left column of \autoref{fig:postprocessing} lists the operations that the routing of \autoref{fig:routing-linear} emits. Besides the ones shown in \autoref{fig:routing-linear}, they include a roundtrip of $q_1$, the translation (3) and its reversal (4). Roundtrip elimination removes both, since no operation between them touches vertices 5 or 6. Afterward, the grouping fills the timesteps in program order. Translation (8) cannot join timestep 4, in which $q_0$ already moves with (7), and opens timestep 5, which (9) joins. In total, the eight remaining operations take six timesteps. These timesteps correspond to the states (2) to (6) of \autoref{fig:routing-linear}, with state (5) spanning the two translation timesteps 4 and 5.
\end{example}

The prompt closes with the acceptance-test matrix of eight test circuits listed in \autoref{tab:linear-tests}. Three carry tight bounds. The first is a working example that must stay within stated operation and timestep bounds, the second a same-pair circuit with all gates on one qubit pair that must produce zero shuttling, and the third a QAOA circuit with chain-like qubit interactions that must produce zero swaps. Each of the remaining five, of varying qubit and gate counts, only needs to execute successfully, with a compile-time guard on one and a swap-count check on another.

\begin{table}[t]
	\centering
	\caption{Acceptance-test matrix for the linear-architecture compiler. Each architecture is a linear path with the listed number of vertices and the gate vertex at its center, and $n$ is the circuit's qubit count. The generated compiler must pass all eight tests. T1--T3 additionally carry the tight schedule bounds shown. The rest need only run to a valid schedule, apart from the compile-time and swap conditions given for T4 and T5.}
	\label{tab:linear-tests}
	\setlength{\tabcolsep}{2pt}
	\scriptsize
	\begin{tabular*}{\columnwidth}{@{\extracolsep{\fill}}llrrrl@{}}
		\toprule
		Test & Circuit & $n$ & Gates & Vertices & Expected \\
		\midrule
		T1 & working example      & 3  & 3   & 7  & $\leq16$ ops, $\leq12$ steps \\
		T2 & \texttt{same\_pair}     & 2  & 10  & 5  & no shuttling \\
		T3 & \texttt{qaoa\_5}        & 5  & 35  & 11 & no swaps \\
		T4 & \texttt{rd32-v0\_66}    & 4  & 42  & 9  & runs, $<1$\,s \\
		T5 & \texttt{4mod5-v1\_23}   & 5  & 102 & 11 & runs, swaps $>0$ \\
		T6 & \texttt{4mod5-bdd\_287} & 7  & 106 & 15 & runs \\
		T7 & \texttt{alu-v2\_31}     & 5  & 640 & 11 & runs \\
		T8 & \texttt{cnt3-5\_179}    & 16 & 230 & 33 & runs \\
		\bottomrule
	\end{tabular*}
\end{table}

\subsection{Emitted compiler}
\label{sec:linear:first}

The emitted compiler structurally implements every block of the prompt, passes all acceptance tests on the first attempt with their bounds preserved, and produces a deterministic schedule. Its initial mapping makes two choices the prompt's placement rule leaves open, both to shorten routing. Within each pair, it turns the qubit that interacts more strongly with one neighboring pair toward that pair, and it centers the chain on the gate vertex. Under the prompt's first-gate exception, the qubits of a first two-qubit gate start on the gate vertex. The emitted compiler applies this exception only when these qubits also form the pair that starts the chain, and it then extends the chain from the gate vertex to one side. Of the two readings of $\mathrm{total\_dag\_depth}$, it takes the number of layers, a choice that barely affects the weights of deep circuits.

In the routing, the emitted compiler deviates from the prompt only in the special case described above, where the endpoint already shares its vertex with its adjacent intermediate and the \textsc{Merge} can be skipped. Rather than take that shortcut, it isolates the two and runs a full \textsc{Merge}-\textsc{Swap}-\textsc{Separate} cycle, which reaches the same end state with more operations.

For the post-processing, the emitted compiler takes the simplest reading of the prompt. Its reduction passes cancel only directly adjacent pairs. Moreover, its grouper fills the timesteps strictly in program order, as in the example of \autoref{fig:postprocessing}, and closes a timestep at the first operation that does not fit.

\subsection{Optimized compiler}
\label{sec:linear:opt}

We initiate every optimization round, here and in the later compiler sections, with a follow-up prompt that asks the LLM to identify code-level improvements reducing timesteps, compile time, and memory use. Four optimizations change the emitted schedule. By far the largest is in the grouping. The optimized grouper opens a timestep with the first operation not yet placed and then scans the following operations within a window that grows with the number of vertices. It adds each one that fits into the timestep and conflicts with none of the operations it skips over, so a later operation may move ahead of earlier operations on other vertices. Then it closes the timestep and opens the next with the first operation still left. Thus, within the scanned window, every operation lands in the earliest timestep its dependencies allow. In the example of \autoref{fig:postprocessing}, the optimized grouper moves (9) ahead of (8) into timestep 4 as a chained translation, since (7) vacates the target vertex 4 of (9) in that timestep. There, the count stays at six timesteps, because (8) still needs a timestep of its own before the final \textsc{Merge}, but on longer operation streams such moves save timesteps.

The second optimization widens the two reduction passes. Beyond directly adjacent pairs, they also remove a roundtrip, a \textsc{Separate}-\textsc{Merge} pair, or a double \textsc{Swap} across a window of intervening operations that touch none of their vertices, as the prompt permits.

The third optimization is a \emph{mirror-mapping trial-compile} on the initial mapping. A placement and its left-right mirror can compile to different lengths even though the architecture is symmetric. The router's tie-breaks favor the lower-numbered vertices rather than following a reflection-invariant rule, so the two orientations are not routed identically. One often yields the shorter schedule, with no way to tell which without compiling. Therefore, the optimized compiler routes both the default mapping and its mirror image with the vertex order reversed and keeps the shorter schedule. This mirror trial is the only strategy the LLM layers on top of the required default, an addition of the kind the prompt invites without naming.

The fourth optimization adopts the prompt's dedicated routine for the special case where an endpoint shares a vertex with its adjacent intermediate, replacing the emitted compiler's isolate-then-cycle handling. The remaining changes are lower-level performance and memory optimizations that leave the emitted schedule unchanged.

\def\TabTotalsLinearLib{%
Specialized, emitted & 14,768k & 7,242k & 1,796k & 1,796k & 1,072k \\
Specialized, optimized & \textbf{12,475k} & \textbf{4,977k} & \textbf{1,783k} & \textbf{1,782k} & \textbf{1,071k} \\
General, emitted & 17,513k & 8,413k & 2,364k & 2,364k & 1,509k \\
General, optimized & 20,243k & 9,727k & 2,836k & 2,835k & 1,984k \\
Baseline \cite{Wagner22} & 16,737k & 8,574k & 2,056k & 2,056k & 1,189k \\}
\def\TabTotalsLinearFam{%
Specialized, emitted & 6k & 110k & 1,076k & 2,196k & 4,066k & 90k & 62k \\
Specialized, optimized & \textbf{4k} & \textbf{79k} & \textbf{806k} & \textbf{1,480k} & \textbf{2,734k} & \textbf{46k} & \textbf{39k} \\
General, emitted & 18k & 209k & 1,090k & 3,151k & 6,185k & 64k & 65k \\
General, optimized & 18k & 299k & 1,144k & 3,907k & 6,651k & 100k & 101k \\
Baseline \cite{Wagner22} & 6k & 267k & 930k & 2,144k & 4,544k & 58k & 67k \\}
\begin{table}[b]
	\centering
	\caption{Linear architecture, circuit library: timesteps, translation timesteps, and separation, merge, and swap operations summed over all 153 library circuits. The lowest value per column is in bold. All values are rounded to thousands.}
	\label{tab:linear-totals}
	\setlength{\tabcolsep}{1.5pt}
	\scriptsize
	\begin{tabular*}{\columnwidth}{@{\extracolsep{\fill}}lrrrrr@{}}
		\toprule
		Compiler & Timesteps & Translate & Separate & Merge & Swap \\
		\midrule
		\TabTotalsLinearLib
		\bottomrule
	\end{tabular*}
\end{table}

\begin{table*}[t]
	\centering
	\caption{Linear architecture, scalable families: timesteps summed over all circuits of each family. The families differ in their tested qubit counts: QAOA and QV range from 5 to 50 qubits in steps of 5, QFT likewise up to 45 qubits and then 49, XEB and XEB\_Sy cover the square numbers of qubits from 4 to 49, and QEC\_Z and QEC\_X only the three sizes 17, 31, and 49. Therefore, the family totals are directly comparable only within these groups. The lowest value per family is in bold. All values are rounded to thousands.}
	\label{tab:linear-large}
	\setlength{\tabcolsep}{4pt}
	\scriptsize
	\begin{tabular*}{\textwidth}{@{\extracolsep{\fill}}lrrrrrrr@{}}
		\toprule
		Compiler & QAOA & QFT & QV & XEB & XEB\_Sy & QEC\_Z & QEC\_X \\
		\midrule
		\TabTotalsLinearFam
		\bottomrule
	\end{tabular*}
\end{table*}

\subsection{Evaluation}
\label{sec:linear:eval}

\begin{figure}[t]
	\centering
	\includegraphics{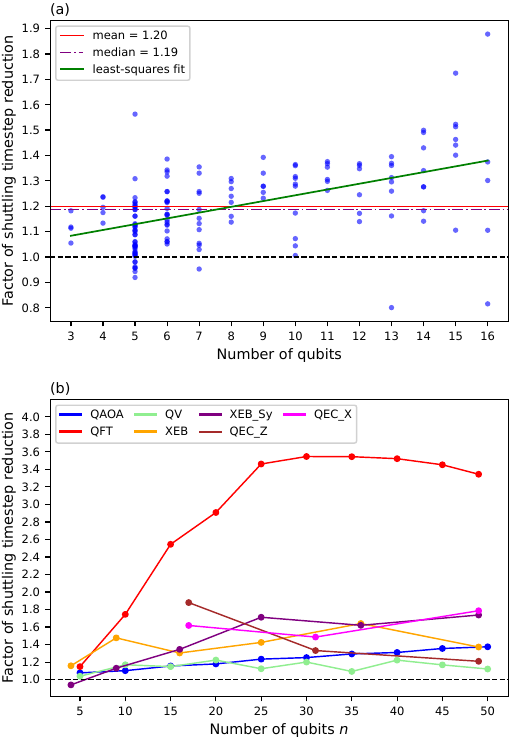}
	\caption{Timestep reduction factor of the optimized LLM-based compiler against the hand-crafted baseline \cite{Wagner22} on the linear architecture. (a) Circuit library, each circuit's factor against its qubit count, with the mean, the median, and a least-squares fit as red, purple, and green lines. (b) Scalable families, one circuit per qubit count and one line per family. Above the dashed line at 1, the LLM-based compiler needs fewer timesteps.}
	\label{fig:lin-final-hand}
\end{figure}

\begin{figure*}[tp]
	\centering
	\includegraphics[width=\textwidth]{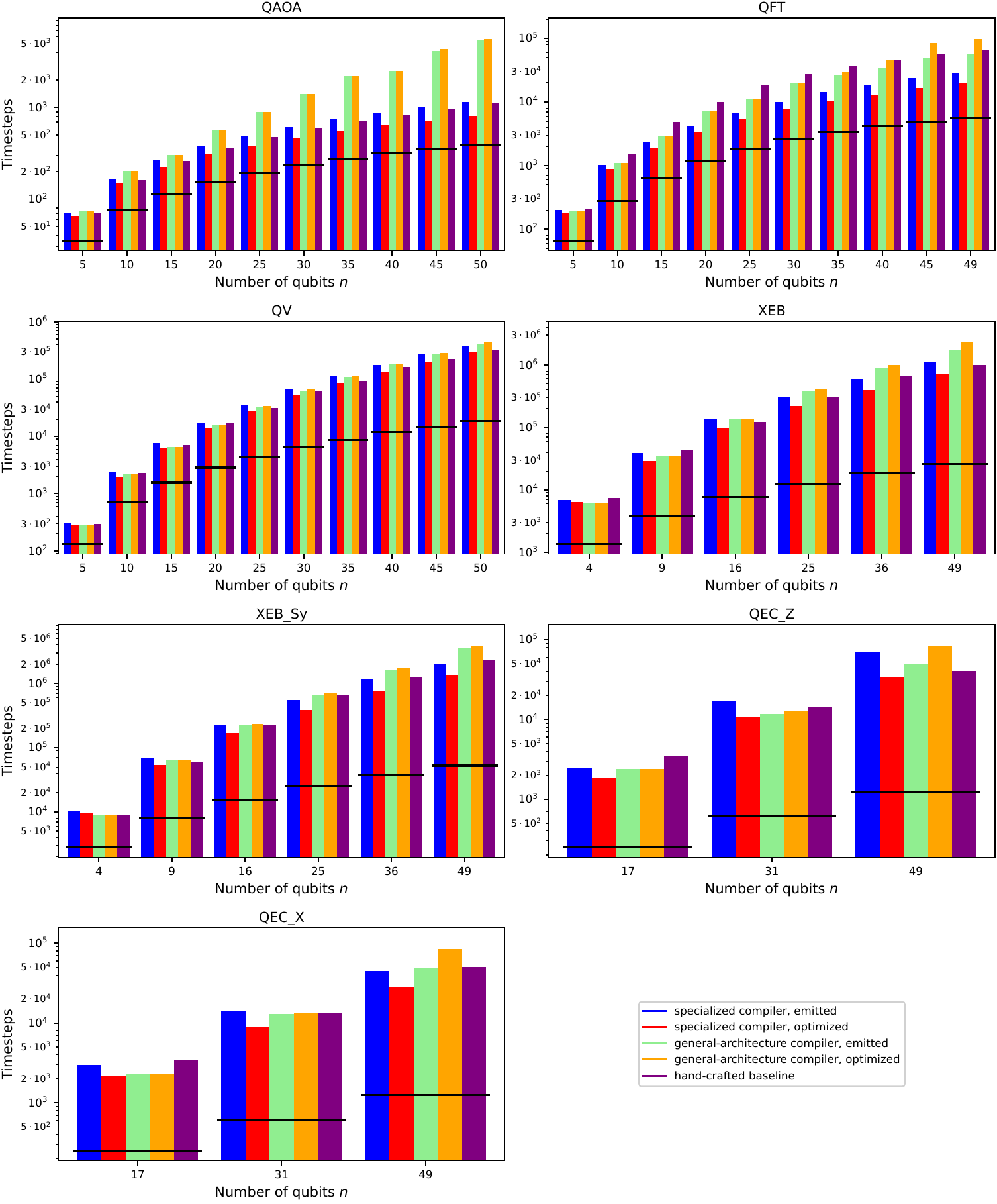}
	\caption{Absolute timesteps of the specialized and general-architecture compilers, each as emitted and optimized, and of the hand-crafted baseline \cite{Wagner22} on the linear architecture for the scalable families, one panel per family and one bar per circuit and compiler. The black line marks the number of gates, a lower bound on the number of timesteps, so the part of each bar above it is the shuttling overhead.}
	\label{fig:abs-linear}
\end{figure*}

\begin{figure}[t]
	\centering
	\includegraphics[width=\columnwidth]{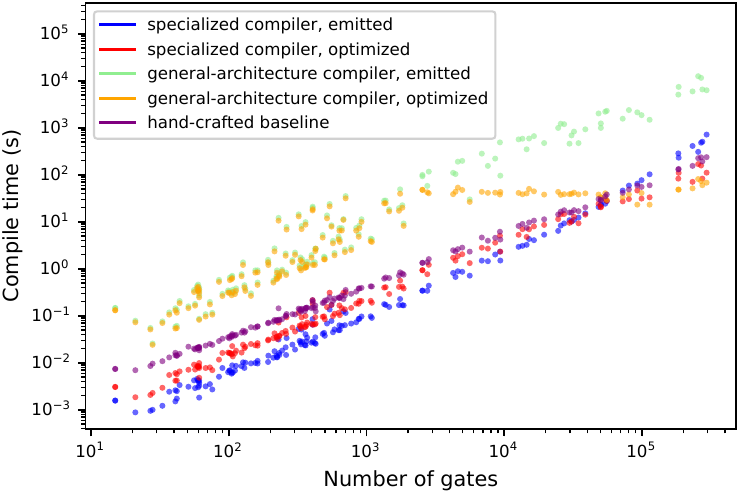}
	\caption{Compile time of the specialized and general-architecture compilers, each as emitted and optimized, and of the hand-crafted baseline \cite{Wagner22} on the linear architecture, one point per library circuit against its number of gates, since the library circuits of one qubit count differ widely in size, as \autoref{fig:circuit-stats} shows. Each value is the mean over ten runs.}
	\label{fig:compile-time-linear}
\end{figure}

The evaluation compares the emitted with the optimized compiler and then the optimized compiler with the hand-crafted baseline \cite{Wagner22}, first in timesteps, whose reduction factors \autoref{fig:lin-final-hand} summarizes, and then in compile time and memory. For a direct comparison with the general-architecture compiler of \autoref{sec:compiler:general}, \autoref{tab:linear-totals}, \autoref{tab:linear-large}, \autoref{fig:abs-linear}, and \autoref{fig:compile-time-linear} also include it, as emitted and as optimized.

\paragraph{Emitted versus optimized compiler.} The follow-up optimizations shorten the library circuits by a median factor of 1.16, with the reduction almost entirely in translations, as the suite totals in \autoref{tab:linear-totals} show. Most of it comes from the improved grouping of \autoref{sec:linear:opt}, and the largest part of the rest from the widened reduction passes, with smaller contributions from the mirror trial and the special-case routine. Per scalable family, whose totals \autoref{tab:linear-large} lists, the median emitted-to-optimized factor is 1.30, 1.28, and 1.28 for QAOA, QFT, and QV, 1.42 and 1.39 for XEB and XEB\_Sy, and 1.60 for QEC\_Z and QEC\_X. The mirror trial doubles the per-circuit routing work, so the optimized compiler usually needs more compile time than the emitted one, by a median factor of 2.25 on the circuit library and 1.97 to 2.83 per scalable family. Only on the largest library circuits do the lower-level performance optimizations outweigh it.

\paragraph{Optimized LLM-based compiler versus the hand-crafted baseline.} The optimized linear-architecture compiler outperforms the baseline on the vast majority of circuits, by a median factor of 1.19 on the circuit library and, per scalable family, 1.24 and 1.16 for QAOA and QV, 1.33 for QEC\_Z, 1.40 for XEB, 1.48 for XEB\_Sy, 1.62 for QEC\_X, and 3.40 for the all-to-all QFT. Across the circuit library, the per-circuit reduction factors, plotted in \autoref{fig:lin-final-hand}(a), run from 0.80 to 1.88, with values below 1 on only 10 of the 153 circuits, and the least-squares fit rises with the qubit count. Most of the saving comes from translations, with smaller reductions in separations, merges, and swaps. This advantage comes almost entirely from the routing, as \autoref{sec:mapping} shows, and within it most likely from where the gate order is decided. Whereas the baseline follows the gate order determined by the circuit compiler, the generated compiler selects each next gate from the ready set during routing. It can thereby adapt the gate order to the current ion placement. The QFT, with its many long-range interactions, is where that predetermined ordering costs the baseline most. Its reduction factor, plotted per family in \autoref{fig:lin-final-hand}(b), climbs from 1.15 at 5 qubits to between 3.45 and 3.55 at 25 to 45 qubits, or up to 72\,\% fewer timesteps, and eases slightly to 3.34 at 49 qubits. QAOA rises more gently, QV and XEB stay at 1.04 to 1.64 regardless of size, and XEB\_Sy grows from 0.94 at its smallest instance to 1.74 at the largest qubit counts. On the two QEC families the factor lies between 1.21 and 1.88.

The number of gates is a lower bound on any schedule, and the absolute timesteps of each circuit of the scalable families in \autoref{fig:abs-linear} show how far the schedules stay above it. Each family reaches its highest ratio of timesteps to gates at its largest size. For the optimized compiler, this ratio is 2.05 on QAOA, 3.51 on QFT, 15.62 on QV, and 22.48 to 28.05 on the QEC and XEB families, against 2.82, 11.75, 17.51, and 32.96 to 44.59 for the baseline. Every gate needs both of its qubits on the single gate vertex, so all ions between them must first be shuttled away. Therefore, the ratios are high on QV, XEB, and QEC, where consecutive gates mostly act on different qubit pairs whose two qubits lie far apart on the chain, and low on the chain-structured QAOA.

\paragraph{Compile time and memory.} As \autoref{fig:compile-time-linear} shows, the optimized compiler is faster than the baseline on all but two library circuits and needs about half the baseline's time at the median. On the largest library circuit, with about 294,000 gates, it takes about 110\,s against the baseline's 240\,s. A similar ratio holds for QAOA and QFT. On the larger QV, XEB, XEB\_Sy, and QEC circuits, the optimized compiler is several times slower, with about 1,950\,s against 270\,s for XEB\_Sy and 49\,s against 4\,s for QEC\_Z at 49 qubits, as the compile times of the scalable families in \autoref{fig:compile-time-linear-fam} in \appref{app:figures} show. The compile time varies little across the ten runs of a compilation, with a median standard deviation of 2.1\,\% of the mean for the LLM-based compilers and 1.1\,\% for the baseline.

Memory is not a limiting factor, as the peak memory of every circuit in \autoref{fig:memory}(b) in \appref{app:figures} shows. Both the emitted and the optimized compiler need less memory than the baseline on every library circuit, on the largest one about 0.9\,GB for the optimized compiler against the baseline's 1.8\,GB. Among the scalable families, the optimized compiler needs more only on QV from 40 qubits on and on the XEB families from 25 qubits on, at most 2.7\,GB on XEB\_Sy with 49 qubits against the baseline's 1.1\,GB.

\section{Branched shuttling architecture with junctions}
\label{sec:compiler:branched}

\subsection{Specification}
\label{sec:branched:spec}

The prompt for the branched-architecture compiler, which appears in \appref{app:branched-prompt}, provides the LLM with the optimized linear-architecture compiler as a starting point and specifies what must change for a tree-structured architecture with junctions, as shown in \autoref{fig:pipeline}. This architecture consists of a single \emph{main axis}, a linear chain of vertices indexed contiguously as in the linear architecture. A \emph{junction} is a main-axis vertex of degree $\geq 3$. In addition to its two axis neighbors, it is the root of one or more \emph{stacks}, each a linear chain of qubit vertices, called \emph{slots}, extending away from the axis. Every stack is accessed only through its junction-adjacent end, the first slot, and behaves as a last-in-first-out buffer. Each gate vertex lies on the main axis and has two non-junction neighbors, so a \textsc{Separate} or \textsc{Merge} there never involves a junction.

Physically, a junction is formed by the intersection of electrodes and lets qubits pass between different linear trap sections. Qubits can neither rest in a junction \cite{BlakestadODWBLW11} nor reverse their direction inside it, since the electric fields there are designed for passing through rather than for oscillatory motion \cite{WrightAFVDHPLDKSH13, MokhberiSW17}. For this reason, the prompt imposes three additional \emph{junction rules}. The \emph{in-transit rule} allows a junction to hold qubits only between two consecutive translations of the same chain, so it is necessarily empty whenever a non-translation operation runs. The \emph{no-roundtrip rule} requires a chain that enters a junction via one edge to leave via a different one, which obliges the routing to plan ahead. Together, these two rules imply the \emph{exit-availability rule}: Before a chain is translated onto a junction, at least one neighbor other than the chain's entry neighbor must be empty, or be made empty by a chained translation in the same timestep. Only then can the chain move off the junction on the next step.

The qubits in a stack shift as a unit. A push translates the chain at the junction into the first slot, while every already-occupied slot shifts one slot deeper in the same timestep. A pop translates the chain at the first slot out through the junction, while every occupied deeper slot shifts one slot toward the first slot. Both push and pop fit in a single timestep regardless of stack height, because the cascading shift of every occupied slot is a chained translation, permitted under the translation grouping rule of the linear architecture.

\begin{figure}[t]
\centering
\resizebox{\columnwidth}{!}{\begin{tikzpicture}[x=11.5mm,y=11.5mm,
  vtx/.style={minimum size=8.0mm,semithick},
  qb/.style={circle,draw=qubitline,fill=qubitcol,text=white,inner sep=0pt,minimum size=3.7mm,font={\fontsize{6.5}{6.5}\selectfont}},
  qg/.style={qb,draw=qgateline,fill=qgatecol},
  ed/.style={draw=black,thick},
  idx/.style={font=\small},
  op/.style={font=\small,anchor=east,align=right}]
\node[op] at (-1.350,0.000) {(1)~pass the junction,\\no roundtrip};
\draw[ed,densely dashed] (0,0.0) -- (-0.6,0.0);
\draw[ed,densely dashed] (4.6,0.0) -- (5.199999999999999,0.0);
\draw[-{Stealth[length=2.2mm]},thick] (0.204,0.354) to[bend left=40] (2.096,0.354);
\draw[-{Stealth[length=2.2mm]},thick] (2.504,0.354) to[bend left=40] (4.396,0.354);
\draw[-{Stealth[length=2.2mm]},thick,draw=gateline,densely dashed] (2.096,-0.354) to[bend left=40] node[midway,text=gateline,font=\bfseries,fill=white,inner sep=0.2pt] {$\times$} (0.204,-0.354);
\draw[ed] (0.000,0.000) -- (2.300,0.000);
\draw[ed] (2.300,0.000) -- (4.600,0.000);
\draw[ed] (2.300,0.000) -- (2.300,1.000);
\draw[ed] (2.300,1.000) -- (2.300,2.000);
\node[storage,vtx] at (0.000,0.000) {};
\node[idx] at (0.000,-0.591) {0};
\node[junc,vtx] at (2.300,0.000) {};
\node[idx] at (2.300,-0.591) {1};
\node[storage,vtx] at (4.600,0.000) {};
\node[idx] at (4.600,-0.591) {2};
\node[storage,vtx] at (2.300,1.000) {};
\node[idx] at (2.891,1.000) {3};
\node[storage,vtx] at (2.300,2.000) {};
\node[idx] at (2.891,2.000) {4};
\node[qg] at ($(0.000,0.000)+(0mm,0)$) {$q_{0}$};
\node[op] at (-1.350,-3.450) {(2)~no free exit,\\no entry};
\draw[ed,densely dashed] (0,-3.45) -- (-0.6,-3.45);
\draw[ed,densely dashed] (4.6,-3.45) -- (5.199999999999999,-3.45);
\draw[-{Stealth[length=2.2mm]},thick,draw=gateline,densely dashed] (0.204,-3.096) to[bend left=40] node[midway,text=gateline,font=\bfseries,fill=white,inner sep=0.2pt] {$\times$} (2.096,-3.096);
\draw[ed] (0.000,-3.450) -- (2.300,-3.450);
\draw[ed] (2.300,-3.450) -- (4.600,-3.450);
\draw[ed] (2.300,-3.450) -- (2.300,-2.450);
\draw[ed] (2.300,-2.450) -- (2.300,-1.450);
\node[storage,vtx] at (0.000,-3.450) {};
\node[idx] at (0.000,-4.041) {0};
\node[junc,vtx] at (2.300,-3.450) {};
\node[idx] at (2.300,-4.041) {1};
\node[storage,vtx] at (4.600,-3.450) {};
\node[idx] at (4.600,-4.041) {2};
\node[storage,vtx] at (2.300,-2.450) {};
\node[idx] at (2.891,-2.450) {3};
\node[storage,vtx] at (2.300,-1.450) {};
\node[idx] at (2.891,-1.450) {4};
\node[qg] at ($(0.000,-3.450)+(0mm,0)$) {$q_{0}$};
\node[qb] at ($(4.600,-3.450)+(0mm,0)$) {$q_{1}$};
\node[qb] at ($(2.300,-2.450)+(0mm,0)$) {$q_{2}$};
\node[qb] at ($(2.300,-1.450)+(0mm,0)$) {$q_{3}$};
\node[op] at (-1.350,-6.900) {(3)~in transit,\\push $q_0$};
\draw[ed,densely dashed] (0,-6.9) -- (-0.6,-6.9);
\draw[ed,densely dashed] (4.6,-6.9) -- (5.199999999999999,-6.9);
\draw[-{Stealth[length=2.2mm]},thick] (1.946,-6.696) to[bend left=40] (1.946,-6.104);
\draw[-{Stealth[length=2.2mm]},thick] (1.946,-5.696) to[bend left=40] (1.946,-5.104);
\draw[ed] (0.000,-6.900) -- (2.300,-6.900);
\draw[ed] (2.300,-6.900) -- (4.600,-6.900);
\draw[ed] (2.300,-6.900) -- (2.300,-5.900);
\draw[ed] (2.300,-5.900) -- (2.300,-4.900);
\node[storage,vtx] at (0.000,-6.900) {};
\node[idx] at (0.000,-7.491) {0};
\node[junc,vtx] at (2.300,-6.900) {};
\node[idx] at (2.300,-7.491) {1};
\node[storage,vtx] at (4.600,-6.900) {};
\node[idx] at (4.600,-7.491) {2};
\node[storage,vtx] at (2.300,-5.900) {};
\node[idx] at (2.891,-5.900) {3};
\node[storage,vtx] at (2.300,-4.900) {};
\node[idx] at (2.891,-4.900) {4};
\node[qg] at ($(2.300,-6.900)+(0mm,0)$) {$q_{0}$};
\node[qb] at ($(2.300,-5.900)+(0mm,0)$) {$q_{1}$};
\end{tikzpicture}}
\caption{Junction rules and a push into a stack at the junction 1 (green), whose neighbors are the vertices 0 and 2 and the first slot 3 of a stack of height 2 (vertices 3 and 4). Dashed lines continue the architecture. The moving qubit $q_0$ is drawn in red, all others in blue. Black arrows are allowed translations, red crossed arrows forbidden ones.}
\label{fig:junction-rules}
\end{figure}

\begin{example}[Junction rules and stack push]
The three states of \autoref{fig:junction-rules} illustrate the junction rules and a push into a stack. In state (1), $q_0$ crosses the junction in consecutive translations and leaves it by a different edge than the one it entered by, whereas returning directly to vertex 0 would violate the no-roundtrip rule. State (2) differs in that vertex 2 and both stack slots are occupied, so $q_0$ would find no exit and may not enter the junction. Finally, state (3) shows the push: $q_0$, in transit on the junction, moves into the first slot of the stack while $q_1$ shifts one slot deeper, and both moves form a single timestep.
\end{example}

Building on these rules, the prompt describes three changes to the compiler. The initial mapping reuses the matching and chain extraction of the linear procedure but places the resulting pairs differently, in three ways. First, qubits are placed only on non-junction vertices, which the prompt adds as a fourth placement rule. Second, instead of forming a contiguous run centered on the gate vertex, the chain is assigned to those vertices in order of increasing graph distance to the nearest gate vertex, filling outward along the main axis in both directions and into the stacks. Because chain extraction starts with the most strongly interacting pair, this pair still lies closest to the gate vertex. Third, the intra-pair orientation generalizes, since the pairs no longer lie in a spatial chain with a definite neighbor on each side. Within each pair, the qubit with the larger total interaction weight to all qubits outside its pair is placed at the chain position that faces the rest of the architecture, where most of the other pairs lie, keeping it close to the qubits it interacts with most.

\begin{example}[Initial mapping]
In \autoref{fig:graph-construction}(e), which shows the branched architecture with $h=d=1$, the pair $(q_1,q_3)$ of the first gate in the circuit of \autoref{fig:graph-construction}(a) again occupies the gate vertex 4, and both of its neighbors stay empty. The nearest remaining non-junction vertices are then the stack slots at the junctions 2 and 6, three edges from the gate vertex and closer to it than the ends of the main axis. Of these two equally close slots, the one with the lower vertex index is taken, so $(q_0,q_2)$ is placed on the stack of junction 2. Finally, within the pair on the gate vertex, $q_3$ faces junction 2 and thus the pair $(q_0,q_2)$, since its interaction weight with qubits outside its pair, $\frac{1}{2}$, exceeds $\frac{1}{4}$ for $q_1$.
\end{example}

The routing again brings each gate's qubits together at the gate vertex, now on a tree with junctions and stacks. Single-qubit routing follows the linear routine and differs only in how a chain moves. Every walk now uses a \emph{move} primitive that follows the unique tree path between source and destination, evacuating any obstacle on its path.

For a two-qubit gate, the intermediates are the qubits sitting on the tree path between the two endpoints' current vertices and their target gate-vertex neighbors. These intermediates are cleared iteratively. In each iteration, the compiler picks whichever of the \emph{swap}, \emph{stack}, and \emph{park-the-endpoint} methods has the lowest estimated timestep cost. Since the swap method always applies, it also serves as the fallback when the other two methods cannot be used, with the \emph{rotor-conveyor} as the guaranteed terminal routine.

The swap method is the \textsc{Merge}-\textsc{Swap}-\textsc{Separate} cycle of the linear architecture, with the endpoint and its adjacent intermediate now routed to the gate vertex's two neighbors along their tree paths rather than along a line. It always removes exactly one intermediate, since a \textsc{Swap} reverses only a single pair.

The stack method applies when a junction on the path has a stack that is not full. This method routes a nearby intermediate to that junction and pushes it onto the stack instead of displacing it past the gate vertex. Each push removes one intermediate from the path, or two when a co-located pair is pushed together.

The park-the-endpoint method parks one endpoint out of the way, on a non-full stack or at an empty end of the main axis. This frees the corridor the endpoint occupied, so the intermediates can be moved through it and off the path. The parked endpoint is then retrieved and delivered to its gate-vertex neighbor. Several intermediates can be cleared in one iteration this way.

The rotor-conveyor is used only when the other methods fail to bring the gate qubits together and is the most costly. It first carries every co-located pair in the architecture to the gate vertex and separates it there. This reshuffles much of the architecture but leaves one qubit per vertex, so that each later \textsc{Merge}-\textsc{Swap}-\textsc{Separate} cycle acts on exactly two qubits. The rotor-conveyor then moves one endpoint, the \emph{anchor}, to the gate-vertex neighbor on the far side from the other endpoint, and carries the qubits between the other endpoint and its target neighbor across the gate vertex one at a time. Each cycle brings the intermediate nearest the target neighbor onto that neighbor and merges it with the anchor at the gate vertex, where a \textsc{Swap} reverses the pair and a \textsc{Separate} sends the two back out with their positions exchanged, leaving the intermediate on the anchor's side and the anchor on the other endpoint's side. The intermediate is then moved off the anchor's neighbor and deep into the anchor's side of the architecture, and the anchor is translated back across the gate vertex to its own neighbor. This repeats until the other endpoint reaches its target neighbor and the gate executes.

In the branched architecture, the rotor-conveyor always succeeds. The gate vertex separates the tree into two sides, so an intermediate stashed on the anchor's side cannot return to the other endpoint's path, and each cycle reduces the number of intermediates by one.

\begin{figure}[t]
\centering
\resizebox{\columnwidth}{!}{\begin{tikzpicture}[x=11.5mm,y=11.5mm,
  vtx/.style={minimum size=8.0mm,semithick},
  qb/.style={circle,draw=qubitline,fill=qubitcol,text=white,inner sep=0pt,minimum size=3.7mm,font={\fontsize{6.5}{6.5}\selectfont}},
  qg/.style={qb,draw=qgateline,fill=qgatecol},
  ed/.style={draw=black,thick},
  idx/.style={font=\small},
  op/.style={font=\small,anchor=east,align=right}]
\node[op] at (-0.550,0.000) {(1)~initial state};
\draw[ed] (0.000,0.000) -- (1.000,0.000);
\draw[ed] (1.000,0.000) -- (2.000,0.000);
\draw[ed] (2.000,0.000) -- (3.000,0.000);
\draw[ed] (3.000,0.000) -- (4.000,0.000);
\draw[ed] (4.000,0.000) -- (5.000,0.000);
\draw[ed] (5.000,0.000) -- (6.000,0.000);
\draw[ed] (1.000,0.000) -- (1.000,1.000);
\draw[ed] (5.000,0.000) -- (5.000,1.000);
\node[storage,vtx] at (0.000,0.000) {};
\node[idx] at (0.000,-0.591) {0};
\node[junc,vtx] at (1.000,0.000) {};
\node[idx] at (1.000,-0.591) {1};
\node[storage,vtx] at (2.000,0.000) {};
\node[idx] at (2.000,-0.591) {2};
\node[gatev,vtx] at (3.000,0.000) {};
\node[idx] at (3.000,-0.591) {3};
\node[idx] at (3.000,0.591) {GS};
\node[storage,vtx] at (4.000,0.000) {};
\node[idx] at (4.000,-0.591) {4};
\node[junc,vtx] at (5.000,0.000) {};
\node[idx] at (5.000,-0.591) {5};
\node[storage,vtx] at (6.000,0.000) {};
\node[idx] at (6.000,-0.591) {6};
\node[storage,vtx] at (1.000,1.000) {};
\node[idx] at (1.591,1.000) {7};
\node[storage,vtx] at (5.000,1.000) {};
\node[idx] at (5.591,1.000) {8};
\node[qg] at ($(0.000,0.000)+(0mm,0)$) {$q_{0}$};
\node[qb] at ($(2.000,0.000)+(0mm,0)$) {$q_{2}$};
\node[qg] at ($(4.000,0.000)+(0mm,0)$) {$q_{1}$};
\node[op] at (-0.550,-2.750) {(2)~push $q_2$ onto\\the stack};
\draw[ed] (0.000,-2.750) -- (1.000,-2.750);
\draw[ed] (1.000,-2.750) -- (2.000,-2.750);
\draw[ed] (2.000,-2.750) -- (3.000,-2.750);
\draw[ed] (3.000,-2.750) -- (4.000,-2.750);
\draw[ed] (4.000,-2.750) -- (5.000,-2.750);
\draw[ed] (5.000,-2.750) -- (6.000,-2.750);
\draw[ed] (1.000,-2.750) -- (1.000,-1.750);
\draw[ed] (5.000,-2.750) -- (5.000,-1.750);
\node[storage,vtx] at (0.000,-2.750) {};
\node[idx] at (0.000,-3.341) {0};
\node[junc,vtx] at (1.000,-2.750) {};
\node[idx] at (1.000,-3.341) {1};
\node[storage,vtx] at (2.000,-2.750) {};
\node[idx] at (2.000,-3.341) {2};
\node[gatev,vtx] at (3.000,-2.750) {};
\node[idx] at (3.000,-3.341) {3};
\node[idx] at (3.000,-2.159) {GS};
\node[storage,vtx] at (4.000,-2.750) {};
\node[idx] at (4.000,-3.341) {4};
\node[junc,vtx] at (5.000,-2.750) {};
\node[idx] at (5.000,-3.341) {5};
\node[storage,vtx] at (6.000,-2.750) {};
\node[idx] at (6.000,-3.341) {6};
\node[storage,vtx] at (1.000,-1.750) {};
\node[idx] at (1.591,-1.750) {7};
\node[storage,vtx] at (5.000,-1.750) {};
\node[idx] at (5.591,-1.750) {8};
\node[qg] at ($(0.000,-2.750)+(0mm,0)$) {$q_{0}$};
\node[qb] at ($(1.000,-1.750)+(0mm,0)$) {$q_{2}$};
\node[qg] at ($(4.000,-2.750)+(0mm,0)$) {$q_{1}$};
\node[op] at (-0.550,-5.500) {(3)~move $q_0$ through\\the junction};
\draw[ed] (0.000,-5.500) -- (1.000,-5.500);
\draw[ed] (1.000,-5.500) -- (2.000,-5.500);
\draw[ed] (2.000,-5.500) -- (3.000,-5.500);
\draw[ed] (3.000,-5.500) -- (4.000,-5.500);
\draw[ed] (4.000,-5.500) -- (5.000,-5.500);
\draw[ed] (5.000,-5.500) -- (6.000,-5.500);
\draw[ed] (1.000,-5.500) -- (1.000,-4.500);
\draw[ed] (5.000,-5.500) -- (5.000,-4.500);
\node[storage,vtx] at (0.000,-5.500) {};
\node[idx] at (0.000,-6.091) {0};
\node[junc,vtx] at (1.000,-5.500) {};
\node[idx] at (1.000,-6.091) {1};
\node[storage,vtx] at (2.000,-5.500) {};
\node[idx] at (2.000,-6.091) {2};
\node[gatev,vtx] at (3.000,-5.500) {};
\node[idx] at (3.000,-6.091) {3};
\node[idx] at (3.000,-4.909) {GS};
\node[storage,vtx] at (4.000,-5.500) {};
\node[idx] at (4.000,-6.091) {4};
\node[junc,vtx] at (5.000,-5.500) {};
\node[idx] at (5.000,-6.091) {5};
\node[storage,vtx] at (6.000,-5.500) {};
\node[idx] at (6.000,-6.091) {6};
\node[storage,vtx] at (1.000,-4.500) {};
\node[idx] at (1.591,-4.500) {7};
\node[storage,vtx] at (5.000,-4.500) {};
\node[idx] at (5.591,-4.500) {8};
\node[qb] at ($(1.000,-4.500)+(0mm,0)$) {$q_{2}$};
\node[qg] at ($(2.000,-5.500)+(0mm,0)$) {$q_{0}$};
\node[qg] at ($(4.000,-5.500)+(0mm,0)$) {$q_{1}$};
\node[op] at (-0.550,-8.250) {(4)~\textsc{Merge},\\gate $g(q_0,q_1)$};
\draw[ed] (0.000,-8.250) -- (1.000,-8.250);
\draw[ed] (1.000,-8.250) -- (2.000,-8.250);
\draw[ed] (2.000,-8.250) -- (3.000,-8.250);
\draw[ed] (3.000,-8.250) -- (4.000,-8.250);
\draw[ed] (4.000,-8.250) -- (5.000,-8.250);
\draw[ed] (5.000,-8.250) -- (6.000,-8.250);
\draw[ed] (1.000,-8.250) -- (1.000,-7.250);
\draw[ed] (5.000,-8.250) -- (5.000,-7.250);
\node[storage,vtx] at (0.000,-8.250) {};
\node[idx] at (0.000,-8.841) {0};
\node[junc,vtx] at (1.000,-8.250) {};
\node[idx] at (1.000,-8.841) {1};
\node[storage,vtx] at (2.000,-8.250) {};
\node[idx] at (2.000,-8.841) {2};
\node[gatev,vtx] at (3.000,-8.250) {};
\node[idx] at (3.000,-8.841) {3};
\node[idx] at (3.000,-7.659) {GS};
\node[storage,vtx] at (4.000,-8.250) {};
\node[idx] at (4.000,-8.841) {4};
\node[junc,vtx] at (5.000,-8.250) {};
\node[idx] at (5.000,-8.841) {5};
\node[storage,vtx] at (6.000,-8.250) {};
\node[idx] at (6.000,-8.841) {6};
\node[storage,vtx] at (1.000,-7.250) {};
\node[idx] at (1.591,-7.250) {7};
\node[storage,vtx] at (5.000,-7.250) {};
\node[idx] at (5.591,-7.250) {8};
\node[qb] at ($(1.000,-7.250)+(0mm,0)$) {$q_{2}$};
\node[qg] at ($(3.000,-8.250)+(-1.95mm,0)$) {$q_{0}$};
\node[qg] at ($(3.000,-8.250)+(1.95mm,0)$) {$q_{1}$};
\end{tikzpicture}}
\caption{Stack method for a two-qubit gate $g(q_0,q_1)$ on a mirror-symmetric branched architecture with the gate vertex 3 (GS, red) at the center. Next to its neighbors lie the junctions 1 and 5 (green), each carrying a stack of height 1 (vertices 7 and 8). Both gate qubits are drawn in red and the other qubit $q_2$ in blue. Each of the states (1) to (4) shows the architecture after the operation named on its left.}
\label{fig:routing-branched}
\end{figure}

\begin{example}[Stack method]
In \autoref{fig:routing-branched}, the intermediate $q_2$ occupies vertex 2, the target of $q_0$. The stack method pushes $q_2$ through junction 1 onto the stack slot 7 and so frees the corridor. Then $q_0$ crosses the junction to vertex 2, and a final \textsc{Merge} brings $q_0$ and $q_1$ together on the gate vertex, where the gate executes. Unlike on the linear architecture of \autoref{fig:routing-linear}, where $q_2$ has to pass through the gate vertex, clearing $q_2$ needs no \textsc{Merge}, \textsc{Swap}, or \textsc{Separate}, so apart from translations, the final \textsc{Merge} is the only shuttling operation. Both $q_2$ and $q_0$ cross junction 1 in transit only, and each leaves it by a different edge than the one it entered by. Since $q_0$ already sits at an end of the main axis, it counts as parked, and the park-the-endpoint method would clear $q_2$ onto slot 7 in the same way.
\end{example}

\begin{figure}[t]
\centering
\resizebox{\columnwidth}{!}{\begin{tikzpicture}[x=11.5mm,y=11.5mm,
  vtx/.style={minimum size=8.0mm,semithick},
  qb/.style={circle,draw=qubitline,fill=qubitcol,text=white,inner sep=0pt,minimum size=3.7mm,font={\fontsize{6.5}{6.5}\selectfont}},
  qg/.style={qb,draw=qgateline,fill=qgatecol},
  ed/.style={draw=black,thick},
  idx/.style={font=\small},
  op/.style={font=\small,anchor=east,align=right}]
\node[op] at (-0.550,0.000) {(1)~initial state};
\draw[ed] (0.000,0.000) -- (1.000,0.000);
\draw[ed] (1.000,0.000) -- (2.000,0.000);
\draw[ed] (2.000,0.000) -- (3.000,0.000);
\draw[ed] (3.000,0.000) -- (4.000,0.000);
\draw[ed] (4.000,0.000) -- (5.000,0.000);
\draw[ed] (5.000,0.000) -- (6.000,0.000);
\draw[ed] (6.000,0.000) -- (7.000,0.000);
\draw[ed] (7.000,0.000) -- (8.000,0.000);
\draw[ed] (2.000,0.000) -- (2.000,1.000);
\draw[ed] (6.000,0.000) -- (6.000,1.000);
\node[storage,vtx] at (0.000,0.000) {};
\node[idx] at (0.000,-0.591) {0};
\node[storage,vtx] at (1.000,0.000) {};
\node[idx] at (1.000,-0.591) {1};
\node[junc,vtx] at (2.000,0.000) {};
\node[idx] at (2.000,-0.591) {2};
\node[storage,vtx] at (3.000,0.000) {};
\node[idx] at (3.000,-0.591) {3};
\node[gatev,vtx] at (4.000,0.000) {};
\node[idx] at (4.000,-0.591) {4};
\node[idx] at (4.000,0.591) {GS};
\node[storage,vtx] at (5.000,0.000) {};
\node[idx] at (5.000,-0.591) {5};
\node[junc,vtx] at (6.000,0.000) {};
\node[idx] at (6.000,-0.591) {6};
\node[storage,vtx] at (7.000,0.000) {};
\node[idx] at (7.000,-0.591) {7};
\node[storage,vtx] at (8.000,0.000) {};
\node[idx] at (8.000,-0.591) {8};
\node[storage,vtx] at (2.000,1.000) {};
\node[idx] at (2.591,1.000) {9};
\node[storage,vtx] at (6.000,1.000) {};
\node[idx] at (6.591,1.000) {10};
\node[qg] at ($(3.000,0.000)+(0mm,0)$) {$q_{0}$};
\node[qb] at ($(5.000,0.000)+(0mm,0)$) {$q_{2}$};
\node[qg] at ($(7.000,0.000)+(0mm,0)$) {$q_{1}$};
\node[qb] at ($(6.000,1.000)+(0mm,0)$) {$q_{3}$};
\node[op] at (-0.550,-2.750) {(2)~\textsc{Merge}, \textsc{Swap},\\\textsc{Separate}};
\draw[ed] (0.000,-2.750) -- (1.000,-2.750);
\draw[ed] (1.000,-2.750) -- (2.000,-2.750);
\draw[ed] (2.000,-2.750) -- (3.000,-2.750);
\draw[ed] (3.000,-2.750) -- (4.000,-2.750);
\draw[ed] (4.000,-2.750) -- (5.000,-2.750);
\draw[ed] (5.000,-2.750) -- (6.000,-2.750);
\draw[ed] (6.000,-2.750) -- (7.000,-2.750);
\draw[ed] (7.000,-2.750) -- (8.000,-2.750);
\draw[ed] (2.000,-2.750) -- (2.000,-1.750);
\draw[ed] (6.000,-2.750) -- (6.000,-1.750);
\node[storage,vtx] at (0.000,-2.750) {};
\node[idx] at (0.000,-3.341) {0};
\node[storage,vtx] at (1.000,-2.750) {};
\node[idx] at (1.000,-3.341) {1};
\node[junc,vtx] at (2.000,-2.750) {};
\node[idx] at (2.000,-3.341) {2};
\node[storage,vtx] at (3.000,-2.750) {};
\node[idx] at (3.000,-3.341) {3};
\node[gatev,vtx] at (4.000,-2.750) {};
\node[idx] at (4.000,-3.341) {4};
\node[idx] at (4.000,-2.159) {GS};
\node[storage,vtx] at (5.000,-2.750) {};
\node[idx] at (5.000,-3.341) {5};
\node[junc,vtx] at (6.000,-2.750) {};
\node[idx] at (6.000,-3.341) {6};
\node[storage,vtx] at (7.000,-2.750) {};
\node[idx] at (7.000,-3.341) {7};
\node[storage,vtx] at (8.000,-2.750) {};
\node[idx] at (8.000,-3.341) {8};
\node[storage,vtx] at (2.000,-1.750) {};
\node[idx] at (2.591,-1.750) {9};
\node[storage,vtx] at (6.000,-1.750) {};
\node[idx] at (6.591,-1.750) {10};
\node[qb] at ($(3.000,-2.750)+(0mm,0)$) {$q_{2}$};
\node[qg] at ($(5.000,-2.750)+(0mm,0)$) {$q_{0}$};
\node[qg] at ($(7.000,-2.750)+(0mm,0)$) {$q_{1}$};
\node[qb] at ($(6.000,-1.750)+(0mm,0)$) {$q_{3}$};
\node[op] at (-0.550,-5.500) {(3)~stash $q_2$,\\move $q_0$ back};
\draw[ed] (0.000,-5.500) -- (1.000,-5.500);
\draw[ed] (1.000,-5.500) -- (2.000,-5.500);
\draw[ed] (2.000,-5.500) -- (3.000,-5.500);
\draw[ed] (3.000,-5.500) -- (4.000,-5.500);
\draw[ed] (4.000,-5.500) -- (5.000,-5.500);
\draw[ed] (5.000,-5.500) -- (6.000,-5.500);
\draw[ed] (6.000,-5.500) -- (7.000,-5.500);
\draw[ed] (7.000,-5.500) -- (8.000,-5.500);
\draw[ed] (2.000,-5.500) -- (2.000,-4.500);
\draw[ed] (6.000,-5.500) -- (6.000,-4.500);
\node[storage,vtx] at (0.000,-5.500) {};
\node[idx] at (0.000,-6.091) {0};
\node[storage,vtx] at (1.000,-5.500) {};
\node[idx] at (1.000,-6.091) {1};
\node[junc,vtx] at (2.000,-5.500) {};
\node[idx] at (2.000,-6.091) {2};
\node[storage,vtx] at (3.000,-5.500) {};
\node[idx] at (3.000,-6.091) {3};
\node[gatev,vtx] at (4.000,-5.500) {};
\node[idx] at (4.000,-6.091) {4};
\node[idx] at (4.000,-4.909) {GS};
\node[storage,vtx] at (5.000,-5.500) {};
\node[idx] at (5.000,-6.091) {5};
\node[junc,vtx] at (6.000,-5.500) {};
\node[idx] at (6.000,-6.091) {6};
\node[storage,vtx] at (7.000,-5.500) {};
\node[idx] at (7.000,-6.091) {7};
\node[storage,vtx] at (8.000,-5.500) {};
\node[idx] at (8.000,-6.091) {8};
\node[storage,vtx] at (2.000,-4.500) {};
\node[idx] at (2.591,-4.500) {9};
\node[storage,vtx] at (6.000,-4.500) {};
\node[idx] at (6.591,-4.500) {10};
\node[qb] at ($(1.000,-5.500)+(0mm,0)$) {$q_{2}$};
\node[qg] at ($(3.000,-5.500)+(0mm,0)$) {$q_{0}$};
\node[qg] at ($(7.000,-5.500)+(0mm,0)$) {$q_{1}$};
\node[qb] at ($(6.000,-4.500)+(0mm,0)$) {$q_{3}$};
\node[op] at (-0.550,-8.250) {(4)~move $q_1$ through\\the junction};
\draw[ed] (0.000,-8.250) -- (1.000,-8.250);
\draw[ed] (1.000,-8.250) -- (2.000,-8.250);
\draw[ed] (2.000,-8.250) -- (3.000,-8.250);
\draw[ed] (3.000,-8.250) -- (4.000,-8.250);
\draw[ed] (4.000,-8.250) -- (5.000,-8.250);
\draw[ed] (5.000,-8.250) -- (6.000,-8.250);
\draw[ed] (6.000,-8.250) -- (7.000,-8.250);
\draw[ed] (7.000,-8.250) -- (8.000,-8.250);
\draw[ed] (2.000,-8.250) -- (2.000,-7.250);
\draw[ed] (6.000,-8.250) -- (6.000,-7.250);
\node[storage,vtx] at (0.000,-8.250) {};
\node[idx] at (0.000,-8.841) {0};
\node[storage,vtx] at (1.000,-8.250) {};
\node[idx] at (1.000,-8.841) {1};
\node[junc,vtx] at (2.000,-8.250) {};
\node[idx] at (2.000,-8.841) {2};
\node[storage,vtx] at (3.000,-8.250) {};
\node[idx] at (3.000,-8.841) {3};
\node[gatev,vtx] at (4.000,-8.250) {};
\node[idx] at (4.000,-8.841) {4};
\node[idx] at (4.000,-7.659) {GS};
\node[storage,vtx] at (5.000,-8.250) {};
\node[idx] at (5.000,-8.841) {5};
\node[junc,vtx] at (6.000,-8.250) {};
\node[idx] at (6.000,-8.841) {6};
\node[storage,vtx] at (7.000,-8.250) {};
\node[idx] at (7.000,-8.841) {7};
\node[storage,vtx] at (8.000,-8.250) {};
\node[idx] at (8.000,-8.841) {8};
\node[storage,vtx] at (2.000,-7.250) {};
\node[idx] at (2.591,-7.250) {9};
\node[storage,vtx] at (6.000,-7.250) {};
\node[idx] at (6.591,-7.250) {10};
\node[qb] at ($(1.000,-8.250)+(0mm,0)$) {$q_{2}$};
\node[qg] at ($(3.000,-8.250)+(0mm,0)$) {$q_{0}$};
\node[qg] at ($(5.000,-8.250)+(0mm,0)$) {$q_{1}$};
\node[qb] at ($(6.000,-7.250)+(0mm,0)$) {$q_{3}$};
\node[op] at (-0.550,-11.000) {(5)~\textsc{Merge},\\gate $g(q_0,q_1)$};
\draw[ed] (0.000,-11.000) -- (1.000,-11.000);
\draw[ed] (1.000,-11.000) -- (2.000,-11.000);
\draw[ed] (2.000,-11.000) -- (3.000,-11.000);
\draw[ed] (3.000,-11.000) -- (4.000,-11.000);
\draw[ed] (4.000,-11.000) -- (5.000,-11.000);
\draw[ed] (5.000,-11.000) -- (6.000,-11.000);
\draw[ed] (6.000,-11.000) -- (7.000,-11.000);
\draw[ed] (7.000,-11.000) -- (8.000,-11.000);
\draw[ed] (2.000,-11.000) -- (2.000,-10.000);
\draw[ed] (6.000,-11.000) -- (6.000,-10.000);
\node[storage,vtx] at (0.000,-11.000) {};
\node[idx] at (0.000,-11.591) {0};
\node[storage,vtx] at (1.000,-11.000) {};
\node[idx] at (1.000,-11.591) {1};
\node[junc,vtx] at (2.000,-11.000) {};
\node[idx] at (2.000,-11.591) {2};
\node[storage,vtx] at (3.000,-11.000) {};
\node[idx] at (3.000,-11.591) {3};
\node[gatev,vtx] at (4.000,-11.000) {};
\node[idx] at (4.000,-11.591) {4};
\node[idx] at (4.000,-10.409) {GS};
\node[storage,vtx] at (5.000,-11.000) {};
\node[idx] at (5.000,-11.591) {5};
\node[junc,vtx] at (6.000,-11.000) {};
\node[idx] at (6.000,-11.591) {6};
\node[storage,vtx] at (7.000,-11.000) {};
\node[idx] at (7.000,-11.591) {7};
\node[storage,vtx] at (8.000,-11.000) {};
\node[idx] at (8.000,-11.591) {8};
\node[storage,vtx] at (2.000,-10.000) {};
\node[idx] at (2.591,-10.000) {9};
\node[storage,vtx] at (6.000,-10.000) {};
\node[idx] at (6.591,-10.000) {10};
\node[qb] at ($(1.000,-11.000)+(0mm,0)$) {$q_{2}$};
\node[qg] at ($(4.000,-11.000)+(-1.95mm,0)$) {$q_{0}$};
\node[qg] at ($(4.000,-11.000)+(1.95mm,0)$) {$q_{1}$};
\node[qb] at ($(6.000,-10.000)+(0mm,0)$) {$q_{3}$};
\end{tikzpicture}}
\caption{Rotor-conveyor for a two-qubit gate $g(q_0,q_1)$ on a mirror-symmetric branched architecture with the gate vertex 4 (GS, red) at the center and the junctions 2 and 6 (green), each carrying a stack of height 1 (vertices 9 and 10). Both gate qubits are drawn in red, all others in blue. Each of the states (1) to (5) shows the architecture after the operation named on its left.}
\label{fig:routing-rotor}
\end{figure}

\begin{table*}[t]
	\centering
	\caption{Acceptance-test matrix for the branched-architecture compiler. The Non-junction vertices column lists each architecture's number of non-junction vertices, with any special structure noted in parentheses, and the Variants column states the number of invocations each test runs, each a combination of a circuit and an architecture variant, the variants differing in stack height and junction distance. Tests T1--T8 reuse the circuits of \autoref{tab:linear-tests} on branched architectures, T9 adds larger circuits, and T10--T14 stress two-qubit delivery on progressively denser architectures. Four tests carry tight schedule bounds, three of them inherited from the linear architecture. The rest need only run to a complete, valid schedule, apart from the compile-time guard on T4.}
	\label{tab:branched-tests}
	\setlength{\tabcolsep}{5pt}
	\scriptsize
	\begin{tabular*}{\textwidth}{@{\extracolsep{\fill}}lllcl@{}}
		\toprule
		Test & Circuit(s) & Non-junction vertices & Variants & Focus / expected \\
		\midrule
		T1  & \texttt{same\_pair}                              & 5 (no junctions)               & 1 & no shuttling \\
		T2  & working example                               & 7                              & 1 & $\leq20$ ops, $\leq12$ steps \\
		T3  & \texttt{qaoa\_5}                                 & 11                    & 4 & no swaps \\
		T4  & \texttt{rd32-v0\_66}                             & 9                     & 2 & runs, $<1$\,s \\
		T5  & \texttt{4mod5-v1\_23}                            & 11                    & 4 & runs \\
		T6  & \texttt{4mod5-bdd\_287}                          & 15                    & 4 & runs \\
		T7  & \texttt{alu-v2\_31}                              & 11                    & 4 & runs \\
		T8  & \texttt{cnt3-5\_179}                             & 33                    & 4 & runs \\
		T9  & \texttt{qaoa\_25}, \texttt{qaoa\_40}, \texttt{qv\_25}, \texttt{qft\_25}, \texttt{xeb\_sy\_25}   & 51, 81             & 5 & runs, no swaps for QAOA \\
		T10 & \texttt{4gt12}, \texttt{4gt4}, \texttt{alu-v2\_30}    & 13 (depth-1 stacks)   & 5 & tight single corridors \\
		T11 & \texttt{4gt4}, \texttt{mod8}, \texttt{rd53}, \texttt{hwb7} & 13, 15, 17         & 5 & occupied stacks, stacked endpoints \\
		T12 & \texttt{xeb\_9}, \texttt{xeb\_sy\_9}          & 19 (4--6 junctions)  & 5 & rotor-conveyor required \\
		T13 & \texttt{qft\_15}                                 & 31               & 8 & congested endpoint corridor \\
		T14 & \texttt{xeb\_16}, \texttt{xeb\_sy\_16}, \texttt{qv\_20}, \texttt{qv\_25}  & 33, 41, 51            & 7 & rotor placement, packed corridors \\
		\bottomrule
	\end{tabular*}
\end{table*}

\begin{example}[Rotor-conveyor]
The state of \autoref{fig:routing-rotor} illustrates the mechanics of the rotor-conveyor, although other methods, such as park-the-endpoint, would also bring the gate qubits together there. The intermediate $q_2$ occupies vertex 5, the target of $q_1$, and cannot be pushed onto a stack, since the stack at junction 6 is already held by $q_3$, and $q_0$ on vertex 3 blocks the way to the empty stack at junction 2. Therefore, the rotor-conveyor takes $q_0$ on vertex 3 as the anchor. One \textsc{Merge}-\textsc{Swap}-\textsc{Separate} cycle at the gate vertex, as in \autoref{fig:routing-linear}, exchanges the positions of $q_0$ and $q_2$ and so carries $q_2$ across the gate vertex. Unlike in the swap method, the intermediate is thereby exchanged with the anchor instead of with $q_1$, whose path it blocks. Afterward, $q_2$ is stashed on vertex 1, deep on the anchor's side, and $q_0$ returns to vertex 3. With its corridor free, $q_1$ passes the junction to vertex 5. The concluding \textsc{Merge} pairs it with $q_0$ on the gate vertex, where the gate executes.
\end{example}

The prompt also specifies two supporting mechanisms. First, a gate qubit may begin deep within a stack, with other qubits between it and the first slot. Because the stack is last-in-first-out, that qubit cannot leave until those qubits are removed, so before the gate qubits are routed, an extraction protocol pops the blocking qubits onto cleared main-axis vertices and then pops the gate qubit out. Second, the stack method needs a stack with a free slot, but on a densely occupied architecture every reachable stack may be full. An eviction routine then frees a slot by popping out a stack qubit whose own next gate lies far ahead. Independently of these mechanisms, every routing attempt is wrapped in a snapshot-and-restore transaction, so a failed attempt is rolled back in full before the next is tried.

The post-processing of the linear architecture is extended with a junction-soundness rule. Removing a roundtrip can leave a chain entering and leaving a junction by the same edge, so each candidate removal is checked by replaying the resulting schedule against the simulated state, and is accepted only if no such violation appears. When a single removal fails this check, the pass retries it together with the smallest enclosing roundtrip pair, since dropping both can clear the junction traversal altogether. Roundtrip elimination and consecutive cancellation then alternate until neither removes any further operation, as in the linear architecture. The prompt also offers a conservative alternative for when the replay is not implemented: refuse any removal in which either vertex of the roundtrip is a junction. This preserves correctness with less implementation effort, but rejects every candidate that might expose a violation, including those the replay would accept.

The acceptance-test matrix consists of the fourteen tests listed in \autoref{tab:branched-tests}, spanning height-1 to height-10 stacks, up to 81 non-junction vertices, and architectures of up to 93 vertices in total. Four carry tight bounds. Three are inherited from the linear architecture: a same-pair test that must produce zero shuttling, a working example that must hit a stated operation and timestep bound, and a chain-structured QAOA test that must produce zero swaps. The fourth extends the same zero-swap requirement to 25- and 40-qubit QAOA circuits. Each of the remaining ten, of varying configurations, only needs to execute successfully, with a compile-time guard on one.

\subsection{Emitted compiler}
\label{sec:branched:first}

The LLM rewrites the linear-architecture seed in full, and the resulting compiler passes all acceptance tests on the first attempt. However, its routing departs from the specification. Where the prompt describes a per-iteration choice among the swap, stack, and park-the-endpoint methods backed by the rotor-conveyor as a guaranteed fallback, the emitted compiler keeps none of that selection. Instead, a \emph{fast attempt} moves each endpoint along its tree path to one of the gate vertex's two neighbors, pushing any obstacle off the path to a nearby free vertex, and hands off to the rotor-conveyor when it fails. Thus, swap, stack, and park-the-endpoint are all omitted as separate methods, with intermediates reaching a stack only incidentally as obstacles pushed off the path, and the \textsc{Merge}-\textsc{Swap}-\textsc{Separate} cycle appearing only inside the rotor-conveyor. This is an outright substitution rather than a re-encoding of the prompt's methods. The fast attempt is not cost-driven either. It tries each combination of which endpoint goes to which neighbor and which one to route first, four in all, in a predetermined order, and accepts the first that succeeds, whereas the prompt specifies first trying the assignment with the shorter total distance from the endpoints to their target neighbors.

Post-processing is adapted in two ways. First, the timestep grouper is tightened for the branched architecture so that it never moves an operation across the boundary between a translation and a non-translation timestep. Otherwise, a non-translation operation could fall between two translations of the same chain, violating the in-transit rule. Second, the junction-soundness rule is implemented in the conservative form the prompt offers as a fallback. Rather than replaying the schedule after a candidate removal and accepting it whenever no junction violation appears, the emitted compiler never removes a roundtrip that touches a junction. This conservative form never violates the junction rules and stays within the specification, but it forgoes the removals that the replay-based rule, which the prompt marks as preferred, would allow.

The emitted compiler also misses the prompt's determinism requirement at one point. To orient each qubit pair of the initial mapping, it compares two sums of floating-point interaction weights. It adds their terms in the iteration order of a Python set, which varies with the hash seed and so changes the rounding of the sums. Therefore, in a near tie, the orientation and with it the schedule can differ from run to run. In our evaluations of this compiler, near ties occur only in QV circuits, and they change the schedule only on the QV circuit with 50 qubits. The prompt's determinism test, two runs under different hash seeds, misses such changes, since the acceptance tests of \autoref{tab:branched-tests} contain QV circuits of at most 25 qubits. For this reason, all evaluations run with one constant hash seed, as stated in \autoref{sec:eval:setup}.

\subsection{Optimized compiler}
\label{sec:branched:opt}

The optimizations mainly change two stages of the pipeline. One change is to the initial mapping, where the single deterministic placement of the prompt is replaced by a search over candidate placements. Among the compiled candidates, the search keeps the one with the shortest schedule, but a swap-free schedule takes precedence over any schedule with swaps, which preserves the zero-swap bounds of the acceptance tests. Its default candidate still comes from the orientation step of the emitted compiler and so keeps the dependence on the hash seed. In our evaluations of the optimized compiler, near ties occur in QV circuits and in one library circuit, and they change the schedule on QV circuits with 30 to 50 qubits.

The search draws its candidates from two sources. Variants of the default construction, for example with a reversed chain or with flipped pairs, are all compiled. A local search generates further placements by repeatedly swapping qubits, but rates them with an inexpensive \emph{proxy} for routing effort instead of compiling them. The proxy multiplies each edge weight of the qubit-interaction graph by the graph distance between the vertices holding its two qubits, and sums the products. Thus, it combines a circuit property, how strongly and how early a pair interacts, with a placement property, how far apart the placement puts that pair. The score is low when strongly or early-interacting qubits start close together and so need little routing, which usually yields a short schedule. Only placements that the proxy rates better than every variant are compiled. Finally, the best placement is refined by exchanging the qubits of two vertices at a time, keeping each exchange that shortens the compiled schedule. On this placement, the search then also tries alternative rules for selecting the next ready gate. A deterministic budget of at most 150 compilations keeps the search time-bounded and reproducible. None of this initial-mapping search appears in the prompt, which invites such additions without naming them.

The routing gains two main changes. The fast attempt now scores all candidate plans and keeps the least costly, instead of accepting the first that succeeds. In addition, the park-the-endpoint method is added as a standalone routine. When the fast attempt fails, both the park-the-endpoint routine and the rotor-conveyor route the gate qubits, and the cheaper delivery is kept. Thus, the routing recovers one of the prompt's three methods, but keeps the substituted structure: the fast attempt is taken whenever it succeeds, and the cost comparison weighs whole gate deliveries rather than choosing a method in each iteration by estimated cost. Further changes make the rotor-conveyor try both endpoints as its anchor and double the scan window of the grouper on architectures of more than 16 vertices. The remaining changes are lower-level performance and memory optimizations that leave the schedule unchanged.

\def\TabTotalsBranchedLib{%
Specialized, emitted & 209,179k & 104,308k & 21,745k & 21,736k & 4,316k \\
Specialized, optimized & \textbf{196,614k} & 100,629k & 18,803k & 18,793k & 1,315k \\
General, emitted & 199,299k & \textbf{98,115k} & 20,530k & 20,521k & 3,060k \\
General, optimized & 199,940k & 98,497k & 20,616k & 20,607k & 3,146k \\
Baseline \cite{KreppelMWHPSB24} & 223,853k & 131,802k & \textbf{17,494k} & \textbf{17,484k} & \textbf{0} \\}
\def\TabTotalsBranchedFam{%
Specialized, emitted & 4,664k & 42,209k & 73,620k & 282,932k & 512,872k & 8,695k & 8,723k \\
Specialized, optimized & 1,199k & \textbf{21,460k} & 50,534k & 184,273k & 328,129k & 4,229k & 4,295k \\
General, emitted & 1,516k & 23,234k & 55,154k & 194,079k & 375,006k & 4,841k & 4,892k \\
General, optimized & 1,553k & 26,445k & 57,774k & 201,135k & 385,556k & 5,575k & 5,548k \\
Baseline \cite{KreppelMWHPSB24} & \textbf{723k} & 22,943k & \textbf{49,324k} & \textbf{143,729k} & \textbf{302,262k} & \textbf{4,189k} & \textbf{4,197k} \\}
\begin{table}[b]
	\centering
	\caption{Branched architecture, circuit library: timesteps, translation timesteps, and separation, merge, and swap operations summed over all 153 library circuits and all tested configurations. The baseline emits no swaps by construction. The lowest value per column is in bold. All values are rounded to thousands.}
	\label{tab:branched-totals}
	\setlength{\tabcolsep}{1.5pt}
	\scriptsize
	\begin{tabular*}{\columnwidth}{@{\extracolsep{\fill}}lrrrrr@{}}
		\toprule
		Compiler & Timesteps & Translate & Separate & Merge & Swap \\
		\midrule
		\TabTotalsBranchedLib
		\bottomrule
	\end{tabular*}
\end{table}

\begin{table*}[t]
	\centering
	\caption{Branched architecture, scalable families: timesteps summed over all circuits of each family and all tested configurations. As in \autoref{tab:linear-large}, the family totals are directly comparable only within the groups of equal qubit counts. The lowest value per family is in bold. All values are rounded to thousands.}
	\label{tab:branched-large}
	\setlength{\tabcolsep}{4pt}
	\scriptsize
	\begin{tabular*}{\textwidth}{@{\extracolsep{\fill}}lrrrrrrr@{}}
		\toprule
		Compiler & QAOA & QFT & QV & XEB & XEB\_Sy & QEC\_Z & QEC\_X \\
		\midrule
		\TabTotalsBranchedFam
		\bottomrule
	\end{tabular*}
\end{table*}

\subsection{Evaluation}
\label{sec:branched:eval}

\begin{figure}[t]
	\centering
	\includegraphics{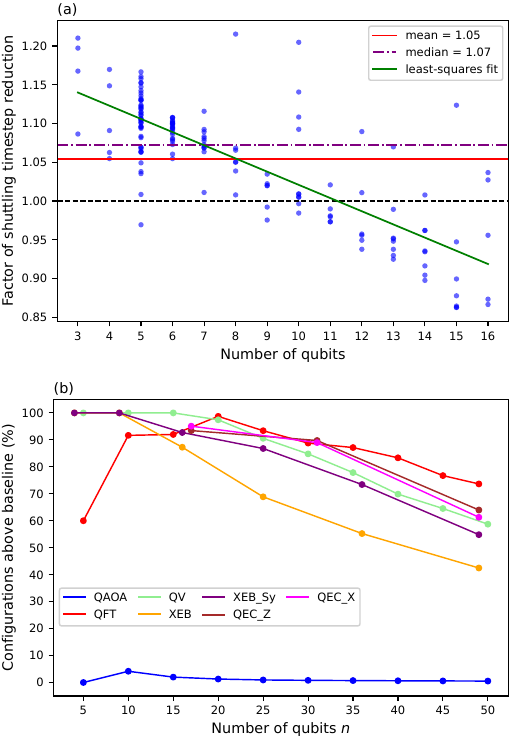}
	\caption{Timestep comparison of the optimized LLM-based compiler with the hand-crafted baseline \cite{KreppelMWHPSB24} on the branched architecture. (a) Circuit library at stack height 1 and junction distance 1, each circuit's timestep reduction factor against its qubit count, with the mean, the median, and a least-squares fit as red, purple, and green lines. Above the dashed line at 1, the LLM-based compiler needs fewer timesteps. (b) Scalable families, per qubit count the proportion of all tested configurations on which the optimized compiler needs fewer timesteps than the baseline, one line per family.}
	\label{fig:branched-lib}
\end{figure}

\begin{figure*}[tp]
	\centering
	\includegraphics[width=\textwidth]{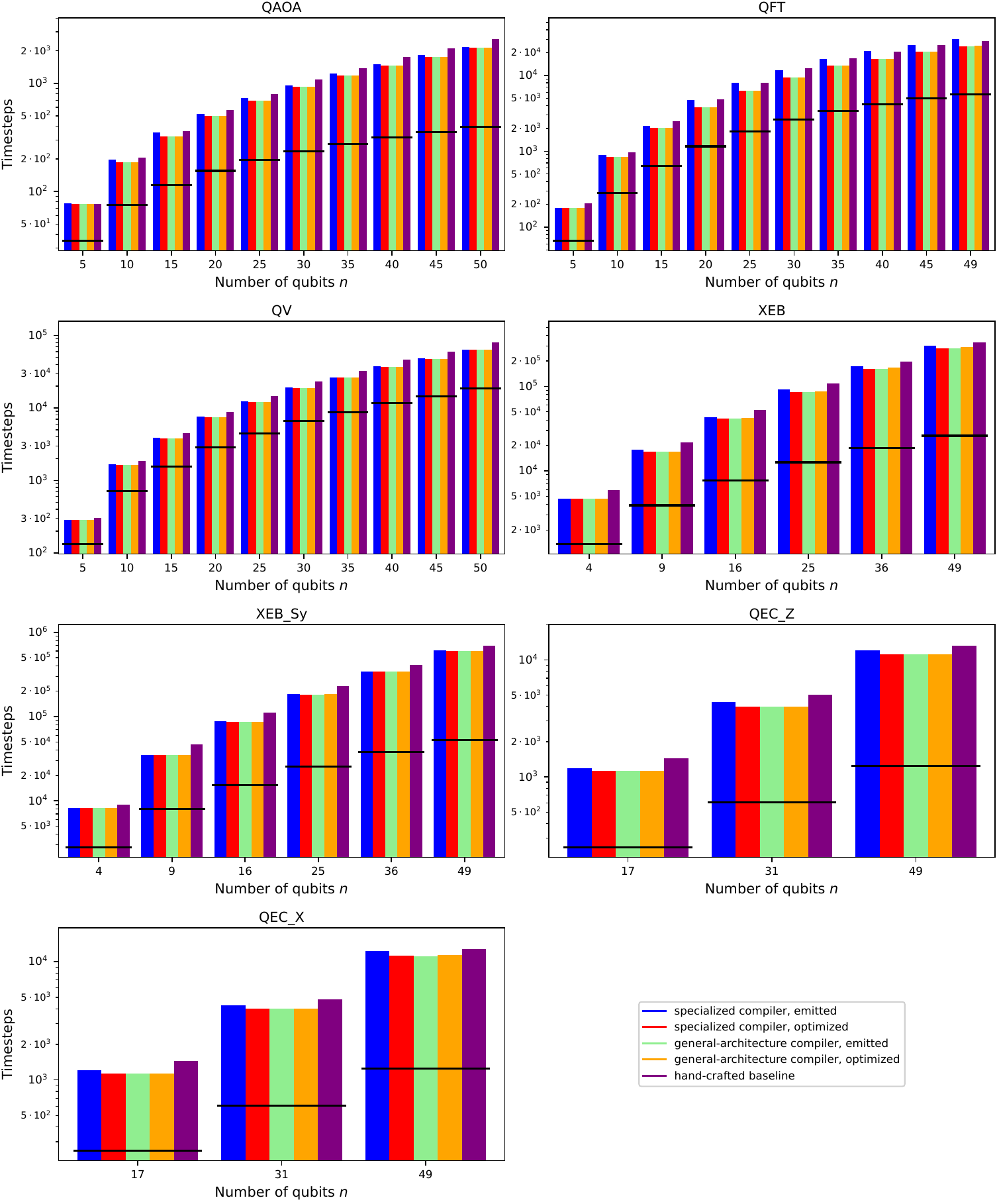}
	\caption{Absolute timesteps of the specialized and general-architecture compilers, each as emitted and optimized, and of the hand-crafted baseline \cite{KreppelMWHPSB24} on the branched architecture at stack height 1 and junction distance 1 for the scalable families, one panel per family and one bar per circuit and compiler. The black line marks the number of gates, a lower bound on the number of timesteps, so the part of each bar above it is the shuttling overhead.}
	\label{fig:abs-branched}
\end{figure*}

\begin{figure}[t]
	\centering
	\includegraphics[width=\columnwidth]{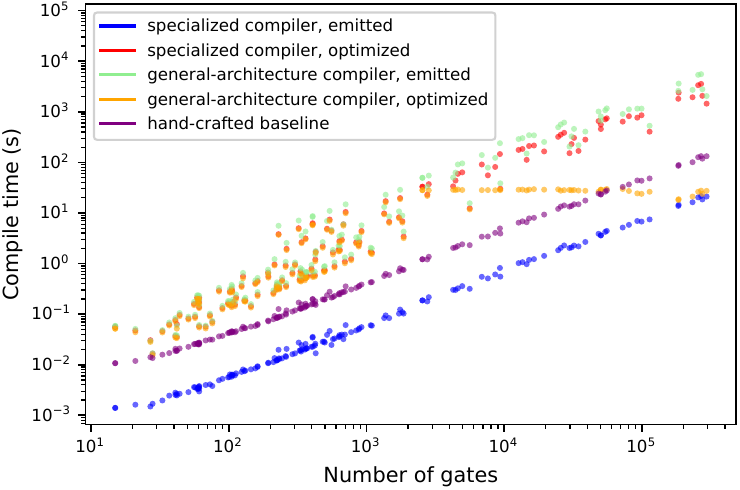}
	\caption{Compile time of the specialized and general-architecture compilers, each as emitted and optimized, and of the hand-crafted baseline \cite{KreppelMWHPSB24} on the branched architecture at stack height 1 and junction distance 1, one point per library circuit against its number of gates. Each value is the mean over ten runs.}
	\label{fig:compile-time-branched}
\end{figure}

The evaluation follows the same order as on the linear architecture, with the hand-crafted baseline \cite{KreppelMWHPSB24} as the reference, whose comparison \autoref{fig:branched-lib} summarizes, and compares the branched with the linear architecture before turning to compile time and memory. As in the linear evaluation, \autoref{tab:branched-totals}, \autoref{tab:branched-large}, \autoref{fig:abs-branched}, and \autoref{fig:compile-time-branched} include the general-architecture compiler, which was generalized from the branched-architecture compiler, as emitted and as optimized.

\paragraph{Emitted versus optimized compiler.} The schedule-affecting changes of \autoref{sec:branched:opt} jointly remove about 70\,\% of the branched-architecture compiler's swaps on the circuit library, as the suite totals over all tested configurations in \autoref{tab:branched-totals} show, and shorten the circuit-library schedules by a median factor of 1.05. Per scalable family, whose totals \autoref{tab:branched-large} lists, the median emitted-to-optimized factor is 1.32 for QV, 1.44 for XEB, 1.41 for XEB\_Sy, 1.67 for QFT, 1.88 for QEC\_Z, 1.85 for QEC\_X, and 3.08 for QAOA.
QAOA improves the most because the optimized compiler removes its swaps almost entirely. Its median grows with the qubit count, from 1.03 at 5 qubits to 4.58 at 50 qubits, and the factor of a single configuration peaks at 9.84 at 45 qubits on an architecture with stack height 7 and junction distance 30.
The search over up to 150 candidate placements of \autoref{sec:branched:opt}, together with the costlier routing of each candidate, raises the compile time, at stack height 1 and junction distance 1 by a median factor of 54.20 on the circuit library and 165.32 on the scalable families, and over all tested configurations by 148.77 and 182.99.

\paragraph{Optimized LLM-based compiler versus the hand-crafted baseline.} Across the tested architectures, the generated branched-architecture compiler outperforms the hand-crafted baseline by a median factor of 1.18 on the circuit library, and per scalable family, the median is 0.66 for the chain-structured QAOA, 1.07 for XEB, 1.17 for QV, 1.21 for QFT, 1.23 for XEB\_Sy, 1.24 for QEC\_X, and 1.26 for QEC\_Z.
The largest advantage of a single configuration is a factor of 1.76, or 43\,\% fewer timesteps, on QEC\_Z at 31 qubits, stack height 9, and junction distance 9. The per-family totals in \autoref{tab:branched-large} sum the timesteps of every circuit, so the largest circuits, with by far the most timesteps, dominate them, and in these totals only QFT remains ahead of the baseline.

By construction, the baseline emits no swaps. At stack height 1 and junction distance 1 the generated compiler emits none either, and on all but one library circuit both emit identical separation and merge counts, so the factor is decided by the translation timesteps alone. The generated compiler's advantage on the circuit library is then slight, a mean of 1.05 and a median of 1.07, and it falls with the qubit count, as \autoref{fig:branched-lib}(a) shows: the least-squares fit drops below the baseline at 11 qubits. The generated compiler needs about 29\,\% fewer translations than the baseline at 3 qubits but about 25\,\% more at 15 and 16 qubits. This advantage at small qubit counts most likely comes from choosing the gate order from the ready set, whereas the baseline follows the input order. However, on a larger architecture the routing must clear more occupied vertices before each gate, and the generated compiler's evacuation of these vertices costs more translations than the baseline's junction-aware routing. Beyond 11 qubits, this extra cost outweighs the ordering benefit. Across all tested architectures, the median factor instead rises to 1.18, even as the compiler begins emitting swaps. Taken per qubit count, this median no longer falls with the qubit count but stays between 1.15 and 1.19.

The scalable families, tested on far larger architectures, diverge. At stack height 1 and junction distance 1, all lie above the baseline, at median factors of 1.15 to 1.26, apart from QAOA at 5 qubits, which ties with it.
The chain-structured QAOA declines most steeply, falling to or below the baseline as soon as the stack height or the junction distance grows, down to 0.32 on single configurations, since the baseline routes its chains more compactly. QFT and QV hold their advantage through most moderate stack heights and junction distances, falling below the baseline mainly at the largest ones, at the largest qubit counts QFT to 0.50 and QV, with its larger swap count, to 0.39. An exception is QV with 40 to 50 qubits at stack height 1, whose factor already falls below 1 from junction distance 6 or 7 on. The grid-based XEB and XEB\_Sy emit by far the most swaps, up to 94,000 and 155,000 at 49 qubits. At 49 qubits, their factors even rise with stack height, to single-configuration peaks of 1.46 and 1.59 at stack height 15, before dropping to 0.24 and 0.29 at the tallest stacks, and they also weaken with junction distance. QEC\_Z and QEC\_X emit at most about 2,300 swaps, yet their factors follow a similar course. Over all tested configurations with a given stack height, their median factors rise to 1.48 and 1.47 at height 15 before dropping to 0.33 and 0.36 at the tallest stacks. With junction distance, they peak at 1.43 and 1.41 at distance 5 and fall below the baseline from about 20 on. Therefore, over all tested configurations at a qubit count, the proportion on which the generated compiler beats the baseline falls with the qubit count, as \autoref{fig:branched-lib}(b) shows, for QFT only after a rise from 60\,\% at 5 qubits to 99\,\% at 20 qubits. At the largest qubit counts, it is 42\,\% for XEB, 55\,\% for XEB\_Sy, 59\,\% for QV, 61\,\% for QEC\_X, 64\,\% for QEC\_Z, and 74\,\% for QFT, while for QAOA it stays near zero at every size.

\paragraph{Branched versus linear architecture.} On the families whose gates act on distant qubit pairs, the junctions bring the schedules much closer to the gate count. At stack height 1 and junction distance 1, for which \autoref{fig:abs-branched} shows the absolute timesteps of each circuit of the scalable families, the optimized compiler needs at most 3.41 times as many timesteps as gates on QV and 8.98 to 11.31 times on the QEC and XEB families, against 15.62 and 22.48 to 28.05 times for the optimized linear-architecture compiler with the same number of qubit segments. Only QAOA and QFT, whose ratios are already low on the linear architecture, end up higher, at 5.40 and 4.35 against 2.05 and 3.51 there.

\paragraph{Compile time and memory.} The optimized compiler routes up to 150 candidate placements per circuit against the baseline's single deterministic one, so at stack height 1 and junction distance 1 it is slower to compile. The median factor is 7.74 on the circuit library, as \autoref{fig:compile-time-branched} shows, and, per scalable family, 15.44 for QV, 18.09 for QFT, 25.72 for XEB, 27.43 for XEB\_Sy, 29.73 for QEC\_Z, 29.99 for QEC\_X, and 30.69 for QAOA, as \autoref{fig:compile-time-branched-fam} in \appref{app:figures} shows. On XEB\_Sy at 49 qubits, it needs about 2,140\,s against the baseline's 80\,s, whereas on QAOA the times stay small despite the large factor, about 7\,s against 0.2\,s at 50 qubits. On the circuit library, the emitted compiler, which lacks this search, compiles 6.87 times faster than the baseline at the median. Across the ten runs of a compilation, the compile time varies by a median standard deviation, over all tested configurations, of 0.9\,\% of the mean for the emitted compiler, 10.1\,\% for the optimized compiler, and 4.8\,\% for the baseline.

Memory is not a limiting factor here either. At stack height 1 and junction distance 1, both the emitted and the optimized compiler need less memory than the baseline on every circuit, as \autoref{fig:memory}(d) in \appref{app:figures} shows, for instance about 0.5\,GB and 0.8\,GB against 1.6\,GB on the largest library circuit. Over all tested configurations, they need more only on the XEB circuit at 49 qubits and the XEB\_Sy circuits at 36 and 49 qubits, the emitted compiler on 3\,\% and the optimized one on 2\,\% of the compilations, and peak at 2.4\,GB for the emitted compiler against the baseline's 1.8\,GB, while the optimized compiler peaks at about the same 1.8\,GB.

\section{General shuttling architecture}
\label{sec:compiler:general}

\subsection{Specification}
\label{sec:general:spec}

The prompt for the general-architecture compiler, which appears in \appref{app:general-prompt}, provides the LLM with the optimized branched-architecture compiler and describes only the differences for an arbitrary connected architecture, which need no longer be a tree. The three junction rules of the branched architecture still apply at every junction. On a tree, a chain that has left a junction by another edge can return to the side it came from only by crossing that junction again. On a cycle, it can instead return along the cycle's other arc. The prompt calls this the main structural gain of cycles and asks the compiler to exploit it whenever a chain transits a junction on a cycle.

The initial mapping is unchanged from the branched architecture. Matching and the chain-extraction procedure make no assumption about the architecture, using only the graph distance and the set of non-junction vertices, so both carry over verbatim.

The routing pipeline structure of the branched architecture carries over to the general graph. Here the prompt describes three changes. First, the tree-specific parts are dropped, namely the selection of park slots on stacks and at main-axis ends and the stack method, since a general graph has neither stacks nor a main axis. Second, in their place, the park-the-endpoint method is generalized to park an endpoint on any empty non-junction vertex that lies off the routing paths, with candidates ranked by the total distance the endpoint must walk, out to the park vertex and then on to its gate-vertex neighbor. Third, a \emph{cycle-detour} method is added. For an intermediate on the direct path from an endpoint to its target gate-vertex neighbor, the router seeks the shortest alternative path to that neighbor that avoids the intermediate. If such a path exists and its extra length, measured against the direct path, costs less than evicting the intermediate, the endpoint takes that alternative arc. The detour needs no \textsc{Separate}, \textsc{Merge}, or \textsc{Swap} and also bypasses any further intermediates on the part of the direct path that it avoids.

In the per-iteration cost comparison, the cycle-detour takes the stack method's place against the park-the-endpoint and swap methods. The prompt keeps both roles of the swap method from the branched architecture, as a candidate in the cost comparison and as the unconditional fallback when neither of the others applies, with the rotor-conveyor as the guaranteed terminal routine. Because the branched-architecture seed routes by the fast attempt described in \autoref{sec:compiler:branched}, with neither a swap method nor a per-iteration cost comparison, the prompt here names features the seed never implemented, which \autoref{sec:compiler:setup} notes can happen.

\begin{figure}[t]
\centering
\resizebox{\columnwidth}{!}{\begin{tikzpicture}[x=9.0mm,y=9.0mm,
  vtx/.style={minimum size=8.0mm,semithick},
  qb/.style={circle,draw=qubitline,fill=qubitcol,text=white,inner sep=0pt,minimum size=3.7mm,font={\fontsize{6.5}{6.5}\selectfont}},
  qg/.style={qb,draw=qgateline,fill=qgatecol},
  ed/.style={draw=black,thick},
  idx/.style={font=\small},
  op/.style={font=\small,anchor=east,align=right}]
\node[op] at (-4.550,0.000) {(1)~initial state,\\cycle-detour\\of $q_0$};
\draw[gateline,line width=12pt,opacity=0.45,dash pattern=on 5pt off 3pt,line cap=round] (4.000,0.600) -- (2.800,1.000) -- (2.000,0.000) -- (1.000,0.000);
\draw[expline,line width=12pt,opacity=0.55,line cap=round,line join=round] (4.000,0.600) -- (4.000,-0.600) -- (2.800,-1.000) -- (2.000,0.000) -- (1.000,0.000);
\draw[ed] (-2.000,0.000) -- (-2.800,1.000);
\draw[ed] (-2.800,1.000) -- (-4.000,0.600);
\draw[ed] (-4.000,0.600) -- (-4.000,-0.600);
\draw[ed] (-4.000,-0.600) -- (-2.800,-1.000);
\draw[ed] (-2.800,-1.000) -- (-2.000,0.000);
\draw[ed] (-2.000,0.000) -- (-1.000,0.000);
\draw[ed] (-1.000,0.000) -- (0.000,0.000);
\draw[ed] (0.000,0.000) -- (1.000,0.000);
\draw[ed] (1.000,0.000) -- (2.000,0.000);
\draw[ed] (2.000,0.000) -- (2.800,1.000);
\draw[ed] (2.800,1.000) -- (4.000,0.600);
\draw[ed] (4.000,0.600) -- (4.000,-0.600);
\draw[ed] (4.000,-0.600) -- (2.800,-1.000);
\draw[ed] (2.800,-1.000) -- (2.000,0.000);
\node[storage,vtx] at (-4.000,0.600) {};
\node[idx] at (-4.000,1.356) {0};
\node[storage,vtx] at (-2.800,1.000) {};
\node[idx] at (-2.800,1.756) {1};
\node[storage,vtx] at (-4.000,-0.600) {};
\node[idx] at (-4.000,-1.356) {2};
\node[storage,vtx] at (-2.800,-1.000) {};
\node[idx] at (-2.800,-1.756) {3};
\node[junc,vtx] at (-2.000,0.000) {};
\node[idx] at (-2.000,-0.756) {4};
\node[storage,vtx] at (-1.000,0.000) {};
\node[idx] at (-1.000,-0.756) {5};
\node[gatev,vtx] at (0.000,0.000) {};
\node[idx] at (0.000,-0.756) {6};
\node[idx] at (0.000,0.756) {GS};
\node[storage,vtx] at (1.000,0.000) {};
\node[idx] at (1.000,-0.756) {7};
\node[junc,vtx] at (2.000,0.000) {};
\node[idx] at (2.000,-0.756) {8};
\node[storage,vtx] at (2.800,1.000) {};
\node[idx] at (2.800,1.756) {9};
\node[storage,vtx] at (4.000,0.600) {};
\node[idx] at (4.000,1.356) {10};
\node[storage,vtx] at (2.800,-1.000) {};
\node[idx] at (2.800,-1.756) {11};
\node[storage,vtx] at (4.000,-0.600) {};
\node[idx] at (4.000,-1.356) {12};
\node[qg] at ($(4.000,0.600)+(0mm,0)$) {$q_{0}$};
\node[qb] at ($(2.800,1.000)+(0mm,0)$) {$q_{2}$};
\node[qg] at ($(-1.000,0.000)+(0mm,0)$) {$q_{1}$};
\node[qb] at ($(-4.000,0.600)+(0mm,0)$) {$q_{3}$};
\node[qb] at ($(-2.800,-1.000)+(0mm,0)$) {$q_{4}$};
\node[op] at (-4.550,-4.300) {(2)~\textsc{Merge},\\gate $g(q_0,q_1)$};
\draw[ed] (-2.000,-4.300) -- (-2.800,-3.300);
\draw[ed] (-2.800,-3.300) -- (-4.000,-3.700);
\draw[ed] (-4.000,-3.700) -- (-4.000,-4.900);
\draw[ed] (-4.000,-4.900) -- (-2.800,-5.300);
\draw[ed] (-2.800,-5.300) -- (-2.000,-4.300);
\draw[ed] (-2.000,-4.300) -- (-1.000,-4.300);
\draw[ed] (-1.000,-4.300) -- (0.000,-4.300);
\draw[ed] (0.000,-4.300) -- (1.000,-4.300);
\draw[ed] (1.000,-4.300) -- (2.000,-4.300);
\draw[ed] (2.000,-4.300) -- (2.800,-3.300);
\draw[ed] (2.800,-3.300) -- (4.000,-3.700);
\draw[ed] (4.000,-3.700) -- (4.000,-4.900);
\draw[ed] (4.000,-4.900) -- (2.800,-5.300);
\draw[ed] (2.800,-5.300) -- (2.000,-4.300);
\node[storage,vtx] at (-4.000,-3.700) {};
\node[idx] at (-4.000,-2.944) {0};
\node[storage,vtx] at (-2.800,-3.300) {};
\node[idx] at (-2.800,-2.544) {1};
\node[storage,vtx] at (-4.000,-4.900) {};
\node[idx] at (-4.000,-5.656) {2};
\node[storage,vtx] at (-2.800,-5.300) {};
\node[idx] at (-2.800,-6.056) {3};
\node[junc,vtx] at (-2.000,-4.300) {};
\node[idx] at (-2.000,-5.056) {4};
\node[storage,vtx] at (-1.000,-4.300) {};
\node[idx] at (-1.000,-5.056) {5};
\node[gatev,vtx] at (0.000,-4.300) {};
\node[idx] at (0.000,-5.056) {6};
\node[idx] at (0.000,-3.544) {GS};
\node[storage,vtx] at (1.000,-4.300) {};
\node[idx] at (1.000,-5.056) {7};
\node[junc,vtx] at (2.000,-4.300) {};
\node[idx] at (2.000,-5.056) {8};
\node[storage,vtx] at (2.800,-3.300) {};
\node[idx] at (2.800,-2.544) {9};
\node[storage,vtx] at (4.000,-3.700) {};
\node[idx] at (4.000,-2.944) {10};
\node[storage,vtx] at (2.800,-5.300) {};
\node[idx] at (2.800,-6.056) {11};
\node[storage,vtx] at (4.000,-4.900) {};
\node[idx] at (4.000,-5.656) {12};
\node[qb] at ($(2.800,-3.300)+(0mm,0)$) {$q_{2}$};
\node[qg] at ($(0.000,-4.300)+(-1.95mm,0)$) {$q_{1}$};
\node[qg] at ($(0.000,-4.300)+(1.95mm,0)$) {$q_{0}$};
\node[qb] at ($(-4.000,-3.700)+(0mm,0)$) {$q_{3}$};
\node[qb] at ($(-2.800,-5.300)+(0mm,0)$) {$q_{4}$};
\end{tikzpicture}}
\caption{Cycle-detour method for a two-qubit gate $g(q_0,q_1)$ on a \texttt{two\_cycles\_bridged} architecture, whose two cycles with the junctions 4 and 8 (green) are joined by a bridge with the gate vertex 6 (GS, red) in its center. Both gate qubits are drawn in red and the other three qubits in blue. State (1) shades the direct path of $q_0$ to vertex 7, blocked by $q_2$, in red and its cycle-detour in blue. State (2) shows the architecture after the final \textsc{Merge}.}
\label{fig:routing-general}
\end{figure}

\begin{example}[Cycle-detour method]\label{ex:cycle-detour}
In \autoref{fig:routing-general}, $q_1$ already sits next to the gate vertex, on vertex 5. The endpoint $q_0$ has to reach vertex 7, the other neighbor of the gate vertex. Its direct path along the upper arc of the right cycle is blocked by the intermediate $q_2$. The lower arc is only one translation longer, so $q_0$ takes it instead of evicting $q_2$, and the routing needs no \textsc{Separate}, \textsc{Merge}, or \textsc{Swap}. A final \textsc{Merge} unites $q_0$ and $q_1$ on the gate vertex, and the gate executes.
\end{example}

One change is made to the post-processing. The conservative guard of the branched-architecture compiler, which refuses every roundtrip deletion at a junction, is replaced with the replay-based junction-soundness check that the branched-architecture prompt marked as preferred.

The acceptance-test matrix has two parts. One retains the fourteen-test suite of \autoref{sec:compiler:branched}, which the general-architecture compiler, seeded from its predecessor, must still pass. The other adds an architecture matrix of fifteen graph families, each specified parametrically at $2n+1$ non-junction vertices for an $n$-qubit circuit and chosen to test particular routing capabilities. None of the tests in this matrix carries a bound, so each only needs to execute successfully. Three are trees, a path like the linear architecture, a caterpillar, and a star, and the remaining twelve add cycles. Of those twelve, the eight drawn in \autoref{fig:pipeline} and carried into the evaluation are:
\begin{itemize}
\item \texttt{cycle}: a single cycle.
\item \texttt{two\_cycles\_bridged}: two cycles connected by a path.
\item \texttt{theta}: two junctions linked by three internally disjoint paths.
\item \texttt{figure8}: two cycles sharing a single vertex.
\item \texttt{cycle\_with\_two\_chords}: a cycle with two non-crossing chords.
\item \texttt{complete\_graph}: a complete graph on four vertices, with edges subdivided by non-junction vertices.
\item \texttt{hypercube}: the three-dimensional hypercube \cite{Leighton91}, subdivided in the same way.
\item \texttt{butterfly}: the two-dimensional butterfly network \cite{Leighton91} with its last level also linked to its first, subdivided in the same way.
\end{itemize}

As \autoref{sec:eval:setup} explains, the subdivision of the last three families provides the qubit and gate vertices that the original networks, consisting predominantly of junctions, do not offer. On every graph the gate vertex meets the constraint of degree two with two non-junction neighbors. The matrix specifies each family by a construction rule rather than as a single given graph, and only specific instances drawn from these rules are tested. Thus, passing the acceptance tests shows that the compiler handles those instances, not every graph the rules admit.

\subsection{Emitted compiler}
\label{sec:general:first}

On the first attempt, the emitted compiler passes both the inherited branched-architecture suite and the architecture matrix. The initial mapping, taken over from the branched-architecture compiler, keeps the dependence on the hash seed described in \autoref{sec:branched:first}. As in the optimized branched-architecture compiler, near ties occur in QV circuits and in one library circuit, and in our evaluations, they change the schedule on QV circuits with 30 to 50 qubits.

The routing follows the prompt only in part. Park-the-endpoint is generalized, but its candidates are ranked by their distance to the gate vertex, and only the nearest three are tried. A cycle's other arc is used only around the gate vertex or the partner endpoint, never around an intermediate. Thus, the cycle-detour method is missing, and in the situation of \hyperref[ex:cycle-detour]{Example~\ref*{ex:cycle-detour}}, the compiler evicts $q_2$ instead. Neither the swap method nor the per-iteration cost comparison is added.

In one respect, the emitted compiler goes beyond the prompt, which, unlike those for the linear and branched architectures, invites no additions. Although the prompt leaves the rotor-conveyor unchanged, the emitted compiler tries a \emph{two-arc} delivery before the rotor-conveyor when the gate vertex lies on a cycle. On a tree, the gate vertex is the only connection between its two neighbors, so an endpoint on the far side has to cross the gate vertex. On a cycle, the two-arc delivery instead routes the endpoints to the two neighbors of the gate vertex along the two arcs of the cycle, so that neither has to cross the gate vertex. The two-arc delivery is the only routing method of the general-architecture compiler that its prompt does not require.

Generality is lost in the rotor-conveyor. The branched-architecture prompt requires it to make routing total, and in the branched architecture it does. That guarantee does not carry over to general graphs. Once the gate vertex lies on a cycle, it is no longer the only connection between the two sides, so a stashed intermediate can reach the other endpoint's path again. The router runs the rotor-conveyor whenever the fast attempt fails, alongside park-the-endpoint, and on some architectures the rotor-conveyor then reports failure even though free qubit vertices remain. Therefore, the compiler does not produce a schedule for every connected graph.

In the post-processing, the emitted compiler applies the replay check to the whole schedule at once. It deletes every roundtrip of the schedule and keeps the result if the replay finds no junction violation, and otherwise falls back to the conservative guard.

\subsection{Optimized compiler}
\label{sec:general:opt}

The optimizations for this compiler are entirely engineering-focused. The prompts ask for reductions in timesteps, compile time, and memory use alike, and here the LLM finds no further way to shorten the schedule, so its changes target compile time and memory use instead. Most are lower-level optimizations that leave the emitted schedule unchanged. Two changes alter it, a \emph{compile-time budget}, which trades schedule length for compilation speed, and a shortcut in the post-processing.

Routing is slow on a dense circuit, so the search caps how many candidates it routes. It estimates how many routings fit in a preset time budget from the operation count of the first routed candidate, a proxy for one routing's cost, and stops there. Deriving the cap from this count rather than from measured compile time keeps it deterministic and the output reproducible. Small circuits never reach the cap. Larger ones reach it, and the compiler accepts a longer schedule in return for a much shorter compile time. A user can raise the budget to compile more candidates, up to the search's preset trial limit. The budget changes only which candidates get compiled, not how the search picks among them. Where the budget compiles the same candidates, the optimized compiler returns the same schedule as the emitted one. Where it compiles fewer, the search mostly returns a longer schedule. Since the search prefers schedules without swaps, it can also return a shorter schedule with swaps when the candidate without swaps lies beyond the cap.

The shortcut in the post-processing skips the replay on trees and applies only the conservative guard there. On a few configurations of the branched architecture, this changes the schedule by a few timesteps. The dependence on the hash seed remains, with the same effect on QV circuits as in the emitted compiler.

\subsection{Evaluation}
\label{sec:general:eval}

The general-architecture compiler handles all ten architectures from a single code base. Our evaluation first compares the emitted with the optimized compiler and then the general-architecture compiler with the specialized ones, in timesteps and then in compile time and memory. Afterward, it compares the ten architectures with each other and closes by explaining why the compilers of related work cannot be compared with ours directly. On the linear and branched architectures, the suite totals, absolute timesteps, and compile times of the general-architecture compiler already appear next to those of the specialized compilers in \autoref{sec:linear:eval} and \autoref{sec:branched:eval}. The absolute timesteps and compile times on the eight cyclic architectures follow in \appref{app:figures}.

\paragraph{Emitted versus optimized compiler.} The general-architecture compiler's follow-up optimizations trade schedule length for compilation speed. On the circuit library, the optimized compiler needs more timesteps than the emitted one on 27\,\% of the circuits on the linear architecture, on 36\,\% of the tested configurations on the branched architecture, and on 23\,\% of the circuits on the cyclic architectures, with at most 36\,\% more timesteps. On the scalable families, these proportions are 58\,\%, 60\,\%, and 46\,\%, with up to 113\,\% more timesteps on QEC\_Z at 49 qubits on the plain \texttt{cycle}. Fewer timesteps occur only on less than 1\,\% of the configurations of the branched architecture, and by at most 6\,\%. Therefore, the following comparisons use the emitted compiler, which produces the shortest schedules.

In return, on the eight cyclic architectures, the optimized compiler is up to 218.09 times faster on the largest circuits, as \autoref{fig:compile-time-general} in \appref{app:figures} shows. The compile time and peak memory of every circuit on these architectures against its number of gates appear in \autoref{fig:memory}(e) and \autoref{fig:memory}(f) in \appref{app:figures}. The two versions take about the same time on small circuits, which never reach the compile-time budget. They diverge once the budget applies: on the XEB families from 9 qubits on, on QFT and QV from about 20 to 25 qubits on, and on the QEC families from 31 qubits on, whereas on the short QAOA circuits they diverge only from about 35 qubits on, by at most a factor of 3.71. On XEB\_Sy at 49 qubits, for instance, the emitted compiler takes about 4,700 to 147,000\,s depending on the architecture, and the optimized one about 64 to 700\,s. As on the other architectures, the compile time varies little across the ten runs of a compilation, with a median standard deviation of 3.4\,\% of the mean, and the peak memory use reaches at most 1.7\,GB as emitted and 0.5\,GB as optimized.

\paragraph{General-architecture compiler versus the specialized compilers.} Each factor here divides the general-architecture compiler's timestep count by the specialized compiler's, so a value above 1 means the specialist emits fewer timesteps. We take the general-architecture compiler as emitted, since that version produces its shortest schedules, and each specialist after optimization.

\begin{figure}[t]
	\centering
	\includegraphics{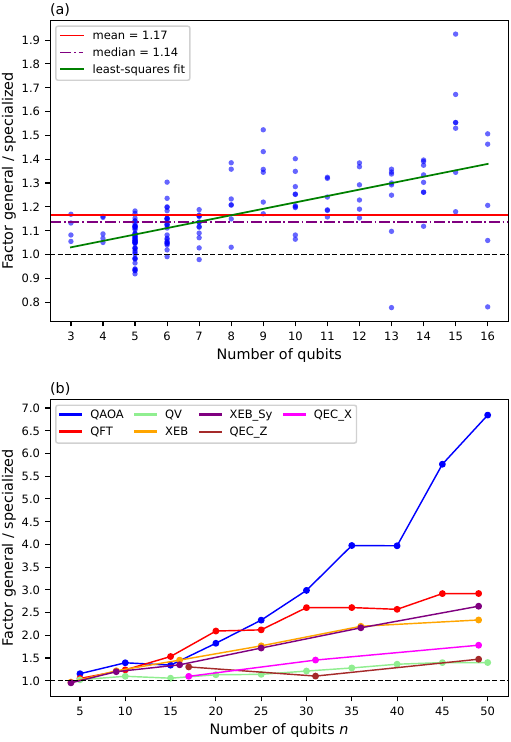}
	\caption{Timesteps of the general-architecture compiler as emitted divided by those of the optimized specialized compiler on the linear architecture. (a) Circuit library, each circuit's factor against its qubit count, with the mean, the median, and a least-squares fit as red, purple, and green lines. (b) Scalable families, one circuit per qubit count and one line per family. Above the dashed line at 1, the specialized compiler needs fewer timesteps.}
	\label{fig:gen-vs-spec}
\end{figure}

On the linear architecture, for which the general-architecture compiler is not tuned, the specialist keeps a clear lead. On the circuit library, the factors spread from 0.78 to 1.93 around a median of 1.14 and rise with the qubit count, as \autoref{fig:gen-vs-spec}(a) shows. On about 8\,\% of the library circuits, the general-architecture compiler needs up to 22\,\% fewer timesteps than the specialized one. Per scalable family, the median factor is 1.18 for QV, 1.30 for QEC\_Z, 1.45 for QEC\_X, 1.53 for XEB\_Sy, 1.61 for XEB, 2.34 for QFT, and 2.66 for QAOA, reaching 6.84 on the largest QAOA circuit, as \autoref{fig:gen-vs-spec}(b) shows.

\begin{figure*}[t]
	\centering
	\includegraphics[width=\textwidth]{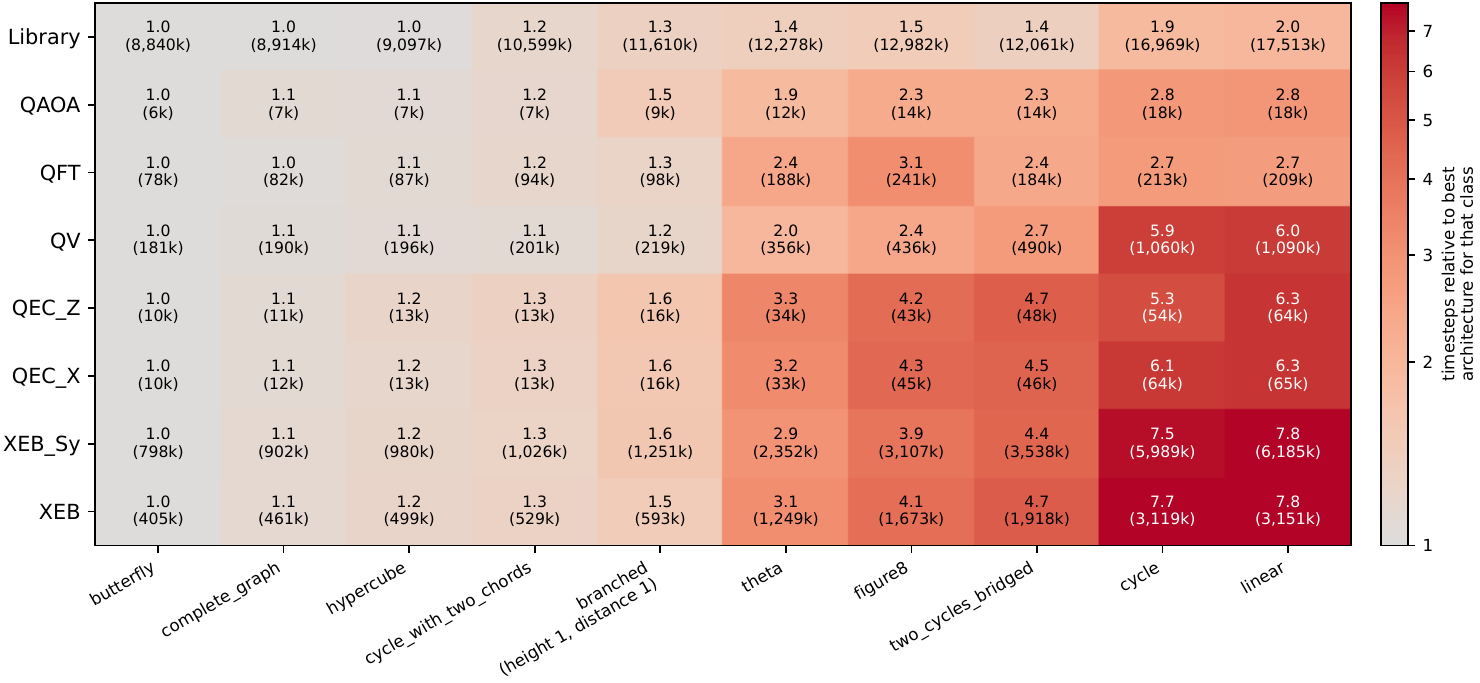}
	\caption{Architecture suitability for the circuit library and each scalable family. Each row is the circuit library or one scalable family, and each column one of the ten architectures, the eight cyclic families plus the linear and branched architectures. Cell values are the total timesteps of the row normalized by its minimum, so 1 marks the best architecture for that row, and the absolute total timesteps are shown in parentheses, rounded to thousands. As in \autoref{tab:linear-large}, the absolute totals of different rows are directly comparable only between families of equal qubit counts. Rows and columns are ordered by the sums of the normalized values.}
	\label{fig:arch-suit}
\end{figure*}

On the branched architecture, from whose compiler it was generalized, the general-architecture compiler nearly matches the specialist. At stack height 1 and junction distance 1, the two compilers emit identical schedules on 144 of the 153 library circuits and differ by at most 4\,\% on the rest. Over all tested configurations, the median factor per scalable family lies between 1.04 and 1.10, though single configurations reach up to 2.48 on QAOA. Taken per family and qubit count, this median grows only slightly with the qubit count, to 1.25 for QAOA at 50 qubits and at most 1.15 for the other families. The timestep totals of both compilers on the two architectures appear in \autoref{tab:linear-totals}, \autoref{tab:linear-large}, \autoref{tab:branched-totals}, and \autoref{tab:branched-large}, and the absolute timesteps of every circuit of the scalable families in \autoref{fig:abs-linear} and \autoref{fig:abs-branched}.

\paragraph{Compile time and memory.} As emitted, the general-architecture compiler is the slowest of the three to compile. It inherits the unbudgeted search over up to 150 candidates of the optimized branched-architecture compiler and, at stack height 1 and junction distance 1, is slower than the optimized branched-architecture compiler only by median factors of 1.20 on the circuit library and 1.11 on the scalable families. After optimization, the compile-time budget caps that search. Against the linear specialist, the optimized general-architecture compiler then compiles the scalable-family circuits with more than 10,000 gates faster, by a median factor of 8.81. For library circuits of that size, it remains 1.57 times slower at the median and is faster only on the largest ones, and for smaller circuits, it is 25.70 times slower at the median. Against the branched specialist, it compiles faster by a median factor of 15.94 on circuits with more than 10,000 gates and of 1.06 on smaller ones, as \autoref{fig:compile-time-linear}, \autoref{fig:compile-time-linear-fam} in \appref{app:figures}, and \autoref{fig:compile-time-branched} show.

Memory is not a limiting factor for the general-architecture compiler either. As emitted, it peaks at 1.8\,GB on the linear and 2.1\,GB on the branched architecture, and after optimization, at 0.5\,GB and 0.7\,GB, against 2.7\,GB and 1.8\,GB for the two specialists.

Thus, on the linear architecture, specializing a compiler still yields a substantial gain in timesteps, and in compile time on all but the largest circuits, whereas on the branched architecture the general-architecture compiler nearly matches the specialist in timesteps and, after optimization, compiles faster, by far on large circuits.

\paragraph{Architecture suitability.} We compare the general-architecture compiler across all ten architectures, the eight cyclic families plus the linear and branched architectures, taking the latter at stack height 1 and junction distance 1, the one configuration that scales across both the circuit library and the scalable families. All ten have $2n$ qubit vertices and a single gate vertex for an $n$-qubit circuit, differing only in how those vertices are connected, so the comparison isolates the effect of connectivity. The heatmap in \autoref{fig:arch-suit} reports the timesteps per architecture, summed over the circuit library and over each scalable family, both in absolute terms and normalized by the best architecture. Summing rather than taking medians weights the largest circuits more heavily. In these, shuttling is heaviest and the choice of architecture matters most.

The densely connected architectures with high-degree junctions and short pairwise distances, \texttt{butterfly}, \texttt{complete\_graph}, and \texttt{hypercube}, lead on the circuit library and on every scalable family, with \texttt{butterfly} the best throughout. Their short distances leave few intermediates on a routing path and put an empty non-junction vertex within easy reach for park-the-endpoint. Both spare the costly \textsc{Merge}-\textsc{Swap}-\textsc{Separate} routine. The two worst architectures overall, the linear architecture and the plain \texttt{cycle}, lack junctions above all. Without them, the only vertices off a routing path lie beyond its endpoints, so parking means long walks along the corridor itself, and the \texttt{cycle}'s single alternative arc spans most of the architecture. Accordingly, the extra arc puts the \texttt{cycle} only slightly ahead of the linear architecture. At the median over the circuits of the circuit library and of each scalable family, it needs 0\,\% to 4\,\% fewer timesteps, except on QEC\_Z with 13\,\% fewer. A single junction pair changes this. The \texttt{theta} family, a cycle with one additional path between two junctions, needs, at the same medians, about 10\,\% fewer timesteps than the plain \texttt{cycle} on the circuit library, QAOA, and QFT, 40\,\% to 47\,\% fewer on the two QEC families, and less than half as many on QV and the two XEB families.

The influence of the architecture scales with a circuit's shuttling demand, which the scalable families vary through their interaction patterns. QAOA is chain-like, QFT and QV are effectively all-to-all, the two XEB families couple only neighboring qubits on a square lattice but run 1,000 cycles, and the QEC circuits couple each measurement qubit to its neighboring data qubits on the surface-code lattice in every round. Accordingly, the spread of the family totals in \autoref{fig:arch-suit} between the best and the worst architecture grows from a factor of 2.84 for the chain-like QAOA to 7.78 for XEB. Per circuit, the spread among the eight cyclic architectures, the factor between the highest and the lowest bar of a circuit in \autoref{fig:abs-general} in \appref{app:figures}, grows overall with the qubit count. At the largest qubit counts, it reaches 3.72 for QFT and up to 9.68 for XEB\_Sy, with \texttt{butterfly} the shortest in every family. On the smaller library circuits, it reaches at most 3.01. Across all ten architectures, the spread widens further, up to a factor of 10.07, or 90\,\% fewer timesteps on the best than on the worst architecture, for the largest XEB\_Sy circuit, on which the linear architecture does even worse than the plain \texttt{cycle}.

The effect is large enough that on the three densest layouts the general-architecture compiler mostly produces shorter schedules than the two specialists do on the architectures they target. Compared by the timesteps summed per scalable family, its schedules are shorter than those of the linear specialist by a factor of 2.79 to 4.54 on QV, the two XEB families, and the two QEC families. On QFT, the two are about even, at factors of 0.91 to 1.01, and on the chain-like QAOA, the general-architecture compiler needs 46\,\% to 61\,\% more timesteps. Against the branched specialist, its schedules are shorter by a factor of 1.12 to 1.59 on each of the seven scalable families. On single circuits, the advantage peaks at 83\,\% over the linear and 41\,\% over the branched specialist, with the branched architecture again at stack height 1 and junction distance 1.

Against the hand-crafted baselines, the general-architecture compiler also comes out ahead on these three layouts. In the same per-family sums, its schedules are shorter than those of the linear-architecture baseline by a factor of 3.07 to 6.44 on each scalable family except QAOA. The linear layout suits QAOA so well that the general-architecture compiler needs 13\,\% to 25\,\% more timesteps there than this baseline. Against the branched-architecture baseline, its schedules are shorter by a factor of 1.38 to 1.94 on each of the seven scalable families. On single circuits, the advantage peaks at 86\,\% and 50\,\% over the two baselines. Thus, the connectivity of the architecture has a first-order effect on shuttling overhead and is worth optimizing alongside the compiler.

\paragraph{Related approaches.} Timestep and operation counts compare directly only between compilers that share both the elementary operations and the architecture, as our hand-crafted baselines do. None of the compilers of \autoref{sec:related} meets both conditions. SHAW and SHAPER \cite{BachSY26, RussonBYS26} and S-SYNC \cite{ZhuWWW25} handle many architectures, like our general-architecture compiler. However, their translation differs from ours. It moves only the outermost ion of a chain into the neighboring segment, which may already hold ions as long as it is not full, whereas ours moves the whole chain of a segment into an empty neighbor. Thus, for a chain of $m$ ions, one of our translations corresponds to $m$ of their operations. With two ions per segment, every schedule of ours can be rewritten in their operations, but not every schedule of theirs in ours, since they also split and join chains outside the gate segment. Therefore, counting operations across both sets would compare the operation sets rather than the compilers.

Other compilers target different hardware. The MQT IonShuttler \cite{SchoenbergerHBW24, SchoenbergerHBW25, SchoenbergerW25} works on a two-dimensional QCCD grid with dedicated processing zones and reorders the ions by rotating them around the loops of the grid. The reinforcement-learning compiler \cite{SchierRSSBHR26} is likewise tied to its two chips, a plus-shaped one, whose four arms meet at a single junction, and a Q-shaped one, in which a junction connects a ring of segments to a short line. Neither uses separations, merges, or swaps. As a consequence, even the schedules that the MQT IonShuttler computes exactly by Boolean satisfiability \cite{SchoenbergerHBW24} are no optimality reference for ours. The compilers for devices in which every segment holds several ions and executes gates \cite{AshSakiTG22, RuanZFLCHHHHPD25, OvideAC26} also rely on operations of their own. Therefore, none of these counts lies on a scale comparable to ours.

\section{LLM comparison}
\label{sec:comparison}

To test whether more recent models can substantially improve the results, we repeat the complete generation and evaluation with Claude Fable 5 \cite{Anthropic26} under the protocol of \autoref{sec:compiler:setup}. The prompts are identical to those of the Opus 4.7 run for all three compilers. For the linear- and branched-architecture compilers, every input file is identical as well, so their results compare directly. Each run seeds each compiler from its own predecessor's optimized code, so the two chains stay independent. The general-architecture compiler differs in two respects. First, its prompt builds on the branched-architecture compiler. Therefore, it contains function names that Opus 4.7 chose. Since the prompt is kept identical for an unbiased comparison, these names are absent from the Fable 5 seed, and Fable 5 matched the prompt's descriptions to its own functions by their role. Second, the acceptance-matrix graphs of the general-architecture prompt were drawn independently from the same construction rules, so the two runs did not necessarily test against identical instances during generation. The evaluation itself uses the same graphs for both runs. Neither difference favors Fable 5.

\subsection{Generation and refinement}
\label{sec:comparison:gen}

\paragraph{Emitted compilers.} All emitted compilers pass every acceptance test of all three stages on the first attempt. Both emitted linear-architecture compilers implement every block of the prompt and fill the timesteps strictly in program order. Fable 5's also implements the special case of an endpoint that shares its vertex with its adjacent intermediate, which Opus 4.7's replaces with a full cycle. However, contrary to the prompt, Fable 5's compiler places the qubit pairs on every second vertex instead of on consecutive ones. Both emitted branched-architecture compilers omit the same part of the specification, the per-iteration choice among the swap, stack, and park-the-endpoint methods, and substitute a simpler fast attempt backed by the rotor-conveyor. In both, intermediates reach a stack only when the routing pushes blocking chains aside, and never through the prompt's stack method. The emitted general-architecture compiler of Fable 5 includes the cycle-detour and weighs it against evicting the intermediates, whereas Opus 4.7's has no cycle-detour, as \autoref{sec:general:first} describes. No compiler of Fable 5 depends on the hash seed, unlike the branched- and general-architecture compilers of Opus 4.7 described in \autoref{sec:branched:first}.

\paragraph{Follow-up refinement.} The follow-up prompts change the compilers of the two LLMs in different ways. On the linear architecture, both optimized compilers first improve the grouping, whose emitted versions only pack consecutive operations into a timestep. Both now place every operation in the earliest timestep its dependencies allow, Opus 4.7's within a limited window of the following operations, as described in \autoref{sec:linear:opt}, and Fable 5's without such a limit. Beyond this, Opus 4.7 adds only the mirror trial, the widened reduction passes, and the special-case routine of \autoref{sec:linear:opt}. On circuits of up to 5,000 gates, Fable 5's optimized compiler runs two variants of its router, which differ in whether a qubit split from its partner for a single-qubit gate is merged back with it afterward. When the default schedule contains swaps, it also tries five further placements and keeps the shortest schedule. On larger circuits, it instead reorders the matched pairs of the chain extraction of \autoref{sec:linear:spec}, so that pairs with many gates between them lie close together along the line.

On the branched architecture, the optimized compilers of both LLMs add the park-the-endpoint method that their emitted versions omit. Beyond this method, both search over placements, but on different scales. Opus 4.7's compiler runs the placement search of \autoref{sec:branched:opt} with up to 150 compilations. Fable 5's routes at most three placements with four variants of its router each and keeps the shortest of these schedules, but prefers a swap-free schedule of the default placement over shorter ones with swaps from the other placements.

The two general-architecture compilers diverge most. Opus 4.7's optimizations add the compile-time budget of \autoref{sec:general:opt}, which trades schedule length for compilation speed. Fable 5's optimizations add a further candidate placement and strengthen the fallback of the routing. The new placement orders the qubits so that frequently interacting ones follow each other and fills the vertices around the gate vertex in this order. When the regular moves cannot clear the path of a gate's qubits, the fallback pushes the blocking chains aside one after another. In its guaranteed-delivery routine, which runs only when all other methods fail, the optimized compiler raises the limit on such pushes from the number of vertices plus four to four times the number of vertices.

\begin{figure*}[t]
	\centering
	\includegraphics{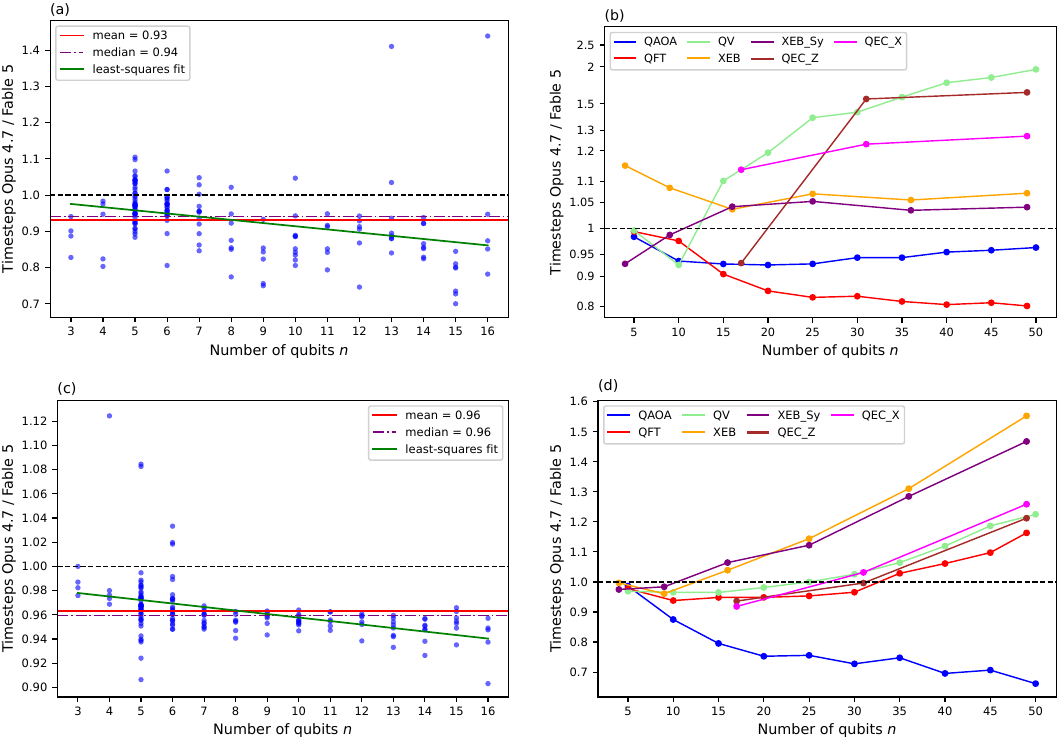}
	\caption{Timesteps of the optimized specialized compilers of Opus 4.7 divided by those of Fable 5. (a) Linear architecture, circuit library, one dot per circuit, with the mean, the median, and a least-squares fit as red, purple, and green lines. (b) Linear architecture, scalable families, one circuit per qubit count. (c) Branched architecture, circuit library, for each circuit the median over all tested configurations, with lines as in (a). (d) Branched architecture, scalable families, the median over all tested configurations at each qubit count. Above the dashed line at 1, Fable 5 needs fewer timesteps. The vertical axis of (b) is stretched around 1.}
	\label{fig:llm-comparison}
\end{figure*}

\begin{table*}[t]
	\centering
	\caption{Timesteps of the optimized linear- and branched-architecture compilers of both LLMs against the hand-crafted baselines, per architecture for the circuit library and each scalable family. Each suite consists of $N$ instances, each a pair of circuit and architecture, and the compilers of both LLMs receive identical inputs. Per LLM, the table lists the median timestep reduction factor, that is, the baseline's timesteps divided by the compiler's, the proportion of instances on which this factor exceeds 1, and the timesteps summed over all instances, rounded to thousands. The better median and proportion per row and the lowest total per row are in bold.}
	\label{tab:model-comparison}
	\setlength{\tabcolsep}{4pt}
	\scriptsize
	\begin{tabular*}{\textwidth}{@{\extracolsep{\fill}}lrrrrrrrr@{}}
		\toprule
		& & \multicolumn{3}{c}{Opus 4.7} & \multicolumn{3}{c}{Fable 5} & Baselines \cite{Wagner22, KreppelMWHPSB24} \\
		\cmidrule(lr){3-5}\cmidrule(lr){6-8}\cmidrule(l){9-9}
		Suite & $N$ & Median & ${>}1$ & Total & Median & ${>}1$ & Total & Total \\
		\midrule
		Linear, library & 153 & \textbf{1.19} & \textbf{93.5\,\%} & \textbf{12,475k} & 1.12 & 79.7\,\% & 15,167k & 16,737k \\
		Linear, QAOA & 10 & \textbf{1.24} & 100.0\,\% & \textbf{4k} & 1.16 & 100.0\,\% & 5k & 6k \\
		Linear, QFT & 10 & \textbf{3.40} & 100.0\,\% & \textbf{79k} & 2.75 & 100.0\,\% & 96k & 267k \\
		Linear, QV & 10 & 1.16 & 100.0\,\% & 806k & \textbf{1.63} & 100.0\,\% & \textbf{463k} & 930k \\
		Linear, XEB & 6 & 1.40 & 100.0\,\% & 1,480k & \textbf{1.50} & 100.0\,\% & \textbf{1,390k} & 2,144k \\
		Linear, XEB\_Sy & 6 & 1.48 & 83.3\,\% & 2,734k & \textbf{1.54} & 83.3\,\% & \textbf{2,631k} & 4,544k \\
		Linear, QEC\_Z & 3 & 1.33 & 100.0\,\% & 46k & \textbf{1.96} & 100.0\,\% & \textbf{30k} & 58k \\
		Linear, QEC\_X & 3 & 1.62 & 100.0\,\% & 39k & \textbf{1.83} & 100.0\,\% & \textbf{31k} & 67k \\
		\midrule
		Branched, library & 1,906 & \textbf{1.18} & 92.5\,\% & \textbf{196,614k} & 1.13 & \textbf{95.1\,\%} & 202,396k & 223,853k \\
		Branched, QAOA & 1,045 & \textbf{0.66} & 0.9\,\% & 1,199k & 0.48 & \textbf{1.2\,\%} & 2,402k & \textbf{723k} \\
		Branched, QFT & 1,042 & 1.21 & 84.4\,\% & 21,460k & \textbf{1.26} & \textbf{98.6\,\%} & \textbf{17,650k} & 22,943k \\
		Branched, QV & 1,045 & 1.17 & 76.3\,\% & 50,534k & \textbf{1.21} & \textbf{89.6\,\%} & \textbf{42,105k} & 49,324k \\
		Branched, XEB & 512 & 1.07 & 58.8\,\% & 184,273k & \textbf{1.26} & \textbf{90.6\,\%} & \textbf{120,752k} & 143,729k \\
		Branched, XEB\_Sy & 512 & 1.23 & 72.7\,\% & 328,129k & \textbf{1.41} & \textbf{93.9\,\%} & \textbf{230,214k} & 302,262k \\
		Branched, QEC\_Z & 374 & 1.26 & 77.5\,\% & 4,229k & \textbf{1.28} & \textbf{89.3\,\%} & \textbf{3,462k} & 4,189k \\
		Branched, QEC\_X & 374 & 1.24 & 76.2\,\% & 4,295k & \textbf{1.28} & \textbf{92.5\,\%} & \textbf{3,399k} & 4,197k \\
		\bottomrule
	\end{tabular*}
\end{table*}

\subsection{Evaluation}
\label{sec:comparison:eval}

The evaluation first compares the emitted compilers of the two LLMs and their gains from the follow-up prompts, then the optimized compilers with each other and with the hand-crafted baselines. Afterward, it turns to the general-architecture compilers on the architecture matrix and to compile time and memory.

\paragraph{Emitted compilers and their refinement.} The emitted compilers of Opus 4.7 produce fewer timesteps on most circuits. The contrast is clearest for the linear-architecture compiler, where Fable 5's emitted compiler needs 16\,\% more timesteps at the median on the circuit library, and more on every scalable family, by median factors from 1.33 on QV up to 3.47 on QFT.

However, on the linear and the general architecture, the compilers of Fable 5 improve more under the follow-up prompts. For the linear architecture, its optimized compiler shortens the schedules of its emitted one by a median factor of 1.25 on the circuit library and of 1.84 to 3.74 across the scalable families, against 1.16 and 1.28 to 1.60 for Opus 4.7. After optimization, Fable 5 thus closes most of the gap of its emitted linear-architecture compiler. For the general architecture, the optimized compiler of Fable 5 needs fewer timesteps than its emitted one, unlike that of Opus 4.7.

\paragraph{Optimized compilers of Opus 4.7 versus Fable 5.} The comparison of the two optimized linear- and branched-architecture compilers depends on the qubit count. The compiler of Opus 4.7 produces the shorter schedule on most of the circuit library, with 3 to 16 qubits. It is ahead on 82\,\% of the library circuits on the linear architecture and on 86\,\% of the pairs of library circuit and configuration on the branched architecture, at median margins of 6\,\% and 5\,\%. The factors per circuit appear in \autoref{fig:llm-comparison}(a) and \autoref{fig:llm-comparison}(c), on the branched architecture as the median over all tested configurations for each circuit.

The compiler of Fable 5 is mostly ahead on the scalable families, with up to 50 qubits. For the linear architecture, it needs fewer timesteps on every circuit of XEB and QEC\_X and on QV, XEB\_Sy, and QEC\_Z apart from their smallest qubit counts. The compiler of Opus 4.7 needs fewer on every circuit of QAOA and QFT, as \autoref{fig:llm-comparison}(b) shows.

For the branched architecture, the compiler of Fable 5 needs fewer timesteps on most configurations for every family apart from QAOA. Its lead builds up with the qubit count. At the smallest qubit counts, the compiler of Opus 4.7 is still slightly ahead on most families, and at the largest ones, the compiler of Fable 5 leads at median factors between 1.16 and 1.55, as \autoref{fig:llm-comparison}(d) shows. On QAOA, the compiler of Opus 4.7 stays ahead throughout. The separation of placement and routing in \autoref{sec:mapping} traces the lead of Fable 5 to the routing.

\paragraph{Against the hand-crafted baselines.} On the linear architecture, neither compiler is ahead of the other throughout, as \autoref{tab:model-comparison} shows. Against the baseline, Opus 4.7's compiler gains more on the circuit library, QAOA, and QFT, and Fable 5's on the other five families. The per-circuit factors on the circuit library appear in \autoref{fig:lin-final-hand}(a) for Opus 4.7 and in \autoref{fig:llm-fable}(a) in \appref{app:figures} for Fable 5. The two compilers differ most on QFT, where Opus 4.7's compiler stays ahead at every qubit count, with peak factors of 3.55 against 2.98 at 30 qubits, and on the largest QV circuits, where Fable 5's compiler reaches 2.19 against 1.12 at 50 qubits, as \autoref{fig:lin-final-hand}(b) for Opus 4.7 and \autoref{fig:llm-fable}(b) for Fable 5 show.

On the branched architecture at stack height 1 and junction distance 1, both stay slightly ahead of the baseline on the circuit library, at median factors of 1.07 and 1.04, as \autoref{fig:branched-lib}(a) and \autoref{fig:llm-fable}(c) show. Over all tested configurations, Fable 5's compiler is the stronger one on the scalable families apart from QAOA and beats the baseline on a larger proportion of the configurations for every family, as \autoref{tab:model-comparison} shows. For QAOA, the baseline is ahead of both compilers on almost every configuration, whereas on QFT, Fable 5's compiler beats the baseline on nearly every configuration at every qubit count. On the other families, and for Opus 4.7's compiler also on QFT, the proportions fall with the qubit count once it exceeds about 15 to 20 qubits, those of Opus 4.7's compiler far more steeply, as \autoref{fig:branched-lib}(b) and, for Fable 5, \autoref{fig:llm-fable}(d) show. At the largest qubit counts, Fable 5's compiler still outperforms the baseline on 80\,\% to 99\,\% of the configurations for each family apart from QAOA, whereas the proportions of Opus 4.7's compiler drop to between 42\,\% and 74\,\%.

\def\TabLlmRuntime{%
Linear, library & 153 & \textbf{2.01} & 98.7\,\% & 1,248 & 1.69 & \textbf{100.0\,\%} & \textbf{104} & 1,980 \\
Linear, QAOA & 10 & 1.88 & 100.0\,\% & 0.5 & \textbf{8.36} & 100.0\,\% & \textbf{0.1} & 0.9 \\
Linear, QFT & 10 & 2.30 & 100.0\,\% & 15.2 & \textbf{2.31} & 100.0\,\% & \textbf{8.6} & 30.1 \\
Linear, QV & 10 & 0.27 & 20.0\,\% & 969 & \textbf{10.34} & \textbf{100.0\,\%} & \textbf{8.4} & 109 \\
Linear, XEB & 6 & 0.34 & 33.3\,\% & 1,322 & \textbf{13.36} & \textbf{100.0\,\%} & \textbf{18.6} & 222 \\
Linear, XEB\_Sy & 6 & 0.36 & 16.7\,\% & 2,881 & \textbf{14.24} & \textbf{100.0\,\%} & \textbf{32.6} & 452 \\
Linear, QEC\_Z & 3 & 0.17 & 0.0\,\% & 54.9 & \textbf{0.91} & 0.0\,\% & 5.9 & \textbf{5.3} \\
Linear, QEC\_X & 3 & 0.22 & 0.0\,\% & 31.1 & \textbf{0.95} & 0.0\,\% & 5.9 & \textbf{5.8} \\
\midrule
Branched, library & 1,906 & 0.046 & 0.0\,\% & 803,173 & \textbf{0.84} & \textbf{42.0\,\%} & 44,425 & \textbf{26,845} \\
Branched, QAOA & 1,045 & 0.0068 & 0.0\,\% & 33,199 & \textbf{0.67} & \textbf{38.0\,\%} & 431 & \textbf{143} \\
Branched, QFT & 1,042 & 0.017 & 0.0\,\% & 283,371 & \textbf{0.68} & \textbf{29.7\,\%} & 10,268 & \textbf{3,151} \\
Branched, QV & 1,045 & 0.020 & 0.0\,\% & 570,284 & \textbf{0.39} & \textbf{5.5\,\%} & 41,876 & \textbf{7,513} \\
Branched, XEB & 512 & 0.0086 & 0.0\,\% & 2,371,380 & \textbf{0.19} & \textbf{1.0\,\%} & 172,197 & \textbf{13,324} \\
Branched, XEB\_Sy & 512 & 0.0094 & 0.0\,\% & 4,528,821 & \textbf{0.23} & \textbf{1.0\,\%} & 307,253 & \textbf{25,670} \\
Branched, QEC\_Z & 374 & 0.0074 & 0.0\,\% & 78,388 & \textbf{0.18} & \textbf{5.6\,\%} & 5,272 & \textbf{373} \\
Branched, QEC\_X & 374 & 0.0074 & 0.0\,\% & 76,926 & \textbf{0.18} & \textbf{4.8\,\%} & 5,126 & \textbf{377} \\}
\begin{table*}[t]
	\centering
	\caption{Compile time of the optimized linear- and branched-architecture compilers of both LLMs against the hand-crafted baselines, per architecture for the circuit library and each scalable family. Each suite consists of $N$ instances, each a pair of circuit and architecture, and the compilers of both LLMs receive identical inputs. Per LLM, the table lists the median compile-time reduction factor, that is, the baseline's compile time divided by the compiler's, the proportion of instances on which this factor exceeds 1, and the compile time summed over all instances in seconds. Each value of an instance is the mean over ten runs. The better median and proportion per row and the lowest total per row are in bold.}
	\label{tab:llm-runtime}
	\setlength{\tabcolsep}{4pt}
	\scriptsize
	\begin{tabular*}{\textwidth}{@{\extracolsep{\fill}}lrrrrrrrr@{}}
		\toprule
		& & \multicolumn{3}{c}{Opus 4.7} & \multicolumn{3}{c}{Fable 5} & Baselines \cite{Wagner22, KreppelMWHPSB24} \\
		\cmidrule(lr){3-5}\cmidrule(lr){6-8}\cmidrule(l){9-9}
		Suite & $N$ & Median & ${>}1$ & Total (s) & Median & ${>}1$ & Total (s) & Total (s) \\
		\midrule
		\TabLlmRuntime
		\bottomrule
	\end{tabular*}
\end{table*}

\paragraph{Architecture matrix.} The qualitative findings of \autoref{sec:compiler:general} hold in the Fable 5 run. For both LLMs the subdivided \texttt{butterfly} is the best on the circuit library and on every scalable family, and the plain \texttt{cycle} and the linear architecture are the worst overall, with best-to-worst spreads of up to 7.78 for Opus 4.7 and 8.78 for Fable 5, so the first-order effect of the architecture's connectivity is not an artifact of one LLM.

For each run we take the general-architecture compiler in the version that produces the shortest schedules: the optimized one for Fable 5 and the emitted one for Opus 4.7, since its optimizations trade schedule length for compilation speed. Compared by the total timesteps on the circuit library and on each scalable family, the two split along connectivity. Fable 5's compiler needs fewer timesteps on most suites of the sparse \texttt{two\_cycles\_bridged}, \texttt{theta}, and \texttt{figure8}. Opus 4.7's leads on most suites of \texttt{cycle\_with\_two\_chords} and on every suite of the three densest architectures and the branched architecture. On the plain \texttt{cycle} and on the linear architecture, each leads on half of the suites.

Against its own optimized specialized compiler on the linear architecture, Fable 5's general-architecture compiler needs about as many timesteps at the median on the circuit library, whereas Opus 4.7's needs 14\,\% more, as \autoref{fig:llm-fable}(e) and \autoref{fig:gen-vs-spec}(a) show. On the scalable families, both need more timesteps than their specialists by similar margins, by median factors of 1.28 to 2.54 for Fable 5 and 1.18 to 2.66 for Opus 4.7, with a wider gap for Fable 5 on QV and the two QEC families, as \autoref{fig:llm-fable}(f) shows. On the branched architecture, over all tested configurations, Fable 5's general-architecture compiler stays even closer to its specialist, within 5\,\% at the median per scalable family, whereas that of Opus 4.7 needs 4\,\% to 10\,\% more.

Against the hand-crafted baselines, both general-architecture compilers stay close to the baseline on the circuit library of the linear architecture, for which neither is tuned, at median timestep reduction factors of 1.03 for Opus 4.7's and 1.10 for Fable 5's compiler. On the scalable families of this architecture, both fall below the baseline on QV and XEB\_Sy, Opus 4.7's also on XEB and Fable 5's also on QEC\_Z and QEC\_X. On the scalable families of the branched architecture apart from QAOA, over all tested configurations, Fable 5's compiler beats the baseline at median factors of 1.17 to 1.33, against 1.04 to 1.19 for Opus 4.7's. On QAOA, both compilers fall below the baseline on both architectures, to median factors of 0.47 to 0.59.

\paragraph{Compile time and memory.} The compile times differ far more between the compilers of Opus 4.7 and Fable 5 than their timesteps do, as \autoref{tab:llm-runtime} shows. The absolute compile times and peak memory of all compilers of both LLMs follow in \autoref{fig:memory} and \autoref{fig:llm-runtime-fable} in \appref{app:figures}.

On the linear architecture, the compiler of Fable 5 is faster than the baseline on the circuit library and on every family apart from QEC\_Z and QEC\_X, whereas the compiler of Opus 4.7 is faster only on the circuit library, QAOA, and QFT. The difference comes from the routing itself. Already as emitted, Fable 5's compiler is 4.71 times faster than that of Opus 4.7 on the scalable families at the median, and its optimized compiler stays faster although it runs up to twelve variants on circuits of up to 5,000 gates, against two for Opus 4.7's. The slowest compilation takes 19\,s for Fable 5's compiler, 1,945\,s for Opus 4.7's, and 269\,s for the baseline, all on the largest XEB\_Sy circuit.

On the branched architecture, the search makes the difference. Opus 4.7's compiler routes up to 150 candidate placements, Fable 5's at most twelve combinations of placement and router variant. Over all tested configurations, both are slower than the baseline at the median, the compiler of Opus 4.7 by one to two orders of magnitude and that of Fable 5 by at most a factor of 5.58, as \autoref{tab:llm-runtime} shows. At stack height 1 and junction distance 1, the compiler of Fable 5 is even faster than the baseline at the median on the circuit library and on every family. There, its slowest compilation takes 179\,s, against 3,565\,s for Opus 4.7's compiler and 133\,s for the baseline.

The optimized general-architecture compiler of Fable 5 is also faster than that of Opus 4.7 at the median, by factors of 3.34, 5.05, and 6.28 on the circuit library of the linear, the branched, and the cyclic architectures, and of 1.34, 4.16, and 5.97 on their scalable families. The exceptions are the largest circuits. On these, the general-architecture compiler of Opus 4.7, whose optimizations trade schedule length for compilation speed, is often the faster one, on the circuit library from about 20,000 gates on the linear and about 50,000 gates on the branched and the cyclic architectures. On the largest circuit, with 294,000 gates, it is faster by a factor of 7.80 on the linear architecture, of up to 9.47 over the tested configurations of the branched architecture, and of 3.07 to 9.43 on the cyclic architectures. Among the scalable families, it is faster at the median on XEB\_Sy on all three, on XEB on the linear and the branched architecture, and on QV and QEC\_Z on the linear architecture.

The peak memory differs much less between the compilers of Opus 4.7 and Fable 5 than the compile time. On the circuit library, the median peak memory of the compilers of Fable 5 lies within 7\,\% of that of Opus 4.7 on the linear and the cyclic architectures and 16\,\% below it on the branched architecture. On the scalable families, it lies 38\,\% to 43\,\% below that of Opus 4.7. The exceptions are QAOA and QFT on the linear architecture, with up to 5\,\% more, and the two XEB families on the branched and the cyclic architectures, with up to 32\,\% more. The optimized specialized compilers of Fable 5 need at most 0.5\,GB on the linear and 1.9\,GB on the branched architecture, and no compiler of Fable 5 needs more than 2.5\,GB, against 2.7\,GB for Opus 4.7.

In sum, the two frontier LLMs show distinct capability profiles under an identical protocol. Opus 4.7 writes the stronger emitted compilers and retains a small advantage on small circuits, whereas the compilers of Fable 5 start weaker, gain more from the follow-up prompts on the linear and the general architecture, and end more competitive against the hand-crafted baselines on the large circuits where shuttling costs dominate. The qualitative claims of the preceding sections hold in both runs, and the two chains deviate from the branched-architecture specification in the same place and recover the same one of its three methods. This indicates that the methodology, rather than one particular LLM, carries the result.

\section{Initial mapping versus routing}
\label{sec:mapping}

Every compiler of this paper combines its own initial mapping with its own routing, so the comparisons of the previous sections measure whole compilers. For Opus 4.7, for instance, the optimized linear-architecture compiler keeps the better of the default placement and its mirror image, and the optimized branched-architecture and general-architecture compilers route the most promising of many candidate placements. Therefore, these comparisons leave open whether a gain over the baselines comes from the placement or from the routing. To separate the two, we run every router from the initial placement of every other compiler on the same architecture and circuit.

\subsection{Experimental design}
\label{sec:mapping:design}

\paragraph{Design.} The experiment is a crossed design of placements and routers. On the linear and the branched architecture, the nine routers are the eight generated compilers, the specialized and the general-architecture compiler of both LLMs, each as emitted and optimized, and the hand-crafted baseline of the architecture \cite{Wagner22, KreppelMWHPSB24}. On the eight cyclic architectures, the routers are the four general-architecture compilers. Each router compiles every circuit from the placement of each of these compilers and from three random placements, which keep the baseline's occupancy of the vertices but assign the qubits uniformly at random. Since the cyclic architectures have no baseline, their random placements keep the occupancy of the emitted Opus 4.7 compiler. This results in a matrix of $9 \times 12$ compilations per circuit on the linear and branched architectures and of $4 \times 7$ on the cyclic ones. The experiment covers the 153 library circuits and the 48 circuits of the scalable families on the linear architecture and on each cyclic architecture, and on every tested configuration of the branched architecture.

\begin{figure*}[t]
	\centering
	\includegraphics{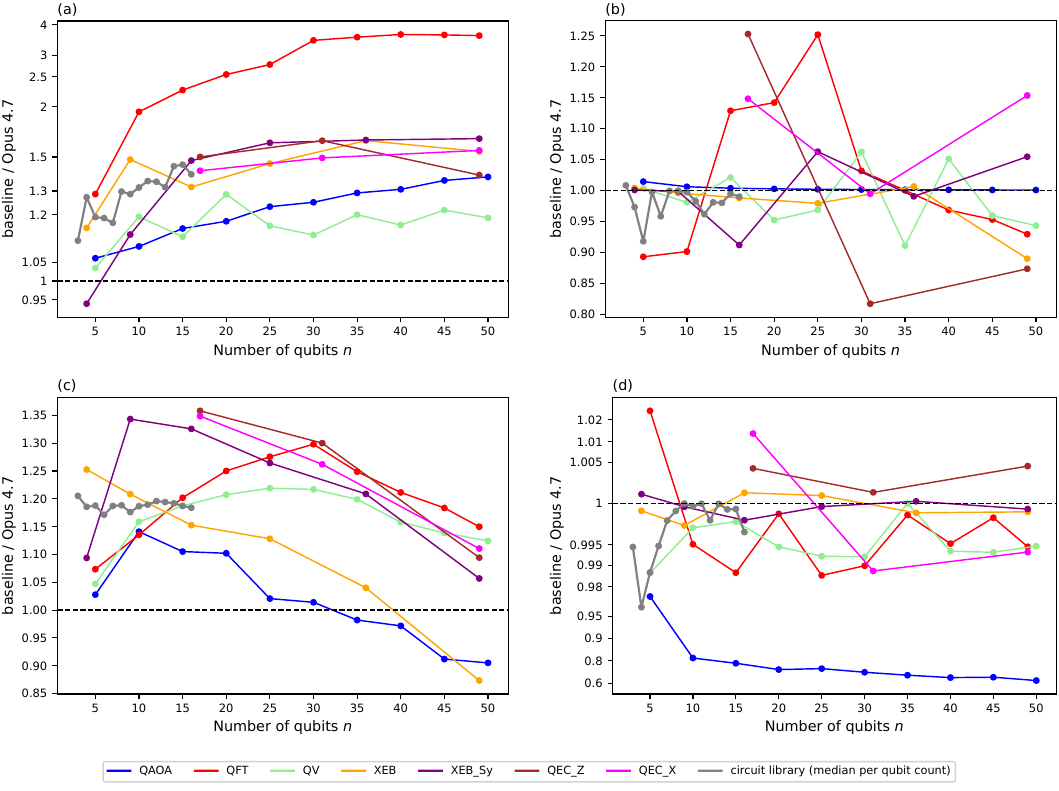}
	\caption{Routing and placement parts of the timestep factor of the hand-crafted baselines \cite{Wagner22, KreppelMWHPSB24} over the optimized specialized compilers of Opus 4.7 per scalable family and qubit count, with the median of the circuit library per qubit count, 3 to 16, as a gray line. (a) Linear architecture, routing part. (b) Linear architecture, placement part. (c) Branched architecture, routing part. (d) Branched architecture, placement part. Each panel divides the timesteps of the baseline by those of the compiler of Opus 4.7, so values above the dashed line at 1 favor the compiler of Opus 4.7. The routing part compares the two routers from the placement of the compiler of Opus 4.7, and the placement part the two placements under the baseline's router. On the branched architecture, each value is the median over all tested configurations. The vertical axes of (a) and (d) are stretched around 1.}
	\label{fig:mapping-split}
\end{figure*}

\paragraph{Procedure.} The placement of each compiler is the one behind its reported result. Both the generated and the hand-crafted compilers receive the given placement instead of computing their own and otherwise run unchanged. Each compilation is a separate deterministic run with the constant hash seed. All factors in this section compare such routes from given placements, including each compiler's own placement. From its own placement, each generated compiler reproduces its own result on all but 44 of 62,520 compilations. In these exceptions, three Fable 5 compilers that otherwise select placement and router variant jointly choose the router variant slightly differently for a single given placement, which changes the schedule in either direction by at most 1.8\,\%. A failure of a router from a foreign placement counts as a result of the experiment and is not discarded, since it shows that the router depends on the kind of placement it was written for. Foreign placements also reveal two special cases, which we treat separately. The optimized linear-architecture compiler of Fable 5 contains a bookkeeping error that makes it fail after a complete routing from some foreign placements. We correct this error, which leaves every other result unchanged. The branched-architecture baseline is designed for stacks that are filled from the junction on without gaps, as its own placements and operations always leave them. A foreign placement with a gap in a stack lies outside this design, so the baseline can misjudge the free space and move qubits onto occupied segments. We stop such a compilation as soon as this happens and count it separately from the routing failures.

\paragraph{Evaluation.} Each comparison divides the timesteps of one compiler by those of another, for instance of a baseline by those of a generated compiler. This timestep factor decomposes into a \emph{routing part}, which compares both routers from the placement of the second compiler, and a \emph{placement part}, which compares both placements under the router of the first one. A part of 1 means that the routing or the placement, respectively, does not change the number of timesteps. Where one compiler or placement is better, the text states its advantage as a factor above 1. We apply this decomposition to four comparisons of the previous sections, the hand-crafted baselines against the generated compilers, the emitted against the optimized compilers, the general-architecture against the specialized compilers, and the compilers of Opus 4.7 against those of Fable 5. The random placements show what each compiler's placement strategy gains over an uninformed one. As elsewhere in the paper, we report the circuit library and each scalable family separately.

\subsection{Results}
\label{sec:mapping:results}

\paragraph{Generated compilers versus baselines.} The advantage of the generated compilers over the hand-crafted baselines comes almost entirely from their routing. Per qubit count, the decomposition of the factor of each baseline over the optimized specialized compiler appears in \autoref{fig:mapping-split} for Opus 4.7 and in \autoref{fig:mapping-split-fable} in \appref{app:figures} for Fable 5, for the circuit library as a median and for each scalable family. On the circuit library, the median routing part is 1.13 to 1.26 for both LLMs on both architectures, and the median placement part 0.98 to 1.00. On the scalable families, the routing part mostly grows with the qubit count on the linear architecture, up to 3.64 on QFT for Opus 4.7 and 2.82 on QEC\_Z for Fable 5, and, apart from the exceptions below, reaches 1.06 to 1.40 at the largest qubit counts on the branched architecture. There, the baseline's router is ahead of that of Opus 4.7 on QAOA and XEB, by factors of 1.10 and 1.14, and ahead of that of Fable 5 only on QAOA, by 1.96.

The placement part stays within about 10\,\% of 1, except on QAOA on the branched architecture and on some families on the linear architecture, namely QFT and QEC and, for Fable 5, the two XEB families. On these linear-architecture families, it varies between 0.69 and 1.25 without a trend. On QFT and QEC, the placement matters, as the random placements below show, but apart from Fable 5's placement on QFT from 15 qubits on, which is better by factors of 1.13 to 1.22, neither the baseline's nor the generated compiler's placement is consistently the better one. QAOA on the branched architecture is the only family on which the baseline consistently gains through its placement. The baseline places these chain-structured circuits increasingly well for its own routing, by a factor that grows with the qubit count to 1.59 against the placements of Opus 4.7 at 50 qubits and to 1.50 against those of Fable 5 at 45 qubits.

\paragraph{Emitted versus optimized compilers.} The follow-up optimizations also gain mostly through the routing. Here, the emitted compiler takes the place of the baseline. On the linear architecture, the median routing part is 1.13 for Opus 4.7 and 1.18 for Fable 5 on the circuit library, and 1.25 to 1.40 and 1.84 to 3.11 on the scalable families. For Opus 4.7, the median placement part is 1.00 on the circuit library and at most 1.08 on the scalable families, apart from two single QEC circuits at 1.18 and 1.60. This confirms that its gain comes from the routing, mainly the grouping, rather than from the mirror trial, as \autoref{sec:linear:eval} states. Only Fable 5's placements add more on a few families, up to 1.51 on QFT.

On the branched architecture over all tested configurations, the median routing part on the scalable families is 1.33 to 2.31 for Opus 4.7 and 1.39 to 1.80 for Fable 5. Under the emitted router, the optimized placements barely help, with median placement parts within 3\,\%, apart from 1.15 on QAOA for Opus 4.7. The search of Opus 4.7's compiler over up to 150 candidate placements instead selects placements that suit its optimized router. Under that router, the emitted placements need 3\,\% to 25\,\% more timesteps on the scalable families than the selected ones. At stack height 1 and junction distance 1, where both routers need the same number of timesteps from the optimized placement on 180 of the 201 circuits, this search accounts for the whole improvement at the median.

For the general-architecture compilers, the decomposition reflects their different follow-up changes. The emitted and the optimized router of Opus 4.7 need the same number of timesteps at the median on every architecture, as expected from a change that only caps the placement search. The placements of the emitted compiler are better by up to 9\,\%, so the cap costs a little. In Fable 5's compiler, the stronger fallback improves the routing by up to 12\,\% on the linear architecture and by less than 1\,\% on the others, and the added placement matters only on QAOA, by factors of 2.69 on the linear and 1.04 on the branched architecture.

\paragraph{General-architecture versus specialized compilers.} Apart from QAOA on the linear architecture, the gap of the general-architecture compiler to the optimized specialized one, which \autoref{sec:compiler:general} reports, also lies in the routing. On the linear architecture, the router of Opus 4.7's specialized compiler is better than that of its general-architecture compiler as emitted, by median factors of 1.20 on the circuit library and 1.18 to 2.08 on the scalable families. For Fable 5, the two routers match on the circuit library, and the specialized one is better by 1.45 to 2.99 on the scalable families. For both LLMs, the median placement part lies between 0.68 and 1.11, so neither placement is consistently better, except on QAOA, where the placement of the specialized compiler is better by a factor of 2.25 for Opus 4.7 and 1.59 for Fable 5.

On the branched architecture over all tested configurations, the router of Opus 4.7's general-architecture compiler as emitted needs 9\,\% to 43\,\% more timesteps on the scalable families than the specialized one from the specialized compiler's placement. Its own placements, better by median factors of 1.02 to 1.21, recover much of this, so that the whole compiler needs only 4\,\% to 9\,\% more.

\paragraph{Opus 4.7 versus Fable 5.} Fable 5's compilers owe their lead on the larger circuits to their routing, as \autoref{fig:mapping-split-llm} in \appref{app:figures} shows. On the linear architecture, the lead of Fable 5 is largest on the QV circuit with 50 qubits, where the router of Opus 4.7 needs twice as many timesteps as that of Fable 5, and the placement of Opus 4.7 hardly changes this. On the branched architecture, at the largest qubit counts of every family apart from QAOA, the router of Opus 4.7 needs 35\,\% to 64\,\% more timesteps from the placement of Fable 5's compiler. Its own placements recover only part of this, by median factors of 1.03 to 1.27, so the routing decides the comparison.

Opus 4.7's router depends more on its own placement than Fable 5's router does. On the scalable families of the branched architecture, it needs 3\,\% to 22\,\% more timesteps from Fable 5's placements at the family median, whereas Fable 5's router needs at most 2\,\% more from those of Opus 4.7. This reflects the search of Opus 4.7's compiler, which selects its placements for its own router.

\paragraph{Foreign and random placements.} The decompositions above rest on routing from foreign placements. Comparing these with random placements shows how much a router depends on an informed placement at all. Per qubit count, the cost of both appears in \autoref{fig:mapping-placement} in \appref{app:figures}, for the circuit library as a median and for each scalable family. From the placement of another compiler, a router needs at most 7\,\% more timesteps at the median on the linear and the branched architecture. The QEC\_Z and QFT circuits on the linear architecture are the only exceptions, at up to 14\,\% and 26\,\%.

Random placements cost more, above all on structured circuits. On the linear architecture, they cost at most 10\,\% on the circuit library, apart from 26\,\% at 16 qubits, and on QV and the two XEB families. On QFT and QEC, they cost up to factors of 1.73 and 1.54, and on QAOA a factor that grows with the qubit count from 1.32 at 5 qubits to 7.35 at 50 qubits. On the branched architecture, where junctions let qubits pass one another, they cost at most 10\,\% on the circuit library and at most 24\,\% on every family.

\paragraph{Cyclic architectures.} On the cyclic architectures, the routing part mostly favors the same compiler as the timestep factor, as \autoref{fig:mapping-cyclic}(a) to \autoref{fig:mapping-cyclic}(c) in \appref{app:figures} show. However, on the sparser ones, the plain \texttt{cycle}, \texttt{two\_cycles\_bridged}, \texttt{theta}, and \texttt{figure8}, Opus 4.7's placements turn a routing advantage of Fable 5 into an overall advantage of Opus 4.7 on QAOA, and on all of them but \texttt{two\_cycles\_bridged} also on some further suites. On \texttt{two\_cycles\_bridged}, the router of Fable 5 is better on every suite apart from QAOA, by median routing parts of 1.05 to 1.85, and Opus 4.7's placements offset part of this advantage. On \texttt{butterfly}, \texttt{complete\_graph}, \texttt{hypercube}, and, apart from the circuit library, \texttt{cycle\_with\_two\_chords}, the router of Opus 4.7 is better, by median routing parts of up to 1.72, and its placements add to this advantage.

Under Opus 4.7's router, its own placements are at least as good as those of Fable 5 on every cyclic architecture, and better most clearly on the QAOA circuits of the plain \texttt{cycle}, by up to a factor of 2.29. As on the branched architecture, junctions let qubits pass one another, so random placements cost at most 24\,\% on every cyclic architecture with junctions. On the plain \texttt{cycle}, which has none, they cost up to a factor of 2.16 on QAOA, as \autoref{fig:mapping-cyclic}(d) shows.

\paragraph{Failures from foreign placements.} Every router handles almost all foreign placements. On the linear and the branched architecture, the generated routers fail on a single one of about 617,000 foreign placements. On the cyclic architectures, the two general-architecture compilers of Opus 4.7 never fail and those of Fable 5 fail on a single one of about 9,600 foreign placements. The linear-architecture baseline never fails. Because of the stack gaps described above, the branched-architecture baseline is stopped on 0.81\,\% of the foreign placements.

Thus, the experiment confirms that the gains of \autoref{sec:compiler:linear} to \autoref{sec:compiler:general} come mainly from the routing that the LLMs generate, whereas a reasonable placement mainly guards against the large losses of an uninformed one on structured circuits such as QAOA.

\section{Conclusion and outlook}
\label{sec:conclusion}

We have shown that frontier LLMs can generate the Python source code of three shuttling compilers for trapped-ion quantum computers from written specifications. The first compiler targets a linear segmented architecture, the second one an architecture with junctions and stack-shaped side branches, and the third one general connected architectures. All three are benchmarked against the published hand-crafted references \cite{Wagner22, KreppelMWHPSB24} on circuits with up to 50 qubits.

For the linear architecture, the compiler generated by Claude Opus 4.7 \cite{Anthropic26_2} emits up to 72\,\% fewer timesteps than the hand-crafted baseline, and for the branched architecture up to 43\,\% fewer, with median reduction factors of 1.19 and 1.18 on the circuit library, respectively. This advantage over the hand-crafted baselines comes almost entirely from the routing, not from the initial placement. A reasonable placement mainly guards against the large losses of a random one on structured circuits such as QAOA. Within the routing, the advantage most likely comes from where gate ordering is decided: the generated compilers select each gate from the ready set during shuttling and so can exploit the current ion placement, whereas the baselines follow a gate order determined by the circuit compiler \cite{KreppelMOWHPSB23}, which never sees that placement. The generated compilers thereby relax the otherwise strict boundary between circuit compilation and shuttling compilation.

The general-architecture compiler covers a broad class of architectures at the cost of more timesteps than the specialists need, up to 6.84 times as many on the linear architecture and 2.48 times as many on the branched one. Run across ten architectures, the general-architecture compiler exposes a first-order effect of connectivity, needing up to 90\,\% fewer timesteps on a dense, junction-rich layout than on a sparse one. The densest layouts are the subdivided \texttt{butterfly}, \texttt{complete\_graph}, and \texttt{hypercube}, and the sparsest are the plain \texttt{cycle} and the linear architecture. On such dense layouts the general-architecture compiler mostly emits fewer timesteps than each specialist does on its own architecture, by up to 83\,\% for the linear one and 41\,\% for the branched one, with similar margins over the hand-crafted baselines.

These results indicate that an unmodified frontier LLM, given a detailed specification, can produce working, correct, and competitive shuttling compilers for distinct trapped-ion architectures without any additional manual algorithmic engineering. Repeating the complete generation and evaluation with a second frontier LLM, Claude Fable 5 \cite{Anthropic26}, reproduces the qualitative findings, indicating that they follow from the methodology rather than from a single model, with the Fable 5 compilers surpassing the hand-crafted baselines more often on the largest circuits. Generating and testing a compiler this way takes a few days instead of the several months needed to hand-craft one. That low cost makes it worthwhile to build a dedicated compiler for each new target architecture, since on its own architecture a specialist compiler mostly outperforms the general one.

For trap design, the results translate into a concrete recommendation. With the same number of qubit segments and a single gate segment, the best junction-rich layout reduces the number of timesteps by up to a factor of 10.07 compared with the worst corridor-like one. Since every saved timestep shortens the run time on the QC and avoids transport-induced motional excitation, connectivity deserves the same attention in trap design as the compiler does.

We see several directions for future work. (i) Most importantly, we plan to extend the architecture to several gate segments with parallel gate execution, as realized on current QCCD processors \cite{PinoDFGMABFHMRN21, MosesB+23, Ransford+26}. (ii) We want to allow separation, merge, and swap at any segment rather than only at a single gate segment. (iii) Moreover, we intend to let a gate segment hold more than two ions, with individually addressed gates on any subset of them, executed in parallel \cite{Pogorelov+21}, and with native multi-qubit gates such as a Toffoli gate in a single laser interaction \cite{ShapiraSMAPGOS20, KatzFRMC23, PelegSNSASBO26}. (iv) Additionally, we would like to make the compilation hardware-aware. Fabricated traps are not homogeneous, and individual segments can differ in operation durations and fidelities, so a compiler may prefer a slower route when that route has a higher fidelity. Specifying such per-segment costs in the prompt is a natural extension of our approach.
(v) Finally, we aim to make the routing complete on every connected architecture. This last direction remains an open problem: the general-architecture compiler covers a broad class of connected architectures but not all of them, and closing the gap would incur the prohibitive cost of a complete search such as A* \cite{HartNR68}.

\begin{acknowledgments}
We thank the data center (ZDV) of Johannes Gutenberg University Mainz for the use of its Mogon NHR cluster for evaluating our generated compilers. We acknowledge funding by the German Federal Ministry of Research, Technology and Space (BMFTR) within the projects IQuAn, ATIQ, and SYNQ, and by the DFG Priority Programme SPP2514.
\end{acknowledgments}

\section*{Author Declarations}

\subsection*{Conflict of Interest}
The authors have no conflicts to disclose.

\subsection*{Author Contributions}
\textbf{Fabian Kreppel}: Conceptualization (equal); Data curation (lead); Formal analysis (lead); Investigation (lead); Methodology (equal); Software (lead); Validation (lead); Writing -- original draft (lead); Writing -- review \& editing (equal). \textbf{Reza Salkhordeh}: Investigation (supporting); Software (supporting). \textbf{Ferdinand Schmidt-Kaler}: Funding acquisition (equal); Supervision (supporting); Writing -- review \& editing (equal). \textbf{Andr\'e Brinkmann}: Conceptualization (equal); Funding acquisition (equal); Methodology (equal); Project administration (lead); Supervision (lead); Writing -- review \& editing (equal).

\section*{Data Availability}
The data that support the findings of this study are available from the corresponding author upon reasonable request.

\renewcommand\refname{References}
\bibliography{references}

\onecolumngrid
\appendix
\vspace*{0.425\textheight}
\section{Supplementary evaluation figures}
\label{app:figures}
This appendix collects the figures that complement the evaluation in \autoref{sec:linear:eval} to \autoref{sec:mapping}. The main text reports their key numbers, and the figures show the underlying data per circuit, scalable family, or architecture. They follow the order in which the main text refers to them.
\newpage

\begin{figure*}[p]
	\centering
	\includegraphics[width=\textwidth]{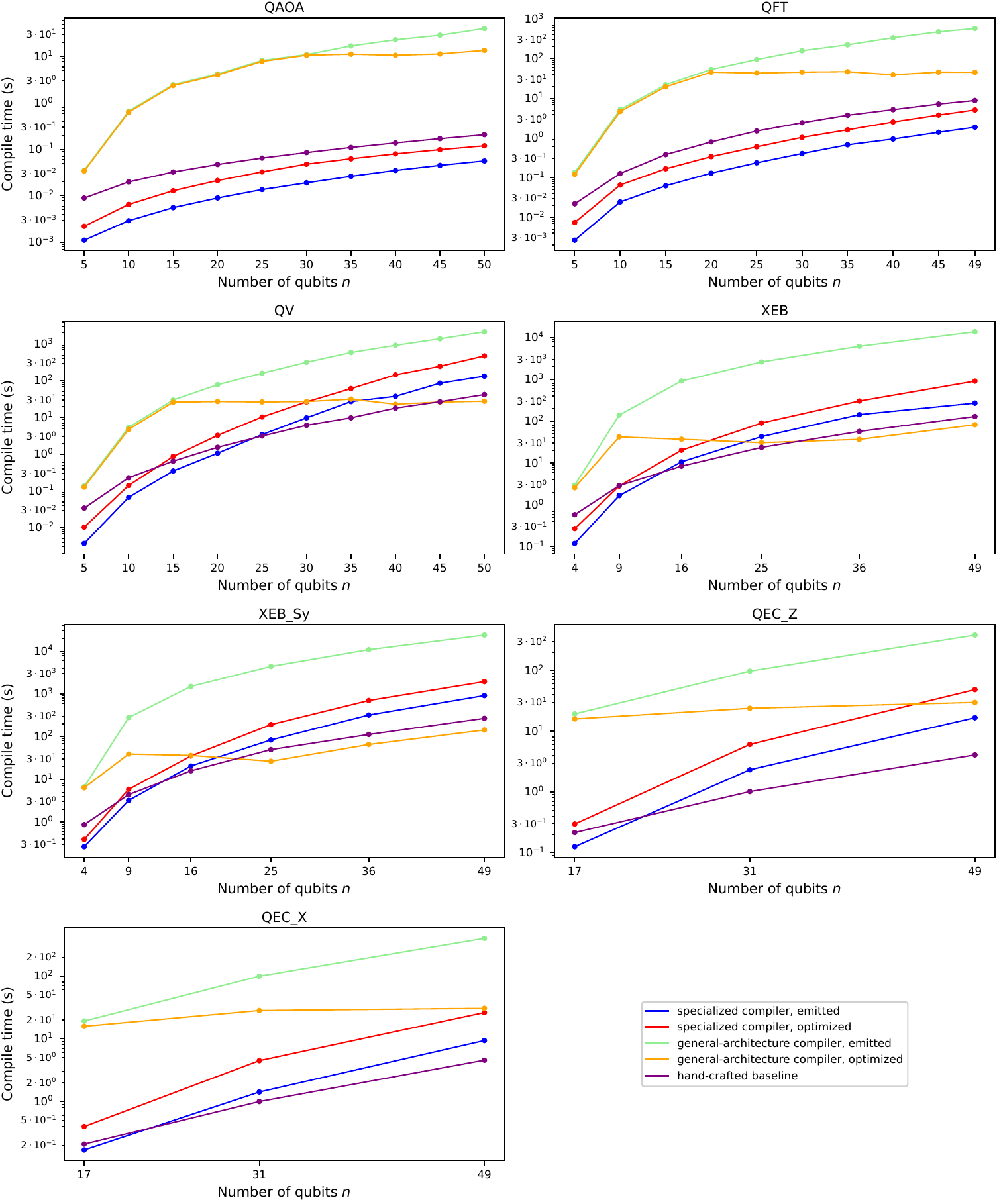}
	\caption{Compile time of the specialized and general-architecture compilers of Opus 4.7, each as emitted and optimized, and of the hand-crafted baseline \cite{Wagner22} on the linear architecture for the scalable families, one panel per family and one circuit per qubit count. Each value is the mean over ten runs.}
	\label{fig:compile-time-linear-fam}
\end{figure*}

\begin{figure*}[tp]
	\centering
	\includegraphics{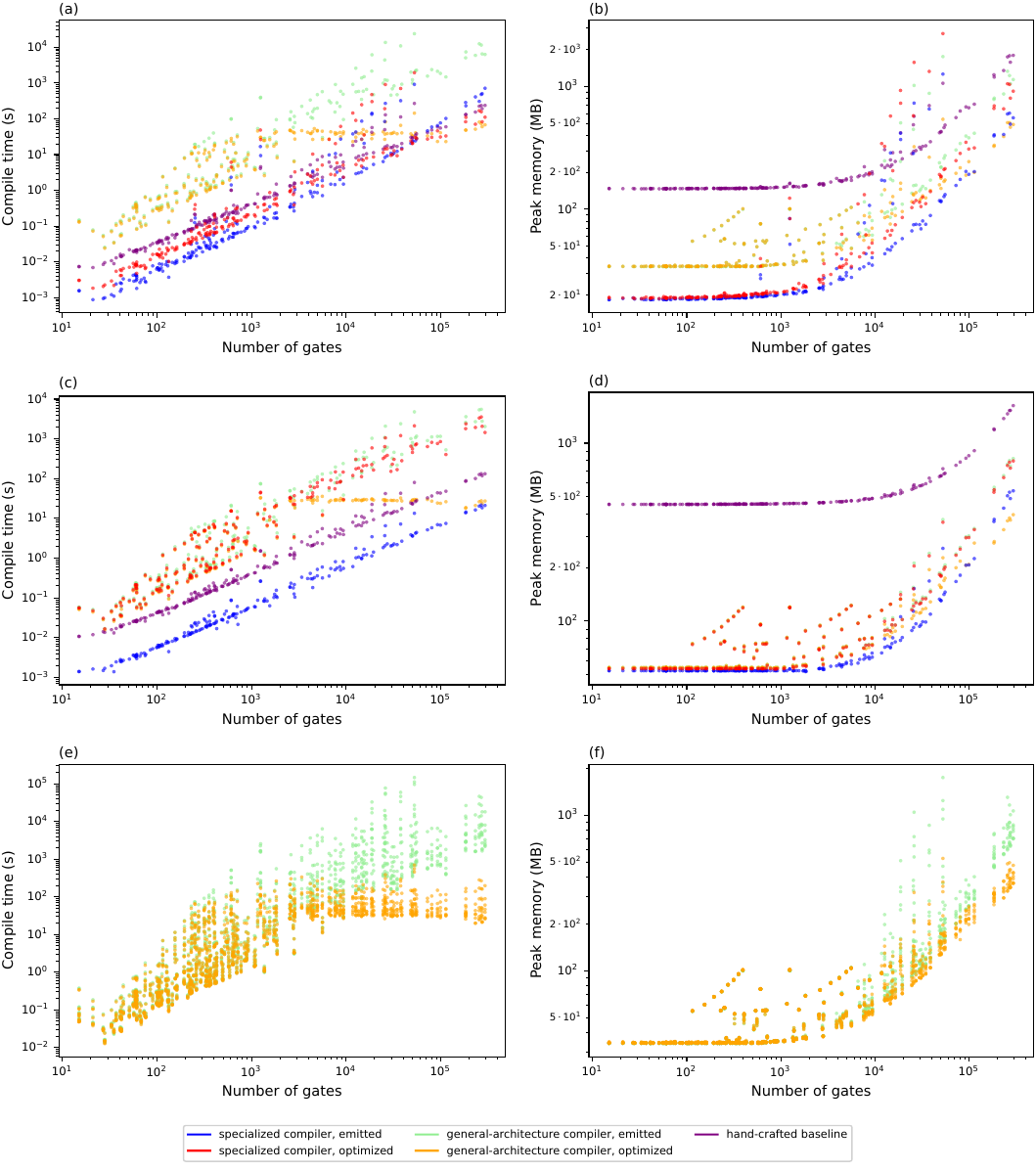}
	\caption{Compile time and peak memory of all compilers of Opus 4.7 and of the hand-crafted baselines \cite{Wagner22, KreppelMWHPSB24} against the number of gates. The left column shows the compile time and the right column the peak memory, in (a) and (b) on the linear architecture, in (c) and (d) on the branched architecture at stack height 1 and junction distance 1, and in (e) and (f) on the eight cyclic architectures. Each point is one circuit on one architecture and the mean over ten runs. The peak memory includes the Python interpreter.}
	\label{fig:memory}
\end{figure*}

\begin{figure*}[p]
	\centering
	\includegraphics[width=\textwidth]{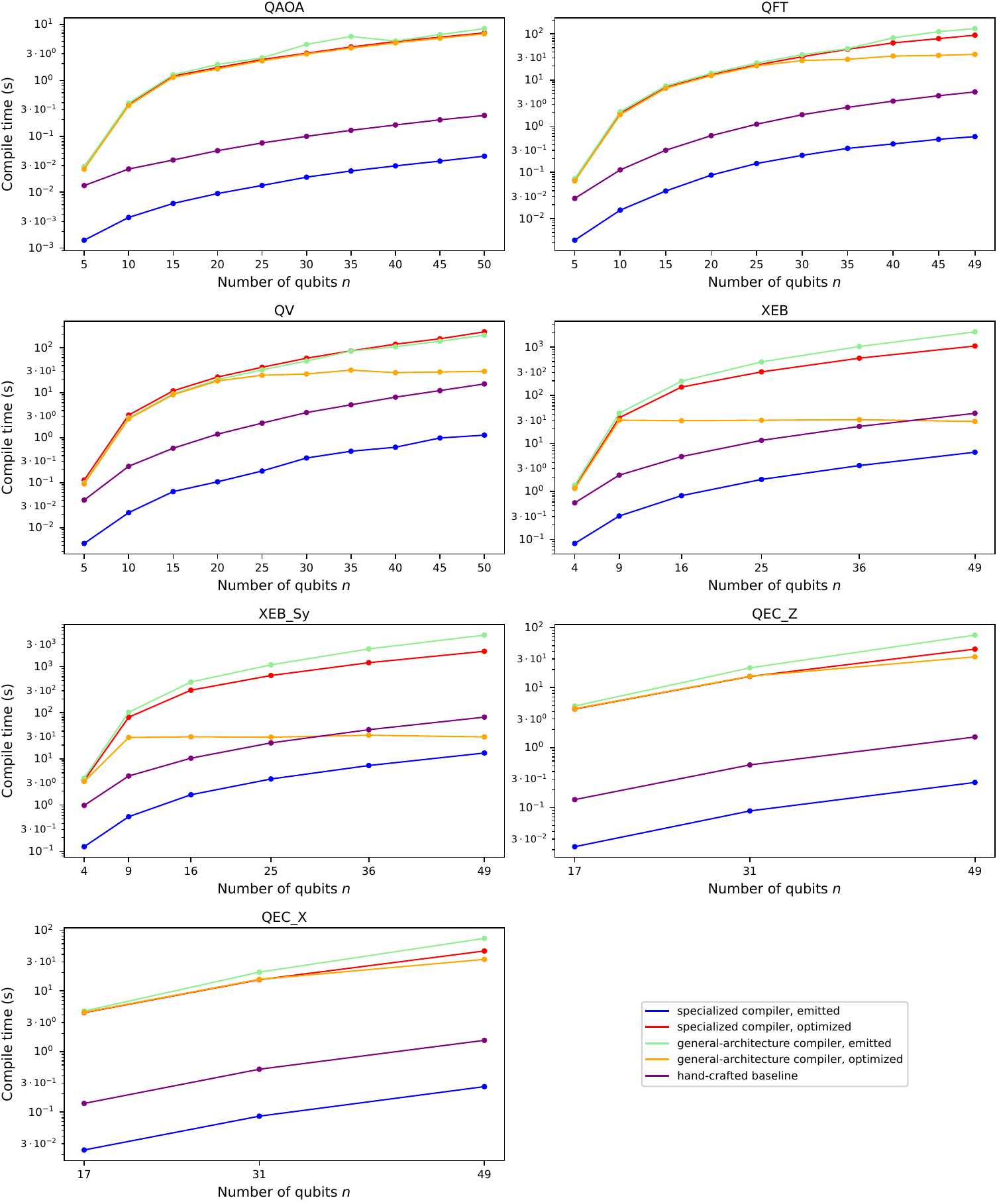}
	\caption{Compile time of the specialized and general-architecture compilers of Opus 4.7, each as emitted and optimized, and of the hand-crafted baseline \cite{KreppelMWHPSB24} on the branched architecture at stack height 1 and junction distance 1 for the scalable families, one panel per family and one circuit per qubit count. Each value is the mean over ten runs.}
	\label{fig:compile-time-branched-fam}
\end{figure*}

\begin{figure*}[p]
	\centering
	\includegraphics[width=\textwidth]{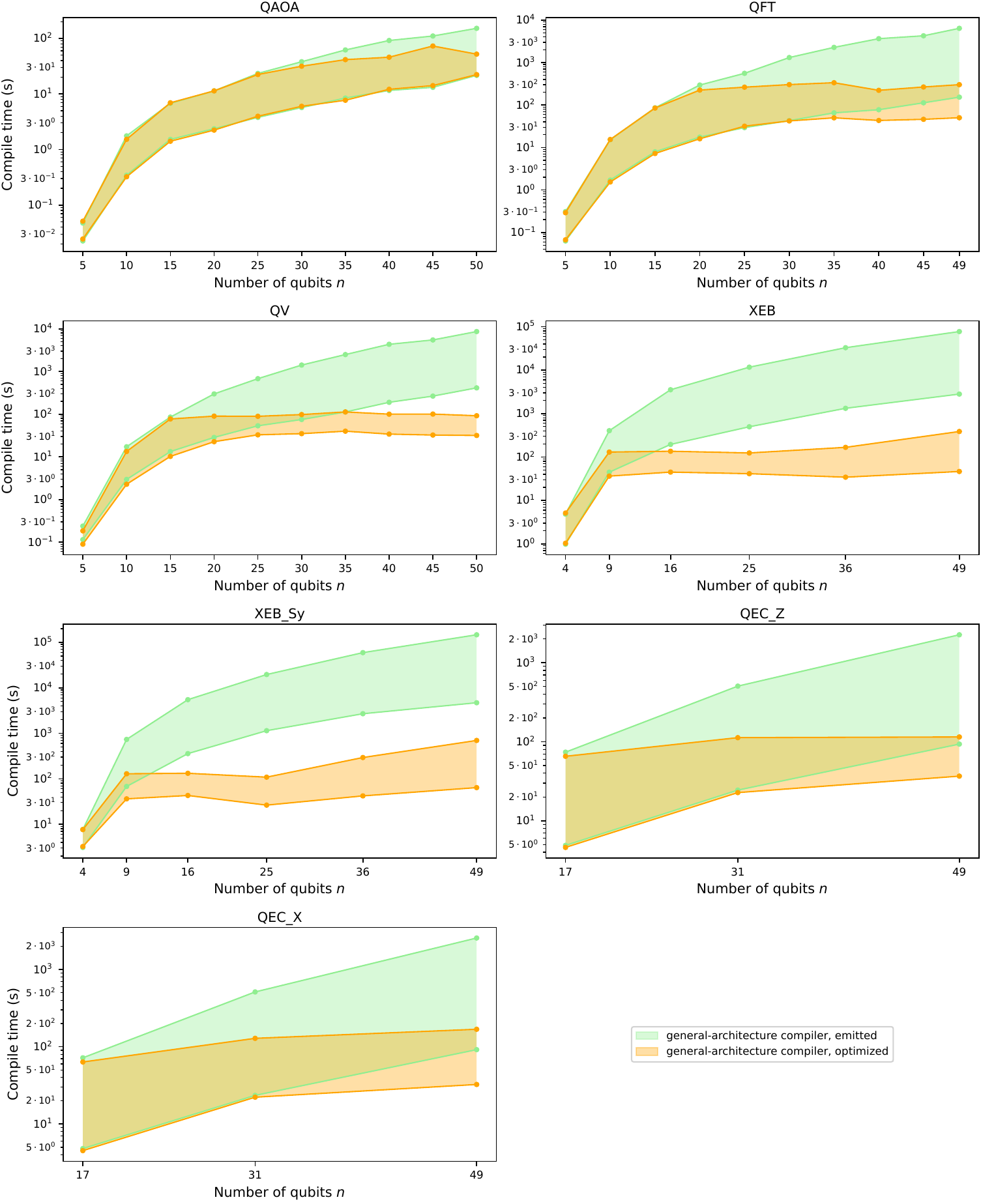}
	\caption{Compile time of the general-architecture compiler of Opus 4.7, as emitted and after optimization, on the eight cyclic architectures of \autoref{fig:pipeline} for the scalable families. Each band spans, per qubit count, from the fastest to the slowest architecture. Each value is the mean over ten runs.}
	\label{fig:compile-time-general}
\end{figure*}

\begin{figure*}[p]
	\centering
	\includegraphics[width=\textwidth]{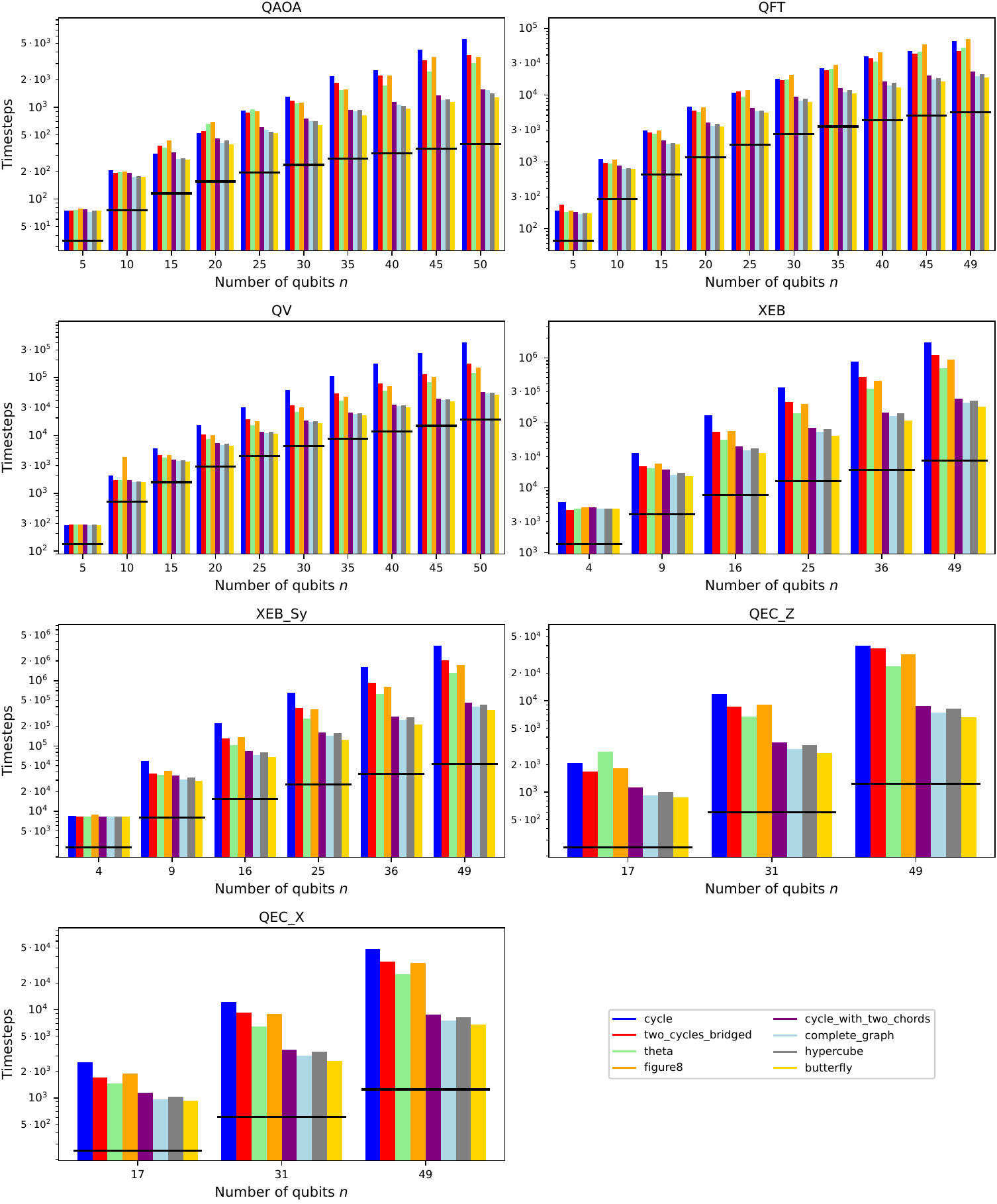}
	\caption{Timesteps of the emitted general-architecture compiler of Opus 4.7 on the eight cyclic architectures for the scalable families, one panel per family and one bar per circuit and architecture. The black line marks the number of gates, a lower bound on the number of timesteps, so the part of each bar above it is the shuttling overhead.}
	\label{fig:abs-general}
\end{figure*}

\begin{figure*}[tp]
	\centering
	\includegraphics{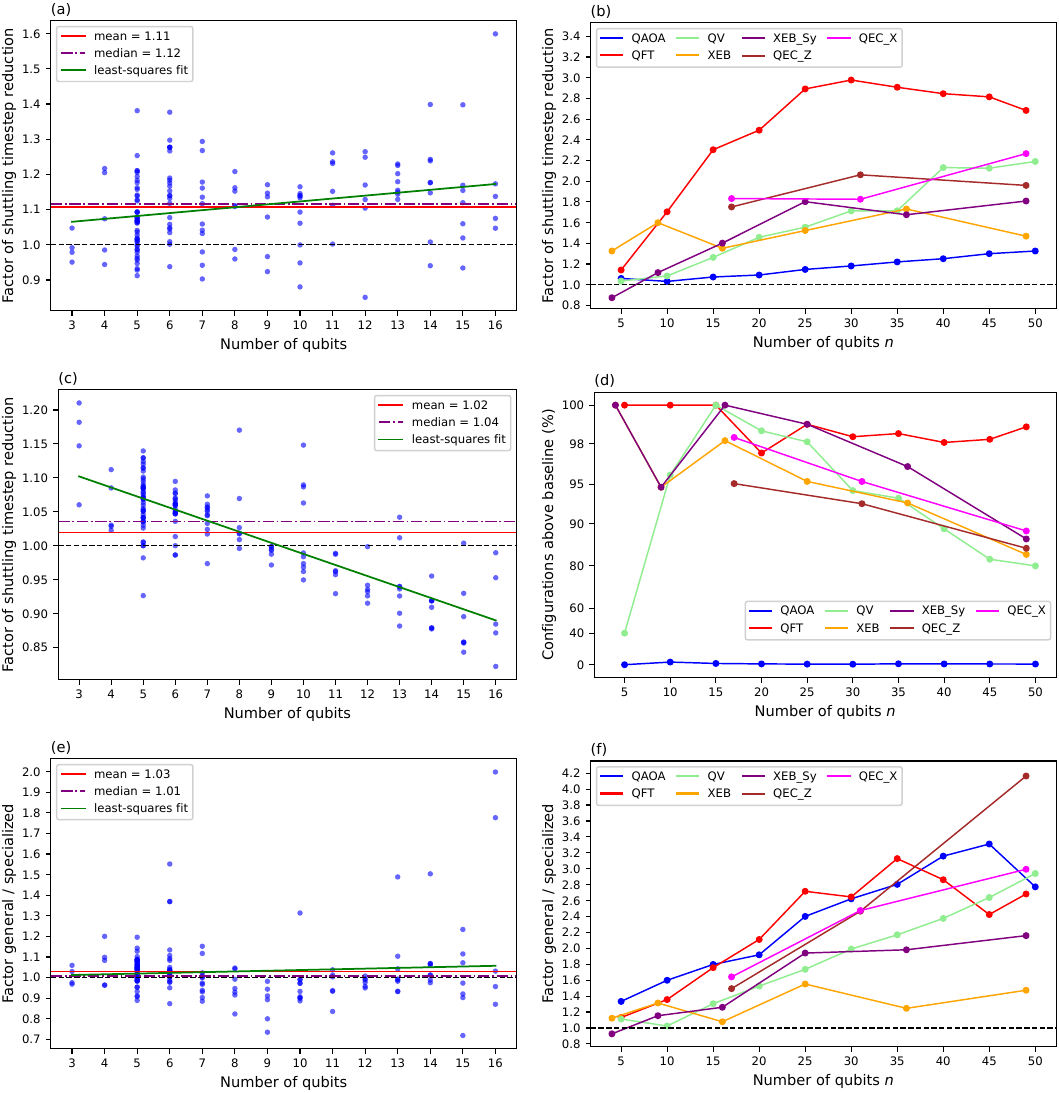}
	\caption{Timesteps of the compilers of Fable 5, the optimized specialized compilers against the hand-crafted baselines \cite{Wagner22, KreppelMWHPSB24} in (a) to (d) and the optimized general-architecture compiler against the optimized specialized one in (e) and (f). (a) Linear architecture, circuit library, reduction factor of the optimized compiler against the baseline per circuit, with the mean, the median, and a least-squares fit as red, purple, and green lines. (b) Linear architecture, scalable families, the factor of (a), one circuit per qubit count. (c) Branched architecture at stack height 1 and junction distance 1, circuit library, as in (a). (d) Branched architecture, scalable families, per qubit count the proportion of the tested configurations on which the optimized compiler needs fewer timesteps than the baseline. (e) Linear architecture, circuit library, timesteps of the optimized general-architecture compiler divided by those of the optimized specialized compiler per circuit, with lines as in (a). (f) Linear architecture, scalable families, the factor of (e), one circuit per qubit count. Above the dashed line at 1, the compiler of Fable 5 needs fewer timesteps than the baseline in (a) to (c), and the specialized compiler fewer than the general-architecture one in (e) and (f). The vertical axis of (d) is stretched around 100\,\%.}
	\label{fig:llm-fable}
\end{figure*}

\begin{figure*}[tp]
	\centering
	\includegraphics{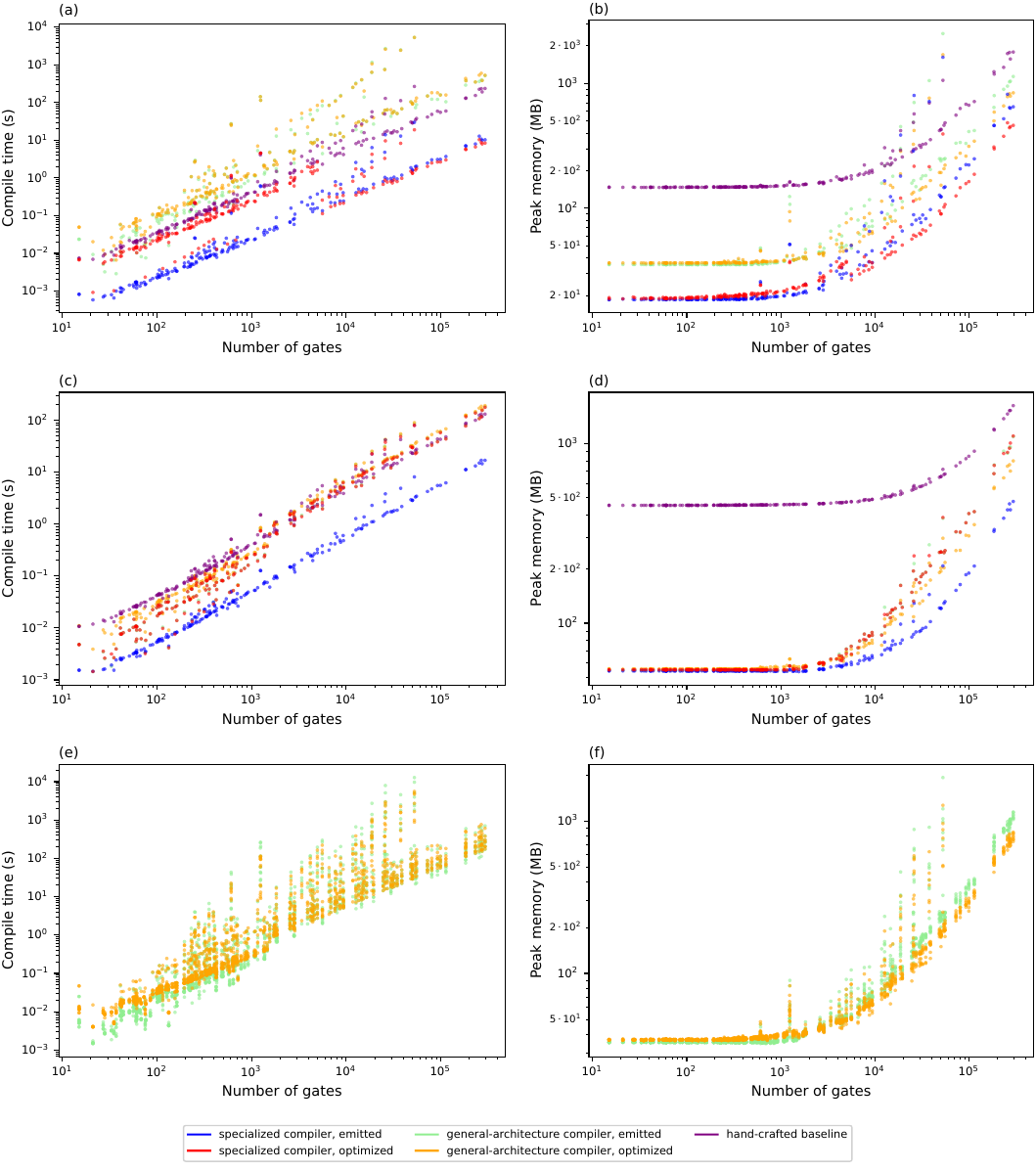}
	\caption{Compile time and peak memory of all compilers of Fable 5 and of the hand-crafted baselines \cite{Wagner22, KreppelMWHPSB24} against the number of gates. The left column shows the compile time and the right column the peak memory, in (a) and (b) on the linear architecture, in (c) and (d) on the branched architecture at stack height 1 and junction distance 1, and in (e) and (f) on the eight cyclic architectures. Each point is one circuit on one architecture and the mean over ten runs. The peak memory includes the Python interpreter.}
	\label{fig:llm-runtime-fable}
\end{figure*}

\begin{figure*}[tp]
	\centering
	\includegraphics{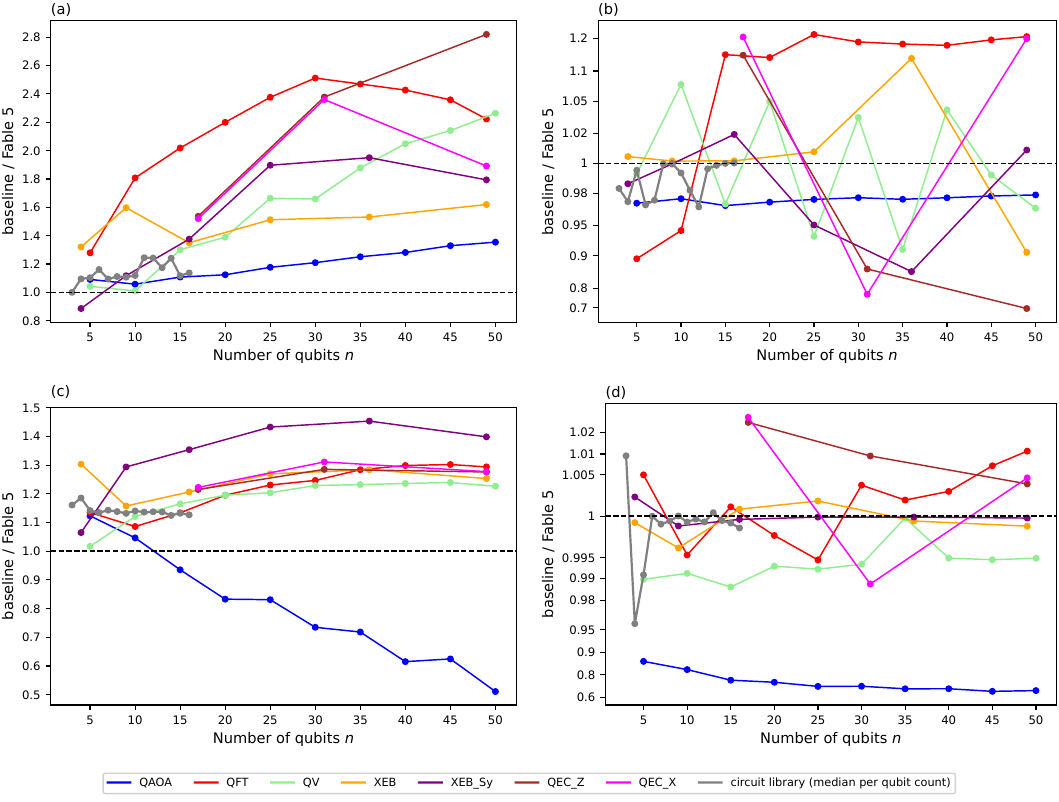}
	\caption{Routing and placement parts of the timestep factor of the hand-crafted baselines \cite{Wagner22, KreppelMWHPSB24} over the optimized specialized compilers of Fable 5 per scalable family and qubit count, with the median of the circuit library per qubit count, 3 to 16, as a gray line. (a) Linear architecture, routing part. (b) Linear architecture, placement part. (c) Branched architecture, routing part. (d) Branched architecture, placement part. Each panel divides the timesteps of the baseline by those of the compiler of Fable 5, so values above the dashed line at 1 favor the compiler of Fable 5. The routing part compares the two routers from the placement of the compiler of Fable 5, and the placement part the two placements under the baseline's router. On the branched architecture, each value is the median over all tested configurations. The vertical axes of (b) and (d) are stretched around 1.}
	\label{fig:mapping-split-fable}
\end{figure*}

\begin{figure*}[tp]
	\centering
	\includegraphics{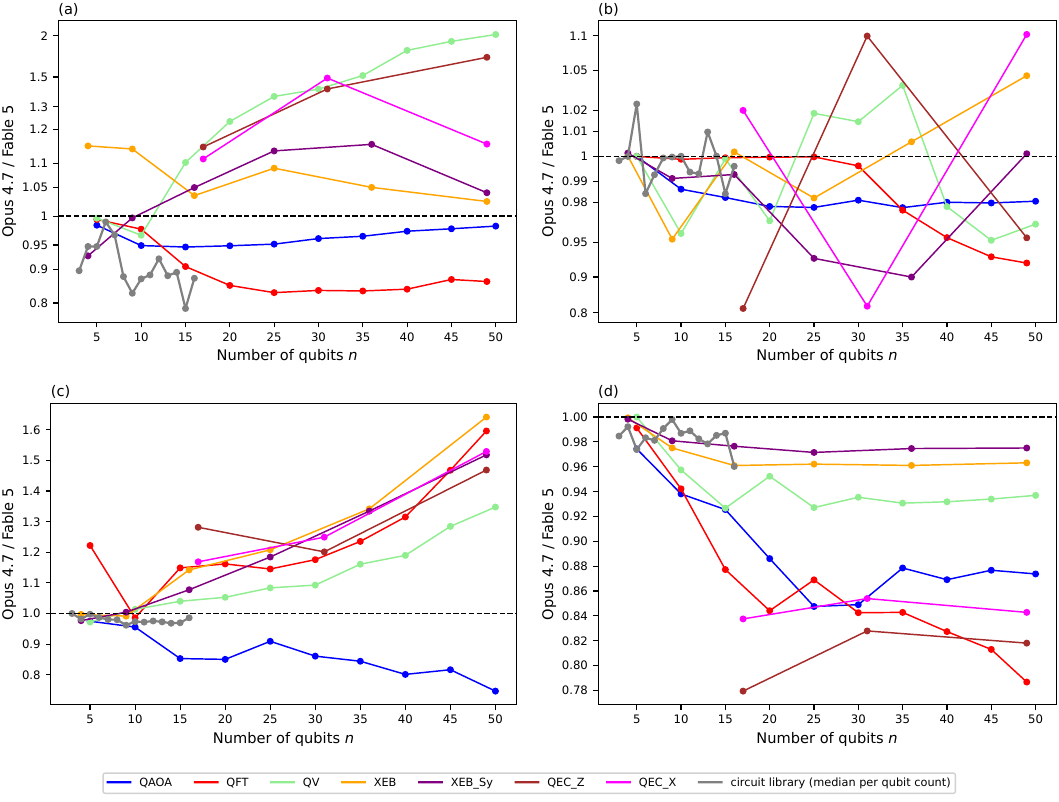}
	\caption{Routing and placement parts of the timestep factor of Opus 4.7 over Fable 5 for the optimized specialized compilers per scalable family and qubit count, with the median of the circuit library per qubit count, 3 to 16, as a gray line. (a) Linear architecture, routing part. (b) Linear architecture, placement part. (c) Branched architecture, routing part. (d) Branched architecture, placement part. Each panel divides the timesteps of Opus 4.7 by those of Fable 5, with both routers from Fable 5's placement in the routing part and both placements under Opus 4.7's router in the placement part, so values above the dashed line at 1 favor Fable 5. On the branched architecture, each value is the median over all tested configurations. The vertical axes of (a) and (b) are stretched around 1.}
	\label{fig:mapping-split-llm}
\end{figure*}

\begin{figure*}[tp]
	\centering
	\includegraphics{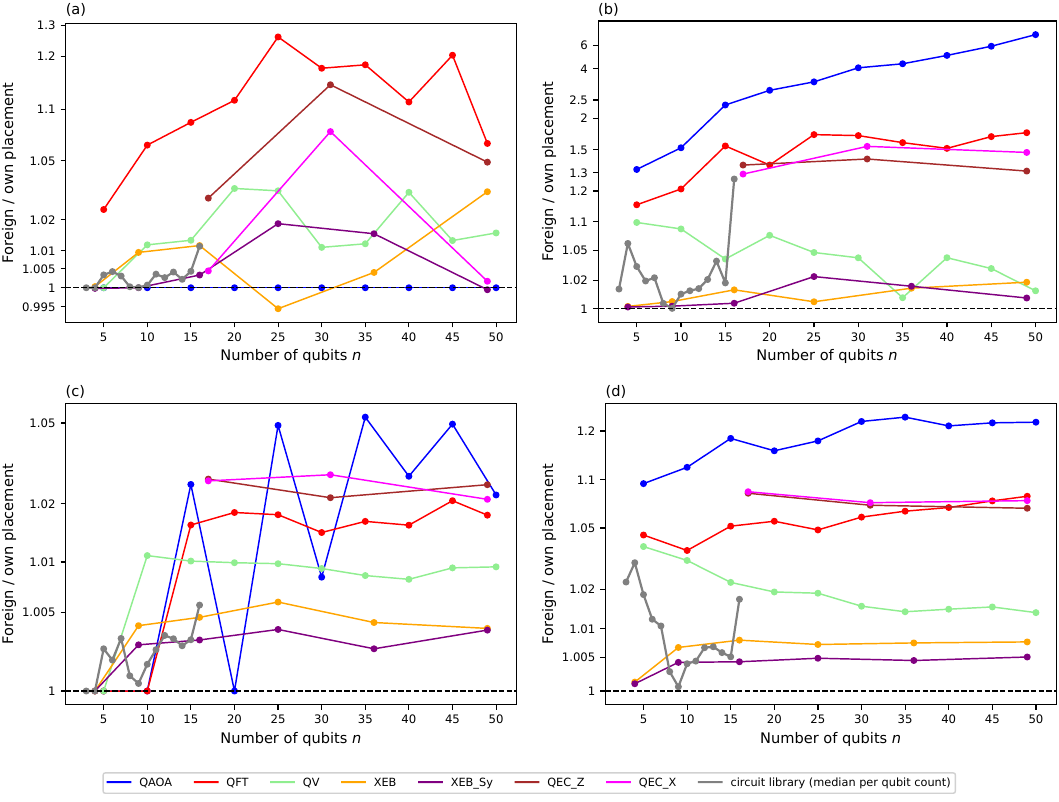}
	\caption{Timesteps of a router from a foreign placement divided by those from its own placement, per scalable family and qubit count and, as a gray line, for the circuit library per qubit count, 3 to 16. (a) Linear architecture, placements of the other compilers. (b) Linear architecture, random placements. (c) Branched architecture, placements of the other compilers. (d) Branched architecture, random placements. Each value is the median over all routers and foreign placements of the respective architecture. On the branched architecture, the median also runs over all tested configurations, and for the circuit library over all circuits of a qubit count. Values above the dashed line at 1 mean that the router needs more timesteps from the foreign placement. All vertical axes are stretched around 1.}
	\label{fig:mapping-placement}
\end{figure*}

\begin{figure*}[tp]
	\centering
	\includegraphics{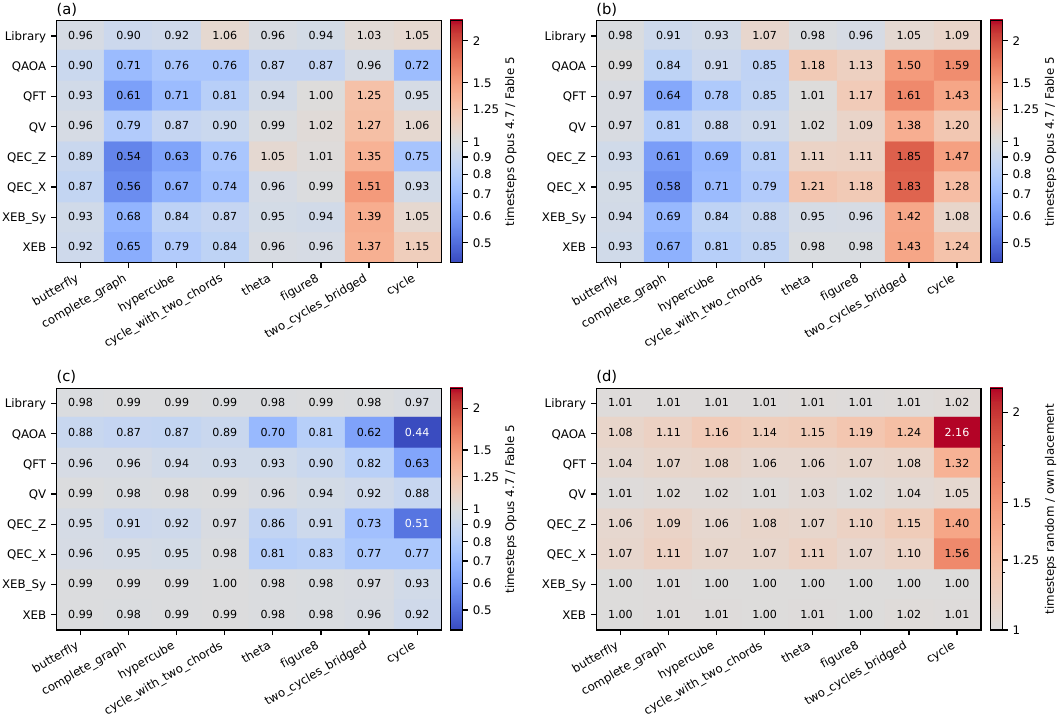}
	\caption{Routing and placement on the eight cyclic architectures of \autoref{fig:pipeline}, one row for the circuit library and each scalable family and one column per architecture, as in \autoref{fig:arch-suit}. Panels (a) to (c) divide the timesteps of Opus 4.7 by those of Fable 5 for the general-architecture compilers in the versions of \autoref{sec:comparison:eval}, emitted for Opus 4.7 and optimized for Fable 5. Panel (a) compares the whole compilers, panel (b) the two routers from the placement of Fable 5's compiler, and panel (c) the two placements under the router of Opus 4.7. Panel (d) divides the timesteps from a random placement by those from the router's own placement, over all four routers. Each cell is a median. The colors run from blue below 1 through gray at 1 to red above 1, so red cells favor Fable 5 in (a) to (c) and mark a loss from the random placement in (d).}
	\label{fig:mapping-cyclic}
\end{figure*}

\clearpage
\section{Prompt for the linear compiler}
\label{app:linear-prompt}
The full prompt used to generate the linear-architecture compiler is reproduced below. The prompt refers to input files by name, concrete files placed in the working directory and not reproduced here. The circuit files, in \texttt{.qasm} or \texttt{.json} form, are the benchmark circuits of \autoref{sec:eval:setup} together with a few small hand-written checks. Each linear trap is a path whose central vertex is the gate vertex, of the length listed in \autoref{tab:linear-tests}.

\begin{lstlisting}[style=prompt]
Implement a complete Ion-Trap Shuttling Compiler in a single Python file. The compiler translates a quantum circuit into a sequence of
physical shuttling operations for a linear ion-trap quantum computer, minimizing the total number of timesteps needed to execute the
circuit.

For the implementation, extend the existing ion_trap_compiler.py file. The file already contains some functions described below, which
can be called directly. Do not re-implement these functions; only their interfaces, as described below, are relevant.

(*@\rule{\linewidth}{0.4pt}@*)
SECTION 1 - HARDWARE MODEL (TrapGraph)
(*@\rule{\linewidth}{0.4pt}@*)

The hardware is represented by the TrapGraph class, whose implementation is already provided in ion_trap_compiler.py.

  class TrapGraph:
    @property def num_vertices(self) -> int                        # Returns the number of vertices in the graph
    @property def vertices(self) -> List[int]                      # Returns all vertices of the graph
    def degree(self, v: int) -> int                                # Returns the degree of vertex v
    def neighbors(self, v: int) -> List[int]                       # Returns the adjacent vertices of vertex v
    @property def qubit_vertices(self) -> List[int]                # Returns all qubit vertices
    @property def gate_vertices(self) -> List[int]                 # Returns all gate vertices
    @property def junctions(self) -> List[int]                     # Returns all vertices with degree > 2
    def bfs_path(self, src: int, dst: int) -> Optional[List[int]]  # Returns the shortest path from src to dst, including both endpoints
    def distance(self, src: int, dst: int) -> int                  # Returns the shortest distance from src to dst

The TrapGraph is a linear graph of trap segments (vertices) numbered 0 to TrapGraph.num_vertices - 1, connected as a path:
0 - 1 - 2 - (*@$\ldots$@*) - (TrapGraph.num_vertices - 1).

Each vertex can hold at most 2 ions, arranged as an ordered linear chain.

Each vertex is either a qubit vertex or a gate vertex. At a gate vertex, the operations SEPARATE, MERGE, SWAP, and GATE can be executed;
a qubit vertex can only store up to 2 ions. During gate execution, the gate vertex must contain exactly the number of ions required by
the gate (no more, no fewer). During the execution of translations, the gate vertices can always be used for transfer.

(*@\rule{\linewidth}{0.4pt}@*)
SECTION 2 - CIRCUIT INPUT
(*@\rule{\linewidth}{0.4pt}@*)

The quantum circuit is represented by two lists:
  - qubits: List[str]         # All qubits in the circuit
  - gates: List[QuantumGate]  # All quantum gates in the circuit

The quantum gates are represented by the QuantumGate class, whose implementation is already provided in ion_trap_compiler.py.

  @dataclass
  class QuantumGate:
      name:    str                      # Name of the unitary operation
      qubits:  List[str]                # Qubits on which the gate acts
      params:  List[Tuple[str, float]]  # Parameters of the unitary operation
      gate_id: int                      # Sequential index in program order

Every gate acts on exactly one or two qubits. Gates acting on the same qubit must be executed in the order in which they appear in the
gates list. Gates acting on disjoint qubits may be executed in any order.

(*@\rule{\linewidth}{0.4pt}@*)
SECTION 3 - PHYSICAL OPERATIONS
(*@\rule{\linewidth}{0.4pt}@*)

There are five elementary operations. Each consumes one timestep unless grouped (see Sec. 6.3).

  TRANSLATE(src: int, dst: int)
    Move the entire contents of vertex src to adjacent vertex dst.
    Preconditions: src and dst are adjacent; dst is empty.
    After: src is empty, dst holds what src held.

  SEPARATE(vertex: int)
    Split the qubit chain at vertex into two halves. Since each vertex holds at most two qubits, the left qubit goes to the left neighbor
    and the right qubit to the right neighbor.
    Preconditions: vertex (*@$\in$@*) TrapGraph.gate_vertices; vertex contains 2 qubits; both neighbors are empty; left neighbor has a smaller
    index than right neighbor.
    After: vertex is empty.

  MERGE(vertex: int)
    Combine the chains from the two neighbors of vertex into vertex, in the order [left_chain, right_chain].
    Preconditions: vertex (*@$\in$@*) TrapGraph.gate_vertices; each neighbor contains exactly one qubit; left neighbor has a smaller index than
    right neighbor.
    After: both neighbors are empty.

  SWAP(vertex: int)
    Reverse the order of the qubit chain at vertex. Not permitted on a single-qubit chain.
    Preconditions: vertex (*@$\in$@*) TrapGraph.gate_vertices; vertex contains exactly two qubits.

  GATE(vertex: int, gate: QuantumGate)
    Execute gate at vertex.
    Preconditions: vertex (*@$\in$@*) TrapGraph.gate_vertices; the vertex contains exactly the qubits required by the gate (no more, no fewer);
    all per-qubit program-order dependencies of this gate are satisfied.
    Note: The qubit order within the vertex chain need not match the order in QuantumGate.qubits; only the sets of qubits must be
    identical.

Implementation requirement:
  Record operations as compact tuples:
    - ("T", src: int, dst: int)                              for TRANSLATE
    - ("S", v: int, left_neighbor: int, right_neighbor: int) for SEPARATE
    - ("M", v: int, left_neighbor: int, right_neighbor: int) for MERGE
    - ("W", v: int)                                          for SWAP
    - ("G", v: int, gate: QuantumGate)                       for GATE

(*@\rule{\linewidth}{0.4pt}@*)
SECTION 4 - INPUT PARSING
(*@\rule{\linewidth}{0.4pt}@*)

At program start, call the following function to parse all inputs. Its implementation is already provided in ion_trap_compiler.py.

  def parse_input() -> Tuple[TrapGraph, List[str], List[QuantumGate], float, Optional[str], Optional[str]]
  # Return values (positional):
  # 1. TrapGraph:         hardware graph (see Sec. 1)
  # 2. List[str]:         qubits (see Sec. 2)
  # 3. List[QuantumGate]: gates (see Sec. 2)
  # 4. float:             time_begin  (pass through to validate_and_output, Sec. 7)
  # 5. Optional[str]:     output_json (pass through to validate_and_output, Sec. 7)
  # 6. Optional[str]:     stats_file  (pass through to validate_and_output, Sec. 7)

(*@\rule{\linewidth}{0.4pt}@*)
SECTION 5 - INITIAL PLACEMENT
(*@\rule{\linewidth}{0.4pt}@*)

Before shuttling begins, assign each qubit to exactly one vertex such that:
  - exactly (*@$\lceil$@*)len(qubits) / 2(*@$\rceil$@*) vertices hold qubits;
  - no vertex holds more than 2 qubits;
  - all remaining vertices are empty.

Implement the function:

  def initial_mapping(trap_graph: TrapGraph,
                      qubits: List[str],
                      gates: List[QuantumGate]
                      ) -> Dict[int, List[str]]
  # Return value:
  # Dict[int, List[str]]: initial_state
  # Keys are vertex indices; values are the qubit chains at those vertices (at most two qubits per vertex). Vertices with empty chains
  # may be omitted or represented with an empty list.

The function can be implemented as follows:

The circuit-DAG (Directed Acyclic Graph) represents the quantum circuit as a dependency graph. Each gate is a node. A directed edge from
gate A to gate B exists if and only if gate B acts on at least one qubit that gate A also acts on, and gate B appears after gate A in
program order with no intervening gate on that qubit -- meaning gate B depends on gate A and cannot execute before A completes. The depth
of a node is the length of the longest directed path from any source node (a gate with no predecessors) to that node. The total DAG depth
is the length of the longest such path in the entire graph, i.e., the minimum number of sequential layers required if all independent
gates ran in parallel.

Weight every gate g at circuit-DAG depth (*@$\delta$@*) by w = 1 / (1 + (*@$\delta$@*) / (*@$\tau$@*)), where (*@$\tau$@*) (*@$\approx$@*) total_dag_depth / 3 (all float divisions), so that gates
executed early carry more influence on placement decisions than gates deeper in the circuit. If total_dag_depth == 0, treat all gates as
equally weighted. Use this weighted interaction matrix to build a weighted interaction graph whose nodes are qubits and whose edge weight
between two qubits is the sum of weights of all gates acting on that pair.

Default pairing rule:
Compute a maximum-weight matching on the weighted interaction graph using a greedy procedure: repeatedly pick the heaviest unpaired edge
and pair its two endpoint qubits; if len(qubits) is odd, the qubit with the lowest total interaction weight across all its edges is left
unpaired as the singleton. Then perform greedy chain extraction over the matched-vertex interaction graph -- starting from the matched
pair with the highest combined edge weight, repeatedly extend the chain by appending the adjacent matched pair with the highest
connecting weight -- to produce an ordered sequence of pairs (and the singleton, if any). Place pairs and the singleton on consecutive
trap vertices in this order. The resulting initial arrangement co-locates frequently interacting qubits at the same vertex and orders
vertices so that the qubit layout along the trap minimizes required timesteps. For circuits with a chain-like qubit-interaction structure
(as in qaoa_5.json), this placement should yield a mapping that requires zero swaps.

Additionally, the initial mapping can reduce shuttling before the first gate. Notwithstanding the above rule, if the first applicable
gate is a two-qubit gate, placing its qubits at a shared gate vertex eliminates any shuttling needed to make that gate executable.
Whenever qubits are placed at a gate vertex, the two neighbors of that gate vertex must be left empty in the initial mapping if a
separation is required before moving these qubits (see Sec. 9).

This greedy matching and chain-extraction procedure is the required default. Additional strategies (e.g., graph-coloring or other
heuristics) may be layered on top to further reduce timesteps, but must not perform worse than the default on the provided benchmark
circuits.

An explicit nx.DiGraph circuit-DAG object is not required; any equivalent per-qubit dependency tracking is acceptable.

(*@\rule{\linewidth}{0.4pt}@*)
SECTION 6 - COMPILER PIPELINE
(*@\rule{\linewidth}{0.4pt}@*)

Implement the function:

  def compile(trap_graph: TrapGraph,
              qubits: List[str],
              gates: List[QuantumGate],
              initial_state: Dict[int, List[str]]
              ) -> List[List[Tuple]]
  # Return value:
  # List[List[Tuple]]: all_timesteps
  # The outer list is ordered by execution time. Each inner list contains all shuttling operations for that timestep, represented as
  # tuples (see Sec. 3).

The compilation pipeline consists of the following steps:

6.1  GATE SCHEDULING

  Maintain a ready-queue of gates whose per-qubit dependencies are fully satisfied -- i.e., all earlier gates on the same qubits have
  already executed. At each scheduling step, select the next gate using the following priority heuristic, whose core principle is to
  minimize unnecessary qubit movement by batching gates that can run without re-shuttling:

    - If multiple ready gates share the same qubit set (one or two qubits), execute all of them consecutively before moving those qubits
      away from the gate vertex.
    - Among the remaining ready gates, prefer gates whose qubits are already at or closest to a gate vertex, minimizing shuttling needed
      to begin execution.

  Additional scheduling strategies are encouraged; this is an open-ended optimization target.

6.2  ROUTING

  Emit the TRANSLATE / SEPARATE / MERGE / SWAP operations needed to bring the required qubits to a gate vertex before each gate
  execution.

  Note: at any routing step, additional qubits may need to be translated to free vertices required for SEPARATE, MERGE, SWAP, or GATE
  operations. Intermediate qubits blocking a path may also need to be moved.

  Single-qubit gate case:
  Given:
    - current_state: Dict[int, List[str]] (same structure as initial_state)
    - QuantumGate with len(QuantumGate.qubits) == 1; qa = QuantumGate.qubits[0]
  1. If qa shares its vertex with another qubit in current_state:
     - Move qa's vertex to a gate vertex.
     - Execute SEPARATE to isolate qa.
  2. Move qa to the gate vertex.
  3. Execute the gate.

  Two-qubit gate case -- distance reduction via iterative merge-swap-separate: Given:
    - current_state: Dict[int, List[str]]
    - QuantumGate with len(QuantumGate.qubits) == 2; call the qubits qa and qb.
  1. Trivial case:
    - If qa and qb are at the same vertex: move that vertex to a gate vertex and execute the gate.
    - Otherwise go to step 2.
  2. Initialization:
    - Build a linear ordering of all qubits by iterating over current_state in ascending key order and concatenating the values:
      (*@$\cdot$@*) linear_ordering: List[str], len == len(qubits)
    - Let intermediate_qubits = [i1, i2, (*@$\ldots$@*), ik] be the qubits between qa and qb in linear_ordering.
    - By definition:
      (*@$\cdot$@*) qa is directly left of i1 in linear_ordering.
      (*@$\cdot$@*) qb is directly right of ik in linear_ordering.
  3. Iterative Reduction:
     Repeat until intermediate_qubits == []:
     a) Choose direction:
        Select the side (qa or qb) that minimizes the approximate number of future timesteps:
        - qa will be merged with i1, or
        - qb will be merged with ik.
     b) Prepare for merge:
        - If the chosen qubit (qa or qb) shares its vertex with another qubit:
          (*@$\cdot$@*) If qa is chosen:
            (*@$\blacklozenge$@*) If qa is the right qubit in its chain:
              (*@$\blacksquare$@*) Move qa's vertex to a gate vertex and execute SEPARATE.
            (*@$\blacklozenge$@*) If qa is the left qubit in its chain (i.e., qa shares with i1):
              (*@$\blacksquare$@*) Move that vertex to a gate vertex.
              (*@$\blacksquare$@*) Execute SWAP, then SEPARATE.
              (*@$\blacksquare$@*) Update: intermediate_qubits = [i2, i3, (*@$\ldots$@*), ik].
          (*@$\cdot$@*) If qb is chosen:
            (*@$\blacklozenge$@*) If qb is the left qubit in its chain:
              (*@$\blacksquare$@*) Move qb's vertex to a gate vertex and execute SEPARATE.
            (*@$\blacklozenge$@*) If qb is the right qubit in its chain (i.e., qb shares with ik):
              (*@$\blacksquare$@*) Move that vertex to a gate vertex.
              (*@$\blacksquare$@*) Execute SWAP, then SEPARATE.
              (*@$\blacksquare$@*) Update: intermediate_qubits = [i1, i2, (*@$\ldots$@*), i(k-1)].
        - If i1 (when qa is chosen) or ik (when qb is chosen) shares its vertex with another qubit:
          (*@$\cdot$@*) Move that vertex to a gate vertex and execute SEPARATE.
        After this step, the chosen qubit and its neighbor (i1 or ik) are each alone at some vertex.
     c) Merge and propagate:
        - If qa is chosen:
          (*@$\cdot$@*) Move qa to the left neighbor of the gate vertex.
          (*@$\cdot$@*) Move i1 to the right neighbor of the gate vertex.
        - If qb is chosen:
          (*@$\cdot$@*) Move qb to the right neighbor of the gate vertex.
          (*@$\cdot$@*) Move ik to the left neighbor of the gate vertex.
        - Execute MERGE. Chain at gate vertex: [qa, i1] or [ik, qb].
        - Execute SWAP. Chain becomes: [i1, qa] or [qb, ik].
        - Execute SEPARATE.
        - Update intermediate_qubits: remove i1 (if qa chosen) or ik (if qb chosen). The intermediate chain is reduced by exactly one
          qubit.
     d) Dynamic reassignment (optional):
        - Different gate vertices may be used each round.
        - The choice of qa or qb in step a) may be re-evaluated each round.
  4. Finalization:
     When intermediate_qubits == []:
       - If qa shares its vertex with another qubit:
         (*@$\cdot$@*) Move qa's vertex to a gate vertex and execute SEPARATE.
       - If qb shares its vertex with another qubit:
         (*@$\cdot$@*) Move qb's vertex to a gate vertex and execute SEPARATE.
       - Move qa to the left neighbor of the gate vertex.
       - Move qb to the right neighbor of the gate vertex.
       - Execute MERGE.
       - Execute the gate.

  Notes for both cases:
    - Not all gate vertices may be usable at every point in time. Always choose a gate vertex that approximately minimizes the number of
      timesteps.

  Chain-order optimization after two-qubit gate execution:
    After routing, the two target qubits are co-located in the gate vertex. Since QuantumGate.qubits order does not affect gate
    correctness, the qubit order within the vertex chain is irrelevant for the current gate but does affect routing efficiency for
    subsequent gates. Use a one-step lookahead over the next gate of each qubit: immediately after gate execution, consider inserting a
    post-execution SWAP if doing so would reduce the routing cost of the next gate touching either qubit. The preferred side of each
    qubit is the side of the gate vertex where its next 2-qubit-gate partner is currently on (None if the qubit has only single-qubit
    gates in the future or there is no next gate for the qubit). Insert the SWAP only when the post-SWAP arrangement matches more qubits
    on the preferred side than the no-SWAP arrangement. When in doubt, do not insert, as an unnecessary SWAP can cascade into extra
    separations downstream.

6.3  TIMESTEP GROUPING AND POST-PROCESSING

  A timestep is a set of operations executed simultaneously. After the routing emits the raw operation stream, apply the following passes
  in order:

    1. Roundtrip elimination -- remove a TRANSLATE A(*@$\rightarrow$@*)B immediately followed by TRANSLATE B(*@$\rightarrow$@*)A when no intervening operation references
       A or B (as src, dst, or operand vertex). Removing the pair must leave A's and B's contents identical to the no-removal trace at
       every intermediate step.

    2. Consecutive cancellation -- remove:
       - a SEPARATE immediately followed by a MERGE on the same vertex,
       - a MERGE immediately followed by a SEPARATE on the same vertex,
       - a SWAP immediately followed by another SWAP on the same vertex,
       in each case only when the qubits are not involved in any intervening operation.

    Apply passes 1 and 2 alternately to fixpoint. Use always the current, post-removal operation stream. When no further reductions are
    possible, proceed to pass 3.

    3. Grouping -- pack independent operations into the same timestep:

       Translation timesteps: may contain only TRANSLATE operations.
         All translations on pairwise disjoint source/destination vertex pairs may be freely grouped. A "chain of translations" along
         consecutive vertices is also permitted: if TRANSLATE B(*@$\rightarrow$@*)C and TRANSLATE A(*@$\rightarrow$@*)B are both needed, they may share a timestep,
         provided each chained vertex except the rightmost (here C) actually holds a qubit at the start of the timestep. This is because
         B(*@$\rightarrow$@*)C vacates B in the same timestep that A(*@$\rightarrow$@*)B fills it. This chain rule extends to arbitrarily long consecutive sequences,
         where each qubit appears in at most one TRANSLATE.

       Non-translation timesteps: may contain only non-TRANSLATE operations (SEPARATE, MERGE, SWAP, GATE). Two such operations may share
       a timestep only if the vertex sets they involve are pairwise disjoint. The vertex set of an operation is:
           - SWAP(v):       {v}
           - GATE(v, (*@$\ldots$@*)):   {v}
           - SEPARATE(v):   {v, v-1, v+1}
           - MERGE(v):      {v, v-1, v+1}
         In particular, a SEPARATE at v1 and a MERGE at v2 cannot share a timestep when {v1, v1(*@$\pm$@*)1} (*@$\cap$@*) {v2, v2(*@$\pm$@*)1} (*@$\neq$@*) (*@$\emptyset$@*), even if v1 (*@$\neq$@*) v2.

(*@\rule{\linewidth}{0.4pt}@*)
SECTION 7 - VALIDATION AND OUTPUT
(*@\rule{\linewidth}{0.4pt}@*)

At the end of the program, call the following function to validate the results and emit all output. Its implementation is already
provided in ion_trap_compiler.py.

  def validate_and_output(trap_graph: TrapGraph,
                          qubits: List[str],
                          gates: List[QuantumGate],
                          initial_state: Dict[int, List[str]], all_timesteps: List[List[Tuple]],
                          time_begin: float,
                          output_json: Optional[str] = None,
                          stats_file: Optional[str] = None)
  # All parameters are the unchanged return values of parse_input (Sec. 4), initial_mapping (Sec. 5), and compile (Sec. 6).

(*@\rule{\linewidth}{0.4pt}@*)
SECTION 8 - PERFORMANCE REQUIREMENTS
(*@\rule{\linewidth}{0.4pt}@*)

The compiler must minimize runtime and peak RAM usage. Complexity should scale roughly linearly in all parameters: number of vertices,
qubits, and gates.

The compiler must handle large circuits (200 qubits, several hundred thousand gates, 400+ vertices) within an acceptable runtime and
memory budget. Prefer greedy/heuristic approaches over exact optimization.

(*@\rule{\linewidth}{0.4pt}@*)
SECTION 9 - WORKING EXAMPLE (VERIFICATION)
(*@\rule{\linewidth}{0.4pt}@*)

The following example must produce a valid (not necessarily identical) shuttling sequence of at most 16 individual operations,
completable in 12 timesteps or fewer.

  Command:
    python ion_trap_compiler.py --circ example.qasm --graph graph_7.json --gate 3

  Trap graph: vertices 0-6 in a path; vertex 3 is the only gate vertex.

Initial mapping:
  Map q[0],q[1] (*@$\rightarrow$@*) Vertex 3
  Map q[2]      (*@$\rightarrow$@*) Vertex 5
  
Reference 16-step sequence with 12 timesteps:
 Timestep 1:
  1.  GATE  cx q[0],q[1] at Vertex 3
 Timestep 2:
  2.  SEPARATE Vertex 3 (*@$\rightarrow$@*) q[0](*@$\rightarrow$@*)V2, q[1](*@$\rightarrow$@*)V4
 Timestep 3:
  3.  TRANSLATE q[0]: V2(*@$\rightarrow$@*)V1
  4.  TRANSLATE q[1]: V4(*@$\rightarrow$@*)V3
  5.  TRANSLATE q[2]: V5(*@$\rightarrow$@*)V4
 Timestep 4:
  6.  TRANSLATE q[1]: V3(*@$\rightarrow$@*)V2
 Timestep 5:
  7.  MERGE V2+V4 (*@$\rightarrow$@*) Vertex 3  (contains q[1],q[2])
 Timestep 6:
  8.  GATE  cx q[1],q[2] at Vertex 3
 Timestep 7:
  9.  SWAP  Vertex 3  (chain becomes q[2],q[1])
 Timestep 8:
  10. SEPARATE Vertex 3 (*@$\rightarrow$@*) q[2](*@$\rightarrow$@*)V2, q[1](*@$\rightarrow$@*)V4
 Timestep 9:
  11. TRANSLATE q[1]: V4(*@$\rightarrow$@*)V5
  12. TRANSLATE q[2]: V2(*@$\rightarrow$@*)V3
  13. TRANSLATE q[0]: V1(*@$\rightarrow$@*)V2
 Timestep 10:
  14. TRANSLATE q[2]: V3(*@$\rightarrow$@*)V4
 Timestep 11:
  15. MERGE V2+V4 (*@$\rightarrow$@*) Vertex 3  (contains q[0],q[2])
 Timestep 12:
  16. GATE  cx q[0],q[2] at Vertex 3

(*@\rule{\linewidth}{0.4pt}@*)
SECTION 10 - CODE QUALITY
(*@\rule{\linewidth}{0.4pt}@*)

  - Single file, Python 3.12+. All Python libraries are permitted; however, only libraries needed to compute the requested output may be
    installed.
  - Determinism: the compiler must produce byte-identical output for identical inputs across runs, independent of PYTHONHASHSEED or any
    other process-level entropy. Iterations over sets, dicts, or any other unordered containers whose order can affect the emitted
    operation stream must be explicitly sorted (e.g., by qubit name, vertex index, or gate id). The random module must not be used
    without a fixed, in-source seed.
  - All graph traversals (DAG depth, routing reductions, cycle detection in timestep grouping, etc.) must be iterative -- using an
    explicit stack or queue -- never recursive. Recursive traversal exceeds Python's default recursion limit on 200-qubit / 100k-gate
    inputs and is forbidden.

(*@\rule{\linewidth}{0.4pt}@*)
SECTION 11 - ACCEPTANCE TEST MATRIX
(*@\rule{\linewidth}{0.4pt}@*)

The implementation must pass every test below; otherwise it must be automatically debugged until all tests pass. Each test fixes the
input, invocation, and expected bound; the schedule itself need not match any reference.

  T1. Sec. 9 cx-triangle.
      Input/invocation: as in Sec. 9.
      Expected: (*@$\leq$@*) 16 individual operations, (*@$\leq$@*) 12 timesteps.

  T2. Same-pair batching.
      Input: 10 gates, each on the same qubit pair (q0, q1).
      Invocation: python ion_trap_compiler.py --circ same_pair.qasm --graph graph_5.json --gate 2
      Expected: q0 and q1 are mapped to vertex 2 in the initial state; the 10 GATE operations execute back-to-back on that vertex;
      SEPARATE, MERGE, SWAP, and TRANSLATE counts are all 0.

  T3-T8 exercise the .json benchmark circuits.

  T3. qaoa_5.json -- chain-structured zero-SWAP target.
      Input: 5 qubits, 35 gates, two-qubit interaction graph is the path q0-q1-q2-q3-q4 (only 4 RZZ gates, all on adjacent pairs).
      Invocation: python ion_trap_compiler.py --circ qaoa_5.json --graph graph_11.json --gate 5
      Expected: 0 SWAP operations. This is the canonical regression target for the initial-placement heuristic described in Sec. 5.

  T4. rd32-v0_66.json -- small reversible circuit, 4 qubits, 42 gates.
      Invocation: python ion_trap_compiler.py --circ rd32-v0_66.json --graph graph_9.json --gate 4
      Expected: schedule verifies; runtime under one second on commodity hardware. The smallest non-trivial benchmark; serves as a fast
      smoke test.

  T5. 4mod5-v1_23.json -- 5 qubits, 102 gates, two-qubit interaction graph is
      non-chain.
      Invocation: python ion_trap_compiler.py --circ 4mod5-v1_23.json --graph graph_11.json --gate 5
      Expected: schedule verifies; SWAP count > 0 (this circuit cannot be routed without swaps, testing that the routing emits them when
      needed).

  T6. 4mod5-bdd_287.json -- 7 qubits, 106 gates.
      Invocation: python ion_trap_compiler.py --circ 4mod5-bdd_287.json --graph graph_15.json --gate 7
      Expected: schedule verifies.

  T7. alu-v2_31.json -- 5 qubits, 640 gates (largest gate count in the set).
      Invocation: python ion_trap_compiler.py --circ alu-v2_31.json --graph graph_11.json --gate 5
      Expected: schedule verifies.

  T8. cnt3-5_179.json -- 16 qubits, 230 gates (largest qubit count in the set).
      Invocation: python ion_trap_compiler.py --circ cnt3-5_179.json --graph graph_33.json --gate 16
      Expected: schedule verifies.

  Cross-cutting requirements over T3-T8:
    - For each circuit, the operation stream is byte-identical across two runs with different PYTHONHASHSEED values.
    - Total wall-clock time and peak RSS scale roughly linearly with number of vertices, qubits, and gates; no individual circuit takes
      more than a small multiple of the time predicted by linear extrapolation from the smallest circuit.
\end{lstlisting}

\clearpage
\section{Prompt for the branched compiler}
\label{app:branched-prompt}
The prompt for the branched-architecture compiler is reproduced below. It is issued with the optimized linear-architecture compiler as its seed and specifies the changes needed for a branched, tree-structured trap with junctions and stacks. Its circuit files are as in \appref{app:linear-prompt}. The trap-graph files encode branched traps as a JSON edge list of segment pairs. A file named \texttt{graph\_$m$\_$h$\_$d$} has $m$ non-junction segments, side-stacks of height $h$, and junction distance $d$.

\begin{lstlisting}[style=prompt]
Implement a complete Ion-Trap Shuttling Compiler in a single Python file. The compiler translates a quantum circuit into a sequence of
physical shuttling operations for a linear ion-trap quantum computer with junctions, minimizing the total number of timesteps needed to
execute the circuit.

For the implementation, extend the existing ion_trap_compiler.py file. The file already contains a working compiler for a junction-free
linear trap and exposes the helpers described below, which can be called directly. Do not re-implement these helpers; only their
interfaces, as described below, are relevant. Every other function in the file (initial_mapping, compile, and the internal
routing/grouping helpers) must be adapted to the junction topology described in this document.

(*@\rule{\linewidth}{0.4pt}@*)
SECTION 1 - HARDWARE MODEL (TrapGraph)
(*@\rule{\linewidth}{0.4pt}@*)

The hardware is represented by the TrapGraph class, whose implementation is already provided in ion_trap_compiler.py. The
class is junction-aware and exposes structural helpers for the main axis and stacks; routing code should prefer these
helpers over any reasoning based on vertex indices.

  class TrapGraph:
    # Basics
    @property def num_vertices(self) -> int                          # Number of vertices
    @property def vertices(self) -> List[int]                        # All vertices, sorted
    def degree(self, v: int) -> int                                  # Degree of v
    def neighbors(self, v: int) -> Tuple[int, ...]                   # Adjacent vertices, sorted (immutable tuple)
    @property def qubit_vertices(self) -> List[int]                  # All qubit vertices, sorted
    @property def gate_vertices(self) -> List[int]                   # Gate vertices (order from CLI)
    @property def junctions(self) -> List[int]                       # All vertices with degree > 2, sorted

    # Topology helpers
    def is_junction(self, v: int) -> bool                            # O(1)
    def is_on_main_axis(self, v: int) -> bool                        # O(1)
    @property def main_axis(self) -> List[int]                       # Ordered [0, 1, (*@$\ldots$@*), M]
    def stacks_at(self, j: int) -> List[List[int]]                   # For junction j: list of stacks; each stack is
                                                                     # [slot 0, slot 1, (*@$\ldots$@*), bottom]
    def stack_of(self, v: int) -> Optional[Tuple[int, int, int]]     # (junction, stack_index, slot_index) for stack vertices, None for
                                                                     # main-axis vertices

    # Distances and paths (precomputed; O(1) distance, paths are cached)
    def bfs_path(self, src: int, dst: int) -> Optional[List[int]]    # Shortest path from src to dst, including both endpoints
    def distance(self, src: int, dst: int) -> int                    # Shortest distance from src to dst
    def next_hop(self, src: int, dst: int) -> Optional[int]          # First vertex after src on the unique src->dst path
    def path_passes_through(self, src, dst, v: int) -> bool          # True iff v is a strict interior vertex of the src->dst path
    def junctions_on_path(self, src, dst) -> List[int]               # Junctions visited along the src->dst path
    def subtree_through(self, j: int, via: int) -> Set[int]          # All vertices reachable from j by first stepping to via, excluding
                                                                     # j

Topology: The TrapGraph is a tree consisting of one main linear axis (the contiguous index range 0, 1, (*@$\ldots$@*), M, available as
TrapGraph.main_axis) with zero or more side branches ("stacks") attached to it. A junction has degree (*@$\geq$@*) 3 and connects one or more
linear stacks of qubit-only vertices that extend away from the main axis. Each stack is enumerated by stacks_at(j) in slot order (slot 0
is junction-adjacent); for a stack vertex v, stack_of(v) returns its (junction, stack_index, slot_index). A graph with no junctions
reduces to the plain linear case; stacks_at returns [] there and stack_of returns None for every vertex.

Example (M = 6, junctions at vertices 1 and 5, two stacks of height 3, gate vertex at 3):

      9               12
      |               |
      8               11
      |               |
      7               10
      |               |
  0 - 1 - 2 - 3 - 4 - 5 - 6

  TrapGraph.num_vertices     == 13
  TrapGraph.qubit_vertices   == [0, 2, 4, 6, 7, 8, 9, 10, 11, 12]
  TrapGraph.gate_vertices    == [3]
  TrapGraph.junctions        == [1, 5]
  TrapGraph.main_axis        == [0, 1, 2, 3, 4, 5, 6]
  TrapGraph.stacks_at(1)     == [[7, 8, 9]]   # 7 is slot 0
  TrapGraph.stacks_at(5)     == [[10, 11, 12]]
  TrapGraph.stack_of(8)      == (1, 0, 1)     # junction 1, stack 0, slot 1
  TrapGraph.stack_of(3)      == None          # main-axis vertex

All benchmark JSON files in this directory follow this convention. Stacks may have different heights and junctions may sit anywhere on
the main axis.

Each vertex can hold at most 2 ions, arranged as an ordered linear chain.

Vertex types and their constraints:
  - Gate vertex ((*@$\in$@*) TrapGraph.gate_vertices):
    (*@$\cdot$@*) The operations SEPARATE, MERGE, SWAP, and GATE may be executed there.
    (*@$\cdot$@*) During gate execution, the gate vertex must contain exactly the qubits required by the gate (no more, no fewer).
    (*@$\cdot$@*) A gate vertex always lies on the main axis. Its two neighbors are never junctions; both have degree (*@$\leq$@*) 2. This means SEPARATE and
      MERGE well-define a left and right neighbor.
    (*@$\cdot$@*) A gate vertex may always be used as a transit vertex during translations.
  - Junction vertex ((*@$\in$@*) TrapGraph.junctions):
    (*@$\cdot$@*) May hold ions only while they are in transit (i.e., between two consecutive TRANSLATEs of the same chain). During every SEPARATE,
      MERGE, SWAP, and GATE all junctions must be empty.
    (*@$\cdot$@*) No-roundtrip rule: if TRANSLATE(a, j) moves a chain into junction j, the next TRANSLATE involving that junction must be
      TRANSLATE(j, b) with b (*@$\neq$@*) a. The chain must leave the junction through a different edge than the one it entered through.
    (*@$\cdot$@*) Exit-availability rule (planning consequence of the no-roundtrip rule and of the "transit only" restriction above): before emitting
      TRANSLATE(a, j) that lands a chain on junction j, the routing must guarantee that, at the moment the chain arrives, at least one
      neighbor of j other than the entry edge `a` is empty (or will be made empty in the same translation timestep by a chained
      translation), so that the chain can be drained on the next translation step. If no such exit can be secured -- e.g. because every
      other neighbor holds an ion and is itself a dead-end or full stack -- the routing must pick a different plan (a different path, a
      stack push elsewhere, or a different gate vertex). Translating onto a junction whose every non-entry exit is occupied by chains
      that cannot themselves be displaced is a permanent deadlock and is forbidden.
  - Qubit vertex:
    (*@$\cdot$@*) Any other vertex. May hold up to 2 ions; never executes SEPARATE / MERGE / SWAP / GATE.

Stack mechanics (LIFO):
  Each stack attached to a junction j is a Last-In, First-Out buffer. The stack vertex adjacent to j is called slot 0; the next deeper
  vertex is slot 1; and so on. The slot-0 vertex is the only access port -- qubits enter and leave the stack only through that vertex,
  via the junction. Pushes and pops have constant timestep cost regardless of stack depth:

    - Push: a qubit chain at the junction j is translated to slot 0. If slot 0 already holds a chain, every occupied slot shifts one slot
      deeper simultaneously (slot k (*@$\rightarrow$@*) slot k+1 for every occupied k, plus j (*@$\rightarrow$@*) slot 0). All of these translations share a single
      timestep under the chain-of-translations rule (Sec. 6.3), so a push costs exactly one translation timestep as long as the deepest
      occupied slot is not already the bottom of the stack. Empty slots between occupied slots do not block the cascade -- only occupied
      slots emit a shift; the chain-rule requirement (each chained source initially occupied) is satisfied for every emitted shift.

    - Pop: the qubit chain at slot 0 is translated through the junction. Simultaneously, every deeper occupant shifts one slot shallower
      (slot k+1 (*@$\rightarrow$@*) slot k for every occupied k+1) within the same timestep. A pop therefore also costs exactly one translation timestep
      as long as slot 0 is occupied. As with push, empty slots between occupied slots are no obstacle; only occupied slots emit a shift.

  Qubits inside a stack cannot pass each other. The order in which qubits exit a stack is the reverse of the order in which they entered.
  To reach a qubit deeper than another, every shallower qubit must be extracted first; in-place permutation of stack contents is
  impossible. Reordering is only achievable by evacuating qubits out of the stack along its edges (through the junction onto the main
  axis or another stack), holding them at other vertices, and re-pushing in the desired order.

(*@\rule{\linewidth}{0.4pt}@*)
SECTION 2 - CIRCUIT INPUT
(*@\rule{\linewidth}{0.4pt}@*)

The quantum circuit is represented by two lists:
  - qubits: List[str]         # All qubits in the circuit
  - gates: List[QuantumGate]  # All quantum gates in the circuit

The QuantumGate class is provided in ion_trap_compiler.py.

  @dataclass
  class QuantumGate:
      name:    str                      # Name of the unitary operation
      qubits:  List[str]                # Qubits on which the gate acts
      params:  List[Tuple[str, float]]  # Parameters of the unitary operation
      gate_id: int                      # Sequential index in program order

Every gate acts on exactly one or two qubits. Gates acting on the same qubit must be executed in the order in which they appear in the
gates list. Gates acting on disjoint qubits may be executed in any order.

(*@\rule{\linewidth}{0.4pt}@*)
SECTION 3 - PHYSICAL OPERATIONS
(*@\rule{\linewidth}{0.4pt}@*)

There are five elementary operations. Each consumes one timestep unless grouped (see Sec. 6.3).

  TRANSLATE(src: int, dst: int)
    Move the entire contents of vertex src to adjacent vertex dst.
    Preconditions: src and dst are adjacent in TrapGraph; dst is empty; src is non-empty; if src is a junction, dst is not the vertex
    from which the chain currently at src entered (no-roundtrip rule).
    After: src is empty, dst holds what src held.

  SEPARATE(vertex: int)
    Split the qubit chain at vertex into its two halves. Since each vertex holds at most two qubits, the left qubit goes to the left
    neighbor and the right qubit to the right neighbor.
    Preconditions: vertex (*@$\in$@*) TrapGraph.gate_vertices; vertex contains 2 qubits; both neighbors are empty; left neighbor has a smaller
    index than right neighbor; all junctions are empty.
    After: vertex is empty.

  MERGE(vertex: int)
    Combine the chains from the two neighbors of vertex into vertex, in the order [left_chain, right_chain].
    Preconditions: vertex (*@$\in$@*) TrapGraph.gate_vertices; each neighbor contains exactly one qubit; left neighbor has a smaller index than
    right neighbor; all junctions are empty.
    After: both neighbors are empty.

  SWAP(vertex: int)
    Reverse the order of the qubit chain at vertex. Not permitted on a single-qubit chain.
    Preconditions: vertex (*@$\in$@*) TrapGraph.gate_vertices; vertex contains exactly two qubits; all junctions are empty.

  GATE(vertex: int, gate: QuantumGate)
    Execute gate at vertex.
    Preconditions: vertex (*@$\in$@*) TrapGraph.gate_vertices; the vertex contains exactly the qubits required by the gate (no more, no fewer);
    all per-qubit program-order dependencies of this gate are satisfied; all junctions are empty.
    Note: The qubit order within the vertex chain need not match the order in QuantumGate.qubits; only the sets of qubits must be
    identical.

Implementation requirement:
  Record operations as compact tuples:
    - ("T", src: int, dst: int)                              for TRANSLATE
    - ("S", v: int, left_neighbor: int, right_neighbor: int) for SEPARATE
    - ("M", v: int, left_neighbor: int, right_neighbor: int) for MERGE
    - ("W", v: int)                                          for SWAP
    - ("G", v: int, gate: QuantumGate)                       for GATE

(*@\rule{\linewidth}{0.4pt}@*)
SECTION 4 - INPUT PARSING
(*@\rule{\linewidth}{0.4pt}@*)

At program start, call the following function to parse all inputs. Its implementation is already provided in ion_trap_compiler.py.

  def parse_input() -> Tuple[TrapGraph, List[str], List[QuantumGate], float, Optional[str], Optional[str]]
  # Return values (positional):
  # 1. TrapGraph:         hardware graph (see Sec. 1)
  # 2. List[str]:         qubits (see Sec. 2)
  # 3. List[QuantumGate]: gates (see Sec. 2)
  # 4. float:             time_begin  (pass through to validate_and_output, Sec. 7)
  # 5. Optional[str]:     output_json (pass through to validate_and_output, Sec. 7)
  # 6. Optional[str]:     stats_file  (pass through to validate_and_output, Sec. 7)

(*@\rule{\linewidth}{0.4pt}@*)
SECTION 5 - INITIAL PLACEMENT
(*@\rule{\linewidth}{0.4pt}@*)

Before shuttling begins, assign each qubit to exactly one vertex such that:
  - exactly (*@$\lceil$@*)len(qubits) / 2(*@$\rceil$@*) vertices hold qubits;
  - no vertex holds more than 2 qubits;
  - no junction vertex holds any qubit;
  - all remaining vertices are empty.

Implement the function:

  def initial_mapping(trap_graph: TrapGraph,
                      qubits: List[str],
                      gates: List[QuantumGate]
                      ) -> Dict[int, List[str]]
  # Return value:
  # Dict[int, List[str]]: initial_state
  # Keys are vertex indices; values are the qubit chains at those vertices (at most two qubits per vertex). Vertices with empty chains
  # may be omitted or represented with an empty list.

The function can be implemented as follows:

The circuit-DAG (Directed Acyclic Graph) represents the quantum circuit as a dependency graph. Each gate is a node. A directed edge from
gate A to gate B exists if and only if gate B acts on at least one qubit that gate A also acts on, and gate B appears after gate A in
program order with no intervening gate on that qubit -- meaning gate B depends on gate A. The depth of a node is the length of the
longest directed path from any source node to it. The total DAG depth is the maximum such path in the graph.

Weight every gate g at circuit-DAG depth (*@$\delta$@*) by w = 1 / (1 + (*@$\delta$@*) / (*@$\tau$@*)), where (*@$\tau$@*) (*@$\approx$@*) total_dag_depth / 3 (all float divisions), so that gates
executed early carry more influence on placement decisions than gates deeper in the circuit. If total_dag_depth == 0, treat all gates as
equally weighted. Build a weighted interaction graph whose nodes are qubits and whose edge weights are summed gate weights over all
2-qubit gates acting on each pair.

Default pairing rule:
Compute a maximum-weight matching on the weighted interaction graph using a greedy procedure: repeatedly pick the heaviest unpaired edge
and pair its two endpoint qubits. Tie-breaker among edges of equal weight: sort by the lexicographic order of the qubit-name tuple
(smaller name first within each edge, then smaller edge first). If len(qubits) is odd, the qubit with the lowest total interaction weight
is left unpaired as the singleton; ties broken by qubit name (string sort).

Then perform greedy chain extraction over the matched-vertex interaction graph: starting from the matched pair with the highest internal
weight (the weight of the edge between the two qubits of the pair; ties broken by the qubit-name tuple of the pair), repeatedly extend
the chain by appending the adjacent unused matched pair connected by the heaviest cross-pair edge to the current chain's last pair (ties
broken by the qubit-name tuple of the candidate pair). The chain extends only at the most recently appended end. The result is an ordered
sequence of pairs (and the singleton, if any), called the items list below.

Intra-pair qubit orientation (the order in which the two qubits of a pair sit within the vertex chain at placement time): for each pair,
compute for each of its two qubits the sum of edge weights from that qubit to every qubit in other items (other pairs and the singleton,
if any). Place the qubit with the greater such total at chain index 0; the other at chain index 1. Ties broken by qubit name
lexicographically (smaller-name qubit at chain index 0). This rule deterministically orients the qubit with stronger downstream
interactions so it ends up on the side of the pair that the candidate-vertex order (sorted by distance to the nearest gate vertex, then
by vertex index) will fill next -- i.e. closer to the later-placed items, reducing routing distance for the gates that act between this
pair and those later items.

Place pairs and the singleton on a sequence of trap vertices chosen as follows:

  1. Compute the candidate vertex list:
     candidates = sorted(
         [v for v in trap_graph.vertices if not trap_graph.is_junction(v)],
         key=lambda v: (min(trap_graph.distance(v, gv) for gv in trap_graph.gate_vertices), v)
     )
     This places vertices closest to any gate vertex first, with vertex index as a deterministic tie-breaker. Main-axis and stack
     vertices that are equidistant from a gate vertex are interleaved naturally.

  2. Walk candidates in order and assign each ordered item (pair or singleton) to the next available candidate vertex. Junction vertices
     are never used. Override: if the first applicable gate is a two-qubit gate whose qubit set equals items[0], place items[0] at the
     gate vertex that minimizes the sum of distances to every non-junction vertex (i.e. the union of trap_graph.qubit_vertices and
     trap_graph.gate_vertices -- a centrality measure that spreads remaining items evenly; ties broken by gate vertex index). Reserve
     both neighbors of that gate vertex as empty for the SEPARATE that follows the first MERGE. Subsequent items continue to fill
     candidates from the front of the list, skipping the gate vertex itself and the reserved neighbors.

     Note: this gate-vertex-selection rule is distinct from the per-gate "best gv" rule used in Sec. 6.2 Step 1 (which ranks by combined
     distance from the two endpoints). The placement rule above is used only at compile-time initial-state setup; the routing rule of
     Sec. 6.2 is recomputed for every two-qubit gate during scheduling.

This greedy matching and graph-distance-based placement is the required default. Additional strategies (e.g., graph-coloring, simulated
annealing) may be layered on top, but must not perform worse than the default on the acceptance tests.

For circuits with a chain-like qubit-interaction structure (as in qaoa_5.json, qaoa_25.json, qaoa_40.json), this placement should yield a
mapping that requires zero swaps independently of the underlying graph topology.

An explicit graph object for the circuit-DAG is not required; any equivalent per-qubit dependency tracking is acceptable.

(*@\rule{\linewidth}{0.4pt}@*)
SECTION 6 - COMPILER PIPELINE
(*@\rule{\linewidth}{0.4pt}@*)

Implement the function:

  def compile(trap_graph: TrapGraph,
              qubits: List[str],
              gates: List[QuantumGate],
              initial_state: Dict[int, List[str]]
              ) -> List[List[Tuple]]
  # Return value:
  # List[List[Tuple]]: all_timesteps
  # The outer list is ordered by execution time. Each inner list contains all shuttling operations for that timestep, represented as
  # tuples (see Sec. 3).
compile must compile the circuit completely: the returned all_timesteps must execute every gate in gates. Partial compilation is
forbidden -- if any gate cannot be routed, compile must raise an exception instead of returning a schedule that omits gates (see the
no-partial-schedule rule in Sec. 11).

The compilation pipeline consists of the following steps:

6.1  GATE SCHEDULING

  Maintain a ready-queue of gates whose per-qubit dependencies are fully satisfied -- i.e., all earlier gates on the same qubits have
  already executed. At each scheduling step, select the next gate using the following priority heuristic, whose core principle is to
  minimize unnecessary qubit movement by batching gates that can run without re-shuttling:

    - If multiple ready gates share the same qubit set (one or two qubits), execute all of them consecutively before moving those qubits
      away from the gate vertex.
    - Among the remaining ready gates, prefer gates whose qubits are currently closest (by trap_graph.distance) to a gate vertex.

  Additional scheduling strategies are encouraged; this is an open-ended optimization target.

6.2  ROUTING

  Routing is graph-aware: every primitive that walks along the trap must consult trap_graph.next_hop / trap_graph.bfs_path /
  trap_graph.distance instead of integer index arithmetic, since the topology is no longer a single linear path. Use is_on_main_axis /
  is_junction / stack_of / stacks_at to discriminate vertex roles; never infer them from index ranges.

  Internal compiler state (in addition to current_state and qubit(*@$\rightarrow$@*)vertex maps):
    - For each junction j currently holding ions, the entry vertex from which those ions arrived. This entry record must be updated on
      every TRANSLATE: writing the entry on TRANSLATE(*, j) and clearing it on TRANSLATE(j, *). It is consulted to enforce the
      no-roundtrip rule during routing decisions and during timestep grouping (Sec. 6.3).

  Terminology used below:
    - "next_gate_id(q)" -- the gate_id (Sec. 2) of the next gate involving qubit q that has not yet been scheduled, in program order. If
      q has no remaining gates, treat next_gate_id(q) as +(*@$\infty$@*).
    - "keep qubit" / "keep set" -- the qubit(s) currently being routed for the gate under consideration. For a single-qubit gate routing
      of q, the keep set is {q}; for a two-qubit gate of {qa, qb}, the keep set is {qa, qb}. Routing helpers must never displace a keep
      qubit except via the controlled displacements permitted by the Endpoint-protection rule (snapshot-wrapped, with guaranteed return
      before MERGE).

  Movement primitives (graph-aware, replace any linear-index versions):

    Endpoint protection (binding across the entire routing of a single gate):
      Once a routing pass has placed an endpoint qubit (qa or qb of the gate currently being routed) on a gv-neighbor, the vertex holding
      that qubit becomes a "placed" vertex. The placed qubit is protected only in the sense that, by the time MERGE is emitted, qa and qb
      must both sit on the two gv-neighbors. Subsequent routing of other qubits (evacuate / clear_junctions / move / push, and any of the
      Swap-, Stack-, Park-the-endpoint methods invoked while still routing the same gate) may temporarily displace a placed endpoint --
      including routing other qubits through the placed vertex via the Swap-Method, or temporarily moving the placed endpoint aside --
      provided the displacement is followed by a guaranteed re-delivery of that endpoint back to a gv-neighbor before MERGE. In
      particular:

        - A placed endpoint may be used as a swap partner in a Swap-Method step that routes another qubit past it; after the SEPARATE the
          endpoint may end up at a vertex adjacent to gv (possibly the opposite neighbor), and that is acceptable as long as it ends up
          at *one* of the two gv-neighbors when MERGE fires.
        - A placed endpoint may be temporarily relocated by any other method (Stack-Method push onto a free stack, Park-the-endpoint, a
          plain move) when doing so is the cheapest way to clear an obstruction, again provided the routing guarantees a return to a
          gv-neighbor before MERGE.
        - Endpoint chains must never be split: an endpoint must remain on its own (or paired with the other endpoint) -- it must not be
          merged with a non-endpoint qubit and left there. Merging with a non-endpoint is only temporarily allowed, e.g. to execute a
          MERGE+SWAP+SEPARATE for the Swap-Method, after which SEPARATE returns the endpoint to its own vertex within the same routing
          pass.

      To enforce "guaranteed return", the routing should plan the relocation and its reversal as a single transaction wrapped in
      snapshot/restore (see Sec. 12.1 for the canonical pattern), rolling back on failure. If, after the displacement, the placed
      endpoint cannot be routed back to either gv-neighbor without mutual exclusion, the routing must restart with a different gate
      vertex; on restart, the placed endpoint is fully unprotected and free to participate from scratch.

      Strict-blocking of placed vertices (i.e., adding every placed vertex unconditionally to the blocked / protect set of every helper
      call) remains an acceptable conservative strategy and is what the simplest finalize implementation should use first. The
      relaxations above are optimizations that the compiler may apply when a strict-blocked attempt would otherwise force a costly
      restart or an unnecessarily long detour. They are not preconditions for correctness.

    move(src, dst, blocked):
      Step along the unique src(*@$\rightarrow$@*)dst tree path using trap_graph.next_hop (or iterate over trap_graph.bfs_path). At each hop:
        - If the next vertex is occupied, open it iteratively via push (see below) using a work-stack (Sec. 12.4), passing blocked (*@$\cup$@*)
          {src}.
        - Emit TRANSLATE(cur, next).
      Returns False if no obstruction can be cleared. The path passes through junctions naturally; because bfs_path returns a simple
      path, each junction is entered and exited through different edges, so the no-roundtrip rule is respected automatically as long as
      the chain is flushed off the junction before any non-translation operation runs (see Sec. 6.3 grouping). move() must not terminate
      with the chain resting on a junction: if dst happens to be a junction, that is a caller bug (the spec only valid dst for move is a
      non-junction); if a junction is the last reachable vertex due to an obstacle, the routine returns False and the caller picks a
      different strategy.

    push(v, away_from, blocked):
      Move the chain at v one hop, away from away_from. Among neighbors of v not in blocked and not equal to away_from, prefer
      non-junction neighbors over junction neighbors (junctions are last resort, since they impose the no-roundtrip constraint). Break
      ties by vertex index for determinism. If the chosen neighbor is occupied, cascade-push it first via the iterative work-stack
      routine of Sec. 12.4 -- never use Python recursion (Sec. 10). The push() primitive here is the convenient short name used by
      move(); its work-stack implementation is documented in Sec. 12.4 (CASCADE PUSH FOR EVACUATE). Do not confuse it with Sec. 12.11
      (STACK PUSH (LIFO CASCADE)), which is a different routine that inserts a chain into a stack column. No-roundtrip skip rule: if v
      itself is a junction, skip any candidate neighbor that equals v's recorded entry edge -- translating back through the same edge the
      chain arrived on is forbidden by the no-roundtrip rule (Sec. 1). Pushing into a candidate neighbor that already holds a chain is
      separately forbidden by the dst-empty precondition of TRANSLATE (Sec. 3) and is handled by the cascade branch above.

    evacuate(v, blocked, keep=None, extra_avoid=None, force=False):
      Move the chain at v to the closest free non-junction vertex. Implement via BFS from v over the trap graph, applying the
      transit/destination rules of Sec. 12.2 to blocked and extra_avoid. Move the chain along that BFS path with move().
        - blocked: vertices treated according to Sec. 12.2 (occupied non-junc in blocked = off-limits; empty non-junc in blocked =
          transit-only; junction in blocked = transit-only; nothing else affected).
        - keep: optional set of qubit names. If keep is given and the chain at v contains any keep qubit, refuse and return False unless
          force=True (relaxed endpoint-protection rule, see Sec. 6.2). The caller is then expected to wrap the call in snapshot/restore
          and re-deliver.
        - extra_avoid: optional set of vertices that are off-limits as both target and transit (typically the junctions on the outer
          walk_path's path; see Sec. 12.3). Note: this parameter has the same semantics as hard_blocked introduced in Sec. 12.10; an
          implementation may merge them into a single parameter (recommended name: hard_blocked) and treat extra_avoid as an alias for
          backward compatibility.
        - force: when True, allow displacement of keep qubits encountered mid-walk. force=True is propagated through all recursive calls
          of evacuate / walk_path / find_empty_targets (see Sec. 12.10); callers must still pass hard_blocked to protect outer-walk
          source chains from being transited through.
      Returns True on success (chain moved to an empty non-junction target), False otherwise.

    clear_junction(j, blocked):
      If j holds ions, evacuate them off the junction so that the next non-translation operation can run.

  All routing helpers must take blocked as an explicit Set[int] of forbidden vertices and must honor it strictly on their *primary*
  invocation. A helper must not silently re-try itself with a weaker blocked set; if the primary attempt fails it returns False. A
  *caller* may, inside its own snapshot/restore transaction, re-invoke the helper with progressively weaker constraints (e.g. drop the
  path-junction extra_avoid, or fall back to a minimal pblocked containing only the source and final destination of the outer walk). Sec.
  12.3 documents the canonical three-step relaxation used inside walk_path's obstacle evac, and Sec. 12.5 lays out the outer fallback
  chain (declutter, decongest, last-resort case (b)). Every level of relaxation must be wrapped in its own snapshot/restore so a wrong
  choice rolls back cleanly. evacuate is allowed to return False; the caller, not the helper, decides whether to relax and retry or to
  abandon and try a different strategy entirely.

  Single-qubit gate routing:
    Given current_state and a QuantumGate with qubits = [qa]:
      1. If qa shares its vertex with another qubit, move the whole pair to a chosen gate vertex, evacuate the gate vertex's two
         neighbors, emit SEPARATE.
      2. Move qa to a gate vertex via move().
      3. Emit GATE(gv, gate).

    Stack-extraction prefix: if qa currently sits at slot k > 0 of some stack column (precisely: trap_graph.stack_of(qpos[qa]) is not
    None and stack_of(qpos[qa])[2] > 0), run the Sec. 6.2 Stack-reordering protocol before step 1, so qa (and any vertex-mate) ends up
    walked out to a main-axis non-junction before any SEPARATE is attempted. This matches the two-qubit pre-extract step in Sec. 12.5
    step 1 and keeps the single-qubit fast path deterministic. The target hint t to pass to the extraction routine is the gate vertex
    about to be used (Sec. 6.2 Phase 3 tie-breaker). When qa's chain at qpos[qa] has size (*@$\geq$@*) 2 and qa sits at slot k > 0 (the "buried
    2-chain" case), the stack-reordering protocol must separate the chain at qpos[qa] before any pop, parking qa's partner one slot
    deeper (or popping the partner outward first if no deeper slot exists) -- see the [step-1-strengthening] "Buried-2-chain stack
    extraction" block at the end of Sec. 12.5 for the detailed protocol. Without this separation step, a 2-chain cannot pop into a
    shallower single-slot vertex and extraction silently returns False.

    Fallback when isolation fails: if the standard isolation flow above cannot land qa alone at a gate vertex (e.g. because aggressive
    displacement during a force-walk left qa parked in a deep stack), the routing must take a snapshot and run a decongest pass: for
    every non-qa qubit, if it sits in a stack at slot > 0, run stack extraction on it; otherwise walk it toward any empty non-junction
    that is not on the qa-to-gate-vertex corridor. Re-run stack extraction on qa one more time (in case the decongest displaced it
    deeper), then retry the standard isolation flow. If that still fails, restore the snapshot and raise an exception per the
    no-partial-schedule rule (Sec. 11): the compiler must never emit a partially compiled circuit. Every gate of the circuit must be
    routed and emitted; if any gate cannot be routed by any available strategy, the compiler must raise an exception rather than return a
    schedule that omits that gate. Returning a partial schedule (one that the validator would reject for unexecuted gates) is forbidden
    -- the only two permitted outcomes are a complete, valid schedule or a raised exception. Without the second re-extract step, a
    force-walk that pushes qa one slot deeper into a stack leaves the chain unreachable on retry.

  Two-qubit gate routing -- iterative distance reduction:
    Given current_state and a QuantumGate with qubits = [qa, qb]:

    Step 0 -- trivial same-vertex case:
      If qa and qb are at the same vertex: move that vertex to a gate vertex (if not already there) and emit GATE.

    Step 1 -- initialization:
      Choose a gate vertex gv (ranked by the sum of distances from qpos[qa] and qpos[qb] to gv; ties broken by gv index). Let l and r be
      gv's two non-junction main-axis neighbors (l < r), and choose the assignment qa (*@$\rightarrow$@*) l, qb (*@$\rightarrow$@*) r that minimizes total movement --
      concretely distance(qpos[qa], l) + distance(qpos[qb], r) versus distance(qpos[qa], r) + distance(qpos[qb], l), swapping roles when
      the alternative is lower; see Sec. 12.9 for the tie-break.
      Define the "intermediate qubits" as those qubits q' (*@$\notin$@*) {qa, qb} whose current vertex v' lies on trap_graph.bfs_path(qpos[qa], l) or
      trap_graph.bfs_path(qpos[qb], r), excluding the endpoint vertices qpos[qa] and qpos[qb] but including l and r themselves whenever
      they are occupied by a non-endpoint chain. If the gate vertex does not lie on a BFS path, any chain at gv is handled separately by
      Step 3 case (b). This definition is the routing-intermediate set: it is exactly the set of qubits that obstruct the delivery of qa
      to l and qb to r, so clearing it is both necessary and sufficient for Step 3 to proceed.

      Edge cases:
        - If qpos[qa] coincides with l (likewise qpos[qb] with r), the corresponding BFS path is empty and contributes no intermediates
          from that side.
        - If qpos[qa] = qpos[qb] (Step 0 handled it already), skip Step 1 entirely.
        - The two BFS paths may share vertices (in particular when gv lies on a shared trunk). Take the set union; do not double-count.

    Step 2 -- iterative reduction. Repeat until no intermediate qubit remains. Each iteration shrinks the intermediate set, choosing one
    of the methods below. The Swap-Method removes one intermediate per iteration; the Stack-Method removes one or two; the
    Park-the-endpoint variant clears the entire intermediate set in a single iteration. Different iterations may pick different methods.
    The loop is driven by the routing-intermediate set defined in Step 1, so the loop is allowed to operate on qubits sitting at l or r
    or anywhere on a routing path from an endpoint to its target neighbor. Iteration count is bounded at trap_graph.num_vertices (*@$\cdot$@*) 6 + 30
    (same bound as Sec. 12.5 step 3); if the bound is reached without emptying the intermediate set, the inner attempt fails and falls
    through to the outer retries of Sec. 12.5.

      Swap-Method (analog of the linear-trap merge-swap-separate step):
        Choose direction:
          i1 = intermediate qubit closest (graph distance) to qa;
          ik = intermediate qubit closest to qb.
          Pick whichever side minimizes the number of timesteps to set up the merge.
        Prepare:
          If the chosen mover qubit (qa or qb) shares a vertex, isolate it via the same SWAP-then-SEPARATE recipe as in the linear case
          (the qubit's order in the chain dictates whether SWAP is required before SEPARATE; the chosen partner i1 / ik is isolated
          likewise).
        Merge and propagate:
          Move the mover to one neighbor of the chosen gate vertex and the partner to the other neighbor (assignment chosen so that the
          final chain order after MERGE matches the linear-case recipe). Emit MERGE, SWAP, SEPARATE. The intermediate set shrinks by one.

        Universal-fallback usage of the Swap-Method:
          The Swap-Method can always be applied when the Stack-Method and the Park-the-endpoint variant are unavailable or fail. The Step
          1 routing-intermediate set covers obstructors on both bfs_path(qa, l) and bfs_path(qb, r); when the Swap-Method is the only
          feasible reducer, walk those paths from each endpoint toward its target gv-neighbor, one intermediate per merge-swap-separate
          step.
          Equivalently, the consecutive intermediates encountered while shrinking the set are exactly the qubits on
          trap_graph.bfs_path(qpos[qa], qpos[qb]) extended through gv to l/r, consumed one at a time as qa and qb move toward their
          targets. Each step removes one qubit from the intermediate set, so a finite number of iterations clears the entire BFS path.

          Free placement of displaced intermediates: once a qubit has been removed from the intermediate set by a Swap-Method step, it is
          no longer constrained to lie on the qa(*@$\leftrightarrow$@*)qb BFS path. It may therefore be moved anywhere in the graph during or after the
          displacement translation. In particular, when the Swap-Method's SEPARATE deposits the displaced qubit at a gate-vertex
          neighbor, the qubit can subsequently be routed off the BFS path entirely -- across a junction onto a side branch, into a free
          stack slot, or onto any empty qubit vertex anywhere in the trap, before the next Swap-Method iteration begins. The only
          requirements are:
            - the displaced qubit must end up at a location that is not on the (updated) BFS path between qa and qb at the start of the
              next iteration, so that it does not re-enter the intermediate set;
            - the routing of the displaced qubit honors the no-roundtrip rule for every junction it traverses, and respects every other
              active blocked / protect set (endpoint protection, endpoint-stack protection, etc.). This gives the Swap-Method the
              flexibility to "park" a displaced intermediate at a junction by translating it onto one of the junction's stacks or onto
              any branch that is not on the qa(*@$\leftrightarrow$@*)qb path. The combination of repeated Swap-Method iterations plus free off-path placement
              of each displaced intermediate guarantees that the Swap-Method alone can always clear the intermediate set, regardless of
              the topology -- making it the unconditional fallback strategy when Stack-Method, Park-the-endpoint, and stack-slot eviction
              do not apply.

      Stack-Method:
        Iterate j over the junctions on the routing-intermediate paths -- i.e. trap_graph.junctions_on_path(qpos[qa], l) (*@$\cup$@*)
        trap_graph.junctions_on_path(qpos[qb], r). For each such j, inspect trap_graph.stacks_at(j); if any of those stacks has at least
        one free slot (its deepest occupied slot is shallower than the stack length), an intermediate qubit q_inter near j can be pushed
        onto that stack instead of being shuffled past via merge-swap-separate. Pre-condition: q_inter's current vertex contains only
        intermediate qubits -- i.e., neither qa nor qb shares that vertex. If q_inter is alone, only it is pushed. If q_inter shares its
        vertex with another intermediate qubit, both qubits in that chain are pushed together as a single chain (TRANSLATE moves whole
        chains), removing two intermediates at once. If q_inter shares a vertex with qa or qb, the Stack-Method does not apply directly
        -- isolate q_inter first via SEPARATE, or fall back to the Swap-Method this iteration.
        Action: route q_inter (or its 2-qubit chain) to the junction j with move(), then push it onto the stack via a single TRANSLATE of
        j (*@$\rightarrow$@*) slot 0. By the LIFO mechanics in Sec. 1, any qubits already in the stack shift one slot deeper in the same timestep
        automatically; no extra cost. After the push, q_inter no longer lies on either of the routing-intermediate paths bfs_path(qa, l)
        or bfs_path(qb, r), so the intermediate set shrinks by one or two.

        Refinements:
          - LIFO ordering preference: q_inter will sit at slot 0 after the push and must therefore be popped before every qubit currently
            deeper in the stack. The ideal case is therefore that next_gate_id(q_inter) is (*@$\leq$@*) the next_gate_id of every existing
            occupant; when this holds, no evacuation is forced later. In practice this preference is often unattainable -- the trap
            rarely has enough free non-junction qubit to keep order strictly. The rule is therefore a tie-breaker, not a hard
            precondition: prefer stacks that satisfy the LIFO order; otherwise still push, and account for the eventual cost of having to
            pop and re-park occupants when their gates come due. Pushing onto an order-violating stack is acceptable as long as the cost
            estimator (see "Method choice each iteration" below) accounts for the future evacuation overhead and the chosen method still
            beats the alternatives.
          - Look-ahead skip: do not push q_inter when its next gate is scheduled to fire within fewer timesteps than one push + one pop
            ((*@$\approx$@*) 2 translation timesteps plus the junction-traversal hops to and from the stack). The retrieval would land before any
            saving materializes and would block the main-axis path during a later gate's routing.
          - Multi-stack parallelism: if several intermediates lie near distinct junctions, push them onto different stacks. Translations
            into disjoint stacks involve pairwise disjoint vertex pairs and therefore share a single translation timestep (Sec. 6.3).
          - Pre-stacking before scheduling reaches the gate: when the scheduler is idle on some main-axis stretch (no ready gate
            currently uses those qubits), opportunistically push qubits whose next gate is far in the future onto nearby stacks. This is
            equivalent to advancing future Stack-Method steps and frees the main axis for in-between gates.
          - Co-routing with SEPARATE: when an intermediate pair must first be isolated via SEPARATE on a gate vertex, route one of the
            resulting qubits toward the stack entry while the other is moved toward the opposite side of the gate vertex; this overlaps
            the stacking move with the routing of the partner qubit.

      Park-the-endpoint variant:
        Instead of pushing an intermediate aside, an entire path can be cleared by pushing qa or qb itself onto a "park slot" -- either
        (a) a stack s connected to a junction j, or (b) a far non-junction main-axis vertex that is not on the routing-intermediate
        paths. Once qa (or qb) is at the park slot, the intermediate qubits can be translated freely past the previously blocking
        endpoint; afterward qa is popped from s (or walked back from the main-axis park) and moved to its target gv-neighbor (l for qa, r
        for qb). For stack parking the LIFO mechanics guarantee constant-time push/pop; for main-axis parking the cost is simply
        distance(qpos[endpoint], park_vertex). During this routing episode no intermediate qubit may be pushed onto a stack that holds a
        parked endpoint (that would land the intermediate above qa and force qa to be popped before it -- LIFO order).

        Refinements:
          - Park slots include both stack slots and far main-axis non-junction vertices. For "very linear" traps -- a long main axis with
            one or two long stacks branching off, plus main-axis ends that extend past the junctions (graph_33_10_1 is the canonical
            example: V0-V4 to the left of junction V5 and V10-V14 to the right of junction V9 are "linear arms" five hops deep,
            equivalent in capacity to the 10-deep stacks at V15+ and V25+) -- every non-junction vertex on the main axis at distance (*@$\geq$@*) 3
            from gv is also a valid park slot. The implementation must enumerate both stack-slot candidates and far-main-axis candidates
            when scheduling Park-the-endpoint. When all stack slot 0s are already occupied with non-keep chains (Stack slot eviction
            would be required otherwise), main-axis-end parking is often the cheaper recovery.
          - Endpoint parking is flexible. qa and qb can be parked independently in any combination, and the choice of which stacks (or
            which main-axis vertices) are used is irrelevant for correctness -- only for cost. The compiler may:
              (*@$\cdot$@*) Park only qa (or only qb), in any stack with a free slot or at any far main-axis non-junction vertex.
              (*@$\cdot$@*) Park qa and qb in the same stack: push qa first, then push qb on top (the roles of qa and qb are exchangeable). After the
                second push, qb sits at slot 0 and qa at slot 1. Since both will be popped and routed to gv-neighbors afterward, the LIFO
                order is fine -- pop qb, route it to one neighbor of gv, pop qa, route it to the other neighbor, MERGE, GATE. While both
                endpoints share the stack, no intermediate may be pushed onto it (that would separate qa from qb in LIFO order).
                Retrieval costs two consecutive pop timesteps instead of one.
              (*@$\cdot$@*) Park qa and qb in different stacks, attached to the same junction or to different junctions -- the routing is identical
                in either case. Pushing qa and qb onto stacks at different junctions can also be done in parallel within a single
                translation timestep when the two routes are disjoint, which is often cheaper than serial pushes onto one junction's
                stacks.
              (*@$\cdot$@*) Park qa at a main-axis end, qb in a stack (or vice versa), or park both at opposite main-axis ends -- whichever
                combination clears the most of the routing-intermediate paths at lowest cost.
            In all variants, once the endpoints are parked the intermediates can be translated freely along the cleared stretch of the
            main axis, often collapsing several Swap-Method iterations into a single translation timestep batch.
          - Endpoint choice by post-gate consumer: when only one of qa, qb is parked, prefer to park whichever has the more distant next
            gate (the parked qubit must be popped again; if its next gate is far, the extra hop is usually amortized by other movement).
          - Linger across multiple gates: if the parked endpoint's next gate is also reachable via the same junction, the qubit may
            remain in the stack across several gates rather than being popped immediately. Track per-qubit "extraction deadline" against
            next_gate_id and only pop when the deadline forces it.
          - Combine with Stack-Method: park one endpoint in a stack at one junction and push one or more intermediates onto another stack
            (different junction) in parallel; the two stacking translations can share the same translation timestep.

      Method choice each iteration: estimate cost using actual graph distances and the constant-time stack push/pop mechanic:
        - Swap-Method (*@$\approx$@*) 4 + max(0, distance(qa, gv) + distance(i1, gv) - 2) (MERGE + SWAP + SEPARATE plus loading the two neighbors of
          gv).
        - Stack-Method (*@$\approx$@*) distance(qpos[q_inter], j) + 1 per push (the +1 is the j (*@$\rightarrow$@*) slot 0 translation; existing stack contents shift
          in the same timestep at no extra cost). A second intermediate co-pushed as part of the same chain costs nothing extra. Eventual
          extraction adds 1 + distance(j, target) per popped chain.
        - Park-the-endpoint (stack park) (*@$\approx$@*) distance(qpos[qa or qb], j) + 1 for the push, plus 1 + distance(j, gv-neighbor) for the
          eventual pop and delivery to the gate vertex.
        - Park-the-endpoint (main-axis-end park) (*@$\approx$@*) 2(*@$\cdot$@*)distance(qpos[qa or qb], park_vertex), i.e. one walk to the park vertex plus one
          walk back to the gv-neighbor when MERGE is due. Main-axis parking has no LIFO constraint and never blocks future stack pushes;
          prefer it over stack parking when the cost is equal or smaller, and always when the candidate stack already holds chains whose
          next gates are sooner than the parked endpoint's next gate. Pick the cheapest method. Different iterations may pick different
          methods; gate-vertex choice may also vary between iterations. Tie-break order when two methods yield equal estimated cost: (1)
          Stack-Method, (2) Park-the-endpoint, (3) Swap-Method (Stack first because it has no SEPARATE/SWAP/MERGE overhead). When two
          candidate vertices within the same method tie, break by the vertex that the method's cost formula refers to -- the intermediate
          vertex's index for Stack-Method and Swap-Method, the junction's index for Park-the-endpoint -- ascending.

      Other strategies are possible (e.g., mixing swap and stack steps within one reduction; opportunistically using any free
      junction-adjacent slot for a cheap detour; allowing intermediates to overshoot through the junction onto the far side of the main
      axis when both endpoints have already cleared it). Two-qubit routing is therefore an open-ended optimization target; the
      Swap-Method, Stack-Method, and park-the-endpoint variant above are the required baselines.

      Swap-Method same-side handling: when both the chosen mover and its intermediate partner currently sit on the same side of gv (so
      the mover's path to its target gv-neighbor passes through the other gv-neighbor, which holds the partner), the naive sequence "move
      target_l to l with blocked = {gv, r}", then "move target_r to r with blocked = {gv, l}" deadlocks: the first move's strict-blocked
      invocation refuses to displace the partner sitting at r. The Swap-Method must handle this by temporarily relocating the blocking
      partner to a safe holding vertex (an empty non-stack, non-junction main-axis vertex not on the mover's path), completing the
      mover's move, then restoring the blocking partner to its destination. With this two-step relocation, the Swap-Method always
      succeeds for a single intermediate regardless of same-side / different-side configuration -- making it the unconditional backup
      whenever Stack-Method, Park-the-endpoint, or stack-slot eviction do not apply (see Universal-fallback usage of the Swap-Method
      above for the off-path displacement rule).

      Stack-Method atomicity: each Stack-Method candidate push is a multi-step procedure (route the intermediate to the junction, then
      translate to slot 0). The traversal itself may displace blockers via cascading evacuations. If the final push step then fails (slot
      0 became occupied during the traversal, or no-roundtrip blocks the push, or move couldn't reach the junction), the partial state
      changes from that candidate must be rolled back via snapshot/restore before the next candidate is tried. A partial push that left
      an endpoint or a non-keep qubit displaced and then declared "failure" without rollback leaves the trap in a worse configuration
      than it started: e.g. an endpoint shoved into a stack while the original target slot 0 becomes blocked, parking the endpoint where
      it cannot subsequently be routed. Snapshot before each candidate; restore on any failure.

      Preferred displacement target for routing-intermediates: when an iteration relocates a qubit out of the routing-intermediate set
      (Step 1), the preferred destination is a free stack slot (slot 0 of a stack whose slot 0 and junction are both empty), not a
      one-hop vacate to an adjacent main-axis vertex. A single-hop shift often leaves the chain on an adjacent junction or on another
      vertex still on the qa(*@$\rightarrow$@*)l or qb(*@$\rightarrow$@*)r path, immediately reintroducing it as an intermediate. A stack push terminates the displacement
      in one translation timestep regardless of how many other stacks are populated. Use Stack-Method first; if no free stack is
      reachable, fall back to Swap-Method or generic evacuate.

      Endpoint-stack protection (cascade-push avoidance): if an endpoint qa or qb currently sits in a stack column (slot k of some stack
      S, k (*@$\geq$@*) 0), the routing must never push another qubit onto slot 0 of S while that endpoint is still parked there. Doing so triggers
      the LIFO cascade and shoves the endpoint one slot deeper (slot k (*@$\rightarrow$@*) slot k+1), after which retrieving the endpoint requires first
      popping the new arrival back out. Therefore: at every point during the routing of the gate (qa, qb) -- Step 2 iterations and Step 3
      finalization alike -- every slot of S other than the endpoint's own slot must be added to the blocked set of every evacuate / move
      / Stack-Method / Park-the-Endpoint call. Free-stack searches must skip S entirely. The endpoint's own slot remains unblocked so the
      endpoint can pop out itself.

      Endpoint chain protection (implementation note): evacuate naturally moves the chain at its target vertex one hop forward as part of
      its reverse-shift. If the chain at that vertex contains qa or qb, this displaces an endpoint, which can cascade the endpoint into
      an arbitrary far vertex. To keep this controlled, evacuate should by default refuse when the chain at its target contains a keep
      qubit, returning False so the caller can pick a different strategy. The relaxed Endpoint-protection rule (Sec. 6.2 movement
      primitives) lists which displacements are nonetheless permitted -- always paired with snapshot/restore so the trap state is
      recoverable on failure.

      Endpoint-on-junction at MERGE/SEPARATE/SWAP/GATE: see Sec. 12.7's hard invariant -- between non-translation ops every junction must
      be empty, and primitives are responsible for draining any transiting chain off a junction before returning.

      Stack-slot eviction (cascading stack rearrangement): on densely-packed traps (graph_33_*, graph_15_5_1, graph_11_2_1, graph_11_3_1)
      every slot 0 of every stack may be occupied at the time Stack-Method tries to park an intermediate. Strict Stack-Method ("only push
      when slot 0 is empty") then has no candidate and returns False, leaving the intermediate to block the merge path. In that case the
      routing must attempt **stack-slot eviction**: pick a stack whose slot 0 occupant is not a keep qubit and whose next gate is far in
      the future, pop that occupant onto a free non-stack vertex (or another free stack), then push the intermediate onto the freed slot
      0. The pop must respect the no-roundtrip rule for the junction; the destination of the evicted qubit must not lie on either
      routing-intermediate path (bfs_path(qa, l) or bfs_path(qb, r)) nor be a gv-neighbor. Without stack-slot eviction, fully-loaded
      traps remain unsolvable. This is the extension of the Stack-Method baseline that Sec. 6.2's "Stack reordering" prose alludes to:
      "pop the relevant prefix of the stack out through the junction onto temporary holding vertices (*@$\ldots$@*) and re-push". Implement that
      explicitly, not just as a future tweak.

      Stack-reordering for trapped endpoints (Sec. 1 mechanics applied to endpoints): if at the start of routing an endpoint qa or qb
      sits at slot k > 0 of some stack column -- i.e. one or more other qubits occupy slots 0(*@$\ldots$@*)k-1 above it -- the endpoint cannot be
      popped directly. The routing must first extract the trapping qubits, then walk the endpoint out to the main axis. The extraction
      routine has three phases and must run before iterative reduction begins (the Swap-Method cannot pull an endpoint up through a
      stack):

        Phase 1 -- pre-clear the junction-adjacent main-axis vertices, and reserve at least one of them as a Phase-3 landing pad. Each
        popped occupant has to land somewhere; the natural landing pad is a main-axis neighbor of the junction. For each main-axis
        neighbor nb of the junction j that is currently non-empty, evacuate nb with blocked equal to the entire stack we are about to
        drain. Blocking the whole stack is essential: it prevents the displaced main-axis chain from being pushed back into the very
        stack we want to empty.

          Reserve-one-landing rule: pick one main-axis neighbor of j (the first one already empty, or -- if all are non-empty -- the
          first one that successfully evacuates) and treat it as the dedicated Phase 3 landing vertex. During Phase 2, that vertex must
          be added to the per-slot blocked set so that displaced shallower-slot chains do not re-occupy it. Without this rule, the
          natural displacement order in Phase 2 fills both main-axis neighbors of j with the displaced shallower-slot occupants, and
          Phase 3 then has nowhere to land (both gv-junction-neighbors blocked, both forbidden by no-roundtrip for some routings, etc.).
          The reserved landing vertex must remain empty until Phase 3 emits TRANSLATE(j (*@$\rightarrow$@*) landing).

        Phase 2 -- pop each shallower slot top-down. For each occupied slot k in 0(*@$\ldots$@*)slot-1, evacuate the slot vertex to anywhere outside
        the stack, using blocked equal to {all other slots of the same stack} (*@$\cup$@*) {the reserved Phase-3 landing vertex from Phase 1}. If
        that fails, retry with the relaxed blocked set containing only the endpoint's own slot vertex. The cascade-shift translations
        described in Sec. 1 (every occupied k+1 shifts to k within the same timestep) are emitted as discrete TRANSLATE operations one
        per slot; the chain-of-translations grouping rule of Sec. 6.3 will pack them into a single timestep when the grouping pass runs.

        Phase 3 -- walk the endpoint out. The caller passes a "target vertex" hint t to this phase (the gate vertex about to be used for
        the single-qubit gate, or the best gate vertex by combined-distance order for a two-qubit pre-extract -- see Sec. 12.5 step 1).
        Starting from the endpoint's current slot, while the endpoint is not yet on the main axis: if it is in the stack at slot > 0,
        translate it one slot shallower (the previous slot is empty because Phase 2 emptied it); if it is at slot 0, translate it onto
        the junction; if it is on the junction, translate it onto an empty main-axis neighbor (evacuating that neighbor first if still
        occupied, which Phase 1 should already have ensured). The choice of main-axis neighbor is governed by the Non-junction landing
        rule and the forced-junction tie-breaker below -- both consult t.

          Non-junction landing rule: when picking which main-axis neighbor of the stack's junction j to land the chain on after the j (*@$\rightarrow$@*)
          main-axis translation, prefer a neighbor that is NOT itself a junction. If a chain lands on a junction-on-the-main-axis (j2),
          then j2 records its entry from j, and the next translation involving j2 cannot exit back to j (no-roundtrip). Almost any
          subsequent walk toward a gate vertex starts by attempting precisely that "back through j" translation, so landing on a junction
          routinely creates an unrecoverable no-roundtrip stall. Only fall back to a junction main-axis neighbor when no non-junction
          neighbor is available.

          Tie-breaker for the forced-junction fallback: when both of j's main-axis neighbors are themselves junctions (j sits in a run of
          back-to-back junctions on the trunk), the extraction caller must pass a "target vertex" hint t. For a single-qubit gate this is
          the gate vertex about to be used; for a two-qubit gate's pre-extract this is the gate vertex of the gate (not the partner
          endpoint), because the chain must ultimately reach a gv-neighbor regardless of where the partner sits -- landing toward the
          partner instead biases the chain away from gv and creates a no-roundtrip stall the moment the routing tries to walk to gv.

          With the hint t, filter the candidates as follows. For each candidate landing neighbor n of j, compute trap_graph.bfs_path(n,
          t): if that path passes through j itself, n is "unsafe" -- landing there forces the very next translation to attempt T(n, j),
          which is forbidden by no-roundtrip at n (n just received the chain from j). Prefer "safe" candidates (path to t does not
          revisit j) over unsafe ones, then break ties within each class by trap_graph.distance(n, t), then by vertex index. Only when
          every candidate is unsafe does the fallback land on an unsafe neighbor; in that case the subsequent routing is likely to fail
          and must unwind to an outer fallback level. This hint and filter matter only in the all-junctions-fallback branch -- when at
          least one non-junction main-axis neighbor exists, the primary rule already gives a no-roundtrip-safe landing.

      The two non-obvious points are: (a) the Phase 1 pre-clear pass blocks the entire stack so the evacuee never returns to it via a
      different slot, and (b) Phase 1 runs before Phase 2 so each pop has a reachable empty destination near the junction. Without both,
      routing fails on traps with depth-5+ stacks fully packed at one end.

      Slot-0 special case: when the endpoint is at slot 0 of a stack (no shallower-slot occupants), Phase 2 is a no-op; Phase 1 and Phase
      3 still apply (pre-clear the junction-adjacent main-axis neighbor, then walk slot 0 (*@$\rightarrow$@*) junction (*@$\rightarrow$@*) main axis).

    Step 3 -- finalization:
      After Step 2's iterative reduction, the routing-intermediate set is empty. By construction this means qpos[qa] = l and qpos[qb] = r
      (or vice versa) -- both endpoints sit at the chosen gv-neighbors. The only remaining obstruction can be the gate vertex gv itself,
      if it still holds a non-endpoint chain. Finalization runs as follows:

        (a) If gv is empty: clear_junctions(blocked = {gv, l, r}), emit MERGE(gv), emit GATE(gv, gate). Done.

        (b) If gv holds a chain (necessarily non-endpoint, since qa and qb are at l and r), the chain at gv is a 1- or 2-qubit obstacle
            sandwiched directly between qa at l and qb at r. Apply one Swap-Method step with qa and qb as the two endpoints and the
            gv-occupants as the (single) intermediate: MERGE qa with the gv-chain at l (or r, whichever side the SWAP-then-SEPARATE
            recipe dictates), SWAP, SEPARATE -- the gv-occupant now lands on one of gv's neighbors opposite to where it came from, and qa
            moves through. Because gv's two neighbors are both non-junction and lie on the main axis, this Swap-Method step is always
            feasible. After the gv-occupant is displaced, evacuate it off l/r via Stack-Method (push onto a free stack at a nearby
            junction) or via Swap-Method's free-placement rule (move it anywhere off the qa(*@$\leftrightarrow$@*)qb path). Re-deliver qa to l and qb to r,
            then run case (a).

            Explicit case (b) protocol -- try every (anchor, anchor_target, other, other_target) combination in turn, wrapping each
            attempt in snapshot/restore. The four combinations are:
              (qa at l, qb to r), (qa at r, qb to l),
              (qb at l, qa to r), (qb at r, qa to l).
            For each combination, perform the following steps in order; on any failure restore the per-attempt snapshot and continue with
            the next combination.

              Step 1. Move the anchor qubit to its anchor_target (the gv-neighbor we want it on going into the M-W-S), with blocked =
              {other_target}.

              Step 2. If other_target is currently non-empty, evacuate it somewhere off the (anchor_target, gv, other_target) triangle --
              use blocked = {anchor_target, gv}.

              Step 3. Translate the chain currently at gv into other_target.
                Special case: if gv is already empty by this point (which means the gv-chain was the anchor itself, an endpoint that we
                just moved out in step 1), skip the M-W-S entirely; instead, deliver other to anchor_target (rotation-through-gv may
                apply, see step 6), clear junctions, emit MERGE and GATE, return success.

              Step 4. Clear junctions with blocked = {l, r, gv}, verify that l and r each hold exactly one qubit, then emit MERGE at gv.
              The chain at gv after MERGE is in (l-side, r-side) order; emit SWAP at gv to reverse it; emit SEPARATE at gv. After
              SEPARATE, the anchor qubit sits at other_target and the displaced gv-chain sits at anchor_target.

              Step 5. Evacuate the displaced gv-chain off anchor_target, with blocked = {other_target, gv}.

              Step 6. Deliver the other qubit to anchor_target. Compute its tree-path to anchor_target via bfs_path. If that path passes
              through other_target (which now holds the anchor), use the rotation-through-gv subroutine: translate from other_target to
              gv, then from gv to anchor_target -- this walks the anchor across gv to anchor_target and frees other_target -- then move
              the other qubit into the freed other_target with blocked = {anchor_target, gv}. Otherwise just move the other qubit to
              anchor_target with blocked = {other_target, gv}.

              Step 7. Clear junctions and emit MERGE, then GATE.

            Why all four anchor combinations: depending on which endpoint occupies which gv-neighbor at entry and which way the gv-chain
            should be displaced for the cleanest subsequent evac, exactly one of the four attempts is usually feasible. Trying them in
            turn is required to handle dense traps. Why the rotation-through-gv in step 6: when other is on the opposite side of gv from
            anchor_target, the tree-unique path passes through other_target (where anchor sits). On a tree there is no alternate route,
            so the only way to deliver other to anchor_target without ejecting anchor permanently is to swap their positions via gv's
            two-step T relay.

      Endpoint relocation during finalization: by default, once qa is at l and qb is at r, no further relocation should be needed except
      the controlled Swap-Method of case (b) (which itself uses qa or qb as a swap-partner -- a controlled displacement allowed by the
      relaxed Endpoint protection rule). If for any other reason an already-placed endpoint must move (e.g. an intermediate that cannot
      be cleared without routing through l or r), the routing must either:
        - plan the displacement and its reversal as a snapshot/restore transaction (see Endpoint chain protection), or
        - restart finalization with a different gate vertex, in which case both endpoints are unprotected again.
      An endpoint must never be left at a vertex other than l or r when MERGE is emitted (see the "Endpoint-on-junction is forbidden at
      MERGE" rule above for the junction-empty part).

  Stack choice priority:
    Slot depth is fully determined by insertion order (Sec. 1), so the compiler chooses only which stack a qubit is pushed onto, not
    where in the stack it lands. Enumerate candidate stacks via trap_graph.stacks_at(j) for each junction of interest; rank as follows,
    with ties broken by junction vertex index, then by stack_index returned by stacks_at:
      1. Free space. The stack must have at least one free slot (i.e., its deepest occupied slot is shallower than the bottom;
         equivalently, the number of currently occupied slots is strictly less than len(trap_graph.stacks_at(j)[stack_index])). This is
         the only hard requirement.
      2. LIFO compatibility (preference, not requirement). The new arrival becomes the next qubit to be popped. Prefer stacks where the
         new arrival's next_gate_id is (*@$\leq$@*) the next_gate_id of every qubit already in that stack, so existing occupants are not blocked
         when their gates come due. When no such stack exists, pushing onto an order-violating stack is still allowed; the cost estimator
         must account for the eventual pop-and-re-park overhead. This relaxation is necessary because the trap rarely has enough free
         non-junction vertices to enforce strict LIFO order.
      3. Distance to next consumer. Among eligible stacks, prefer the one whose junction minimizes trap_graph.distance from the new
         arrival's likely future location (gate vertex of its next gate).
      4. Distance to current location. As a secondary tie-breaker, prefer the stack whose junction is closest to the qubit's current
         vertex, to minimize the cost of the push itself.

    Stack reordering is possible but only by full-or-partial evacuation, never in place: qubits inside a stack cannot pass each other. To
    change the order, pop the relevant prefix of the stack out through the junction onto temporary holding vertices (main axis or another
    stack), realize the desired ordering, and re-push. The pop and re-push must each respect the junction no-roundtrip rule. Reordering
    only pays off when its translation cost is less than the extraction cost it saves; estimate before emitting. The chosen ordering must
    remain consistent with the deterministic-ordering requirement of Sec. 10. This pop-and-re-push mechanic is the same machinery used by
    the Stack-slot eviction rule and by Stack-reordering for trapped endpoints (both above) -- three different applications of the same
    primitive.

  Notes for both single- and two-qubit cases:
    - When several gate vertices are available, rank candidates by trap_graph.distance to the qubit's current location, not by index
      difference, and honor any "avoid" set imposed by the routing context.

  Chain-order optimization after two-qubit gate execution:
    After routing, the two target qubits are co-located at the gate vertex. Since QuantumGate.qubits order does not affect gate
    correctness, but qubit order within the chain affects routing of subsequent gates, use a one-step lookahead over the next gate of
    each qubit and insert a post-execution SWAP only when doing so brings more qubits to their preferred side. Use trap_graph.distance to
    determine the preferred side. When in doubt, do not insert the SWAP -- an unnecessary SWAP can cascade into extra separations
    downstream.

6.3  TIMESTEP GROUPING AND POST-PROCESSING

  A timestep is a set of operations executed simultaneously. After the routing emits the raw operation stream, apply the following passes
  in order:

    1. Roundtrip elimination -- remove a TRANSLATE A(*@$\rightarrow$@*)B immediately followed by TRANSLATE B(*@$\rightarrow$@*)A when no intervening operation references
       A or B (as src, dst, or operand vertex). Removing the pair must leave A's and B's contents identical to the no-removal trace at
       every intermediate step.

       General junction soundness rule: let the candidate removal be a set R of (possibly nested) (X_i(*@$\rightarrow$@*)Y_i, Y_i(*@$\rightarrow$@*)X_i) pairs. R is legal
       iff replaying the trace with all pairs in R deleted yields, for every junction j, a sequence of entries and exits at j in which no
       exit reuses the immediately preceding entry edge. The pair (X(*@$\rightarrow$@*)Y, Y(*@$\rightarrow$@*)X) by itself is never the junction violation (it is a normal
       back-and-forth on Y); a violation can only appear in what the removal exposes around it. The rewriter therefore replays the trace
       after each candidate removal, tracks each junction's current entry vertex per chain, and accepts only when the residual is legal.
       When a singleton R fails the check, try a combined R that also drops the smallest enclosing roundtrip -- a combined removal can be
       legal even when no proper subset is. The basic non-junction case is recovered by |R| = 1.

       Six illustrative consequences of the general rule (B is a junction in all):

         Example 1 -- fully nested roundtrip, smallest legal R is the pair:
           TRANSLATE(A(*@$\rightarrow$@*)B)
           TRANSLATE(B(*@$\rightarrow$@*)C)
           TRANSLATE(C(*@$\rightarrow$@*)B)
           TRANSLATE(B(*@$\rightarrow$@*)A)
         Removing only the inner pair (B(*@$\rightarrow$@*)C, C(*@$\rightarrow$@*)B) leaves A(*@$\rightarrow$@*)B, B(*@$\rightarrow$@*)A, which is a junction roundtrip violation at B (entered from A,
         exits back to A). Removing only the outer pair (A(*@$\rightarrow$@*)B, B(*@$\rightarrow$@*)A) leaves B(*@$\rightarrow$@*)C, C(*@$\rightarrow$@*)B with no source for the chain at B and is
         therefore unsound for chain contents as well. Removing both pairs together leaves an empty residual, which is trivially legal.
         Apply this combined removal recursively: arbitrary nested roundtrips around a junction collapse simultaneously when every layer
         pairs up.

         Example 2 -- outer translation continues to a different exit:
           TRANSLATE(A(*@$\rightarrow$@*)B)
           TRANSLATE(B(*@$\rightarrow$@*)C)
           TRANSLATE(C(*@$\rightarrow$@*)B)
           TRANSLATE(B(*@$\rightarrow$@*)D) with (D (*@$\neq$@*) A)
         Removing only the inner pair (B(*@$\rightarrow$@*)C, C(*@$\rightarrow$@*)B) leaves A(*@$\rightarrow$@*)B, B(*@$\rightarrow$@*)D. The chain enters B from A and exits to D (*@$\neq$@*) A -- legal. The outer
         translations stay. Recursively, sequences of multiple inner detours through B can be peeled off one at a time as long as each
         successive exit from B remains different from the chain's most recent entry into B.

         Example 3 -- multiple stacked detours, all-or-none:
           TRANSLATE(A(*@$\rightarrow$@*)B)
           TRANSLATE(B(*@$\rightarrow$@*)C)
           TRANSLATE(C(*@$\rightarrow$@*)B)
           TRANSLATE(B(*@$\rightarrow$@*)D)
           TRANSLATE(D(*@$\rightarrow$@*)B)
           TRANSLATE(B(*@$\rightarrow$@*)A) with (D (*@$\neq$@*) A, D (*@$\neq$@*) C)
         The chain enters B from A, makes two detours (to C and to D), and leaves back to A. Each individual exit in the original trace
         alternates entry edges (legal). Removing only one inner pair, e.g. (B(*@$\rightarrow$@*)C, C(*@$\rightarrow$@*)B), yields A(*@$\rightarrow$@*)B, B(*@$\rightarrow$@*)D, D(*@$\rightarrow$@*)B, B(*@$\rightarrow$@*)A -- still legal,
         since each exit from B continues to alternate. But removing both inner pairs alone yields A(*@$\rightarrow$@*)B, B(*@$\rightarrow$@*)A -- a junction violation.
         The minimal R that collapses everything is therefore the pair (A(*@$\rightarrow$@*)B, B(*@$\rightarrow$@*)A) together with every inner detour pair
         simultaneously; or, equivalently, peel off the inner pairs one at a time first and stop before exposing the A(*@$\rightarrow$@*)B, B(*@$\rightarrow$@*)A
         roundtrip. The general rule subsumes both choices.

         Example 4 -- recursive Pattern 2:
           TRANSLATE(A(*@$\rightarrow$@*)B)
           TRANSLATE(B(*@$\rightarrow$@*)C)
           TRANSLATE(C(*@$\rightarrow$@*)B)
           TRANSLATE(B(*@$\rightarrow$@*)D)
           TRANSLATE(D(*@$\rightarrow$@*)B)
           TRANSLATE(B(*@$\rightarrow$@*)E) with (E (*@$\neq$@*) A, D (*@$\neq$@*) C, E (*@$\neq$@*) D)
         Apply Example-2 reasoning to (B(*@$\rightarrow$@*)C, C(*@$\rightarrow$@*)B): residual A(*@$\rightarrow$@*)B, B(*@$\rightarrow$@*)D, D(*@$\rightarrow$@*)B, B(*@$\rightarrow$@*)E is legal (D (*@$\neq$@*) A, E (*@$\neq$@*) D). Apply Example-2 again to
         (B(*@$\rightarrow$@*)D, D(*@$\rightarrow$@*)B): residual A(*@$\rightarrow$@*)B, B(*@$\rightarrow$@*)E with E (*@$\neq$@*) A -- legal. Inner pairs peel off one at a time as long as each successive exit
         avoids the chain's current entry edge.

         Example 5 -- non-adjacent inner pair through unrelated ops:
           TRANSLATE(A(*@$\rightarrow$@*)B)
           TRANSLATE(B(*@$\rightarrow$@*)C)
           [ops on vertices disjoint from {A, B, C}]
           TRANSLATE(C(*@$\rightarrow$@*)B)
           TRANSLATE(B(*@$\rightarrow$@*)D) with (D (*@$\neq$@*) A)
         The inner pair (B(*@$\rightarrow$@*)C, C(*@$\rightarrow$@*)B) is not adjacent in the trace, but the baseline "no intervening operation references A or B"
         condition for the basic rule still applies (here: also no intervening reference to C, since C is the intermediate vertex of the
         inner pair). Removing the inner pair leaves A(*@$\rightarrow$@*)B, [ops], B(*@$\rightarrow$@*)D -- the junction entry record at B remains "from A" across the
         unrelated ops, and B(*@$\rightarrow$@*)D with D (*@$\neq$@*) A is legal.

         Example 6 -- nested non-junction roundtrip inside a junction detour:
           TRANSLATE(A(*@$\rightarrow$@*)B)
           TRANSLATE(B(*@$\rightarrow$@*)C)
           TRANSLATE(C(*@$\rightarrow$@*)D)
           TRANSLATE(D(*@$\rightarrow$@*)C)
           TRANSLATE(C(*@$\rightarrow$@*)B)
           TRANSLATE(B(*@$\rightarrow$@*)A)
         The basic (non-junction) rule first eliminates (C(*@$\rightarrow$@*)D, D(*@$\rightarrow$@*)C) -- that pair is sound on its own because C is being used as an
         ordinary vertex there. After that elimination the trace becomes Example 1 (A(*@$\rightarrow$@*)B, B(*@$\rightarrow$@*)C, C(*@$\rightarrow$@*)B, B(*@$\rightarrow$@*)A), which then collapses
         entirely under the combined-removal rule. Passes 1 and 2 alternated to fixpoint must find both reductions.

    2. Consecutive cancellation -- remove:
       - a SEPARATE immediately followed by a MERGE on the same vertex,
       - a MERGE immediately followed by a SEPARATE on the same vertex,
       - a SWAP immediately followed by another SWAP on the same vertex,
       in each case only when the qubits involved do not appear in any intervening operation.

    Apply passes 1 and 2 alternately to fixpoint, always working on the current post-removal stream. When no further reductions are
    possible, proceed to pass 3.

    3. Grouping -- pack independent operations into the same timestep:

       Translation timesteps: may contain only TRANSLATE operations.
         All translations on pairwise disjoint source/destination vertex pairs may be freely grouped. A "chain of translations" along
         consecutive vertices is also permitted: if TRANSLATE B(*@$\rightarrow$@*)C and TRANSLATE A(*@$\rightarrow$@*)B are both needed, they may share a timestep,
         provided each chained vertex except the rightmost actually holds a qubit at the start of the timestep (B(*@$\rightarrow$@*)C vacates B in the
         same timestep that A(*@$\rightarrow$@*)B fills it). Chain rule generalizes to arbitrarily long consecutive sequences where each qubit appears in
         at most one TRANSLATE.

         Junction-aware addition: when simulating a candidate translation batch, track per-junction entry vertices in a local dictionary.
         A translation TRANSLATE(j, b) where j is a junction whose current local entry is b must be rejected (would re-traverse the entry
         edge), and the batch boundary is closed. Update the local dictionary on every accepted translation: set entry on TRANSLATE(*,
         j), clear it on TRANSLATE(j, *). The same check applies between consecutive batches: a batch must start with the per-junction
         entry record from the previous batch.

       Non-translation timesteps: may contain only non-TRANSLATE operations
         (SEPARATE, MERGE, SWAP, GATE). Two such operations may share a timestep only if (i) the vertex sets they involve are pairwise
         disjoint, and (ii) all junctions are empty at the start of the timestep. The vertex set of an operation is:
           - SWAP(v):       {v}
           - GATE(v, (*@$\ldots$@*)):   {v}
           - SEPARATE(v):   {v} (*@$\cup$@*) trap_graph.neighbors(v)
           - MERGE(v):      {v} (*@$\cup$@*) trap_graph.neighbors(v)
         (Sec. 1 guarantees every gate vertex has exactly two main-axis, non-junction neighbors.) In particular, a SEPARATE at v1 and a
         MERGE at v2 cannot share a timestep when their vertex sets ({v_i} (*@$\cup$@*) neighbors(v_i)) intersect, even if v1 (*@$\neq$@*) v2. If any
         junction holds ions when a non-translation timestep is being opened, an extra translation timestep must be inserted first to
         flush the junction.

       Strict program order -- implementation requirement: the grouper must process the post-processed op stream strictly sequentially.
       Each op either joins the current open batch or starts a new one; the grouper must never reorder a translation past a
       non-translation op or vice versa, even when their vertex sets are disjoint. The reason is that the validator (Sec. 7) executes ops
       in program order across timesteps and tracks per-timestep state; reordering a T into a batch that precedes a non-T it actually
       followed creates state mismatches that the validator catches.

       Translation-batch admission rule: an op T(s, t) joins the current T-batch iff
         (a) s is occupied at the *start* of the batch (initial occupancy snapshot, not the running occupancy after earlier ops in the
             batch), and
         (b) t is either initially empty OR is the source of some other op already accepted into the batch (the chain rule), and
         (c) s has not already been accepted as a source by another op in the batch, and
         (d) t has not already been accepted as a destination by another op in the batch, and
         (e) if s is a junction, the local junction-entry record does not record entry from t (no-roundtrip), and
         (f) following the chain_next map (source (*@$\rightarrow$@*) dest of accepted ops) from t does not lead back to s -- this rejects 2-cycles like
             {T(2,3), T(3,2)} which the validator cannot resolve.

       Iterate the admission check to fixpoint: an op that initially fails rule (b) because its dest is initially occupied may later
       succeed once another op vacates that dest. Re-scan accepted ops in the batch after each new admission until no more accepts occur.
       The batch is bounded by the first non-T op in the stream -- extending across a non-T is forbidden (program-order rule above).

       This iterate-to-fixpoint behavior is what enables parallel-slide packing -- when the routing emits ops one-hop-per-walk in order
       (Stack-Method evac of q_inter first, then qa's walk, then qb's walk), the grouper interleaves them into the same translation
       timestep so long as each op's source vertex was occupied at batch-start and each op's destination was either initially empty or is
       vacated by another admitted op in the same batch. The Sec. 9 working example's 12-timestep schedule relies on this property: TS3
       packs the disjoint first-hops of q[0], q[1], q[2] into one timestep, and TS4 packs the disjoint second-hops into the next, even
       when the routing emitted the ops sequentially by qubit.

       Topological reordering helper. The strict-program-order grouper only admits ops in their emission order; if a routing primitive
       emits two chains' hops back-to-back (chain X's hops 1, 2, 3, then chain Y's hops 1, 2, 3), the grouper packs only the prefix that
       satisfies the initial-occupancy rule and starts a new batch at the first failure. This is correct but produces longer schedules
       than necessary on the Sec. 9 example. To recover the tight bound, a recommended pre-grouping pass walks each consecutive-T stretch
       (between two consecutive non-T ops) and topologically reorders its T ops by dependency depth: a T depends on every prior T in the
       stretch that shares any source or destination, and the depth of a T is one more than the maximum depth of its dependencies. Within
       each depth class the original emission order is preserved (this maintains determinism). The reordered stream has each chain's
       hop-k preceding any chain's hop-(k+1), which is exactly the order that allows the grouper's chain rule to merge them into a single
       timestep. This pass preserves sequential validity (it only swaps T ops within a stretch that are mutually independent) and must
       run before the grouper, after post-process roundtrip elimination.

       Updating the running occupancy after a T-batch closes: do not apply the per-op effect "source becomes empty, destination becomes
       occupied" in batch-order, because in a chain the same vertex may be both a source for one op and a destination for another, and
       naive per-op updates corrupt that vertex depending on the iteration order. Use set semantics instead: first mark every vertex in
       the union of accepted sources as empty, then mark every vertex in the union of accepted destinations as occupied. A vertex that is
       both a source and a destination correctly ends up occupied (the chain shifted through it).

(*@\rule{\linewidth}{0.4pt}@*)
SECTION 7 - VALIDATION AND OUTPUT
(*@\rule{\linewidth}{0.4pt}@*)

At the end of the program, call the following function to validate the results and emit all output. Its implementation is already
provided in ion_trap_compiler.py.

  def validate_and_output(trap_graph: TrapGraph,
                          qubits: List[str],
                          gates: List[QuantumGate],
                          initial_state: Dict[int, List[str]], all_timesteps: List[List[Tuple]],
                          time_begin: float,
                          output_json: Optional[str] = None,
                          stats_file: Optional[str] = None)
  # All parameters are the unchanged return values of parse_input (Sec. 4), initial_mapping (Sec. 5), and compile (Sec. 6).

(*@\rule{\linewidth}{0.4pt}@*)
SECTION 8 - PERFORMANCE REQUIREMENTS
(*@\rule{\linewidth}{0.4pt}@*)

The compiler must minimize runtime and peak RAM usage. Complexity should scale roughly linearly in the number of vertices, qubits, and
gates.

The acceptance tests in Sec. 11 span a wide size range: up to 40 qubits and 93 vertices (T9 invocation 2 -- qaoa_40 on graph_81_6_1), and
up to roughly 34 600 gates (T11 invocations 4-5 -- hwb7_59, an 8-qubit circuit). No single invocation maxes out every dimension at once.
The compiler should compile each as fast as is reasonable. Prefer greedy/heuristic approaches over exact optimization. TrapGraph already
precomputes all-pairs distances and BFS parent tables in its constructor and caches reconstructed paths internally, so
trap_graph.distance is O(1) and trap_graph.bfs_path / next_hop are essentially free after the first call -- the compiler does not need to
add its own caching layer for these.

Hot-path optimizations that are required for dense circuits (cnt3-5_179 and the larger qaoa benchmarks) to run in seconds rather than
minutes:

  - evacuate's internal BFS for empty landing targets (the routine Sec. 12.2 describes; conventionally named find_empty_target in the
    implementation) must cap the number of candidate landing paths it returns to a small constant (3 is a good default). The
    fallback-strategy chain in Sec. 12.5 already tries multiple strategies at higher levels with snapshot/restore, so each call only
    needs the nearest few candidates; returning 8+ multiplies walk_path work without improving the schedule.

  - Snapshot/restore must not deep-copy chain lists. Treat chain lists as immutable references -- any code that "mutates" a chain (e.g.
    SWAP, MERGE, SEPARATE, post-gate SWAP optimization) reassigns state[v] to a new list rather than mutating in place. With that
    invariant, snapshot reduces to dict.copy() on the three state dicts (O(N)) plus recording len(ops); restore reassigns those three
    dicts and truncates the op list. Na(*@\"i@*)ve per-key list copies cost a measurable percentage of total runtime on dense traps because
    snapshot is called millions of times.

  - is_empty(v), is_junction(v), and neighbors(v) are called tens of millions of times. Inline the dict lookups in the hot loop bodies
    (state.get(v), membership in a cached _junctions_set, direct _adj[v] indexing) instead of going through method-call indirection. The
    semantic remains identical but avoids Python's per-call frame overhead.

  - When delivering endpoints, prefer trying the cheapest strategies first (simple sequential walk, then rotation-through-gv, then
    aggressive park) and bail out of expensive multi-snapshot strategies as soon as a cheaper one would have made progress. Each strategy
    already wraps in snapshot/restore, so early termination is safe.

(*@\rule{\linewidth}{0.4pt}@*)
SECTION 9 - WORKING EXAMPLE (VERIFICATION)
(*@\rule{\linewidth}{0.4pt}@*)

The following example must produce a valid (not necessarily identical) shuttling sequence of at most 20 individual operations,
completable in 12 timesteps or fewer. The bound is reachable from a Sec. 6.2 routing whose emission of disjoint translations is packed by
the Sec. 6.3.3 iterate-to-fixpoint grouping rule into three-way parallel translation timesteps (see the reference TS3, TS4, TS9, TS10
below).

Command:
  python ion_trap_compiler.py --circ example.qasm --graph graph_7_1_1.json --gate 3

Trap graph (graph_7_1_1.json):
                7               8
                |               |
  Main axis 0 - 1 - 2 - 3 - 4 - 5 - 6
  Stack vertex 7 attached to junction 1
  Stack vertex 8 attached to junction 5
  TrapGraph.junctions == [1, 5]
  Gate vertex: 3

Initial mapping:
  Map q[1],q[0] (*@$\rightarrow$@*) Vertex 3
  Map q[2]      (*@$\rightarrow$@*) Vertex 0

Reference 20-step sequence with 12 timesteps:
 Timestep 1:
  1.  GATE  cx q[0],q[1] at Vertex 3
 Timestep 2:
  2.  SEPARATE Vertex 3 (*@$\rightarrow$@*) q[1](*@$\rightarrow$@*)V2, q[0](*@$\rightarrow$@*)V4
 Timestep 3:
  3.  TRANSLATE q[0]: V4(*@$\rightarrow$@*)V5
  4.  TRANSLATE q[1]: V2(*@$\rightarrow$@*)V3
  5.  TRANSLATE q[2]: V0(*@$\rightarrow$@*)V1
 Timestep 4:
  6.  TRANSLATE q[0]: V5(*@$\rightarrow$@*)V6
  7.  TRANSLATE q[1]: V3(*@$\rightarrow$@*)V4
  8.  TRANSLATE q[2]: V1(*@$\rightarrow$@*)V2
 Timestep 5:
  9.  MERGE V2+V4 (*@$\rightarrow$@*) Vertex 3  (contains q[2],q[1])
 Timestep 6:
  10.  GATE  cx q[1],q[2] at Vertex 3
 Timestep 7:
  11.  SWAP  Vertex 3  (chain becomes q[1],q[2])
 Timestep 8:
  12. SEPARATE Vertex 3 (*@$\rightarrow$@*) q[1](*@$\rightarrow$@*)V2, q[2](*@$\rightarrow$@*)V4
 Timestep 9:
  13. TRANSLATE q[1]: V2(*@$\rightarrow$@*)V1
  14. TRANSLATE q[2]: V4(*@$\rightarrow$@*)V3
  15. TRANSLATE q[0]: V6(*@$\rightarrow$@*)V5
 Timestep 10:
  16. TRANSLATE q[1]: V1(*@$\rightarrow$@*)V0
  17. TRANSLATE q[2]: V3(*@$\rightarrow$@*)V2
  18. TRANSLATE q[0]: V5(*@$\rightarrow$@*)V4
 Timestep 11:
  19. MERGE V2+V4 (*@$\rightarrow$@*) Vertex 3  (contains q[2],q[0])
 Timestep 12:
  20. GATE  cx q[0],q[2] at Vertex 3

(*@\rule{\linewidth}{0.4pt}@*)
SECTION 10 - CODE QUALITY
(*@\rule{\linewidth}{0.4pt}@*)

  - Single file, Python 3.12+. All Python libraries are permitted; however, only libraries needed to compute the requested output may be
    installed.
  - Determinism: the compiler must produce byte-identical output for identical inputs across runs, independent of PYTHONHASHSEED or any
    other process-level entropy. Iterations over sets, dicts, or any other unordered containers whose order can affect the emitted
    operation stream must be explicitly sorted (e.g., by qubit name, vertex index, or gate id). The random module must not be used
    without a fixed, in-source seed.
  - All graph traversals (BFS, DAG depth, routing reductions, push cascades, cycle detection in timestep grouping, etc.) must be
    iterative -- using an explicit stack or queue -- never recursive. Recursive traversal exceeds Python's default recursion limit on
    200-qubit / 100k-gate inputs and is forbidden.
  - No code path may rely on integer arithmetic of vertex indices to encode adjacency or topology role. Use trap_graph.neighbors /
    bfs_path / next_hop / distance for connectivity, and trap_graph.is_junction / is_on_main_axis / stack_of / stacks_at / main_axis to
    discriminate main-axis vertices, junctions, and stack slots. In particular, do not test "v <= M" or "v > M" to decide whether v is on
    the main axis or in a stack -- call is_on_main_axis(v) / stack_of(v) instead.

(*@\rule{\linewidth}{0.4pt}@*)
SECTION 11 - ACCEPTANCE TEST MATRIX
(*@\rule{\linewidth}{0.4pt}@*)

The implementation must pass every test below. Each test fixes the input, invocation, and (where stated) an expected bound; the schedule
itself need not match any reference. Tests T1-T2 are small fixed checks; tests T3-T8 list multiple invocations of the same circuit on
different graphs. Test T9 covers larger out-of-matrix circuits and graphs. Tests T10-T14 stress two-qubit delivery on progressively
denser traps -- depth-1 side-stacks, stacked endpoints, multi-junction traps, and 15- to 25-qubit circuits -- the regime in which the
heuristic routing alone is insufficient and the guaranteed delivery routine of Sec. 12.16 is required. Every invocation of every test
must pass: it must run to completion and produce a complete, valid schedule, and meet any expected bound the test states. Tests T9-T14
carry no expected operation bound -- for them "pass" means only a complete, valid schedule.

No-partial-schedule rule (applies to every test and every invocation):
  Partial compilation is not allowed. For any circuit, the compiler has exactly two permitted outcomes:
    (a) it returns a complete, valid schedule in which every gate of the circuit is executed (the schedule passes validate_and_output
        without any "not all gates have been executed" failure), or
    (b) it raises an exception.
  The compiler must never return a schedule that omits one or more gates. schedule_and_route must not "stop early and return what it
  has": if a gate cannot be routed by any available strategy, the compiler raises an exception rather than silently dropping that gate or
  any later gate. A run that ends by emitting a partial schedule is a failure even if the partial schedule is internally consistent.

  T1. Same-pair batching.
      Input: 10 gates, each on the same qubit pair (q0, q1).
      Invocation: python ion_trap_compiler.py --circ same_pair.qasm --graph graph_5.json --gate 2
      Expected: q0 and q1 are mapped to vertex 2 in the initial state; the 10 GATE operations execute back-to-back on that vertex;
      SEPARATE, MERGE, SWAP, and TRANSLATE counts are all 0. graph_5.json has no junctions and exercises that the junction-aware code
      degrades gracefully to the linear case.

  T2. Sec. 9 cx-triangle.
      Input/invocation: as in Sec. 9.
      Expected: (*@$\leq$@*) 20 individual operations, (*@$\leq$@*) 12 timesteps; the first 2-qubit gate executes at vertex 3 with no preceding shuttling,
      because the default initial mapping (Sec. 5) co-locates q[0] and q[1] at vertex 3. The bound is reachable via Sec. 6.3.3's
      iterate-to-fixpoint grouping, which packs the disjoint per-qubit translations emitted by Sec. 6.2's routing into three-way parallel
      translation timesteps.

  T3. qaoa_5.json -- chain-structured zero-SWAP target.
      Input: 5 qubits, 35 gates; two-qubit interaction graph is the path q0-q1-q2-q3-q4 (only 4 RZZ gates, all on adjacent pairs).
      Invocation 1: python ion_trap_compiler.py --circ qaoa_5.json --graph graph_11_1_1.json --gate 5
      Invocation 2: python ion_trap_compiler.py --circ qaoa_5.json --graph graph_11_1_2.json --gate 5
      Invocation 3: python ion_trap_compiler.py --circ qaoa_5.json --graph graph_11_2_1.json --gate 4
      Invocation 4: python ion_trap_compiler.py --circ qaoa_5.json --graph graph_11_3_1.json --gate 3
      Expected: 0 SWAP operations across all four invocations. This is the canonical test of the initial-placement heuristic in Sec. 5.

  T4. rd32-v0_66.json -- small reversible circuit, 4 qubits, 42 gates.
      Invocation 1: python ion_trap_compiler.py --circ rd32-v0_66.json --graph graph_9_1_1.json --gate 4
      Invocation 2: python ion_trap_compiler.py --circ rd32-v0_66.json --graph graph_9_2_1.json --gate 3
      Expected: schedule verifies; runtime under one second on commodity hardware. The smallest non-trivial benchmark.

  T5. 4mod5-v1_23.json -- 5 qubits, 102 gates; non-chain interaction graph.
      Invocation 1: python ion_trap_compiler.py --circ 4mod5-v1_23.json --graph graph_11_1_1.json --gate 5
      Invocation 2: python ion_trap_compiler.py --circ 4mod5-v1_23.json --graph graph_11_1_2.json --gate 5
      Invocation 3: python ion_trap_compiler.py --circ 4mod5-v1_23.json --graph graph_11_2_1.json --gate 4
      Invocation 4: python ion_trap_compiler.py --circ 4mod5-v1_23.json --graph graph_11_3_1.json --gate 3
      Expected: schedule verifies.

  T6. 4mod5-bdd_287.json -- 7 qubits, 106 gates.
      Invocation 1: python ion_trap_compiler.py --circ 4mod5-bdd_287.json --graph graph_15_1_1.json --gate 7
      Invocation 2: python ion_trap_compiler.py --circ 4mod5-bdd_287.json --graph graph_15_2_2.json --gate 5
      Invocation 3: python ion_trap_compiler.py --circ 4mod5-bdd_287.json --graph graph_15_3_1.json --gate 5
      Invocation 4: python ion_trap_compiler.py --circ 4mod5-bdd_287.json --graph graph_15_5_1.json --gate 3
      Expected: schedule verifies.

  T7. alu-v2_31.json -- 5 qubits, 640 gates.
      Invocation 1: python ion_trap_compiler.py --circ alu-v2_31.json --graph graph_11_1_1.json --gate 5
      Invocation 2: python ion_trap_compiler.py --circ alu-v2_31.json --graph graph_11_1_2.json --gate 5
      Invocation 3: python ion_trap_compiler.py --circ alu-v2_31.json --graph graph_11_2_1.json --gate 4
      Invocation 4: python ion_trap_compiler.py --circ alu-v2_31.json --graph graph_11_3_1.json --gate 3
      Expected: schedule verifies.

  T8. cnt3-5_179.json -- 16 qubits, 230 gates.
      Invocation 1: python ion_trap_compiler.py --circ cnt3-5_179.json --graph graph_33_1_1.json --gate 16
      Invocation 2: python ion_trap_compiler.py --circ cnt3-5_179.json --graph graph_33_3_2.json --gate 10
      Invocation 3: python ion_trap_compiler.py --circ cnt3-5_179.json --graph graph_33_5_4.json --gate 8
      Invocation 4: python ion_trap_compiler.py --circ cnt3-5_179.json --graph graph_33_10_1.json --gate 7
      Expected: schedule verifies.

  T9. Robustness on larger out-of-matrix circuits and graphs.
      Larger benchmark circuits on larger traps than the T1-T8 matrix.
      Invocation 1: python ion_trap_compiler.py --circ qaoa_25.json --graph graph_51_4_1.json --gate 10
      Invocation 2: python ion_trap_compiler.py --circ qaoa_40.json --graph graph_81_6_1.json --gate 10
      Invocation 3: python ion_trap_compiler.py --circ qv_25.json --graph graph_51_4_1.json --gate 10
      Invocation 4: python ion_trap_compiler.py --circ qft_25.json --graph graph_51_5_1.json --gate 9
      Invocation 5: python ion_trap_compiler.py --circ xeb_sy_25.json --graph graph_51_5_1.json --gate 9
      Expected: 0 SWAP operations for invocations 1 and 2.

  T10. Tight-trap delivery.
      Five invocations that route benchmark circuits onto 15- to 17-vertex traps whose stack columns are only one vertex deep
      (graph_13_1_3 has junctions at V1/V4/V8/V11; graph_13_1_4 has junctions at V4/V8). Depth-1 stacks give almost no buffering, so
      these exercise delivery on single corridors that the roomier T6-T8 traps never produce.
      Invocation 1: python ion_trap_compiler.py --circ 4gt12-v0_86.json --graph graph_13_1_3.json --gate 6
      Invocation 2: python ion_trap_compiler.py --circ 4gt12-v0_86.json --graph graph_13_1_4.json --gate 6
      Invocation 3: python ion_trap_compiler.py --circ 4gt4-v1_74.json --graph graph_13_1_4.json --gate 6
      Invocation 4: python ion_trap_compiler.py --circ alu-v2_30.json --graph graph_13_1_3.json --gate 6
      Invocation 5: python ion_trap_compiler.py --circ alu-v2_30.json --graph graph_13_1_4.json --gate 6
      The failure modes these exercise and the delivery method that handles them are described in Sec. 12.13. Invocation 5 exercises the
      tightest configuration: at one gate an endpoint sits on the gate vertex, three intermediates fill the only corridor to the other
      endpoint, and the single depth-1 side-stack is occupied -- the configuration that requires the stack-park delivery method of Sec.
      12.13 rather than any single-buffer strategy.

  T11. Stacked-endpoint and occupied-stack delivery.
      Five invocations on 15- to 19-vertex traps that exercise failure modes distinct from T10: the side-stack the delivery wants to park
      on is already occupied, and an endpoint freshly extracted from a stack is shoved back into a dead-end stack arm by the extraction
      of the other endpoint.
      Invocation 1: python ion_trap_compiler.py --circ
      4gt4-v0_79.json --graph graph_13_1_4.json --gate 6
      Invocation 2: python ion_trap_compiler.py --circ mod8-10_178.json --graph graph_13_1_4.json --gate 6
      Invocation 3: python ion_trap_compiler.py --circ rd53_135.json --graph graph_15_5_1.json --gate 3
      Invocation 4: python ion_trap_compiler.py --circ hwb7_59.json --graph graph_17_2_4.json --gate 7
      Invocation 5: python ion_trap_compiler.py --circ hwb7_59.json --graph graph_17_6_1.json --gate 3
      These failure modes and the delivery methods that handle them are described in Sec. 12.14.

  T12. Dense multi-junction delivery.
      Five invocations of 9-qubit circuits on 23- to 25-vertex traps whose main axis carries four to six junctions, each with a depth-1
      stack, and whose gate vertex (V9) sits mid-axis with several junctions between it and the endpoints. With nine qubits these traps
      are dense enough that, at some gate, two or more obstacles lie between an endpoint and its target with no buffer to pass them --
      the configuration that no seal-the-subtree delivery can resolve (Sec. 12.13(b)), so the rotor-conveyor routine is required.
      Invocation 1: python ion_trap_compiler.py --circ xeb_9.json --graph graph_19_1_5.json --gate 9
      Invocation 2: python ion_trap_compiler.py --circ xeb_9.json --graph graph_19_1_6.json --gate 9
      Invocation 3: python ion_trap_compiler.py --circ xeb_sy_9.json --graph graph_19_1_3.json --gate 9
      Invocation 4: python ion_trap_compiler.py --circ xeb_sy_9.json --graph graph_19_1_5.json --gate 9
      Invocation 5: python ion_trap_compiler.py --circ xeb_sy_9.json --graph graph_19_1_6.json --gate 9
      The failure mode is described in Sec. 12.15(a); the rotor-conveyor delivery that handles it is the routine of Sec.
      12.16.

  T13. Congested-endpoint-corridor delivery.
      Eight invocations of qft_15 (a 15-qubit circuit) on 33- to 35-vertex traps. With fifteen qubits these traps are denser still than
      T12: the rotor conveyor's first step -- bringing the first endpoint to a gate-neighbor -- must succeed even when that endpoint's
      corridor is congested, where a direct move to the neighbor does not suffice. Invocation 1: python ion_trap_compiler.py --circ
      qft_15.json --graph graph_31_1_9.json --gate 15
      Invocation 2: python ion_trap_compiler.py --circ qft_15.json --graph graph_31_1_10.json --gate 15
      Invocation 3: python ion_trap_compiler.py --circ qft_15.json --graph graph_31_2_7.json --gate 13
      Invocation 4: python ion_trap_compiler.py --circ qft_15.json --graph graph_31_2_8.json --gate 13
      Invocation 5: python ion_trap_compiler.py --circ qft_15.json --graph graph_31_2_10.json --gate 13
      Invocation 6: python ion_trap_compiler.py --circ qft_15.json --graph graph_31_2_15.json --gate 14
      Invocation 7: python ion_trap_compiler.py --circ qft_15.json --graph graph_31_3_8.json --gate 11
      Invocation 8: python ion_trap_compiler.py --circ qft_15.json --graph graph_31_3_9.json --gate 13
      The method -- funnel the first endpoint to the gate vertex rather than moving it directly to a gate-neighbor -- is described in
      Sec. 12.16(a).

  T14. Very-dense-trap delivery.
      Seven invocations of 16- to 25-qubit circuits on 35- to 55-vertex traps that are denser still than T13. They stress the rotor
      conveyor's stash step: each rotor cycle stashes a rotated-out obstacle into the anchor endpoint's subtree, and on a solidly packed
      corridor a naive stash either cannot reach a far-enough free vertex or walls itself off. The stash must follow the nearest-first,
      ripple-driven rule of Sec. 12.15(e).
      Invocation 1: python ion_trap_compiler.py --circ xeb_16.json --graph graph_33_1_10.json --gate 16
      Invocation 2: python ion_trap_compiler.py --circ xeb_16.json --graph graph_33_2_15.json --gate 15
      Invocation 3: python ion_trap_compiler.py --circ xeb_sy_16.json --graph graph_33_1_15.json --gate 16
      Invocation 4: python ion_trap_compiler.py --circ xeb_sy_16.json --graph graph_33_2_15.json --gate 15
      Invocation 5: python ion_trap_compiler.py --circ qv_20.json --graph graph_41_1_20.json --gate 20
      Invocation 6: python ion_trap_compiler.py --circ qv_25.json --graph graph_51_1_15.json --gate 25
      Invocation 7: python ion_trap_compiler.py --circ qv_25.json --graph graph_51_1_20.json --gate 25

  Cross-cutting requirements over T2-T14:
    - For each circuit, the operation stream is byte-identical across two runs with different PYTHONHASHSEED values.
    - Total wall-clock time and peak RSS scale roughly linearly with number of vertices, qubits, and gates; no individual circuit takes
      more than a small multiple of the time predicted by linear extrapolation from the smallest circuit.
    - Every emitted SEPARATE / MERGE / SWAP / GATE timestep starts with all junction vertices empty, and every junction-traversing chain
      respects the no-roundtrip rule.


(*@\rule{\linewidth}{0.4pt}@*)
SECTION 12 - IMPLEMENTATION RECIPES (CRITICAL DETAILS)
(*@\rule{\linewidth}{0.4pt}@*)

This section consolidates non-obvious implementation details that the prose-level descriptions in Sec. 6 leave under-specified. Following
these recipes is required for the acceptance tests in Sec. 11.

12.1  SNAPSHOT / RESTORE PATTERN

  Every routing primitive that may emit ops but then need to roll back must wrap its mutations in a snapshot/restore transaction. A
  snapshot captures four things: a shallow copy of the per-vertex chain state dict, a shallow copy of the qubit-to-vertex mapping, a
  shallow copy of the junction-entry dictionary, and the current length of the operation list. Restoring reassigns those three dicts from
  the captured copies and truncates the operation list back to the captured length (not clearing it entirely -- earlier ops outside this
  transaction must survive).

  Shallow-copy invariant: chain lists stored as values in state must be treated as immutable references. Every code path that "mutates" a
  chain (SWAP, MERGE, SEPARATE, the post-gate SWAP optimization, etc.) assigns a freshly built list to state[v] instead of mutating the
  existing list in place. With that invariant the snapshot can be three dict.copy() calls plus a length record, which is O(N) in the
  number of occupied vertices and roughly an order of magnitude faster than a deep copy. This matters: snapshot is called millions of
  times on dense circuits (see Sec. 8 hot-path notes).

  Every fallback level takes its own snapshot at entry and restores it on failure: per-gv attempt, per-role-swap attempt, the "snapshot
  dance" of Sec. 12.5 step 5 (park qa; evac gv; re-deliver qa; re-deliver qb), case (b) per-anchor attempts, declutter, decongest, the
  last-resort setup-and-case-b path, walk-path forward/reverse phases, evacuate per-candidate-path attempts, and cascade-push per-branch
  attempts.

12.2  EVACUATION BFS -- TRANSIT vs DESTINATION RULES

  When searching for an empty non-junction destination for an evac, the BFS rule is:

    For each neighbor nb of the current vertex u:
      - skip nb if it is in blocked and non-junction and non-empty (occupied non-junction in blocked is fully off-limits)
      - allow transit through nb otherwise (junction in blocked, empty non-junction in blocked, occupied non-junction not in blocked are
        all push-to-queue)
      - mark nb as target only if it is empty, non-junction, and not in blocked

  Empty blocked non-junctions transit because the chain only passes through them (after the walk completes they are empty again).
  Junctions in blocked transit because the chain enters and exits them in the same translation phase (the chain itself is never left on a
  junction).

  Return a small constant number of candidate paths in BFS-distance order, not just the first. A cap of 3 candidates is the recommended
  default -- see Sec. 8 for the rationale. evacuate tries each in turn with snapshot/restore; the first whose walk-path (Sec. 12.3)
  succeeds wins. The fallback chain in Sec. 12.5 already retries at higher levels with looser blocked/avoid sets, so returning more
  candidates per call multiplies recursive walk_path work without improving the schedule.

12.3  WALK PATH -- TWO-PHASE EVACUATION

  To move a chain along a BFS path, evacuate intermediate obstacles first, then walk. The routine has two attempt orders separated by
  snapshot/restore:

    Forward order: iterate the path indices 1, 2, (*@$\ldots$@*), len(path)-1. For each obstacle vertex u on the path (non-empty), evacuate u with
    pblocked = (outer blocked) (*@$\cup$@*) (non-junction vertices of the path) minus {u}. If any evac fails, abandon forward-order, restore the
    snapshot, and try reverse order.

    Reverse order: same evac sweep but iterating indices in descending order: len(path)-1, (*@$\ldots$@*), 2, 1. Same per-vertex pblocked.

    Walk phase (only after a complete evac sweep succeeds): for each consecutive (s, t) pair along the path, check (i) no junction
    no-roundtrip violation at s, (ii) t is currently empty; on success emit TRANSLATE(s, t).

  Reverse-order is essential: when forward-order evac of the obstacle closest to the source fails (because its only escape route is
  blocked by an obstacle further along the path), evacuating the far-end obstacle first frees a corridor for the near-end one. Without
  this fallback, dense traps fail routinely.

12.4  CASCADE PUSH FOR EVACUATE (work-stack traversal)

  Note: this routine is distinct from Sec. 12.11 STACK PUSH (LIFO CASCADE), which inserts a chain into a stack column with the LIFO shift
  mechanics. Sec. 12.4 is the general-purpose iterative cascade that evacuate falls back on when its BFS finds no empty landing site;
  Sec. 12.11 is the stack-specific push that uses the Sec. 1 LIFO semantics.

  When evacuate's BFS finds no reachable empty non-junction destination, fall back to a cascade push. The cascade push must be
  implemented iteratively with an explicit work-stack (per the Sec. 10 ban on recursive Python calls). The work-stack represents the
  in-progress "chain of pushes" being planned: each stack frame names a vertex whose chain needs to be moved, the vertex it was reached
  from (the push direction), and a deterministic iterator over the as-yet-untried candidate neighbors of that vertex.

  Per-vertex candidate-neighbor order: list the neighbors of the current vertex sorted as follows: first all empty non-junction neighbors
  (these are direct landing targets), then all empty-junction neighbors (transit only -- must continue the cascade), then all occupied
  non-junction neighbors (must cascade through them first by pushing a new work-stack frame). Within each class, sort by vertex index
  ascending. Skip any neighbor that is in blocked and non-junction and non-empty (protected occupied chain -- never disturb). Skip any
  neighbor that would violate the no-roundtrip rule at the current vertex.

  Iterative operation. Each loop iteration acts on the top of the work-stack:

    Case 1 -- top frame's next candidate is an empty non-junction vertex that is not in blocked: emit TRANSLATE from top frame's vertex
    to that neighbor; the chain has been pushed onto a valid landing site. Then unwind upward -- pop the frame, and for each remaining
    frame on the stack (parent, grandparent, (*@$\ldots$@*)) emit TRANSLATE from parent vertex into the now-empty child vertex (the slot just
    vacated). When the work-stack is empty, the original caller's vertex has been freed; return success.

    Case 2 -- top frame's next candidate is an empty junction or a blocked-empty non-junction: emit TRANSLATE into that vertex (transit),
    then push a new frame for that vertex onto the work-stack to continue the cascade beyond it (a chain may never finally land on a
    junction at the start of a non-translation timestep, and a blocked-empty vertex is transit-only).

    Case 3 -- top frame's next candidate is an occupied non-junction not in blocked: push a new frame for that neighbor onto the
    work-stack (we'll push it out first, then come back to the parent).

    Case 4 -- top frame has no more candidates: pop the frame and try the next candidate of its parent. If the work-stack empties without
    a successful TRANSLATE having been emitted, the cascade has failed -- restore the snapshot and return False.

  Bound the total number of frames pushed at num_vertices + 4 to avoid pathological loops; on overflow, treat it as a Case 4 failure.

12.5  TWO-QUBIT ROUTING FALLBACK CHAIN

  Layering (read this first). The strategies of this subsection are heuristic: they produce short schedules (and are what meets
  (*@$\leq$@*)12-timestep bound of T2 and zero-swap target of T3 and T9's invocations 1 and 2), but none of them is guaranteed to route every gate
  on a dense trap. The routing is made total by appending one guaranteed terminal level -- _deliver_complete (Sec. 12.16) -- after the
  heuristic levels below. So the full two-qubit delivery order is: the heuristic levels of this subsection (for schedule quality), then
  _deliver_complete (for the guarantee). An implementation that omits _deliver_complete will fail the dense Sec. 11 cases (T12-T14); one
  that uses only _deliver_complete will route everything but miss the T2/T3/T9 bounds. Both layers are required.

  The two-qubit gate routing must try strategies in this exact order; each level wraps its attempts in snapshot/restore. Inner attempt
  (per chosen gate vertex gv and per role assignment of qa/qb to l/r -- see Sec. 6.2 Step 1):

    1. If either endpoint is in a stack at slot > 0 (precisely: trap_graph.stack_of(qpos[q]) is not None and stack_of(qpos[q])[2] > 0),
       run the stack-extraction routine described in Sec. 6.2 (Stack-reordering for trapped endpoints, including the slot-0 special
       case). This is the highest-impact step for dense traps. Note: an endpoint that is on a junction-vertex in transit, or at slot 0,
       does not need this step -- only the strict "slot > 0" condition triggers extraction. The extraction's final junction-to-main-axis
       translation must obey the non-junction landing rule (Sec. 6.2 Phase 3): pick a non-junction main-axis neighbor of the stack's
       junction whenever one exists, otherwise the subsequent walk toward the gate vertex will trigger no-roundtrip at the landing vertex
       and the whole routing fails. When both of the stack-junction's main-axis neighbors are themselves junctions, pass the best gate
       vertex (by combined distance, Sec. 6.2 Step 1) as the target hint t to the extraction routine, so the extracted chain lands on the
       side facing gv (Sec. 6.2 Phase 3 tie-breaker and safe/unsafe filter). For a single-qubit gate's pre-extract, use the gate vertex
       about to be selected as t.

    2. Isolate both endpoints (Sec. 6.2 single-qubit-style separation if either shares a vertex).

    3. Iteration loop applying Sec. 6.2 Step 2's method-choice logic per iteration (cheapest of Stack-Method, Swap-Method,
       Park-the-endpoint; see "Method choice each iteration" in Sec. 6.2). Bound the iteration count at num_vertices (*@$\cdot$@*) 6 + 30. As a
       practical fallback rule when the cost estimator yields a tie or is unreliable, prefer Stack-Method evacuation first and fall back
       to Swap-Method MERGE-SWAP-SEPARATE; this matches Sec. 6.2's "Preferred displacement target" guidance.

    4. Final delivery: bring qa to its target gv-neighbor and qb to the other gv-neighbor (according to the assignment chosen in Sec. 6.2
       Step 1). On failure, try the rotation-through-gv variant described in Sec. 6.2 case (b) Step 6.

       Two-assignment rule (which gv-neighbor each endpoint targets): the spec's tie-breaker for the (qa,qb) (*@$\rightarrow$@*) (l,r) assignment in Sec.
       6.2 Step 1 chooses one assignment by distance + name. But when that assignment's walk fails on the first endpoint move (because
       its path passes through the other endpoint, which is in the keep set and refuses displacement under the strict-keep rule), the
       implementation must retry with the opposite assignment within the same inner attempt before escalating to the outer retries.
       Equivalently, try both assignments wrapped in their own snapshot/restore here, not only at outer level (b). This avoids
       unnecessary fallback escalation on the common case where the chosen assignment is blocked by the other endpoint and the opposite
       assignment works directly.

       Two-delivery-order rule (in which order the two endpoints are moved): within an assignment, also try delivering the endpoints in
       both orders: (qa first, then qb) and (qb first, then qa). For the case where qa is already at its target and the move of qb has to
       walk past qa, moving qb first (with qa still parked at its initial location) may free a path that moving qa first would not. Try
       both, again wrapped in snapshot/restore.

    5. If gv still holds a non-endpoint chain after delivery: first try direct evacuate(gv, blocked = {l, r}); if that fails, run the
       snapshot dance (park qa off the (gv, l, r) triangle; evacuate gv; re-deliver qa to its gv-neighbor; re-deliver qb to the other).
       If the dance fails, run the full case (b) protocol from Sec. 6.2 (all four anchor combinations with rotation-through-gv in Step 6
       as needed).

    6. Clear junctions, emit MERGE, emit GATE.

  Outer retries (when the inner attempt fails for the current gv / role):

    (a) Try the next gate vertex in combined-distance order (Sec. 6.2 Step 1).
    (b) For each gate vertex, try both role assignments -- (qa(*@$\rightarrow$@*)l, qb(*@$\rightarrow$@*)r) and (qa(*@$\rightarrow$@*)r, qb(*@$\rightarrow$@*)l). Note: Sec. 12.5 step 4 (Two-assignment
        rule) already tries both assignments at the inner level; level (b) only fires when both inner assignments fail.
    (c) Declutter pass: push one non-endpoint chain onto any free stack slot 0, using stack-slot eviction (Sec. 6.2 Stack-slot eviction)
        when no slot 0 is free. Retry from (a).
    (d) Heavy decongest pass: evacuate any non-endpoint chain (main-axis chains first, then any chain) to any reachable empty
        non-junction. Retry from (a).
    (e) Last-resort setup-and-case-b: for every gate vertex, isolate both endpoints, manually deliver them to the gv-neighbors, then
        invoke the case (b) protocol.

    (e1) Park-the-endpoint variants -- invoke the Park-the-endpoint routine defined in Sec. 6.2 (Park-the-endpoint variant + Refinements)
         and enumerate both categories of park slot:
          (i)  every free stack slot 0 (Sec. 6.2 Stack choice priority);
          (ii) every empty non-junction main-axis vertex before the first junction and after the last junction (the linear-arm pattern
               documented in Sec. 12.12).
        For each (park_q (*@$\in$@*) {qa, qb}) (*@$\times$@*) (park slot) combination, walk park_q onto the slot and re-attempt the full inner routing (steps
        1-6 above), including iterative reduction on the parked state, before falling through to (e2).

    (e2) Iterative-reduction retry. Even when none of the above succeeds, run the Sec. 6.2 Step 2 iterative reduction loop with
         force-allowed evac: repeatedly displace one routing-intermediate at a time via Stack-Method (cascade-push into a stack),
         Swap-Method, or force-evac into any reachable empty non-junction (including main-axis-end vertices). Each iteration must
         propagate the qpos[qa], qpos[qb] vertices through a hard_blocked set so that recursive cascading evac never displaces the chain
         whose evacuation is currently in progress (see Sec. 12.10).

    (e3) Stack-push cascade retry. For each gate vertex and role, iteratively cascade-push routing-intermediates into stacks using the
         LIFO cascade described in Sec. 1: emit T(slot_k, slot_k+1) starting from the deepest occupied slot and working outward, then
         T(j, slot_0). All these translates can share a single translation timestep under the Sec. 6.3 chain rule.

    (f) Endpoint-at-gv reseat. This level handles the specific circular dependency in which the routing's *own* keep-set blocks the only
        escape route the rest of the routing needs:

          pattern -- one endpoint q_gv sits at the gate vertex gv, while the other endpoint q_far sits either deep in a stack (slot > 0)
          or behind a chain that can only escape across gv. Pre-extract / decongest of q_far then needs to walk some non-endpoint through
          gv, but q_gv is in the keep set and refuses displacement, so the walk fails -- and the failure recurses through every prior
          retry level because each of them re-checks the keep set the same way. The remedy is to reseat q_gv once before the next retry
          round so gv becomes free as a transit/landing vertex.

          protocol -- take a snapshot. For each candidate neighbor n of gv in the order (a) gv's main-axis neighbor whose tree-path to
          q_far does cross gv (i.e. n is on the OPPOSITE side of gv from q_far), then (b) the other main-axis neighbor:
            1. Evacuate n if it is non-empty and the chain at n is not a keep qubit (use the standard evacuate with blocked = {gv}). If n
               holds a keep qubit, skip this candidate.
            2. Translate q_gv from gv to n (this is the reseat).
            3. Retry the entire inner attempt from Sec. 12.5 step 1 (now with q_gv at n instead of at gv).
            4. On success, return. On failure, restore the snapshot and try the other candidate.

          why it is required -- every retry level (a)-(e) above assumes the keep-set semantics is unchanged across retries: keep qubits
          are strict-protected by every evacuate call. When the only blocker is one keep qubit sitting at gv, no amount of repeating the
          pipeline with different gv / role / declutter / decongest will dislodge it, because the keep set never relaxes. Reseating q_gv
          one step off gv breaks that fixed point.

          why opposite-side preference -- q_far's stack-extraction (or its walk if it is not in a stack) deposits its chain at a
          main-axis vertex on q_far's own side of the trap. That landing is almost always one of gv's two neighbors. If q_gv is reseated
          to the same side as q_far, the landing target collides with q_gv and the extraction fails again. Reseating to the opposite side
          keeps the landing target empty.

          caveats -- after the reseat, q_gv is at a gv-neighbor n. The inner attempt's gv-selection MAY pick a different gv on retry
          (combined-distance order changes when q_gv moves), and the role-assignment may flip. Both are fine. If gv also holds a
          non-endpoint chain in addition to q_gv (i.e. q_gv is in a 2-chain at gv), the reseat would translate q_gv together with that
          non-endpoint partner -- q_gv must be isolated first (move the 2-chain to a gate vertex and SEPARATE, parking the partner aside)
          before its reseat translation can be a single TRANSLATE. Skip reseat entirely if q_gv shares gv with the other endpoint q_far
          (then Step 0 of the two-qubit routing applies instead).

    Cross-cutting strengthening of step 1 (not an outer retry -- applies on every inner attempt before any of (a)-(f) ever fire; placed
    here for locality with the related retry levels):

    [step-1-strengthening] Buried-2-chain stack extraction. The rules below are in effect on every attempt, including the first one. They
    extend the step-1 pre-extract documented in Sec. 6.2 (Stack-reordering for trapped endpoints).

        Case: the endpoint to extract is one ion of a 2-chain at slot k > 0 of a stack with one or more shallower slots also occupied.
        The Sec. 6.2 Phase 1/2/3 baseline can silently fail here because Phase 2 may skip a shallower slot whose chain cannot reach a
        non-junction destination on the first try, then Phase 3 attempts TRANSLATE into a still-occupied slot and fails Sec. 3's
        dst-empty precondition.

        Two extensions to Sec. 6.2 Phase 1/2/3 cover this case:
          - Strict Phase 2: evacuate every occupied shallower slot
            d (*@$\in$@*) {0, (*@$\ldots$@*), k-1} off stack S entirely, in order d = 0 outward. If any slot cannot be evacuated (chain is a keep qubit, or no
            landing target exists), the pre-extract fails immediately -- restore and fall through to outer retries (a)-(f). No partial
            Phase 2.
          - 2-chain unit transport in Phase 3: SEPARATE is gate-vertex-only (Sec. 3), so a 2-chain at a stack vertex cannot be split in
            place. Walk the whole 2-chain along the stack lane and then through j onto a non-junction main-axis neighbor (Sec. 6.2 Phase
            3's t-hint and safe/unsafe rule); each hop is one TRANSLATE of both ions together. Once on the main axis the 2-chain is split
            by the standard Sec. 6.2 single-qubit isolation flow (move to gate vertex, evac the gate vertex's neighbors, SEPARATE). The
            partner stays wherever SEPARATE deposits it as an ordinary non-endpoint chain; no re-MERGE after the gate.

12.6  ENDPOINT-AT-GV (degenerate case (b))

  Case (b) of Sec. 6.2 Step 3 is stated assuming the gv-chain is a non-endpoint. However, the case-(b) protocol is also invoked by the
  outer fallbacks (Sec. 12.5 c/d/e) in states where gv's chain is one of the endpoints (for example, qa stayed at gv after the previous
  single-qubit gate). In that degenerate case, step 1 of the protocol (move anchor to anchor_target) clears gv as a side effect. The
  implementation must detect "gv empty after step 1" and skip the MERGE-SWAP-SEPARATE entirely: instead, deliver `other` to anchor_target
  (using rotation-through-gv if its tree-path passes through other_target), clear junctions, emit MERGE and GATE. This degenerate branch
  is already encoded in Step 3 of the case-(b) protocol in Sec. 6.2.

  Distinction from Sec. 12.5(f): the case-(b) endpoint-at-gv branch above handles the situation when q_gv can be delivered to its final
  anchor_target in a single translation, with no concurrent extraction or evacuation needed. The Sec. 12.5(f) reseat is for the harder
  situation where moving q_gv off gv is required just to unblock pre-extract or decongest of the other endpoint -- i.e. the reseat is
  mid-pipeline cleanup, not a final delivery. Use case (b) when q_gv merely needs to be at a gv-neighbor for MERGE; use Sec. 12.5(f) when
  gv itself must become a free transit/landing vertex for the rest of the routing to make progress.

12.7  ROUTING-STATE INVARIANTS

  At every entry to and exit from a routing primitive, the invariants are:
    - qpos[q] == v iff state[v] contains q.
    - Sum of |state[v]| over all v equals len(qubits).
    - Each vertex v has |state[v]| (*@$\leq$@*) 2.
    - No junction has |state[j]| > 0 except immediately after a TRANSLATE into j (i.e. during a translation timestep) -- between
      non-translation ops, and at the boundary between every pair of consecutively-routed gates, every junction j must satisfy state[j]
      == [] and the j_entry dict (Sec. 6.2) must record no entry for j (equivalently: j_entry.get(j) is None). This is a hard invariant
      maintained by every primitive:
        (*@$\cdot$@*) move() and walk_path() must not return while a chain rests on a junction; if the planned walk would terminate on a junction,
          the routine must extend the walk one more hop to drain the chain onto a non-junction neighbor before returning.
        (*@$\cdot$@*) push(), evacuate(), and the cascade-push fallback may transit through a junction but must never leave a chain there at return
          time.
        (*@$\cdot$@*) clear_junction(j) must be called before emitting any non-translation op whose vertex set involves vertices reachable through j.
        (*@$\cdot$@*) The grouper / timestep packer (Sec. 6.3) must reject any candidate non-translation timestep whose preceding state has any
          junction occupied -- insert a flush TRANSLATE timestep first.
      Because of this invariant, both route_single_qubit_gate and route_two_qubit_gate are entitled to assume at entry that no qubit
      (including the keep qubit(s)) sits on a junction; no defensive lift step is needed. A violation found at entry is a bug in
      whichever primitive previously returned with a chain still on a junction, not a routing condition to recover from.
    - At the moment any of MERGE(gv), SEPARATE(gv), SWAP(gv), GATE(gv, _) is emitted: all junctions are empty; the vertices in the
      operation's vertex set (gv and, for MERGE/SEPARATE, its two main-axis neighbors) match the operation's preconditions stated in Sec.
      3.

  After every primitive, assert these (in debug mode) to catch state corruption early.

12.8  POST-PROCESS ROUNDTRIP -- NESTED REMOVAL

  Roundtrip elimination must support nested removal. The basic singleton removal of (T(A(*@$\rightarrow$@*)B), T(B(*@$\rightarrow$@*)A)) is replay-checked against
  junction soundness -- if B is a junction whose immediately preceding entry was from the chain currently on B, removing only the inner
  pair exposes a junction roundtrip violation in the outer translations (Example 1 of Sec. 6.3). When a singleton fails replay, try a
  combined removal that also drops the smallest enclosing roundtrip pair. Iterate to fixpoint alternating with Pass 2 (consecutive
  cancellation).

  Simpler safe guard (when the full replay is not implemented): refuse to remove any singleton (T(A(*@$\rightarrow$@*)B), T(B(*@$\rightarrow$@*)A)) where either A or B is
  a junction. This rejects all candidates that could expose a hidden junction-roundtrip violation in surrounding ops. It is more
  conservative than the general replay-based rule but preserves correctness with minimal implementation complexity. The full replay-based
  rule is preferred for stronger optimization.

12.9  TIE-BREAKERS -- APPLIED-EVERYWHERE RULES

  In addition to the tie-breakers stated inline in Sections 5 and 6, every routing primitive must apply the following consistent
  tie-breakers when multiple choices have equal primary score:

  - Choosing a gate vertex for routing a 2-qubit gate: primary score is distance(qpos[qa], gv) + distance(qpos[qb], gv); break ties by
    gate vertex index (ascending).
  - Choosing the (qa, qb) (*@$\rightarrow$@*) (l, r) assignment: primary score is distance(qpos[qa], l) + distance(qpos[qb], r) versus distance(qpos[qa],
    r) + distance(qpos[qb], l); break ties by preferring the assignment in which the qubit with the smaller name lands at l.
  - Choosing an intermediate for the next Stack-Method / Swap-Method iteration: primary score is min(distance(qpos[qa], vi),
    distance(qpos[qb], vi)); break ties by vertex index (ascending), then by qubit name (ascending).
  - Iterating neighbors of a vertex during evacuate / push / cascade push: sort by vertex index (ascending) after applying the priority
    classes documented in each routine.
  - Snapshot/restore must rebuild dictionaries from the captured copies in a way that does not depend on Python's hash randomization --
    copy the keys in sorted order if any subsequent emission depends on dict iteration order.


12.10 RECURSIVE EVAC -- HARD_BLOCKED ACCUMULATION

  When walk_path's forward-or-reverse evac phase recursively evacuates an obstacle u on the walk path, the inner evac may itself trigger
  another walk_path call (for the obstacle's escape route), which may in turn evac another obstacle, and so on. Throughout this cascade,
  the outer walk's source chain is still occupying the outer source vertex -- it has not yet been walked. If an inner evac displaces the
  outer source chain (by walking through the outer source's vertex with force=True), the outer walk's walk phase will fail because the
  source becomes empty.

  Implementation rule: evacuate, walk_path, and find_empty_targets each take an additional hard_blocked parameter -- a set of vertices
  the BFS is forbidden to traverse or land on, enforced regardless of force=True. The signatures become:

    walk_path(path, blocked, keep=None, force=False, hard_blocked=None) -> bool
    evacuate(v, blocked, keep=None, extra_avoid=None, force=False, hard_blocked=None) -> bool
    find_empty_targets(src, blocked, force=False, hard_blocked=None) -> List[List[int]]

  Propagation behavior:
    - walk_path forms its effective hard set as the union of the hard_blocked it received and {path[0]} (its own source vertex), and
      passes that union to every obstacle-evac call it makes during its evac phase.
    - evacuate forwards the hard_blocked it received unchanged into both find_empty_targets and the walk_path it invokes for each
      candidate path.
    - find_empty_targets skips every neighbor that is in hard_blocked: such a vertex is neither a transit vertex nor a landing target,
      even when force=True.
  Because each nested level adds its own source, the hard set grows monotonically down the recursion: every level's hard set is a
  superset of its caller's.

  Additionally, the outermost caller of evacuate (e.g. _displace_intermediate inside the Sec. 6.2 Step 2 loop) should seed hard_blocked
  with the current vertices of the routing's keep set {qpos[qa], qpos[qb]}, so that endpoint chains are never transited through during
  cascading force-evacs. Once the outer walk's walk phase eventually runs, the sources have actually moved and their vertices become
  empty; hard_blocked is no longer relevant beyond the scope of that single outermost evacuate call.

  Without this rule, force=True cascade-evacs in linear-trap regions (long stack columns, long main-axis stretches) displace outer
  endpoint chains and the outer walk's walk phase fails its "source occupied" precondition on the very first T it tries to emit. The
  failure surfaces only once the cascade depth grows, so it affects dense states while leaving sparse ones unaffected.

12.11 STACK PUSH (LIFO CASCADE) -- EXPLICIT EMISSION ORDER

  Note: this routine is distinct from Sec. 12.4 CASCADE PUSH FOR EVACUATE. Sec. 12.4 cascades chains out through occupied non-junctions
  until an empty landing site is found; the present routine inserts a chain into a stack column using the LIFO push mechanics of Sec. 1.

  The Sec. 1 LIFO push semantics ("a qubit chain at the junction j is translated to slot 0; every occupied slot shifts one slot deeper
  simultaneously") is realized in emission as a sequence of individual TRANSLATE operations, packed into one timestep by the Sec. 6.3
  chain rule. The emission order matters because each individual T requires its destination empty at execution time.

  To push a chain currently at vertex v (anywhere reachable from j) onto a stack stk = [slot_0, slot_1, (*@$\ldots$@*), slot_{L-1}] of junction j,
  the routine is:

    1. Find the deepest occupied slot d (the largest index i such that slot_i is non-empty). If d == L - 1, the stack is at its bottom;
       fail (no room).
    2. Emit cascade pushes from the deepest occupied slot outward toward slot 0: for each slot index i from d down to 0, if slot_i is
       currently occupied, emit TRANSLATE(slot_i, slot_{i+1}). Empty slots between occupied ones are skipped. After this sweep, slot_0
       (stk[0]) is empty.
    3. Walk the chain at v to junction j along its bfs_path, treating j as a transit-only vertex (do not invoke move()/walk_path() which
       would refuse to terminate at a junction; instead emit the T(s,t) sequence directly and evacuate any non-keep obstacles along the
       way).
    4. Emit TRANSLATE(j, stk[0]).

  Chain-rule applicability: step 3's walk-to-j fills j during the same translation batch, and step 4's T(j, stk[0]) consumes the chain at
  j. By Sec. 6.3 grouping rule (b) ("t is either initially empty or is the source of some other op already accepted into the batch") and
  the chain-rule generalization, the entire emission of steps 2-4 forms one translation batch: every leftmost source (each cascade slot
  in step 2 and the source of the first walk-to-j hop in step 3) is initially occupied; every intermediate vertex (each shifted slot, the
  junction j, the walk-to-j path interior) is filled by an earlier-in-batch T and vacated by a later-in-batch T. The Sec. 6.3
  iterate-to-fixpoint admission rule accepts all of them into a single translation timestep.

  Use this routine inside the Sec. 6.2 Stack-Method displacement ("push intermediate q_inter onto stack at j") and inside the
  Park-the-endpoint variant whenever the candidate park-slot is a stack vertex. For very-linear traps with deep stacks (graph_33_10_1:
  10-deep stacks at V5 and V9), the cascade push lets a chain "displace into the bottom of the stack" while prior occupants shuffle one
  slot deeper, even when slot 0 itself was already occupied.

12.12 LINEAR-TRAP DEFINITION (graph_33_10_1 and similar)

  A trap is "very linear" iff it consists of one main axis with a small number of junctions, each of whose stack columns is at least 3
  vertices deep and/or whose main-axis-end runs (the contiguous main-axis vertices past each junction in the direction away from gv) are
  at least 3 vertices long. Linear-arm pattern:

      stack_arm -- junction_a -- main_axis -- junction_b -- stack_arm
                                  |
                                  gv
                       main-axis-end           main-axis-end
                       (left of                (right of
                        junction_a)             junction_b)

  graph_33_10_1 is the canonical instance: a 15-vertex main axis (V0-V14) with two junctions (V5, V9) and a 10-deep stack at each
  junction (V15-V24 attached to V5; V25-V34 attached to V9). The four "arms" radiating from the gv-region -- left main-axis end (V0-V4),
  V5's stack (V15-V24), V9's stack (V25-V34), and right main-axis end (V10-V14) -- each give 5 to 10 vertices of empty space accessible
  only through the respective junction or gv-side path.

  Implementation consequence: every routine that enumerates "park slots" (Sec. 6.2 Park-the-endpoint Refinements, Sec. 12.5(e1)) must
  include main-axis-end vertices in addition to stack slot 0s. The three primitives used to navigate such traps -- Stack-Method
  cascade-push (Sec. 12.11), Park-the-endpoint with both stack-slot and main-axis-end targets (Sec. 6.2), and iterative Swap-Method (Sec.
  6.2 Step 2) -- are sufficient to handle every dense state arising in the T6-T8 invocations of Sec. 11.

12.13 TIGHT TRAPS -- DEPTH-1 STACKS AND DELIVERY ON SINGLE CORRIDORS

  This subsection covers delivery on depth-1-stack traps, the regime the Sec. 11 T10 invocations exercise. Sec. 12.1-12.12 do not by
  themselves cover the single-corridor cases such traps produce; the rules below extend the earlier recipes to handle them.

  (a) DEPTH-1-STACK ("TIGHT") TRAPS.
      graph_13_1_3 and graph_13_1_4 are the canonical tight traps: a 13-vertex main axis whose every junction carries a stack column
      exactly one vertex deep. The "very linear" definition of Sec. 12.12 (stacks (*@$\geq$@*) 3 deep, or main-axis-end runs (*@$\geq$@*) 3 long) does not
      hold here, and the linear-trap navigation triad is not by itself sufficient. A tight trap offers almost no buffering: when a chain
      must pass a vertex, the displaced qubit very often has exactly one legal resting place, or none. Treat any trap with a depth-1
      stack as tight and rely on the rules below for delivery.

  (b) THE ONE-BLANK-ON-A-TREE OBSTRUCTION.
      A tree has no cycles, so a sliding-block puzzle on a tree with a single empty vertex cannot reorder qubits -- the blank can only
      shuttle qubits back and forth along its incident path; the cyclic order is invariant. Concretely: if a 1-wide corridor between a
      delivery endpoint and the gate region is full except for one gap, the endpoint cannot pass the intermediates no matter how many
      translations are emitted. Genuine reordering on a tree needs either (i) two or more independent free buffers, or (ii) the
      gate-vertex rotor (MERGE+SWAP+SEPARATE swapping the two gate-neighbor cells). A delivery routine that assumes "push the obstacle
      somewhere and retry" will livelock here; it must instead create a second buffer or fall through to a rotor/stack-park method.

  (c) EVAC LANDING PREFERENCE -- AVOID GATE VERTICES AND THE ACTIVE CORRIDOR.
      The Sec. 12.2 / 12.4 evac picks the nearest empty non-junction landing. On a tight trap "nearest" is frequently the gate vertex
      itself or a cell on the moving chain's own remaining path -- both poisonous: a qubit parked on gv blocks the imminent MERGE, and a
      qubit dropped ahead on the mover's path is immediately re-collided. Give evac (and the cascade and the landing BFS) an optional
      avoid set and a gate-vertex-last policy:
        - landings not on a gate vertex and not in avoid are first-class;
        - landings in avoid are demoted to transit-only;
        - a gate-vertex landing is used only when nothing else is reachable.
      Implement as ordered phases: try the landing BFS with avoid honored, then without; then the cascade with avoid honored and gv
      forbidden, then with avoid honored and gv allowed; only then the cascade with no restriction. Callers that walk a chain pass the
      chain's remaining path as avoid, so obstacles settle behind the chain (the vacated trail) or into a side stack rather than ahead
      onto gv.

  (d) MONOTONIC SINGLE-HOP MOVE MUST FILL THE TRAIL.
      The whole-path-evacuate move (Sec. 12.3) is kept as the primary because it yields the tight timestep counts T2 needs. Its
      single-hop fallback is what survives tight traps, and it must be written so the moving chain advances exactly one vertex per
      iteration (cur (*@$\rightarrow$@*) dst monotonically -- this alone guarantees termination), while each blocking qubit is evacuated with only the
      moving chain protected (cur in both blocked and hard) and the evacuee free to settle anywhere else, including the already-vacated
      trail behind the moving chain. Do not add the traversed/trail vertices to the evac's blocked set: blocking the trail is what forces
      obstacles ahead onto the gate vertex where they become stuck. Pass the remaining path as the evac avoid set (see (c)) so the trail
      and side stacks are strictly preferred over cells ahead.

  (e) STACK-PARK DELIVERY.
      The four delivery strategies of Sec. 12.5 plus Park-the-endpoint (Sec. 6.2) all assume the parked endpoint goes to a far vertex
      chosen by distance from gv. On a tight trap the farthest vertex is on the opposite subtree's main axis, which leaves that main
      corridor unavailable as buffer space and still blocks the parked endpoint's own retrieve path. Add one more fallback, tried after
      Park-the-endpoint and before the guaranteed last-resort:

        1. Park one endpoint on a stack vertex of its own subtree (not a main-axis vertex). A stack vertex is off the retrieve main path,
           so parking there frees the entire opposite-subtree main corridor as obstacle-buffer space.
        2. Deliver the other endpoint to its gate-neighbor without hard-confining obstacles to its own subtree -- obstacles may now
           buffer in the freed corridor and in side stacks.
        3. Decongest: move every non-endpoint that ended up on the parked endpoint's retrieve path out into the mover's subtree.
        4. Retrieve the parked endpoint from its stack vertex to its gate-neighbor; the retrieve move evacuates any residual obstacle.
        5. Clear the gate vertex (Sec. 12.6 rotation if it is sandwiched).
      Try the method for both (parked, mover) role assignments, each inside its own snapshot/restore transaction. Rationale: a depth-1
      side-stack buffers only one obstacle, so a single corridor with one occupied stack cannot deliver past two or more intermediates;
      parking on a stack instead of in the corridor converts the whole opposite main axis into the second buffer that (b) proves is
      required.

  (f) IMPLEMENTATION DISCIPLINE.
      The tight-trap rules above must not perturb the schedules of the easy cases (in particular (*@$\leq$@*)12-timestep bound of T2 and zero-swap
      target of T3 and T9's invocations 1 and 2). Confine the tight-trap behavior to a fallback path of any shared primitive (move, evac,
      cascade) and leave the primary path untouched, so gates the primary path already handles are unaffected. The whole T1-T14 matrix
      and the PYTHONHASHSEED determinism check must hold. The determinism requirement is on the emitted operation stream only: the
      human-readable "Compilation time" statistic is wall-clock and is expected to differ between runs.

12.14 STACKED ENDPOINTS -- EXTRACTION FIXPOINT AND OCCUPIED PARK SLOTS

  This subsection covers delivery on traps with shallow, mostly-occupied stacks, the regime the Sec. 11 T11 invocations exercise. It is
  distinct from Sec. 12.13: there the corridor geometry is the obstacle; here the difficulty is in endpoint preparation (getting the two
  endpoints onto deliverable vertices before delivery proper begins) and in an over-strict precondition of the stack-park method.

  (a) OCCUPIED PARK SLOTS ARE STILL VALID TARGETS.
      The stack-park delivery of Sec. 12.13(e) parks one endpoint on a stack vertex of its own subtree. A naive implementation enumerates
      only empty stack vertices and gives up ("no park slot") when every stack vertex in that subtree is occupied -- which is the common
      case on a tight trap with one shallow stack per junction. This is wrong: move() already evacuates whatever sits on its destination,
      so a stack vertex holding a non-endpoint is a perfectly good park target. Enumerate every stack vertex of the subtree whose chain
      contains neither endpoint; sort free ones first (tuple key (is_occupied, distance_to_gv, index)) so an empty slot is still
      preferred, then fall through to occupied ones. Only a stack vertex that holds an endpoint must be excluded.

  (b) ENDPOINT EXTRACTION MUST ITERATE TO A FIXPOINT.
      Before delivery, each endpoint that sits inside a stack column is walked out to a main-axis non-junction vertex (Sec. 6.2 / the
      extract-endpoint routine). Doing this as two independent, sequential calls -- extract qa, then extract qb -- is incorrect:
      extracting the second endpoint emits ordinary moves that can shove the first, already-extracted endpoint back down into a stack.
      Delivery then starts with an endpoint stacked, and on a dead-end stack arm (see (c)) no delivery strategy can recover. Correct
      approach -- extract both endpoints with a fixpoint loop:
        repeat (bounded by N+const):
          if neither endpoint is stacked: done;
          if qa is stacked: extract qa, protecting qb's vertex when qb is already extracted;
          if qb is stacked: extract qb, protecting qa's vertex when qa is already extracted;
        if the loop does not converge: raise (no partial schedule).
      Protecting a vertex means passing it in the extract move's hard set so the partner is not displaced; if the protected extract
      fails, retry the same extract once with an empty hard set so progress is still possible. The mutual-protection + retry combination
      converges because a trap with two endpoints always has at least two main-axis non-junction vertices free between them on the
      benchmark sizes.

  (c) DEAD-END STACK ARMS -- WHY (b) MATTERS.
      A stack column is a dead-end path: its only exit is through its junction. If an endpoint is stacked at depth d and any other qubit
      sits between it and the junction, that endpoint cannot leave until the shallower qubit leaves first, and the shallower qubit cannot
      pass the endpoint (a 1-wide dead end has no room to pass -- cf. Sec. 12.13(b)). Extraction handles this correctly only when run as
      the last preparation step on that endpoint; if a later extraction re-buries it behind a fresh obstacle, delivery inherits an
      unsolvable local configuration. The fixpoint of (b) guarantees the final state has both endpoints on the main axis, so no delivery
      method ever has to extract from a dead-end arm itself.

  (d) ORDER OF PREPARATION STEPS.
      The required order inside the two-qubit delivery driver is: isolate qa (split it out of any shared chain), isolate qb, then the
      extract-both fixpoint of (b) -- never interleave isolate and extract per-endpoint, and never extract before both isolations are
      done, because isolate emits its own moves and can itself re-stack an endpoint. Each step is wrapped in the usual snapshot/restore
      transaction.

12.15 ROTOR-CONVEYOR -- THE GUARANTEED DELIVERY MECHANISM

  The Sec. 11 T12-T14 invocations require a delivery the heuristic chain cannot provide: the family of "deliver one endpoint, then the
  other" methods of Sec. 12.5/12.13/12.14 is fundamentally incomplete on dense traps, so a correct compiler must also provide one
  guaranteed delivery routine built on the gate-vertex rotor. This subsection describes the mechanism; Sec. 12.16 is the routine to
  implement and to wire into the fallback chain.

  (a) WHY THE SEAL-THE-SUBTREE METHODS CANNOT BE PATCHED.
      Every delivery in Sec. 12.5/12.13/12.14 ultimately moves one endpoint to a gate-neighbor and then the other. The moment an endpoint
      occupies a gate-neighbor, its subtree is sealed: that subtree connects to the rest of the trap only through that one gate-neighbor
      vertex, now occupied. An obstacle that must cross from one side to the other can no longer do so. On a dense trap (e.g. 9 qubits on
      a 23-vertex, multi-junction trap) there is always some gate where two or more obstacles sit between an endpoint and its target, the
      side-stacks cannot buffer them all, and at least one obstacle must cross the gate region -- which the seal makes impossible. No
      amount of corridor-sweeping, stack-parking, or delivery reordering removes this; it is a structural dead end. Counting confirms it:
      an endpoint N hops deep with k obstacles ahead and only s reachable side-stacks strands k-s obstacles when k > s.

  (b) THE GATE VERTEX IS A 3-CELL ROTOR.
      With single qubits at l and r, gv empty and all junctions empty, the sequence MERGE(gv), SWAP(gv), SEPARATE(gv) exchanges the
      qubits at l and r. This is the only primitive that moves a qubit from one subtree to the other without either qubit needing free
      space to pass the other -- exactly what defeats the obstruction of (a).

  (c) THE CONVEYOR LOOP.
      Given qa already on gate-neighbor ga, gv empty, and qb in the subtree of the other neighbor gb, this loop delivers qb to gb,
      draining every obstacle between qb and gb across the rotor into qa's subtree:

        Repeat until qb is the qubit of gb's subtree closest to gb, then move qb straight to gb:
          - pick x = a qubit in gb's subtree strictly closer to gb than qb (an obstacle between qb and its target);
          - move x to gb (x is closest, so its path carries no other qubit -- a clean translation);
          - rotor: MERGE/SWAP/SEPARATE at gv -- now x is at ga and qa is at gb (the rotor flips qa to the far side);
          - stash x off ga, deep into qa's subtree (see (e));
          - move qa back from gb to ga.
        Each cycle drains exactly one obstacle out of gb's subtree, so the loop is bounded by the qubit count. Succeed when qa and qb sit
        on the two gate-neighbors, gv empty and no junction occupied.

  (d) CONVEYOR INVARIANTS (each is mandatory for correctness).
      - Isolate every qubit (not just the endpoints) before the conveyor: after a previous gate two non-endpoints can still be a merged
        2-chain, and MERGE/SWAP/SEPARATE on a 3-chain drops a qubit (SEPARATE re-emits only chain[0] and chain[1]).
      - gv must be empty before every rotor cycle: clear it explicitly once before the loop (an endpoint-delivery move keeps gv off the
        landing set but does not evict a qubit already sitting on gv).
      - "Closest qubit to gb" is well defined: its path to gb carries no other qubit (any would itself be closer), so moving it to gb is
        a clean translation that disturbs nothing.
      - The rotor leaves qa on the far gate-neighbor each cycle; bring it back to ga before the next cycle. Verify the invariant (qa at
        ga, gv empty, junctions empty) before every MERGE/SWAP/SEPARATE and roll back if it does not hold.

  (e) STASHING ROTATED-OUT OBSTACLES -- NEAREST-FIRST, RIPPLE-DRIVEN.
      Every conveyor cycle of (c) rotates an obstacle x to ga and must stash it somewhere in ga's subtree before the next cycle. On the
      very dense traps of Sec. 11 T14 (16-25 qubits) the naive stash fails two ways:
        - stashing x at the farthest free vertex needs a move across the whole subtree corridor, which is impossible once that corridor
          is packed (x cannot pass a wall of qubits -- (a) again); and
        - stashing only at currently-empty vertices fills the near-ga corridor first and seals off the far free space behind a wall, so
          the conveyor jams after a handful of cycles.
      Two measures make the stash robust:

        (i) Pre-decongest at the routine's entry. Before the conveyor, spread every non-endpoint toward the trap extremities (the
            existing decongest routine, best-effort, never raising). This frees the gate-vertex corridor so the conveyor has somewhere to
            stash.

        (ii) Stash nearest-first, and allow an occupied target. Enumerate the stash vertices of ga's subtree ordered by distance to ga
             ascending (nearest first), excluding only junctions, gate vertices, ga itself, qb's retrieve path, and any vertex holding an
             endpoint -- but not excluding occupied vertices. move(ga, u) to an occupied u is fine: move() evacuates u, and that
             evacuation ripples the local corridor one vertex toward the subtree's free far end. Because the stash always targets the
             nearest spot and lets the ripple carry the displacement outward, the stashes never wall themselves off -- the conveyor keeps
             draining for as many cycles as the gate needs.

      General principle: when you must repeatedly park qubits into a packed region, push each one to the nearest slot and rely on the
      evacuation cascade to ripple the crowd outward; do not target far slots directly and do not restrict targets to empty vertices.

12.16 COMPLETE DELIVERY -- THE GUARANTEED ROUTINE

  _deliver_complete is the single delivery routine that makes routing total on dense traps. Implement it; the heuristic chain of Sec.
  6.2/12.5/12.13/12.14 is then only an optimization that yields shorter schedules on the easy gates. Do not rely on the heuristic chain
  alone -- on dense traps it leaves some gates unroutable.

  (a) THE KEY ASYMMETRY: ROUTING TO gv IS ROBUST, ROUTING TO A NEIGHBOR IS NOT.
      Every routing failure on a dense trap reduces to one thing: an endpoint could not pass an obstacle inside a 1-wide corridor.
      That problem disappears when the destination is the gate vertex itself. Walk an endpoint q toward gv: q and every obstacle ahead of
      it travel the same direction, and the front-most obstacle simply exits across gv into the opposite subtree. No qubit ever has to
      pass another. So an unconstrained move(q, gv) -- blocked and hard both empty -- succeeds whenever the trap has free space anywhere,
      because the cascade evacuation always finds a hole on one side or the other. This is the "funnel". Routing to a gate-neighbor has
      no such property (a neighbor offers obstacles only one corridor direction once gv is taken), which is why a constrained delivery to
      a gate-neighbor can fail where the funnel does not.

      Corollary, and the key principle: a heuristic method fails not because the trap is unroutable but because its moves are
      over-constrained. The same move(q, gv) that fails with blocked={gv}/hard={subtree} succeeds with blocked=set()/hard=set(). When a
      delivery sub-step must simply succeed, pass it the emptiest possible blocked/hard sets and route through gv.

  (b) THE ROUTINE.
      _deliver_complete(qa, qb, gv, l, r):
        1. Isolate every qubit (single-qubit chains -- Sec. 12.15(d)).
        2. Funnel qa to gv with an unconstrained move(qa, gv). By (a) this succeeds whenever the trap has free space.
        3. Read qb's side; let gb be the gate-neighbor on qb's side and ga the other. Drop qa on ga (evacuate ga, then T(gv, ga)).
           Choosing ga opposite qb's side makes qb's subtree the one to be drained.
        4. Run the rotor conveyor of Sec. 12.15(c) (with its pre-decongest and nearest-first stash, Sec. 12.15(e)) to deliver qb to gb.
        5. Succeed iff qa and qb sit on the two gate-neighbors, gv empty, no junction occupied.
      Step 2's funnel is robust by construction; step 4's rotor carries each obstacle past qb with neither needing room to pass. Together
      they make delivery total: no congestion pattern defeats this.

  (c) ORDERING AND COST.
      Keep the heuristic levels of Sec. 12.5 ahead of _deliver_complete in the fallback chain -- they produce shorter schedules on the
      gates they can handle, and _deliver_complete is the guaranteed catch-all that ends the chain. The rotor cost is O(corridor-length)
      per hard gate, so total wall-clock stays linear in gate count; a large circuit of many thousands of gates takes minutes to compile,
      which is the expected cost rather than a failure, and the no-partial-schedule rule means such a run must complete rather than be
      abandoned. Tests with a swap bound (only T3 and T9's invocations 1 and 2, zero swaps) are unaffected because their gates are routed
      by the structured fast path and never reach the rotor; confirm this after wiring _deliver_complete into the fallback chain.
\end{lstlisting}

\clearpage
\section{Prompt for the general compiler}
\label{app:general-prompt}
The prompt for the general-architecture compiler is reproduced below. It is issued with the optimized branched-architecture compiler as its seed and specifies the changes needed for an arbitrary connected trap graph. Its circuit files are as in \appref{app:linear-prompt}. The trap-graph files encode arbitrary connected graphs as a JSON edge list of vertex-identifier pairs, with the gate vertex given on the command line.

\begin{lstlisting}[style=prompt]
Adapt the existing Ion-Trap Shuttling Compiler in ion_trap_compiler.py from a linear trap with junctions (a tree consisting of a main
axis with attached side-stacks) to a general trap graph (an arbitrary connected undirected graph, which may or may not contain cycles).
The compiler translates a quantum circuit into a sequence of physical shuttling operations, minimizing the number of timesteps.

ion_trap_compiler.py already contains a complete and working compiler for the linear-with-junctions case. Read it: it is the source of
truth for every rule and primitive that does not change in the adaptation. This prompt only describes what needs to change for the
general-graph case; everything else -- the operation tuple format, the snapshot/restore pattern, the routing primitives' contracts, the
post-processing passes, the timestep-grouping admission rules, the iterative-traversal coding requirements, the no-partial-schedule rule,
etc. -- is documented by the existing code and must be inferred from it.

In particular, every hardware rule that the validator enforces is in the not-allowed-to-modify code. If anything in this prompt appears
to conflict with what those functions enforce, the code is the spec.

(*@\rule{\linewidth}{0.4pt}@*)
SCOPE OF THE ADAPTATION
(*@\rule{\linewidth}{0.4pt}@*)

May be modified: initial_mapping, compile, and every internal helper function that one of them transitively calls. These must be
rewritten so they handle general graphs correctly -- trees, graphs with one or more cycles, and mixes of the two.

Must not be modified: the TrapGraph class, the QuantumGate class, parse_input, validate_and_output, and every helper they call.

The TrapGraph class already exposes every topology helper the router needs. Do not extend it.

(*@\rule{\linewidth}{0.4pt}@*)
INPUT FORMAT
(*@\rule{\linewidth}{0.4pt}@*)

The TrapGraph constructor accepts an edge-list JSON whose entries are pairs of vertex identifiers given as strings:

    [["0", "1"], ["1", "2"], ["2", "3"], (*@$\ldots$@*)]

Vertex identifiers are converted to int internally, so all comparisons -- including the validator's u1 < u2 assertion for SEPARATE /
MERGE neighbor pairs at gate vertices -- use numeric ordering. Identifiers need not start at 0, need not be contiguous, and need not be
assigned in any particular pattern, but strings that parse as integers.

The --gate command-line argument is a comma-separated list of vertex identifiers (e.g. --gate 3 or --gate 2,7), also parsed as ints.

The graph itself has no spatial embedding -- there is no inherent "left" or "right" of a gate vertex. The validator nevertheless requires
a canonical order on the gate vertex's two neighbors, and it uses numeric comparison to define that order: in every SEPARATE / MERGE
operation tuple (gv, u1, u2), the smaller-numbered neighbor is u1 and the larger-numbered neighbor is u2. The "left" / "right" language
used in routing prose is just shorthand for this convention:

  - When a 2-chain [a, b] sits at gv and SEPARATE is applied, the validator places a at u1 (the smaller-numbered neighbor) and b at u2
    (the larger-numbered neighbor). See apply_ts_non_trans's 'S' branch.
  - When the two empty neighbors u1, u2 hold single ions and MERGE is applied, the validator forms the chain [state[u1], state[u2]] at gv
    -- i.e. the smaller-numbered neighbor's ion is the first element of the resulting chain. See apply_ts_non_trans's 'M' branch.

The router that builds the operation stream must respect this convention when reasoning about chain order at gv: which qubit ends up at
the smaller-numbered neighbor after SEPARATE is determined by which element is index 0 of the chain at gv, which in turn is determined by
how the chain was assembled (e.g. by the order of two MERGEs combined with intervening SWAPs). The graph-builder does not need to align
"left" and "right" with anything -- the numbering of the vertices alone fixes the convention.

(*@\rule{\linewidth}{0.4pt}@*)
TRAPGRAPH METHODS
(*@\rule{\linewidth}{0.4pt}@*)

TrapGraph exposes the standard graph primitives (vertices, neighbors, degree, distance, bfs_path, next_hop, is_junction,
path_passes_through, junctions_on_path, subtree_through) and the general-graph topology helpers needed for cycle-aware routing:

  - articulation_points -> List[int]
    Vertices whose removal disconnects the graph. Useful as a coarse classifier: an articulation-point junction has at least one neighbor
    that becomes unreachable from its other neighbors when the junction is removed. Note: "is an articulation point" and "lies on a
    cycle" are independent properties (e.g. the shared vertex of a figure-8 is both on two cycles and an articulation point). For
    per-pair-of-edges reasoning, use per_junction_components below, not articulation_points.

  - bridge_edges -> List[Tuple[int, int]]
    Edges whose removal disconnects the graph. Returned as sorted (u, v) tuples with u < v numerically. An edge is a bridge iff it lies
    on no cycle.

  - per_junction_components(j: int) -> Tuple[frozenset, (*@$\ldots$@*)]
    For the junction j, the connected components of V \ {j}. This is the routing-relevant helper: given a chain that enters j via edge
    (a, j) and exits via edge (j, b), the chain can later reach a's side without re-entering j via (a, j) iff a and b lie in the same
    component (i.e. an alternative arc through V \ {j} connects them). When they lie in different components, the chain is committed to
    b's side until j is emptied. Consult this helper at every junction-transit decision.

  - alternative_path(src: int, dst: int, forbidden: Set[int]) -> Optional[Tuple[int, (*@$\ldots$@*)]]
    Shortest path from src to dst that avoids every vertex in forbidden, or None if no such path exists. The endpoints src and dst are
    never excluded. Underpins the cycle-detour routing method: passing the obstacles on the canonical bfs_path as forbidden returns the
    alternative arc around them, if one exists.

All four helpers are precomputed once in the TrapGraph constructor. articulation_points and bridge_edges are iterative Tarjan
computations, O(|V| + |E|). per_junction_components is also O(|V| + |E|) per junction total O(|V|(*@$\cdot$@*)(|V| + |E|)) for the whole table, with
the results cached. alternative_path runs a fresh BFS on each call, O(|V| + |E|).

(*@\rule{\linewidth}{0.4pt}@*)
WHAT IS DIFFERENT ON A GENERAL GRAPH
(*@\rule{\linewidth}{0.4pt}@*)

A general graph may be a tree, may contain cycles, or may be a mix. The adaptation must handle all of these. Where the existing
implementation assumes a tree structure or stack semantics, replace that assumption with one of the alternatives below.

1.  Topology. A junction is still any vertex of degree > 2, a non-junction vertex has degree 1 or 2, and the gate-vertex constraints are
    unchanged (degree exactly 2; neither neighbor is a junction). Every junction rule still applies on the general graph: junctions hold
    ions only in transit; during SEPARATE / MERGE / SWAP / GATE every junction must be empty; the no-roundtrip rule holds at every
    junction the chain transits (a chain that entered j from a must leave j via some b != a).

2.  No stacks, no main axis. Qubit vertices are simply non-junction non-gate vertices that hold up to 2 ions each, with no LIFO or
    ordering constraint. The existing tree-specific routing structures (stack-shaped side arms with depth-ordered slots, main-axis
    indexing, per-stack endpoint protection rules) are no longer meaningful on a general graph and must be replaced by general-graph
    routing during the rewrite. The specific code-visible structures to remove are listed under "REMOVE OR REPLACE" in the ROUTING
    section.

3.  Cycles relax the no-roundtrip rule at cycle-member junctions. The structural difference between a tree and a cycle-containing graph
    is not "obstacles can be detoured around for free" -- a pure cycle is topologically a linear chain with the endpoints joined, so the
    1-wide-corridor obstruction "ions cannot pass each other along a path" still holds on every edge. What changes is the freedom a chain
    has when transiting a junction that sits on a cycle.

    On a tree-junction j, the no-roundtrip rule forces every chain entering via edge a to leave via some other edge b; once committed to
    b, the chain cannot reach a's side of j without first emptying j and re-entering. On a cycle-member junction j, a chain that needs to
    reach a's side after exiting via b can take the cycle's alternative arc back around, satisfying no-roundtrip via topology rather than
    via eviction. This is the main structural win.

    A secondary, situational win is obstacle bypass on a cycle: if one arc of a cycle is heavily packed and the other arc is mostly
    empty, routing along the empty arc avoids evicting the packed arc's obstacles. The win is real only when the asymmetric packing makes
    the alternative arc strictly shorter than (eviction cost + direct arc); on uniformly packed cycles the alternative arc has roughly
    the same number of obstacles as the direct arc and the cycle-detour is a wash or a loss.

    The adaptation should exploit (1) whenever a chain transits a cycle-member junction (consult per_junction_components to know whether
    the entry and intended exit lie in the same component), and (2) opportunistically when alternative_path reports a strictly shorter
    obstacle-free path than the direct arc.

4.  The set of routing methods that must remain after the adaptation:
      (a) cycle-detour (only applies when alternative_path returns a path and its cost beats the direct-arc-plus-eviction cost -- often
          loses on uniformly packed cycles);
      (b) park-the-endpoint at any empty non-junction vertex anywhere in the graph;
      (c) the standard MERGE / SWAP / SEPARATE Swap-Method (works on any graph, used as the unconditional fallback);
      (d) the gate-vertex rotor (MERGE / SWAP / SEPARATE on gv) as the terminal level of the fallback chain, since the rotor never relies
          on cycles -- it relies only on the gate-vertex having two non-junction neighbors, which is true on every legal trap graph.
    On a graph with no cycles at all, method (a) is unavailable on every iteration and the router falls through to (b), (c), (d)
    automatically. On a pure cycle (no junctions), method (a) is available on every iteration but rarely wins on cost, so the schedule
    length is similar to the linear-chain case. The structural cycle benefit (item 3) only materializes on graphs that have both cycles
    and junctions on those cycles.

5.  Park-the-endpoint generalizes. In the existing linear-trap code, "park slot" meant either a stack column head or a main-axis-end
    vertex. On the general graph it means any empty non-junction vertex that lies off the routing-intermediate paths between the two
    endpoints' current positions and their target gv-neighbors.
    Enumerate candidates by:

      candidates = [v for v in trap_graph.qubit_vertices
                      if v not in routing_intermediate_paths and current_state.get(v, []) == []]

    Rank by total cost distance(qpos[endpoint], park_vertex) + distance(park_vertex, gv_neighbor); tie-break by vertex identifier
    ascending. Park-the-endpoint always has at least one candidate on a 2 * n + 1 non-junction-vertex trap because the placement leaves
    one empty non-junction vertex free.

(*@\rule{\linewidth}{0.4pt}@*)
INITIAL MAPPING
(*@\rule{\linewidth}{0.4pt}@*)

The four placement rules must remain in force unchanged (they are also enforced by check_state in the validator):
  - exactly (*@$\lceil$@*)len(qubits) / 2(*@$\rceil$@*) vertices hold qubits;
  - no vertex holds more than 2 qubits;
  - no junction vertex holds any qubit;
  - all remaining vertices are empty.

The placement heuristic in the existing initial_mapping (circuit-DAG depth-weighted interaction graph (*@$\rightarrow$@*) greedy maximum-weight matching
(*@$\rightarrow$@*) greedy chain extraction (*@$\rightarrow$@*) intra-pair orientation by total cross-pair weight (*@$\rightarrow$@*) greedy assignment to candidate vertices sorted by
minimum distance to any gate vertex, with the items[0]-equals-first-2q-gate override that places items[0] at the most-central gate
vertex) is topology-agnostic: it uses only trap_graph.distance and the set of non-junction vertices. Carry it over unchanged.

(*@\rule{\linewidth}{0.4pt}@*)
ROUTING
(*@\rule{\linewidth}{0.4pt}@*)

The existing compile uses a multi-level routing pipeline (gate scheduling (*@$\rightarrow$@*) routing-intermediate-set construction (*@$\rightarrow$@*) iterative
reduction (*@$\rightarrow$@*) finalization (*@$\rightarrow$@*) post-processing (*@$\rightarrow$@*) timestep grouping). The pipeline structure carries over to the general graph. Read the
existing code for the contract of every primitive (move, push, evacuate, walk_path, clear_junction, snapshot/restore, the routing-state
invariants on junctions and endpoint protection, etc.).

The changes required for the general-graph adaptation:

REMOVE OR REPLACE in the existing code:

  - The park-candidate scaffolding at the top of _Compiler.__init__ -- the local_main_axis / local_on_main_axis / local_stacks_at
    recomputation and the _park_candidates / _stack_heads / _PARK_MAX state it produces. This entire block hardcodes the assumption that
    the graph is a linear main axis with stack-shaped side arms, and is used by _fast_two_park / _try_park_deliver to enumerate park
    slots only at stack-column heads and main-axis-end vertices. Rewrite the park-slot enumeration to use any empty non-junction vertex
    of the graph (see point 5 of "What is different").

  - The conservative roundtrip-elimination guard inside _post_process (the rule that refuses to remove (A(*@$\rightarrow$@*)B, B(*@$\rightarrow$@*)A) whenever either A or
    B is a junction). Replace it with the replay-based junction-soundness check: replay the trace with the candidate removal applied and
    accept the removal iff no junction's exit edge matches its immediately preceding entry edge in the residual. The replay-based check
    accepts strictly more removals, which matters on cycle graphs where an A(*@$\rightarrow$@*)B detour at a cycle-member junction B can be sound (the
    chain entered B from somewhere else via the alternative arc before re-emerging at A).

ADD: the cycle-detour routing method. For each routing-intermediate q_inter on the BFS path from an endpoint's current vertex to its
target gv-neighbor, call trap_graph.alternative_path with q_inter (and any already-placed endpoint that should stay stationary) in the
forbidden set. If the call returns a path, the endpoint can walk along the alternative arc; q_inter stays where it is, and no SEPARATE /
MERGE / SWAP is emitted for it. The bypass cost is just the extra translations the endpoint walks (alternative_arc_length minus
direct_arc_length); see METHOD CHOICE below for the full cost comparison.

  The cycle-detour method is applicable only when the graph contains a suitable cycle. On trees and on dense tree-like subregions of a
  general graph, alternative_path returns None for any non-trivial obstacle set, and routing falls through to Park-the-endpoint or
  Swap-Method. Even when an alternative path exists, cycle-detour can lose on cost because the alternative arc may be much longer than
  the direct arc plus an eviction. The method is an optimization, not a correctness primitive -- never rely on it as the only
  displacement method.

KEEP (these methods work on every general graph, with or without cycles):

  - Park-the-endpoint, with the generalized park-slot enumeration of point 5 of "What is different."
  - The Swap-Method intermediate displacement (MERGE (*@$\rightarrow$@*) SWAP (*@$\rightarrow$@*) SEPARATE on a gate vertex, shifting one intermediate one position
    outward per step). This is the unconditional fallback when neither cycle-detour nor Park-the-endpoint applies. The same-side handling
    of the existing implementation carries over verbatim.
  - The two-qubit finalization protocol when the gate vertex still holds a non-endpoint chain (_deliver_complete + _deliver_rotor in the
    existing code): try each of the two role assignments (qa-anchor / qb-anchor) wrapped in snapshot/restore, keep the cheaper.
  - The guaranteed-delivery terminal routine (_deliver_rotor in the existing code): funnel one endpoint to gv with an unconstrained move;
    drop it on the gv-neighbor opposite the other endpoint; run the MERGE / SWAP / SEPARATE rotor to drain obstacles across gv with
    nearest-first ripple-driven stashing of rotated-out obstacles. This routine relies only on the gate-vertex's two non-junction
    neighbors and on the funnel-move-to-gv property, both of which hold on every legal trap graph regardless of cycle structure.

METHOD CHOICE per iteration of the intermediate-set reduction loop. Estimate the cost of each applicable method in elementary
translations. Both the bypass cost (extra distance walked along the alternative arc) and the eviction cost must be charged honestly:
  - cycle-detour: alternative_arc_length minus direct_arc_length, i.e. the extra translations the endpoint walks along the longer arc to
    avoid the intermediate. Free only when the alternative arc happens to be the same length as the direct one. Often the alternative arc
    is much longer, in which case cycle-detour loses on cost.
  - Park-the-endpoint: distance(qpos[endpoint], park_vertex) + distance(park_vertex, gv_neighbor).
  - Swap-Method: 4 + max(0, distance(endpoint, gv) + distance(intermediate, gv) - 2). The +4 accounts for the SEPARATE + SWAP + MERGE
    non-translation overhead.
Pick the cheapest applicable method. Tie-break order when two methods yield equal estimated cost: (1) cycle-detour (no non-translation
overhead), (2) Park-the-endpoint (also no SEPARATE/SWAP/MERGE), (3) Swap-Method. Within the same method, tie-break by the relevant vertex
identifier ascending.

Special case worth checking explicitly: when the intermediate sits on a cycle-member junction's "wrong side" (i.e. between the endpoint
and the junction, on the direct arc), cycle-detour around the junction may be the cheapest method even when the alternative arc is
longer, because evicting an intermediate at a junction often requires the intermediate to first be moved off the junction via
no-roundtrip-compliant routing, which can cascade. Consult per_junction_components to determine whether the alternative arc exists at
all.

WIRING. The multi-level routing fallback chain in the existing implementation remains in the same shape. The level-method names in the
existing code (_route_two, _fast_two, _fast_two_park, _try_deliver_pair, _deliver_complete, _deliver_rotor and the snapshot/restore
wrapping between them) form the skeleton; the general-graph adaptation changes which method-choice options are available at each level,
not the level structure.

(*@\rule{\linewidth}{0.4pt}@*)
CODE QUALITY
(*@\rule{\linewidth}{0.4pt}@*)

The existing implementation's coding standards apply unchanged. The ones most likely to bite during a rewrite:

  - Single file, Python 3.12+.
  - Determinism: byte-identical output across PYTHONHASHSEED values. Sort every iteration over a set or dict whose order could leak into
    the emitted ops.
  - All graph traversals iterative -- explicit stack/queue, no recursion. This applies to BFS, DAG depth, evacuation cascades, and the
    timestep-grouping admission loop.
  - No integer-arithmetic-on-vertex-identifiers to encode topology role or position. Vertex identifiers are opaque ints: do not assume
    any structural meaning from the numbering (consecutive indices need not be neighbors; mid-range indices need not be interior
    vertices, etc.). Use trap_graph.is_junction together with trap_graph.articulation_points / per_junction_components for topology
    classification, and trap_graph.neighbors / trap_graph.next_hop / trap_graph.bfs_path for adjacency and routing.
  - The tree-specific scaffolding identifiers listed in "REMOVE OR REPLACE" (local_main_axis, local_on_main_axis, local_stacks_at,
    _stack_heads, _park_candidates, _PARK_MAX, _fast_two_park, _try_park_deliver) should not survive in the rewritten code.

(*@\rule{\linewidth}{0.4pt}@*)
TESTING
(*@\rule{\linewidth}{0.4pt}@*)

No-partial-schedule rule (applies to every (circuit, graph) test and every T1-T14 invocation below): the compiler has exactly two
permitted outcomes for any circuit:
  (a) it returns a complete, valid schedule in which every gate is executed (the schedule passes validate_and_output without any "not all
      gates have been executed" failure), or
  (b) it raises an exception.
The compiler must never return a schedule that omits one or more gates. If a gate cannot be routed by any available strategy, the
compiler raises an exception rather than silently dropping that gate or any later gate. A run that ends by emitting a partial schedule is
a failure even if the partial schedule is internally consistent.

Test focus is single-gate-vertex architectures sized at 2 * n + 1 non-junction vertices for an n-qubit circuit. Single-gate-vertex
architectures are the hardest because every two-qubit interaction must funnel through one gv, removing the ability to absorb obstacles at
a second gate vertex.

THE BENCHMARK CIRCUITS supplied in the working directory (example.qasm, same_pair.qasm, and the .json files: 4gt12-v0_86, 4gt4-v0_79,
4gt4-v1_74, 4mod5-bdd_287, 4mod5-v1_23, alu-v2_30, alu-v2_31, cnt3-5_179, hwb7_59, mod8-10_178, qaoa_5, qaoa_25, qaoa_40, qft_15, qft_25,
qv_20, qv_25, rd32-v0_66, rd53_135, xeb_9, xeb_16, xeb_sy_9, xeb_sy_16, xeb_sy_25) must each compile to a valid schedule on a suitably
sized general graph. The number of qubits in each circuit is the number of distinct qubit names appearing in its gates list.

GRAPHS -- the test graphs are provided in the working directory as JSON edge-list files. The architecture-matrix graphs (groups A-E
below) are named graph_<shape>_n<n>.json (one per shape per circuit qubit-count n, e.g. graph_cycle_n5.json,
graph_subdivided_K4_n16.json); arch_graph_gates.json maps each (shape, n) to its gate vertex. The linear-with-junctions graphs used by
the explicit T1-T14 invocations are named graph_<V>_<junction_depth>_<junction_distance>.json -- a main axis of V = 2 * n + 1 segments
with the gate vertex at its center, junction segments spaced junction_distance apart along the axis, each carrying a side-stack of
junction_depth vertices (e.g. graph_13_1_3 has depth-1 stacks every third segment; graph_19_1_5 has depth-1 stacks spaced 5 apart on a
19-segment axis) -- plus the plain path graph_5.json (no junctions; gate vertices 2). Every file uses the JSON edge-list format
documented under "INPUT FORMAT" above. The descriptions below document how each provided graph is built and what routing capability it
exercises.

Every provided architecture graph satisfies:
  (i)   gate-vertex degree exactly 2 (asserted by TrapGraph.__init__);
  (ii)  neither gate-vertex neighbor is a junction (asserted by TrapGraph.__init__ and by check_sep_merge_asserts);
  (iii) exactly 2 * n + 1 non-junction vertices for the n-qubit test circuit. The total vertex count is 2 * n + 1 plus the number of
        junctions, which varies by architecture (zero for pure cycles and paths, one or more for the mixed and dense families below). The
        placement fills (*@$\lceil$@*)n / 2(*@$\rceil$@*) non-junction vertices with qubits and needs at least one empty non-junction vertex of headroom;
  (iv)  connectivity (asserted by TrapGraph.__init__).


REQUIRED ARCHITECTURE COVERAGE. For each circuit size n, a graph is provided for each of the fifteen test cases below
(graph_<shape>_n<n>.json). The groups are chosen so each exercises a distinct routing capability; passing on one group does not imply
passing on the others. Every entry's construction is stated for general n so the same pattern scales to any circuit. The non-junction
vertex count in every entry is exactly 2 * n + 1.

GROUP A -- Tree shapes (no cycles, baseline). These graphs verify that the routing still works correctly when no alternative arc ever
exists. The cycle-detour method is unavailable at every iteration; the router must compile every benchmark using only Park-the-endpoint,
Swap-Method, and the gate-vertex rotor terminal.

  A1. path. A single path of 2 * n + 1 vertices, no junctions at all. Gate vertex mid-path. Total vertex count: 2 * n + 1.

  A2. caterpillar. A path of 2 * n + 1 vertices with k additional pendant vertices attached to k distinct interior vertices. Each
      pendant-bearing vertex has degree 3 and is a junction. All junctions are articulation points (every edge is a bridge). Non-junction
      vertices: (2 * n + 1 - k) interior + k pendants = 2*n + 1. Total vertex count: 2*n + 1 + k.

  A3. star. A central vertex of degree k connected to k arms whose lengths sum to 2 * n + 1. If k >= 3 the central vertex is a
      high-degree junction (degree 4, 5, (*@$\ldots$@*)). Gate vertex on one arm. Total vertex count: 2 * n + 2 (the 2 * n + 1 arm vertices plus the
      central junction).

GROUP B -- Cycle-only shapes (one cycle, no junctions). A pure cycle is topologically a linear chain with one extra edge joining the
endpoints. Routing on it is mostly the same as on a linear path: the 1-wide corridor constraint still holds on every edge, and the
cycle-detour method usually loses on uniformly packed cycles because the alternative arc carries roughly as many obstacles as the direct
arc. These graphs verify that the cycle-detour method does not regress the schedule when it does not help, and that the BFS-based
alternative-arc search is correct when it does help (asymmetric obstacle packing). Expect schedule lengths similar to a linear-chain
shape of the same vertex count, not dramatically better.

  B1. cycle. A single cycle of length 2 * n + 1, no junctions. Gate vertex anywhere on the cycle (a pure cycle is vertex-transitive, so
      the gate-vertex position is irrelevant up to relabeling). Total vertex count: 2 * n + 1.

GROUP C -- Mixed-topology shapes (cycles and bridges in the same graph). These are the primary stress tests for the cycle benefit: each
contains at least one cycle-member junction (where the no-roundtrip rule relaxes via the alternative arc) and at least one bridge (where
the routing must fall back to tree-style methods). The router must pick the right method per junction based on whether the junction lies
on a cycle.

  C1. cycle_with_pendant. A cycle of length 2 * n + 1 plus k pendant vertices attached to k distinct cycle vertices. Each pendant-bearing
      cycle vertex is a junction; it is not an articulation point (the cycle survives its removal), so the no-roundtrip rule relaxes at
      it. The pendant edges are bridges. Non-junction vertices: (2 * n + 1 - k) cycle non-junctions + k pendant vertices = 2 * n + 1.
      Total vertex count: 2 * n + 1 + k.

  C2. lollipop. A cycle of length m + 1 plus a path-tail of length 2 * n + 1 - m attached to one cycle vertex. The attachment vertex has
      degree 3 and is an articulation point of the whole graph (removing it disconnects the tail from the cycle). Routing within the
      cycle benefits from the relaxed no-roundtrip at the attachment vertex; routing along the tail has no such relaxation. Non-junction
      vertices: m cycle vertices (all cycle vertices except the attachment) + (2 * n + 1 - m) tail vertices = 2 * n + 1. Total vertex
      count: 2 * n + 2 (the 2 * n + 1 non-junctions plus the attachment junction).

  C3. two_cycles_bridged. Two disjoint cycles connected by a path. The path endpoints (one in each cycle) are junctions; every edge on
      the connecting path is a bridge. The connecting path itself has zero, one, or more interior non-junction vertices. Non-junction
      vertices: the cycle non-junction vertices in cycle 1 plus those in cycle 2 plus the connecting-path interior vertices, summing to 2
      * n + 1. Total vertex count: 2 * n + 3 if the connecting path has zero interior vertices (just one edge between the two junctions),
      more otherwise.

GROUP D -- Dense / multi-cycle shapes (multiple overlapping cycles). Most or all junctions are cycle-member junctions, so the
no-roundtrip relaxation applies almost everywhere. These verify that the router correctly handles junctions belonging to several cycles
simultaneously.

  D1. theta. Two vertices joined by three internally disjoint paths. Both degree-3 vertices are junctions on the same pair of cycles. The
      gate vertex is on the interior of one of the three paths. Non-junction vertices: the path interiors, summing to 2 * n + 1. Total
      vertex count: 2 * n + 3.

  D2. figure8. Two cycles sharing one vertex. The shared vertex has degree 4 and is an articulation point of the whole graph: removing it
      splits the two cycles into two disjoint paths. Cycle-detour applies within each cycle but cannot cross the shared vertex
      (alternative_path through forbidden = {shared} returns None for any pair on different sides). Tests that the router does not
      mistakenly assume an alternative arc exists across the shared vertex. Non-junction vertices: (cycle1_length - 1) + (cycle2_length -
      1) = 2 * n + 1. Total vertex count: 2 * n + 2.

  D3. cycle_with_chord. A cycle of length 2 * n + 3 plus one chord edge connecting two non-adjacent cycle vertices. The chord adds no new
      vertices, but its two endpoints (already cycle vertices) become junctions of degree 3. Non-junction vertices: 2 * n + 3 - 2 = 2 * n
      + 1. The chord creates two additional cycles (the chord plus each of the two arcs of the original cycle that it separates), giving
      three cycles total that share pairs of vertices. The gate vertex sits on the cycle interior at a non-junction position whose two
      neighbors are also non-junctions. Total vertex count: 2 * n + 3.

  D4. cycle_with_two_chords. A cycle of length 2 * n + 5 plus two non-crossing chord edges. The four chord endpoints become junctions of
      degree 3. Non-junction vertices: 2 * n + 5 - 4 = 2 * n + 1. Place the chord endpoints so that at least one consecutive run of three
      non-junction cycle vertices remains, then put the gate vertex in the middle of that run. (A naive placement that scatters all four
      chord endpoints uniformly leaves only length-2 non-junction runs, and no such vertex has two non-junction neighbors -- the
      construction then fails the gate-vertex constraint.) Total vertex count: 2 * n + 5.

GROUP E -- Structured / parameterized families. These are adaptations of classical network topologies. The unadapted forms typically have
every vertex at degree 3 or higher (so every vertex would be a junction and no qubit or gate vertex would exist). Each entry below
describes both the base topology and the adaptation that introduces enough non-junction vertices to host the qubits and the gate vertex.

The adaptation principle is the same in every case: subdivide selected edges by inserting degree-2 intermediate vertices, until the
non-junction count reaches exactly 2 * n + 1 and a gate-vertex candidate (degree-2, both neighbors non-junctions) exists.

  E1. subdivided_K4. Base: the complete graph K4 has 4 vertices, each of degree 3, with 6 edges. Subdivide each edge by inserting one
      degree-2 vertex in the middle of the edge. After subdivision: 4 original vertices (junctions, still degree 3) + 6 subdivision
      vertices (non-junctions, degree 2). For larger n, subdivide each edge with more than one vertex -- k subdivision vertices per edge
      yields 6k non-junctions; tune the per-edge subdivision counts (they need not all be equal) so the total non-junction count equals 2
      * n + 1. Pick a subdivision vertex on a sufficiently subdivided edge as the gate vertex so that its two neighbors (other
      subdivision vertices on the same edge) are non-junctions. Tests dense multi-cycle routing with junctions belonging to multiple
      short cycles each.

  E2. subdivided_Q3. Base: the 3-dimensional hypercube Q3 has 8 vertices (the corners of a cube), each of degree 3, with 12 edges. The 8
      original vertices are all junctions in any subdivision. To get a viable gate vertex, at least one edge must be subdivided with 3 or
      more intermediate vertices (so the middle subdivision vertex has two subdivision-vertex neighbors, both non-junctions). The
      remaining edges may be subdivided with any number of vertices (including zero -- leaving them as direct Q3 edges between
      junctions). Tune the per-edge subdivision counts so the total subdivision-vertex count equals 2 * n + 1. For example, for n = 5
      (need 11 non-junctions): one edge with 3 subdivisions for the gate vertex, plus 8 edges with 1 subdivision and 3 edges left direct
      = 3 + 8 = 11 non-junctions. Tests regularly-structured multi-cycle routing with every junction equivalent under graph automorphism.

  E3. subdivided_butterfly_B2. Base: the butterfly network B2 has 3 levels of 4 vertices each (12 vertices total) with edges following
      the butterfly bit-flip pattern between adjacent levels. The standard butterfly is constructed so every vertex has degree 4 (each
      level-i vertex connects to two vertices in level i-1 and two in level i+1, with end levels wrapping or terminating depending on the
      variant). Every vertex is a junction in the unadapted form, so no qubit or gate vertex exists. Adapt by subdividing every edge that
      lies on a candidate gate-vertex chain: a chain of three subdivision vertices in series on one edge (call them A-X1-X2-X3-B where A
      and B are original butterfly vertices) gives X2 as a viable gate vertex (its neighbors X1 and X3 are both non-junctions). For n = 4
      (need 9 non-junctions), subdivide one edge with 3 vertices and another 6 edges with 1 vertex each (3 + 6 = 9 non-junctions). For
      larger n, distribute additional single-subdivision vertices across more edges. The level structure and the multiple short cycles
      between adjacent levels make this a stress test for routing through junctions belonging to many short cycles each.

  E4. ladder. Two parallel paths of P + 1 vertices each (so P edges along each path), connected by P + 1 rungs (edges between
      corresponding path-vertices). Subdivide each rung by inserting one degree-2 vertex in the middle. Junctions: the 2 * P - 2 internal
      path vertices (each has two path neighbors and one mid-rung neighbor, so degree 3). Non-junctions: 4 corner path vertices (degree 2
      each, with one path neighbor and one mid-rung neighbor) plus P + 1 mid-rung vertices (degree 2 each, with two path-vertex
      neighbors). Total non-junctions: 4 + (P + 1) = P + 5. Set P = 2 * n - 4 so non-junctions = 2 * n + 1. Total vertex count 2 * (P +
      1) + (P + 1) = 3 * (P + 1) = 6 * n - 9. Gate vertex: an end-rung mid-vertex (rung 0 or rung P), whose two neighbors are corner
      path-vertices, both non-junctions. (Interior-rung mid-vertices do not qualify because their neighbors are internal path junctions.)
      Tests routing on a two-rail topology with a regular cycle structure (every face is a 4-cycle).

VERIFICATION (already satisfied by the provided files):
  - TrapGraph instantiates on each graph with len(tg.qubit_vertices) + len(tg.gate_vertices) == 2 * n + 1.
  - The smallest benchmark (same_pair.qasm or example.qasm) compiles on each graph; if that fails, the larger benchmarks on the same
    shape fail too.

EXPLICIT INVOCATIONS (T1-T14). Beyond the architecture matrix above, the compiler must pass the explicit invocation suite below. It is
the original linear-with-junctions test suite -- a linear main axis with depth-k side-stacks is a tree, a special case of a general
graph, so every invocation the linear compiler passed must still pass after the general-graph adaptation. Each test fixes the input,
invocation, and (where stated) an expected bound; the schedule itself need not match any reference. Tests T1-T2 are small fixed checks;
T3-T8 list multiple invocations of one circuit on different graphs; T9 covers larger out-of-matrix circuits and graphs; T10-T14 stress
two-qubit delivery on progressively denser traps (depth-1 side-stacks, stacked endpoints, multi-junction traps, 15- to 25-qubit circuits)
-- the regime in which heuristic routing alone is insufficient and the guaranteed-delivery rotor is required. Tests T9-T14 carry no
expected operation bound -- for them "pass" means only a complete, valid schedule.

  T1. Same-pair batching. Input: 10 gates, each on the same qubit pair (q0, q1). Invocation: python ion_trap_compiler.py --circ
      same_pair.qasm --graph graph_5.json --gate 2 Expected: q0 and q1 are mapped to vertex 2 in the initial state; the 10 GATE
      operations execute back-to-back on that vertex; SEPARATE, MERGE, SWAP, and TRANSLATE counts are all 0. graph_5.json has no
      junctions and exercises that the junction-aware code degrades gracefully to the linear case.

  T2. cx-triangle. Input: example.qasm -- three qubits, the gates cx q0,q1; cx q1,q2; cx q0,q2. Invocation: python ion_trap_compiler.py
      --circ example.qasm --graph graph_7_1_1.json --gate 3 Expected: (*@$\leq$@*) 20 individual operations, (*@$\leq$@*) 12 timesteps; the first 2-qubit
      gate executes at vertex 3 with no preceding shuttling, because the default initial mapping co-locates q[0] and q[1] at vertex 3.

  T3. qaoa_5.json -- chain-structured zero-SWAP target. Input: 5 qubits, 35 gates; two-qubit interaction graph is the path q0-q1-q2-q3-q4
      (only 4 RZZ gates, all on adjacent pairs). Invocation 1: python ion_trap_compiler.py --circ qaoa_5.json --graph graph_11_1_1.json
      --gate 5 Invocation 2: python ion_trap_compiler.py --circ qaoa_5.json --graph graph_11_1_2.json --gate 5 Invocation 3: python
      ion_trap_compiler.py --circ qaoa_5.json --graph graph_11_2_1.json --gate 4 Invocation 4: python ion_trap_compiler.py --circ
      qaoa_5.json --graph graph_11_3_1.json --gate 3 Expected: 0 SWAP operations across all four invocations. This is the canonical test
      of the initial-placement heuristic.

  T4. rd32-v0_66.json -- small reversible circuit, 4 qubits, 42 gates. Invocation 1: python ion_trap_compiler.py --circ rd32-v0_66.json
      --graph graph_9_1_1.json --gate 4 Invocation 2: python ion_trap_compiler.py --circ rd32-v0_66.json --graph graph_9_2_1.json --gate
      3 Expected: schedule verifies; runtime under one second on commodity hardware. The smallest non-trivial benchmark.

  T5. 4mod5-v1_23.json -- 5 qubits, 102 gates; non-chain interaction graph. Invocation 1: python ion_trap_compiler.py --circ
      4mod5-v1_23.json --graph graph_11_1_1.json --gate 5 Invocation 2: python ion_trap_compiler.py --circ 4mod5-v1_23.json --graph
      graph_11_1_2.json --gate 5 Invocation 3: python ion_trap_compiler.py --circ 4mod5-v1_23.json --graph graph_11_2_1.json --gate 4
      Invocation 4: python ion_trap_compiler.py --circ 4mod5-v1_23.json --graph graph_11_3_1.json --gate 3 Expected: schedule verifies.

  T6. 4mod5-bdd_287.json -- 7 qubits, 106 gates. Invocation 1: python ion_trap_compiler.py --circ 4mod5-bdd_287.json --graph
      graph_15_1_1.json --gate 7 Invocation 2: python ion_trap_compiler.py --circ 4mod5-bdd_287.json --graph graph_15_2_2.json --gate 5
      Invocation 3: python ion_trap_compiler.py --circ 4mod5-bdd_287.json --graph graph_15_3_1.json --gate 5 Invocation 4: python
      ion_trap_compiler.py --circ 4mod5-bdd_287.json --graph graph_15_5_1.json --gate 3 Expected: schedule verifies.

  T7. alu-v2_31.json -- 5 qubits, 640 gates. Invocation 1: python ion_trap_compiler.py --circ alu-v2_31.json --graph graph_11_1_1.json
      --gate 5 Invocation 2: python ion_trap_compiler.py --circ alu-v2_31.json --graph graph_11_1_2.json --gate 5 Invocation 3: python
      ion_trap_compiler.py --circ alu-v2_31.json --graph graph_11_2_1.json --gate 4 Invocation 4: python ion_trap_compiler.py --circ
      alu-v2_31.json --graph graph_11_3_1.json --gate 3 Expected: schedule verifies.

  T8. cnt3-5_179.json -- 16 qubits, 230 gates. Invocation 1: python ion_trap_compiler.py --circ cnt3-5_179.json --graph graph_33_1_1.json
      --gate 16 Invocation 2: python ion_trap_compiler.py --circ cnt3-5_179.json --graph graph_33_3_2.json --gate 10 Invocation 3: python
      ion_trap_compiler.py --circ cnt3-5_179.json --graph graph_33_5_4.json --gate 8 Invocation 4: python ion_trap_compiler.py --circ
      cnt3-5_179.json --graph graph_33_10_1.json --gate 7 Expected: schedule verifies.

  T9. Robustness on larger out-of-matrix circuits and graphs. Larger benchmark circuits on larger traps than the T1-T8 matrix. Invocation
      1: python ion_trap_compiler.py --circ qaoa_25.json --graph graph_51_4_1.json --gate 10 Invocation 2: python ion_trap_compiler.py
      --circ qaoa_40.json --graph graph_81_6_1.json --gate 10 Invocation 3: python ion_trap_compiler.py --circ qv_25.json --graph
      graph_51_4_1.json --gate 10 Invocation 4: python ion_trap_compiler.py --circ qft_25.json --graph graph_51_5_1.json --gate 9
      Invocation 5: python ion_trap_compiler.py --circ xeb_sy_25.json --graph graph_51_5_1.json --gate 9 Expected: 0 SWAP operations for
      invocations 1 and 2.

  T10. Tight-trap delivery. Invocation 1: python ion_trap_compiler.py --circ 4gt12-v0_86.json --graph graph_13_1_3.json --gate 6
       Invocation 2: python ion_trap_compiler.py --circ 4gt12-v0_86.json --graph graph_13_1_4.json --gate 6 Invocation 3: python
       ion_trap_compiler.py --circ 4gt4-v1_74.json --graph graph_13_1_4.json --gate 6 Invocation 4: python ion_trap_compiler.py --circ
       alu-v2_30.json --graph graph_13_1_3.json --gate 6 Invocation 5: python ion_trap_compiler.py --circ alu-v2_30.json --graph
       graph_13_1_4.json --gate 6

  T11. Stacked-endpoint and occupied-stack delivery. Invocation 1: python ion_trap_compiler.py --circ 4gt4-v0_79.json --graph
       graph_13_1_4.json --gate 6 Invocation 2: python ion_trap_compiler.py --circ mod8-10_178.json --graph graph_13_1_4.json --gate 6
       Invocation 3: python ion_trap_compiler.py --circ rd53_135.json --graph graph_15_5_1.json --gate 3 Invocation 4: python
       ion_trap_compiler.py --circ hwb7_59.json --graph graph_17_2_4.json --gate 7 Invocation 5: python ion_trap_compiler.py --circ
       hwb7_59.json --graph graph_17_6_1.json --gate 3

  T12. Dense multi-junction delivery. Invocation 1: python ion_trap_compiler.py --circ xeb_9.json --graph graph_19_1_5.json --gate 9
       Invocation 2: python ion_trap_compiler.py --circ xeb_9.json --graph graph_19_1_6.json --gate 9 Invocation 3: python
       ion_trap_compiler.py --circ xeb_sy_9.json --graph graph_19_1_3.json --gate 9 Invocation 4: python ion_trap_compiler.py --circ
       xeb_sy_9.json --graph graph_19_1_5.json --gate 9 Invocation 5: python ion_trap_compiler.py --circ xeb_sy_9.json --graph
       graph_19_1_6.json --gate 9

  T13. Congested-endpoint-corridor delivery. Invocation 1: python ion_trap_compiler.py --circ qft_15.json --graph graph_31_1_9.json
       --gate 15 Invocation 2: python ion_trap_compiler.py --circ qft_15.json --graph graph_31_1_10.json --gate 15 Invocation 3: python
       ion_trap_compiler.py --circ qft_15.json --graph graph_31_2_7.json --gate 13 Invocation 4: python ion_trap_compiler.py --circ
       qft_15.json --graph graph_31_2_8.json --gate 13 Invocation 5: python ion_trap_compiler.py --circ qft_15.json --graph
       graph_31_2_10.json --gate 13 Invocation 6: python ion_trap_compiler.py --circ qft_15.json --graph graph_31_2_15.json --gate 14
       Invocation 7: python ion_trap_compiler.py --circ qft_15.json --graph graph_31_3_8.json --gate 11 Invocation 8: python
       ion_trap_compiler.py --circ qft_15.json --graph graph_31_3_9.json --gate 13

  T14. Very-dense-trap delivery. Invocation 1: python ion_trap_compiler.py --circ xeb_16.json --graph graph_33_1_10.json --gate 16
       Invocation 2: python ion_trap_compiler.py --circ xeb_16.json --graph graph_33_2_15.json --gate 15 Invocation 3: python
       ion_trap_compiler.py --circ xeb_sy_16.json --graph graph_33_1_15.json --gate 16 Invocation 4: python ion_trap_compiler.py --circ
       xeb_sy_16.json --graph graph_33_2_15.json --gate 15 Invocation 5: python ion_trap_compiler.py --circ qv_20.json --graph
       graph_41_1_20.json --gate 20 Invocation 6: python ion_trap_compiler.py --circ qv_25.json --graph graph_51_1_15.json --gate 25
       Invocation 7: python ion_trap_compiler.py --circ qv_25.json --graph graph_51_1_20.json --gate 25

CROSS-CUTTING REQUIREMENTS over every (circuit, graph) test of the architecture matrix and every T1-T14 invocation:
  - The operation stream is byte-identical across two runs with different PYTHONHASHSEED values.
  - Total wall-clock time and peak RSS scale roughly linearly with number of vertices, qubits, and gates; no individual circuit takes
    more than a small multiple of the time predicted by linear extrapolation from the smallest circuit.
  - Every emitted SEPARATE / MERGE / SWAP / GATE timestep starts with all junction vertices empty, and every junction-traversing chain
    respects the no-roundtrip rule (both are enforced by the validator).
\end{lstlisting}

\end{document}